CITY UNIVERSITY OF HONG KONG
香港城市大學

# Phonon Dephasing, Entanglement and Exchange-Only Toffoli Gate Sequence in Quantum Dot Spin Chains
# 量子點自旋鏈系統中的聲子退相干機制、糾纏及僅交換托佛利門序列研究

Submitted to
Department of Physics
物理學系
in Partial Fulfillment of the Requirements
for the Degree of Doctor of Philosophy
哲學博士學位

by

HE GUANJIE
賀冠傑

July 2024
二零二四年七月

# Abstract

The quantum dot spin chain system represents a significant platform for quantum simulation and investigating the collective behaviors of electrons. As such, understanding its mechanisms and control protocols is crucial. Chapter 1 of the thesis introduces the fundamental concepts of the quantum dot chain system, focusing on the extended Hubbard model, the double quantum dot system, and electron-phonon coupling within these systems. Chapter 2 delves into the electron-phonon coupling mechanism within a multielectron double quantum dot system. Here, I examine two distinct scenarios based on the detuning variations of the system: the unbiased case and the biased case. In the unbiased case, the dephasing rate due to electron-phonon coupling generally increases with the number of electrons in the right dot. However, this trend is inconsistent in the biased case, indicating that multielectron quantum dots may offer advantages under certain conditions. Chapter 3 investigates the entanglement entropy in a multielectron quantum dot spin chain system, as the extended Hubbard model describes. Local and pairwise entanglement is influenced by the Coulomb interactions and tunneling strengths settings, shaped by the system's electronic configurations and potential energy of sites. The entanglement diagram exhibits clear phase transitions, significantly impacted by the ratios of coupling strength and potential energy variations. Adjusting the potential energy of a particular dot crucially influences the ground state configurations and, as a result, the entanglement entropy. Chapter 4 explores the possible operation sequences in the quantum dot spin chain system as defined by the Heisenberg model, inspired by the concept of a decoherence-free subspace. This chapter describes a nine-spin system within a nine-quantum-dot arrangement, where the bases are determined by the total angular quantum number. By employing the Krotov method of quantum optimal control, we identify a more efficient pulse-level operation sequence for an exchange-only quantum dot spin chain system, which offers a superior alternative to conventional quantum gate decomposition methods. This approach could enhance the development of more concise quantum algorithm representations.

**CITY UNIVERSITY OF HONG KONG**
**Qualifying Panel and Examination Panel**

| | |
|---|---|
| Surname: | He |
| First Name: | Guanjie |
| Degree: | Doctor of Philosophy |
| College/Department: | Department of Physics |

The Qualifying Panel of the above student is composed of:

*Supervisor(s)*

| | |
|---|---|
| Prof. WANG Xin Sunny | Department of Physics<br>City University of Hong Kong |

*Qualifying Panel Member(s)*

| | |
|---|---|
| Prof. LI Xiao | Department of Physics<br>City University of Hong Kong |
| Prof. CHAI Yu | Department of Physics<br>City University of Hong Kong |

This thesis has been examined and approved by the following examiners:

| | |
|---|---|
| Prof. Liu Qi | Department of Physics<br>City University of Hong Kong |
| Prof. Dai Liang | Department of Physics<br>City University of Hong Kong |
| Prof. Haidong Yuan | Department of Mechanical and Automation Engineering<br>The Chinese University of Hong Kong |
| Prof. WANG Xin Sunny | Department of Physics<br>City University of Hong Kong |

# Acknowledgements

I extend my heartfelt thanks to:

My supervisor, for his invaluable guidance.

My family, for their love and encouragement.

My true friends, for their unwavering friendship.

The genuine people in my life, for their sincerity.

Those who wish me well, for their support.

Your presence and support have been instrumental in my lifelong journey.

# Table of contents

# List of figures

# List of tables

# Chapter 1

# Introduction

## 1.1 Background

Semiconductor quantum dots, often referred to as ”artificial atoms”, have emerged as a powerful platform for studying many-body physics in highly controllable environments [149, 99, 150, 130, 203, 263, 255, 104, 129, 143, 218, 285, 44]. These nanoscale structures confine electrons in all three spatial dimensions, resulting in discrete energy levels and tunable electron-electron interactions. The ability to manipulate these systems with external electric and magnetic fields makes them ideal candidates for exploring fundamental quantum phenomena and developing novel quantum technologies [109, 292, 243, 251, 289, 313, 279, 134, 44].

The study of quantum dots has gained significant momentum in recent years due to their potential applications in quantum computing, quantum simulation, and the investigation of strongly correlated electron systems [296, 280, 276, 140, 88, 23, 141, 109, 71, 162, 144]. In particular, arrays of coupled quantum dots provide a unique opportunity to emulate complex many-body systems, such as those described by the Hubbard model, in a highly controllable solid-state environment [109, 280, 88, 141].

The versatility of semiconductor quantum dots stems from their ability to be engineered with precise control over their electronic properties. This includes the ability to tune the number of electrons in each dot, the coupling between dots, and the energy levels within individual dots [149, 99, 50, 61, 110, 299, 84, 52, 290, 160, 310, 131, 312]. Such control has enabled the observation of a wide range of quantum phenomena, including Coulomb blockade, the Kondo effect, and spin-orbit coupling [150, 99].

Recent advancements in fabrication techniques have allowed for the creation of increasingly complex quantum dot arrays, ranging from double dots to linear chains

and two-dimensional lattices [109, 296, 280, 61, 110, 299, 84, 52, 290, 160, 310, 131, 312]. These structures provide an ideal testbed for studying the physics of interacting electrons in confined geometries, with potential applications in quantum information processing and quantum simulation [29, 276, 23, 141, 71, 162, 144].

Spin qubits in semiconductor quantum dots have emerged as a promising platform for quantum information processing and quantum computation [167, 204, 101]. These qubits leverage the intrinsic angular momentum of individual electrons or nuclei confined within nanometer-scale semiconductor devices, providing a natural two-level system that is relatively insensitive to electric fields and exhibits long quantum coherence times [204]. The ability to precisely control and measure the electronic states in quantum dots has opened up new avenues for studying fundamental quantum mechanical effects. For example, researchers have demonstrated coherent manipulation of electron spins in quantum dots, paving the way for spin-based quantum computation [194, 280, 185]. Additionally, the strong confinement in quantum dots leads to enhanced electron-electron interactions, allowing for the exploration of strongly correlated electron physics in a highly tunable system [139, 65]. First proposed by Loss and DiVincenzo in 1998 [167], spin qubits have since been developed in various semiconductor materials, including gallium arsenide (GaAs), silicon (Si), and germanium (Ge). They offer several advantages, such as long coherence times, compatibility with existing microelectronics industry infrastructure, small size, high density, and versatility in qubit types (e.g., single spin, donor spin, singlet-triplet, and exchange-only qubits) [101]. Recent advancements have focused on improving charge control and readout, coupling spins to other quantum degrees of freedom, and scaling to larger system sizes [204]. Additionally, efforts to hybridize spin qubits with superconducting systems have opened new avenues for long-range interactions and improved readout schemes [204, 101]. More details will be introduced in the remaining chapters.

Furthermore, the scalability of semiconductor quantum dot systems makes them particularly attractive for realizing large-scale quantum simulators [29, 280, 88, 140]. By engineering arrays of coupled quantum dots, researchers aim to simulate complex many-body Hamiltonians that are intractable for classical computers, potentially leading to new insights in condensed matter physics and quantum chemistry [296, 276, 84, 13, 47, 87, 193, 106, 42, 140, 278, 100, 122]. These simulations leverage the precise control over quantum dot properties to explore fundamental quantum phenomena and develop novel quantum technologies.

In summary, semiconductor quantum dots offer a unique and versatile platform for exploring quantum many-body physics, with potential applications ranging from fundamental science to quantum technologies. The ability to precisely control and measure these systems, combined with their scalability and compatibility with existing semiconductor fabrication techniques, positions quantum dots as a promising candidate for realizing practical quantum simulators and quantum computers in the near future [109, 280, 29, 61, 299, 160].

## 1.2 Extended Hubbard Model

The Hubbard model, originally proposed to describe interacting electrons in narrow energy bands of solids, has become a cornerstone in understanding strongly correlated electron systems [120]. In recent years, this model has found exciting new applications in the field of semiconductor quantum dots[161, 280, 140, 156], where it serves as a powerful tool for describing and predicting the behavior of confined electrons in artificial atom-like structures.

The extended Hubbard model, when applied to semiconductor quantum dots, captures the essential physics of these systems through several key parameters and interactions. The Hamiltonian for this model is given by:

$$H = -\sum_{<i,j>,\sigma} t_{ij}(c_{i\sigma}^{\dagger} c_{j\sigma} + \text{h.c.}) + \sum_i U_i n_{i\uparrow} n_{i\downarrow} + \sum_{i,j} V_{ij} n_i n_j - \sum_i \epsilon_i n_i \quad (1.1)$$

where:

- $t_{ij}$ represents the tunneling amplitude between neighboring sites $i$ and $j$. This term quantifies the kinetic energy associated with electron hopping, which is crucial for describing electron mobility and transport properties in the quantum dot lattice.

- $U_i$ denotes the on-site Coulomb repulsion, which is the energy cost for double occupancy of site $i$ by electrons with opposite spins. This parameter is fundamental for modeling electron-electron interactions that lead to phenomena such as the Mott insulator state.

- $V_{ij}$ describes the long-range Coulomb interaction between electrons at different sites $i$ and $j$. This term accounts for the electrostatic repulsion between electrons, which can lead to the formation of charge density waves and other correlated electron states.

- $\epsilon_i$ is the on-site energy, representing the local potential energy experienced by an electron at site $i$. This term can include contributions from external electric fields, local impurities, and other site-specific factors that affect the electron's energy landscape.

- $c_{i\sigma}^\dagger$ and $c_{i\sigma}$ are the creation and annihilation operators, respectively, for an electron with spin $\sigma$ at site $i$. These operators obey the anti-commutation relations characteristic of fermions and are essential for describing the quantum states of the system.

- $n_i = \sum_\sigma c_{i\sigma}^\dagger c_{i\sigma}$ is the number operator, representing the total number of electrons at site $i$. This operator is used to quantify electron occupancy and is critical for evaluating interaction terms in the Hamiltonian.

The extended Hubbard model provides a comprehensive framework for studying the interplay between kinetic energy, on-site interactions, and long-range Coulomb interactions in semiconductor quantum dot systems. By tuning the parameters $t_{ij}$, $U_i$, $V_{ij}$, and $\epsilon_i$, researchers can explore a wide range of physical phenomena, including metal-insulator transitions, magnetic ordering, and the emergence of topological phases. This model is particularly relevant for simulating quantum many-body effects and designing novel quantum devices based on semiconductor quantum dots.

## 1.3 Density Matrices

The density matrix formalism provides a powerful tool for describing quantum systems, especially when dealing with mixed states or subsystems of entangled states. For a pure state $|\psi\rangle$, the density matrix is given by:

$$\rho = |\psi\rangle\langle\psi| \tag{1.2}$$

This representation captures all the information about the quantum state and allows for the computation of expectation values and other statistical properties.

More generally, for a mixed state, which represents a statistical ensemble of pure states, the density matrix is a weighted sum of pure state density matrices:

$$\rho = \sum_i p_i |\psi_i\rangle\langle\psi_i|, \tag{1.3}$$

where $p_i$ are probabilities satisfying $\sum_i p_i = 1$ [229]. Each $|\psi_i\rangle$ represents a possible pure state of the system, and $p_i$ represents the probability of the system being in that state.

The density matrix formalism is particularly useful in quantum statistical mechanics and for describing subsystems of larger quantum systems. It allows for a unified treatment of pure and mixed states and provides a natural way to compute expectation values of observables. For an observable $A$, the expectation value is given by:

$$\langle A \rangle = \text{Tr}(\rho A), \tag{1.4}$$

where Tr denotes the trace operation. This formulation is essential in the study of open quantum systems, where a system interacts with its environment, leading to mixed states.

Furthermore, the density matrix formalism is crucial in quantum information theory. It is used to describe the state of qubits in quantum computing, particularly when dealing with decoherence and noise. The reduced density matrix, obtained by tracing out the degrees of freedom of a subsystem, provides a means to study entanglement and correlations in multipartite systems.

For a bipartite system described by the state $\rho_{AB}$, the reduced density matrix of subsystem $A$ is obtained by tracing out subsystem $B$:

$$\rho_A = \text{Tr}_B(\rho_{AB}), \tag{1.5}$$

where $\text{Tr}_B$ denotes the partial trace over subsystem $B$. The reduced density matrix $\rho_A$ contains all the information about the subsystem $A$'s statistical properties, making it a vital tool for analyzing subsystems' behavior within a larger entangled system.

The purity of a quantum state, which quantifies the degree of mixedness, can also be evaluated using the density matrix. It is defined as:

$$\text{Purity} = \text{Tr}(\rho^2). \tag{1.6}$$

For a pure state, the purity is 1, while for a completely mixed state, it is less than 1. This measure is useful in various contexts, including quantum information processing and the study of decoherence.

The density matrix formalism thus provides a comprehensive framework for analyzing and understanding a wide range of quantum phenomena, from basic quantum mechanics to advanced quantum information science.

## 1.4 Quantum Entanglement, Entropy, and the Fermi-Hubbard Model

Quantum entanglement is a fundamental phenomenon in quantum mechanics that has no classical analogue. It describes non-local correlations between quantum systems that are stronger than any classical correlation. For a pure bipartite quantum state $|\psi\rangle_{AB}$ in a Hilbert space $\mathcal{H}_A \otimes \mathcal{H}_B$, the state is considered entangled if it cannot be decomposed as a tensor product of states in the individual subsystems, i.e., $|\psi\rangle_{AB} \neq |\psi\rangle_A \otimes |\phi\rangle_B$ [114].

Entanglement is a key resource in quantum information theory, underpinning protocols such as quantum teleportation, superdense coding, and entanglement-based quantum cryptography. These applications exploit the unique properties of entangled states to perform tasks that are impossible or less efficient classically.

The degree of entanglement can be quantified using various measures, with entropy being a fundamental concept. The von Neumann entropy, a quantum analogue of the classical Shannon entropy, is defined for a density matrix $\rho$ as:

$$S(\rho) = -\mathrm{Tr}(\rho \log \rho), \tag{1.7}$$

where Tr denotes the trace operation. This entropy measure reflects the amount of quantum uncertainty or mixedness in the state $\rho$. For a pure state $\rho = |\psi\rangle\langle\psi|$, the von Neumann entropy is zero, indicating no uncertainty. However, for a mixed state, the entropy is positive, reflecting the degree of mixture.

For a pure bipartite state $|\psi\rangle_{AB}$, the entanglement entropy is given by the von Neumann entropy of the reduced density matrix of either subsystem:

$$E(|\psi\rangle_{AB}) = S(\rho_A) = S(\rho_B), \tag{1.8}$$

where $\rho_A = \mathrm{Tr}_B(|\psi\rangle_{AB}\langle\psi|)$ and $\rho_B = \mathrm{Tr}_A(|\psi\rangle_{AB}\langle\psi|)$ are the reduced density matrices obtained by tracing out the degrees of freedom of subsystem $B$ and $A$, respectively [192]. The equality $S(\rho_A) = S(\rho_B)$ holds because $|\psi\rangle_{AB}$ is a pure state, and the entropy quantifies the entanglement between the subsystems $A$ and $B$.

Entanglement entropy has significant implications in various fields of physics. In condensed matter physics, it is used to study quantum phase transitions and topological orders. In high-energy physics, it is related to black hole thermodynamics and the holographic principle. The scaling of entanglement entropy with the size of a subsys-tem is a powerful tool for characterizing different phases of matter, including critical points and topologically ordered phases [30, 69].

The study of entanglement in the Fermi-Hubbard model provides profound insights into the nature of quantum correlations in many-body systems. This model, which describes interacting fermions on a lattice, is a cornerstone of condensed matter physics and has been extensively used to investigate phenomena such as magnetism, superconductivity, and metal-insulator transitions.

For a bipartition of the lattice into subsystems $A$ and $B$, the entanglement entropy $S_A$ is a key quantity that characterizes the quantum correlations between these subsystems. It is defined as:

$$S_A = -\mathrm{Tr}(\rho_A \log \rho_A), \tag{1.9}$$

where $\rho_A$ is the reduced density matrix of subsystem $A$, obtained by tracing out the degrees of freedom of subsystem $B$:

$$\rho_A = \mathrm{Tr}_B(\rho_{AB}), \tag{1.10}$$

with $\rho_{AB}$ being the density matrix of the entire system. The entanglement entropy $S_A$ measures the amount of quantum information shared between subsystems $A$ and $B$.

The behavior of $S_A$ as a function of subsystem size, interaction strength $U/t$, and filling factor provides crucial information about the system's quantum phase transitions and correlation structure. For example, in the case of a one-dimensional Hubbard model at half-filling, $S_A$ exhibits a logarithmic scaling with subsystem size, characteristic of critical systems described by conformal field theory [4, 230]. In higher dimensions or away from half-filling, different scaling behaviors can emerge, reflecting the diverse phases and transitions of the model [30, 69].

The entanglement entropy is particularly useful for identifying and characterizing quantum phase transitions. At a quantum critical point, $S_A$ often shows non-trivial scaling behavior, which can be used to extract critical exponents and other universal properties. For example, in the presence of strong interactions (large $U/t$), the system may undergo a transition from a metallic to a Mott insulating phase, with a corresponding change in the entanglement entropy [146, 135].

Moreover, the study of entanglement in the Hubbard model extends to various modifications and generalizations, such as the extended Hubbard model with long-range interactions, the Hubbard model on different lattice geometries, and models incorpo-rating spin-orbit coupling. These studies reveal rich and complex entanglement structures, providing a deeper understanding of correlated electron systems [212, 95, 168, 199]. For instance, the extended Hubbard model can describe phenomena like charge density waves and superconductivity, which are not captured by the standard Hubbard model. Spin-orbit coupling introduces additional complexity, leading to the emergence of topologically non-trivial states [212].

In recent years, advanced numerical methods such as tensor network techniques and quantum Monte Carlo simulations have been employed to study entanglement in the Hubbard model, offering detailed insights into the ground state properties and dynamics of these systems [273, 103]. These approaches have proven particularly effective in capturing the intricate entanglement patterns and their evolution under various conditions. Tensor network methods, such as the density matrix renormalization group (DMRG) and its higher-dimensional extensions, are particularly powerful for one-dimensional and quasi-one-dimensional systems [230]. Quantum Monte Carlo simulations, on the other hand, are well-suited for studying finite-temperature properties and phase transitions in higher dimensions [135].

The entanglement properties of the Fermi-Hubbard model are also crucial for understanding the dynamics of quantum systems. For example, the growth of entanglement entropy following a quantum quench and a sudden change in the system's parameters provides valuable information about thermalization and the spread of correlations [30]. In systems with strong interactions, the entanglement growth can be highly non-trivial and is influenced by conserved quantities and integrability [146].

## 1.5 Electron-Phonon Coupling in Quantum Dots

Semiconductor quantum dots (QDs) have emerged as a promising platform for quantum information processing and cavity quantum electrodynamics (QED) due to their discrete energy levels and strong light-matter interactions [195]. However, unlike isolated atomic systems, QDs are embedded in a solid-state environment, leading to intrinsic interactions with their surroundings, particularly with phonons. These electron-phonon interactions play a crucial role in determining the optical and electronic properties of QDs, including their coherence times, emission spectra, and spin dynamics [287, 152, 216, 153, 148, 147, 115].

### 1.5.1 Electron-Phonon Coupling in Double Quantum Dots

Electron-phonon coupling in quantum dots can be understood by considering the perturbation of the electronic states due to the lattice vibrations. Phonons can be classified into two main types: acoustic phonons and optical phonons. Acoustic phonons correspond to the low-energy, long-wavelength vibrations where atoms move in phase, while optical phonons involve higher-energy vibrations where atoms move out of phase.

The total Hamiltonian describing the system, including electron-phonon interactions, is given by [198]:

$$H = H_{\text{e}} + H_{\text{ph}} + H_{\text{ep}}, \tag{1.11}$$

where $H_{\text{e}}$ represents the system Hamiltonian for electrons in a double quantum dot (DQD) (see Eq. (2.1)), $H_{\text{ph}}$ denotes the environment Hamiltonian characterized by phonon modes, expressed as

$$H_{\text{ph}} = \sum_{\mathbf{q},\lambda} \omega_{\mathbf{q},\lambda} a^{\dagger}_{\mathbf{q},\lambda} a_{\mathbf{q},\lambda}, \tag{1.12}$$

and $H_{\text{ep}}$ describes the electron-phonon interaction.

For a singlet-triplet qubit, $H_{\text{e}}$ is written in the basis of the lowest singlet, $|\text{S}\rangle$, and lowest triplet state, $|\text{T}\rangle$, i.e.

$$H_{\text{e}} = \frac{J}{2}\sigma_z, \tag{1.13}$$

where $\sigma_z = |\text{T}\rangle\langle\text{T}| - |\text{S}\rangle\langle\text{S}|$ (see Sec. 2.3.2 for details).

In a semiconductor, $H_{\text{ep}}$ is the Hamiltonian that describes the effective electron-phonon interaction, taking the form [115]:

$$H_{\text{ep}} = \sum_{\mathbf{q},\lambda} M_{\lambda}(\mathbf{q})\rho(\mathbf{q})(a_{\mathbf{q},\lambda} + a^{\dagger}_{\text{-}\mathbf{q},\lambda}), \tag{1.14}$$

where $a_{\mathbf{q},\lambda}$ and $a^{\dagger}_{\text{-}\mathbf{q},\lambda}$ are phonon annihilation and creation operators respectively, $\mathbf{q}$ the lattice momentum, and $\lambda$ the branch index. $\rho(\mathbf{q})$ is the electron density operator, taking the form $\rho(\mathbf{q}) = \sum_{i=1}^{2N} e^{i\mathbf{q}\cdot\text{R}_\text{i}}$ in a $2N$ electron DQD system. $M(\mathbf{q})$ represents different kinds of electron-phonon interactions.

### 1.5.2 Effects on Optical Properties

The electron-phonon coupling in quantum dots significantly influences their optical properties, particularly in processes such as photoluminescence (PL) and Raman scattering. When an electron in a quantum dot recombines with a hole, it can emit a photon. However, due to electron-phonon coupling, this recombination can also involve the emission or absorption of phonons, leading to a broadening and shifting of the PL spectrum.

The linewidth of the photoluminescence peaks provides information about the strength of the electron-phonon coupling. A stronger coupling results in broader peaks, indicating a higher degree of interaction between the electronic states and the lattice vibrations. This interaction can be quantified by the Huang-Rhys factor, $S$, which is a dimensionless parameter representing the average number of phonons involved in the electronic transition.

### 1.5.3 Electron-Phonon Interactions in GaAs Double Quantum Dots

In GaAs double quantum dots (DQDs), the deformation potential (DP) and piezoelectric (PE) interaction provide the main contributions to phonon dephasing, while contributions from other interactions are negligible [115, 38]. The DP and PE have the forms [115]:

$$M^{\mathrm{DP}}_{\mathrm{GaAs}}(\mathbf{q}) = D\left(\frac{\hbar}{\rho V \omega_{\mathbf{q}}}\right)^{\frac{1}{2}} |\mathbf{q}|, \tag{1.15}$$

$$M^{\mathrm{PE}}_{\mathrm{GaAs}}(\mathbf{q}) = i\left(\frac{\hbar}{\rho V \omega_{\mathbf{q}}}\right)^{\frac{1}{2}} 2ee_{14}(\hat{q}_x\hat{q}_y\hat{\xi}_z + \hat{q}_y\hat{q}_z\hat{\xi}_x + \hat{q}_z\hat{q}_x\hat{\xi}_y), \tag{1.16}$$

and one should note that $M^{\mathrm{DP}}_{\mathrm{GaAs}}(\mathbf{q})$ only couples electrons to longitudinal acoustic phonons, while $M^{\mathrm{PE}}_{\mathrm{GaAs}}(\mathbf{q})$ can couple electrons to both longitudinal acoustic (LA) and transverse acoustic (TA) phonons. Here, $D = 8.6$ eV is the deformation potential constant, $\rho = 5.3 \times 10^3$ kg/m$^3$ the mass density, $e$ is the elementary electric charge, $e_{14} = 1.38 \times 10^9$ V/m is the piezoelectric constant, $\hat{\xi}$ is the polarization vector, and $\omega_{\mathbf{q}}$ the angular frequency of the phonon mode $\mathbf{q}$. We further define $\gamma_{\mathbf{q}}$ as the population relaxation rate of the phonon mode $\mathbf{q}$, which is assumed to have the form $\gamma_{\mathbf{q}} = \gamma_0 q^n$.

## 1.6 Exchange-Only Qubits and Decoherence-Free Subspaces in Quantum Dots

### 1.6.1 Exchange-Only Qubits

Exchange-only qubits are a type of multi-spin qubit that can be fully controlled using only the exchange interaction between neighboring spins, without the need for individual addressing or magnetic field gradients [166, 63]. The simplest implementation of an exchange-only qubit uses three electron spins in a triple quantum dot, where the logical qubit states are encoded in the two-dimensional subspace of total spin $S = 1/2$ and total $z$-component of spin $S_z = 1/2$.

The Hamiltonian for a triple-dot exchange-only qubit can be written as:

$$H = J_{12}(\epsilon_1)\mathbf{S}_1 \cdot \mathbf{S}_2 + J_{23}(\epsilon_2)\mathbf{S}_2 \cdot \mathbf{S}_3 \tag{1.17}$$

where $J_{ij}(\epsilon_k)$ is the exchange coupling between dots $i$ and $j$, controlled by the detuning parameter $\epsilon_k$, and $\mathbf{S}_i$ is the spin operator for the electron in dot $i$ [166].

The key advantage of exchange-only qubits is that all single-qubit operations can be performed by simply modulating the exchange couplings $J_{12}$ and $J_{23}$, which can be achieved through fast electrical control of the dot potentials. This eliminates the need for oscillating magnetic fields or $g$-factor engineering, simplifying the experimental implementation [166, 63].

### 1.6.2 Decoherence-Free Subspaces

Decoherence-free subspaces (DFS) provide a passive error correction mechanism by encoding quantum information in subspaces that are inherently protected against certain types of noise. These subspaces are defined such that the encoded states are immune to specific interactions with the environment [166, 298].

A DFS is mathematically defined as a subspace $\mathcal{H}_{\text{DFS}}$ of the system's Hilbert space $\mathcal{H}$ spanned by states $|\phi\rangle$ that are simultaneous eigenstates of all system operators $S_\alpha$ that couple to the environment, with corresponding eigenvalues $\lambda_\alpha$:

$$\mathcal{H}_{\text{DFS}} = \{|\phi\rangle \in \mathcal{H} : S_\alpha|\phi\rangle = \lambda_\alpha|\phi\rangle, \forall\alpha\}. \tag{1.18}$$

For a system-environment interaction Hamiltonian of the form:

$$H_{\text{SE}} = \sum_{\alpha} S_{\alpha} \otimes B_{\alpha}, \tag{1.19}$$

where $S_{\alpha}$ and $B_{\alpha}$ are system and bath operators, respectively, states in the DFS evolve unitarily under the action of $H_{\text{SE}}$:

$$e^{-iH_{\text{SE}}t}|\phi\rangle \otimes |B\rangle = |\phi\rangle \otimes e^{-i\sum_{\alpha}\lambda_{\alpha}B_{\alpha}t}|B\rangle, \tag{1.20}$$

where $|B\rangle$ represents the state of the environment. This unitary evolution implies that the quantum information encoded in $|\phi\rangle$ is preserved, as the environment-induced decoherence effects cancel out.

In semiconductor quantum dots, decoherence-free subspaces (DFS) have been proposed as a robust method to protect against various sources of decoherence, including fluctuations in uniform magnetic fields for spin qubits and electric fields for charge qubits [83, 205, 99]. For instance, a DFS immune to uniform magnetic field fluctuations $\delta\mathbf{B}$ can be constructed from states $|\phi\rangle$ satisfying:

$$(\mathbf{S}_1 + \mathbf{S}_2) \cdot \delta\mathbf{B}|\phi\rangle = 0, \tag{1.21}$$

where $\mathbf{S}_1$ and $\mathbf{S}_2$ are the spin operators for two electron spins. The singlet state $\frac{1}{\sqrt{2}}(|\uparrow\downarrow\rangle - |\downarrow\uparrow\rangle)$ satisfies this condition and forms a one-dimensional DFS, making it robust against uniform magnetic field fluctuations [154, 18, 164].

In addition to magnetic field fluctuations, DFS can also be designed to protect against electric field noise. For charge qubits in semiconductor quantum dots, where electric field fluctuations can cause significant decoherence, encoding information in states that are symmetric with respect to charge displacement can help mitigate decoherence effects [256, 201].

The practical implementation of DFS in quantum dot systems requires precise control over the quantum states and their interactions with the environment. Techniques such as dynamical decoupling, which involves applying sequences of control pulses to refocus the system's state, can be used in conjunction with DFS to further enhance coherence times [274, 28, 249]. The combination of DFS and dynamical decoupling provides a powerful strategy for error correction and noise suppression in quantum computing [155, 17].

# Chapter 2

# Theory on electron-phonon spin dephasing in GaAs multi-electron double quantum dots

## 2.1 Overview

Recent studies have demonstrated that a double-quantum-dot system hosting more than two electrons may exhibit superior characteristics compared to the traditional two-electron singlet-triplet qubit configuration. Our research focuses on the electron-phonon dephasing in a GaAs multi-electron double-quantum-dot system, examining both a biased scenario where the singlet state is hybridized and an unbiased scenario where hybridization is absent.

We have observed that, in the unbiased case, the electron-phonon dephasing rate increases with the number of confined electrons. However, this trend does not persist in the biased case. To quantify the performance, we introduce a figure of merit defined as the ratio between the exchange energy and the dephasing rate. Our analysis reveals that within an experimentally relevant range of exchange energies, the figure of merit increases with the number of electrons in the biased case.

These findings indicate that multi-electron quantum-dot systems possess an additional advantage in mitigating electron-phonon dephasing effects, an aspect that has been previously underappreciated in the literature.

## 2.2 Background

Semiconductor quantum-dot spin qubits, serving as platforms for the physical realization of quantum computation, have garnered significant research interest due to their tunability, scalability, and high-fidelity gate operations [167, 307, 293, 25, 227, 202, 109, 288, 12, 73, 76, 102, 45, 56, 10, 245, 246, 37, 36, 250, 177, 268, 297, 35, 55, 158]. Traditional spin qubits typically host no more than two electrons per quantum dot. However, recent research suggests that multi-electron qubits, where certain dots host more than two electrons, may offer distinct advantages [174, 183, 281, 252, 242, 275, 163, 209, 181, 142, 173, 176, 151, 112, 8, 180, 169, 57].

For instance, a multi-electron quantum dot can act as a mediator for fast spin exchange [174] or as a tunable coupling mechanism between nearby dots [242]. Additionally, it has been demonstrated that multi-electron quantum-dot devices may exhibit greater resilience to noise compared to traditional single or double-electron systems, owing to the screening effect provided by core electrons [112, 8, 180].

Experiments indicate that in certain asymmetric multi-electron triple-quantum-dot systems, the dependence of the exchange energy on the absolute value of detuning can be non-monotonic, suggesting the existence of a ”sweet spot” [173]. Additionally, in similar systems, it has been observed that the sign of the exchange energy can reverse, eliminating a longstanding constraint in constructing dynamically corrected exchange gates [176].

From a theoretical perspective, calculations using Configuration Interaction (CI) techniques on few-electron multi-quantum-dot systems have demonstrated negative exchange interactions, highlighting their implications for robust quantum control [57, 38, 39]. Other studies have revealed these systems' potential for tunable couplings [242], robust quantum gates [180], and other intriguing properties [169]. These findings underscore the promise of multi-electron quantum-dot systems in achieving noise-resilient quantum information processing.

Environmental noise and strategies to mitigate it have been extensively studied in conventional two-electron singlet-triplet qubits [311, 148, 147, 115, 187, 60, 117, 295, 238, 214, 119, 257, 85, 170, 14, 294, 86, 221, 21, 190, 215, 178, 189, 153]. Among these noise sources, electron-phonon dephasing is a significant channel leading to decoherence [311, 147, 148, 115, 187]. In GaAs, the deformation potential interaction, polar optical interaction, and piezoelectric interaction are phonon couplings that contribute to decoherence [85].

In double quantum dots (DQDs) hosting two electrons, the deformation potential and piezoelectric interactions are the primary contributors to electron-phonon dephasing, with all phonon coupling channels diminishing as the distance between dots increases [115, 147]. An open question remains on how electron-phonon dephasing behavior changes with an increasing number of electrons in DQDs [39].

In this chapter, we investigate electron-phonon dephasing in a GaAs multi-electron double quantum dot (DQD) system. The electron configurations, definitions of singlet and triplet states, and their hybridization are more complex compared to systems with only two electrons. We examine phonon-mediated dephasing in two scenarios: the unbiased case and the biased case. In the unbiased case, the hybridization between the lowest singlet states is minimized. In the biased case, the hybridization is enhanced because the first excited singlet states are energetically brought closer to the lowest singlet state by increasing the relative detuning between the two dots in a DQD.

We define a figure of merit as the ratio between the exchange energy and the dephasing rate in the biased case. Our analysis shows that within an experimentally relevant range of exchange energies, this figure of merit increases with the number of electrons. These findings suggest that multi-electron quantum-dot systems have advantages in reducing noise from electron-phonon interactions, an aspect previously underappreciated in the literature.

The remainder of the chapter is organized as follows. In Sec. 2.3, we present our model of the multi-electron DQD system and the methods used to address the electron-phonon interaction problem. In the subsection, we provide the results on dephasing rates, exchange energies, and the figures of merit in different scenarios. Finally, we conclude in Sec. 2.4.

## 2.3 Model

### 2.3.1 Hamiltonian

In this study, we consider an asymmetric double-quantum-dot (DQD) system where the right dot (R) is larger than the left dot (L), with the center-to-center distance between the two dots being $2x_0$. We retain the lowest $N$ orbitals in the right dot, denoted as $\mathrm{R}_1$ through $\mathrm{R}_N$, as illustrated in Fig. 2.1. The system hosts a total of $2N$ electrons, with one electron in the left dot (L) and $2N-1$ electrons in the right dot (R). The Hamiltonian of the system is given by:

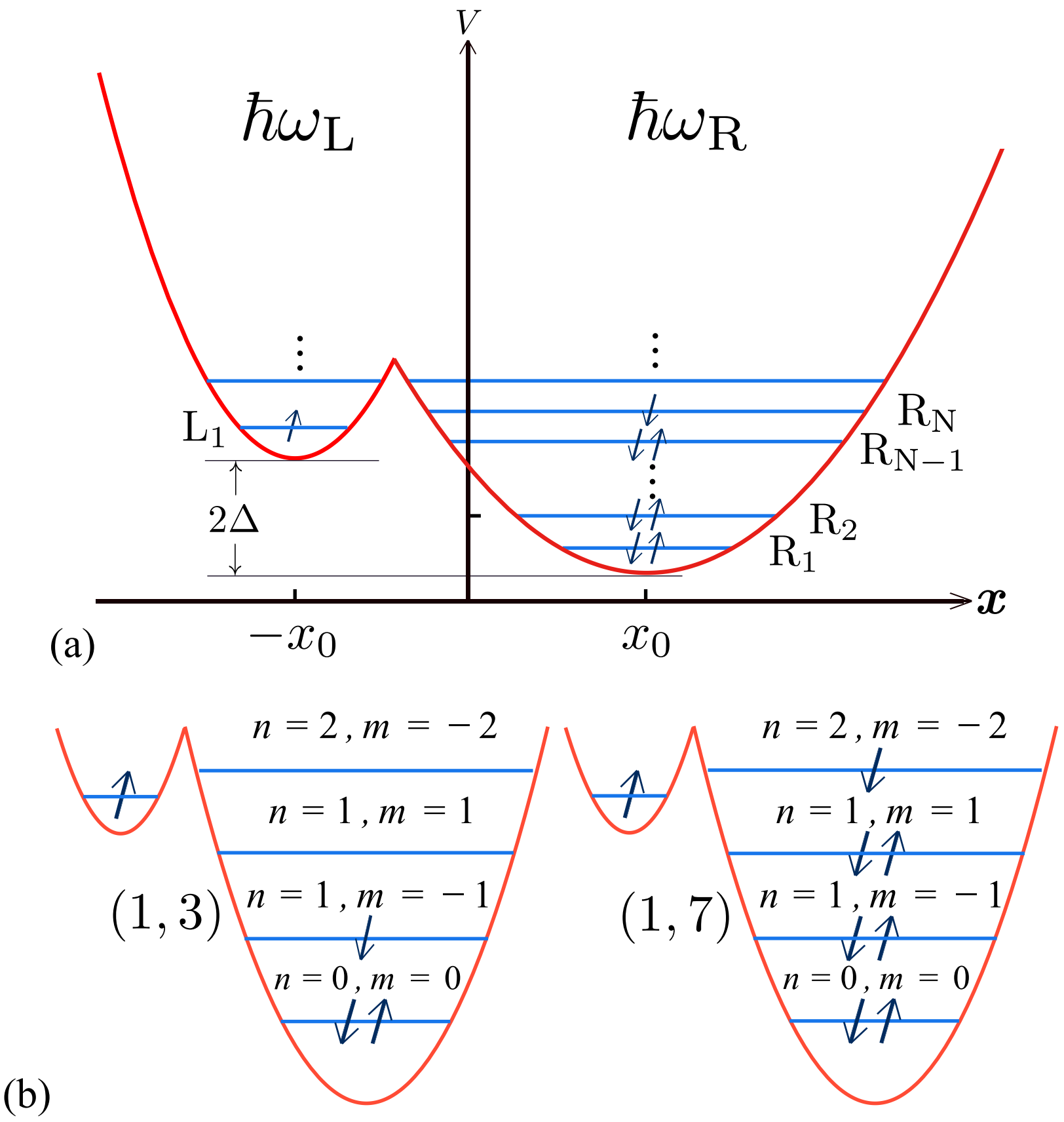

Fig. 2.1 (a) Schematic illustration of a double-quantum-dot system hosting $2N$ electrons. One electron occupies the $L_1$ orbital of the left dot, while $2N - 1$ electrons occupy the $R_1$ through $R_N$ orbitals of the right dot. (b) Cases with electron configuration $(1, 3)$ and $(1, 7)$ considered in this chapter. Here, $n$ is the principal quantum number of the relevant Fock-Darwin state, and $m$ the magnetic quantum number.

$$H_{\mathrm{e}} = \sum_{j=1}^{2N} h_j + \sum_{j,k=1}^{2N} \frac{e^2}{\epsilon \left|\mathbf{R}_j - \mathbf{R}_k\right|}, \tag{2.1}$$

where

$$h_i = \frac{(-i\hbar\nabla_i + e\mathbf{A}/c)^2}{2m^*} + V(\mathbf{r}) + g^*\mu_B \mathbf{B}\cdot\mathbf{S}. \tag{2.2}$$

The confinement potential in the $xy$ plane is:

$$V(\mathbf{r}) = \frac{1}{2}m^* \min\left[\omega_{\mathrm{R}}^2(\mathbf{r}-\mathbf{r}_0)^2 - \Delta, \omega_{\mathrm{L}}^2(\mathbf{r}+\mathbf{r}_0)^2 + \Delta\right], \tag{2.3}$$

where $\omega_{\mathrm{L}}$ ($\omega_{\mathrm{R}}$) represents the confinement strengths in dot L (R), $\mathbf{r} = (x, y)$, $\mathbf{r}_0 = (x_0, 0)$, and $x_0$ is half the distance between the centers of the two dots. The effective mass of the electron is $m^* = 0.067m_e$, and $\Delta$ is the detuning, as shown in Fig. 2.1. The last term in $h_i$ is the Zeeman energy, where $g^*$ is the electron g-factor, $\mu_B$ is the Bohr magneton, $\mathbf{B} = B\hat{\mathbf{z}}$ is the perpendicular magnetic field, and $\mathbf{S}$ is the total electron spin. The magnetic field is set to $B = 0.7$ T. In our model, we have $\hbar\omega_{\mathrm{L}} > \hbar\omega_{\mathrm{R}}$, with $\hbar\omega_{\mathrm{L}} = 2.838$ meV, and $\hbar\omega_{\mathrm{R}}$ varying between $\hbar\omega_{\mathrm{L}}/4$ and $\hbar\omega_{\mathrm{L}}/2$.

We consider a singlet-triplet (ST) qubit in the detuning regime where the electron occupancy is approximately $(n_{\mathrm{L}}, n_{\mathrm{R}}) \approx (1, 2N-1)$. Here, $(n_{\mathrm{L}}, n_{\mathrm{R}})$ indicates the number of electrons in the left ($n_{\mathrm{L}}$) and right ($n_{\mathrm{R}}$) dots. We denote a singlet (triplet) state formed by $n_{\mathrm{L}}$ electrons in the left dot and $n_{\mathrm{R}}$ electrons in the right dot as $|\mathrm{S}(n_{\mathrm{L}}, n_{\mathrm{R}})\rangle$ ($|\mathrm{T}(n_{\mathrm{L}}, n_{\mathrm{R}})\rangle$).

In this work, we study dephasing in two scenarios:

1. Unbiased case: The logical bases are $|\mathrm{S}(1, 2N-1)\rangle$ and $|\mathrm{T}(1, 2N-1)\rangle$.

2. Biased case: The logical bases are a hybridized singlet $|\mathrm{S}(0, 2N)\rangle/|\mathrm{S}(2, 2N-2)\rangle$ and $|\mathrm{T}(1, 2N-1)\rangle$.

The hybridized singlet state depends on the detuning direction $\Delta$. For large positive detuning, the hybridized singlet is $|\mathrm{S}_{\mathrm{mix}}^{(0,2N)}\rangle$. For small positive or negative detuning, the hybridized singlet is $|\mathrm{S}_{\mathrm{mix}}^{(2,2N-2)}\rangle$.

In the biased case, the exchange energy between the singlet and triplet states is a key quantity. To evaluate the exchange energy as a function of detuning, we consider an effective Hamiltonian in the basis states $|\mathrm{S}(1, 2N-1)\rangle$, $|\mathrm{T}(1, 2N-1)\rangle$, and $|\mathrm{S}(0, 2N)\rangle$ for the large positive detuning regime, and $|\mathrm{S}(1, 2N-1)\rangle$, $|\mathrm{T}(1, 2N-1)\rangle$, and $|\mathrm{S}(2, 2N-2)\rangle$ for the small positive or negative detuning regime. We have verified that other electron configurations have much higher energy and therefore do not significantly affect the exchange energy (see Appendix 2.5.1).

By diagonalizing the effective Hamiltonian, the exchange energy $J$ is defined as:

$$J = E_{|\mathrm{T}\rangle} - E_{|\mathrm{S}\rangle}, \tag{2.4}$$

where $|\mathrm{S}\rangle$ and $|\mathrm{T}\rangle$ are the lowest singlet and triplet states, respectively, and $E_{|\mathrm{S}\rangle}$, $E_{|\mathrm{T}\rangle}$ are their corresponding eigenvalues.

### 2.3.2 Singlet and Triplet in Multi-Electron Double Quantum Dot

Similar to the two-electron case, singlet-triplet (ST) spin qubits can be well defined in multi-electron double quantum dots (DQDs) [242, 180, 169, 173, 176, 142, 57, 38, 39].

In the unbiased case, the singlet state $|\mathrm{S}(1, 2N-1)\rangle$ and the triplet state $|\mathrm{T}(1, 2N-1)\rangle$ can be expressed as:

$$|\mathrm{S}(1, 2N-1)\rangle = \frac{1}{\sqrt{2(1+\mathcal{I}_{N,\mathrm{S}})}}(|\uparrow_{\mathrm{L}_1}\downarrow_{\mathrm{R}_N}\rangle + |\uparrow_{\mathrm{R}_N}\downarrow_{\mathrm{L}_1}\rangle), \tag{2.5}$$

$$|\mathrm{T}(1, 2N-1)\rangle = \frac{1}{\sqrt{2(1-\mathcal{I}_{N,\mathrm{T}})}}(|\uparrow_{\mathrm{L}_1}\downarrow_{\mathrm{R}_N}\rangle - |\uparrow_{\mathrm{R}_N}\downarrow_{\mathrm{L}_1}\rangle), \tag{2.6}$$

where

$$|\uparrow_{\mathrm{L}_1}\downarrow_{\mathrm{R}_N}\rangle = |\uparrow_{\mathrm{L}_1}, \downarrow_{\mathrm{R}_N}, \uparrow_{\mathrm{R}_{N-1}}, \downarrow_{\mathrm{R}_{N-1}}, \cdots, \uparrow_{\mathrm{R}_1}, \downarrow_{\mathrm{R}_1}\rangle, \tag{2.7}$$

$$|\uparrow_{\mathrm{R}_N}\downarrow_{\mathrm{L}_1}\rangle = |\uparrow_{\mathrm{L}_1}, \downarrow_{\mathrm{R}_N}, \uparrow_{\mathrm{R}_{N-1}}, \downarrow_{\mathrm{R}_{N-1}}, \cdots, \uparrow_{\mathrm{R}_1}, \downarrow_{\mathrm{R}_1}\rangle, \tag{2.8}$$

are $2N$-electron Slater determinants with different electron configurations in the DQD. Here, $\mathrm{L}_1$ and $\mathrm{R}_i$ $(i = 1 \ldots N)$ label the orbital states occupied by electrons in the left and right dots, respectively, as shown in Fig. 2.1. The symbols $\uparrow$ and $\downarrow$ represent the spin states, while $\mathcal{I}_{N,\mathrm{S}}$ and $\mathcal{I}_{N,\mathrm{T}}$ are normalization factors provided in Appendix 2.5.2. Thus, in the multi-electron case, the singlet state is symmetric and the triplet state is antisymmetric. When $N = 1$, this reduces to the conventional two-electron ST qubit [115].

In the biased DQD, due to varying detuning values, the singlet states hybridize into combinations of Slater determinants. For large positive detuning, the hybridized singlet is:

$$|\mathrm{S}^{(0,2N)}_{\mathrm{mix}}\rangle = \frac{|\mathrm{S}(1, 2N-1)\rangle + \beta|\mathrm{S}(0, 2N)\rangle}{\sqrt{1+\beta^2}}, \tag{2.9}$$

whereas for small positive or negative detuning, the hybridized singlet is:

$$|\mathrm{S}_{\mathrm{mix}}^{(2,2N-2)}\rangle = \frac{|\mathrm{S}(1,2N-1)\rangle + \beta|\mathrm{S}(2,2N-2)\rangle}{\sqrt{1+\beta^2}}, \tag{2.10}$$

where $1/\sqrt{1+\beta^2}$ and $\beta/\sqrt{1+\beta^2}$ are functions of detuning, and

$$|\mathrm{S}(0,2N)\rangle = |\uparrow_{\mathrm{R}_N}\downarrow_{\mathrm{R}_N}\rangle, \tag{2.11}$$

$$|\mathrm{S}(2,2N-2)\rangle = |\uparrow_{\mathrm{L}_1}\downarrow_{\mathrm{L}_1}\rangle, \tag{2.12}$$

where

$$|\uparrow_{\mathrm{R}_N}\downarrow_{\mathrm{R}_N}\rangle = |\uparrow_{\mathrm{R}_N}, \downarrow_{\mathrm{R}_N}, \uparrow_{\mathrm{R}_{N-1}}, \downarrow_{\mathrm{R}_{N-1}}, \cdots, \uparrow_{\mathrm{R}_1}, \downarrow_{\mathrm{R}_1}\rangle, \tag{2.13}$$

$$|\uparrow_{\mathrm{L}_1}\downarrow_{\mathrm{L}_1}\rangle = |\uparrow_{\mathrm{L}_1}, \downarrow_{\mathrm{L}_1}, \uparrow_{\mathrm{R}_{N-1}}, \downarrow_{\mathrm{R}_{N-1}}, \cdots, \uparrow_{\mathrm{R}_1}, \downarrow_{\mathrm{R}_1}\rangle, \tag{2.14}$$

are Slater determinants representing the electron configurations considered in this work.

### 2.3.3 Multi-Electron Dephasing of Electron-Phonon Interaction

The total Hamiltonian is given by [198]:

$$H = H_{\mathrm{e}} + H_{\mathrm{ph}} + H_{\mathrm{ep}}, \tag{2.15}$$

where $H_{\mathrm{e}}$ represents the electron system Hamiltonian in a double quantum dot (DQD) (see Eq. (2.1)), $H_{\mathrm{ph}}$ denotes the phonon environment Hamiltonian, expressed as $H_{\mathrm{ph}} = \sum_{\mathbf{q},\lambda} \omega_{\mathbf{q},\lambda} a^\dagger_{\mathbf{q},\lambda} a_{\mathbf{q},\lambda}$, and $H_{\mathrm{ep}}$ describes the electron-phonon interaction.

For a singlet-triplet (ST) qubit, $H_{\mathrm{e}}$ can be written in the basis of the lowest singlet, $|\mathrm{S}\rangle$, and the lowest triplet state, $|\mathrm{T}\rangle$, as:

$$H_{\mathrm{e}} = \frac{J}{2}\sigma_z, \tag{2.16}$$

where $\sigma_z = |\mathrm{T}\rangle\langle\mathrm{T}| - |\mathrm{S}\rangle\langle\mathrm{S}|$ (see Sec. 2.3.2 for details).

In a semiconductor, the Hamiltonian $H_{\text{ep}}$, which describes the effective electron-phonon interaction, takes the form [115]:

$$H_{\text{ep}} = \sum_{\mathbf{q},\lambda} M_\lambda(\mathbf{q})\rho(\mathbf{q})(a_{\mathbf{q},\lambda} + a^\dagger_{\text{-}\mathbf{q},\lambda}), \tag{2.17}$$

where $a_{\mathbf{q},\lambda}$ and $a^\dagger_{\text{-}\mathbf{q},\lambda}$ are the phonon annihilation and creation operators, respectively, $\mathbf{q}$ is the lattice momentum, and $\lambda$ is the phonon branch index. The electron density operator $\rho(\mathbf{q})$ is given by $\rho(\mathbf{q}) = \sum_{i=1}^{2N} e^{i\mathbf{q}\cdot\mathrm{R}_i}$ in a $2N$-electron DQD system. $M(\mathbf{q})$ represents various types of electron-phonon interactions.

In GaAs DQDs, the deformation potential (DP) and piezoelectric (PE) interactions are the main contributors to phonon-induced dephasing, while other interactions are negligible [115, 38]. The DP and PE interactions are given by [115]:

$$M^{\text{DP}}_{\text{GaAs}}(\mathbf{q}) = D\left(\frac{\hbar}{\rho V \omega_{\mathbf{q}}}\right)^{\frac{1}{2}} |\mathbf{q}|, \tag{2.18}$$

$$M^{\text{PE}}_{\text{GaAs}}(\mathbf{q}) = i\left(\frac{\hbar}{\rho V \omega_{\mathbf{q}}}\right)^{\frac{1}{2}} 2ee_{14}(\hat{q}_x\hat{q}_y\hat{\xi}_z + \hat{q}_y\hat{q}_z\hat{\xi}_x + \hat{q}_z\hat{q}_x\hat{\xi}_y), \tag{2.19}$$

where $M^{\text{DP}}_{\text{GaAs}}(\mathbf{q})$ couples electrons only to longitudinal acoustic phonons, while $M^{\text{PE}}_{\text{GaAs}}(\mathbf{q})$ couples electrons to both longitudinal and transverse acoustic phonons. Here, $D = 8.6$ eV is the deformation potential constant, $\rho = 5.3 \times 10^3$ kg/m$^3$ is the mass density, $e$ is the elementary charge, $e_{14} = 1.38 \times 10^9$ V/m is the piezoelectric constant, $\hat{\xi}$ is the polarization vector, and $\omega_{\mathbf{q}}$ is the angular frequency of the phonon mode $\mathbf{q}$. We define $\gamma_{\mathbf{q}}$ as the population relaxation rate of the phonon mode $\mathbf{q}$, which is assumed to follow the form $\gamma_{\mathbf{q}} = \gamma_0 q^n$ in our calculations. We fix $\gamma_0 = 10^8$ Hz and consider cases where $n = 2$ or $n = 3$ [115], confirming that varying $\gamma_0$ and $n$ does not significantly alter our main findings.

For a two-level system (qubit), the off-diagonal element of the effective electron-phonon interaction Hamiltonian leads to decay as [67, 291]:

$$\rho_{\text{ST}}(t) = \rho_{\text{ST}}(0)e^{-B^2(t)}, \tag{2.20}$$

where $B^2(t)$ is the dephasing factor. For an ST qubit in a $2N$-electron system, the logical eigenstates are defined as the lowest singlet and triplet states. Written in the basis of $|\mathrm{S}\rangle$ and $|\mathrm{T}\rangle$:

$$H_{\mathrm{ep}} = \begin{pmatrix} \langle \mathrm{T}|H_{\mathrm{ep}}|\mathrm{T}\rangle & \langle \mathrm{T}|H_{\mathrm{ep}}|\mathrm{S}\rangle \\ \langle \mathrm{S}|H_{\mathrm{ep}}|\mathrm{T}\rangle & \langle \mathrm{S}|H_{\mathrm{ep}}|\mathrm{S}\rangle \end{pmatrix}. \tag{2.21}$$

In Eq. (2.21), $H_{\mathrm{ep}}$ can be expanded in the basis of Pauli matrices:

$$H_{\mathrm{ep}} = H^z_{\mathrm{ep}}\sigma_z + H^x_{\mathrm{ep}}\sigma_x + H^y_{\mathrm{ep}}\sigma_y, \tag{2.22}$$

where $H^z_{\mathrm{ep}} = (\langle \mathrm{T}|H_{\mathrm{ep}}|\mathrm{T}\rangle - \langle \mathrm{S}|H_{\mathrm{ep}}|\mathrm{S}\rangle)/2$, with a global shift of $(\langle \mathrm{S}|H_{\mathrm{ep}}|\mathrm{S}\rangle + \langle \mathrm{T}|H_{\mathrm{ep}}|\mathrm{T}\rangle)/2$. Since $H_{\mathrm{ep}}$ has no spin dependence and no imaginary part, $H^x_{\mathrm{ep}}$ and $H^y_{\mathrm{ep}}$ are zero. Thus, we can rewrite $H_{\mathrm{ep}}$ as:

$$H_{\mathrm{ep}} = \sum_{\mathbf{q},\lambda} M_\lambda(\mathbf{q}) A_\phi \sigma_z (a_{\mathbf{q},\lambda} + a^\dagger_{\mathbf{q},\lambda}), \tag{2.23}$$

where $A_\phi$ is defined as:

$$A_\phi = \frac{1}{2}\left[\langle \mathrm{T}|\rho(\mathbf{q})|\mathrm{T}\rangle - \langle \mathrm{S}|\rho(\mathbf{q})|\mathrm{S}\rangle\right], \tag{2.24}$$

see Appendix 2.5.2 for details.

For a dissipative phonon reservoir with finite $\gamma_{\mathbf{q}}$, the main contribution to $B^2(t)$ can be calculated as [115, 67]:

$$B^2_{\mathrm{Decay}}(t) = \frac{V}{2\pi^3\hbar^2}\int d^3\mathbf{q} \frac{|M(\mathbf{q})A_\phi(\mathbf{q})|^2}{\omega^2_{\mathbf{q}} + \gamma^2_{\mathbf{q}}/4}\frac{\gamma_{\mathbf{q}}}{2} t \equiv \Gamma_{\mathrm{ST}} t, \tag{2.25}$$

where $\Gamma_{\mathrm{ST}}$, the dephasing rate, is the key quantity considered in this work.

### 2.3.4 Dephasing rate of unbiased case

According to Eq. (2.25), the electron-phonon dephasing rate can be expressed as:

$$\Gamma_{\mathrm{ST}} = \frac{V}{2\pi^3\hbar^2}\int d^3\mathbf{q} \frac{|M(\mathbf{q})A_\phi(\mathbf{q})|^2}{\omega^2_{\mathbf{q}} + \gamma^2_{\mathbf{q}}/4}\frac{\gamma_{\mathbf{q}}}{2}. \tag{2.26}$$

In the unbiased case, $A_\phi$ depends on the singlet state $|\mathrm{S}(1, 2N-1)\rangle$ and the triplet state $|\mathrm{T}(1, 2N-1)\rangle$ (Eqs. (2.5) and (2.6)), indicating that $\Gamma_{\mathrm{ST}}$ varies with the number of electrons.

We consider three unbiased cases with electron configurations $(1, 1)$, $(1, 3)$, and $(1, 7)$, where the first entry represents the number of electrons in dot L and the second entry represents the number of electrons in dot R. A schematic of the latter two cases is shown in Fig. 2.1(b). Details on the evaluation of $A_\phi$ for these configurations are provided in Appendix 2.5.2.

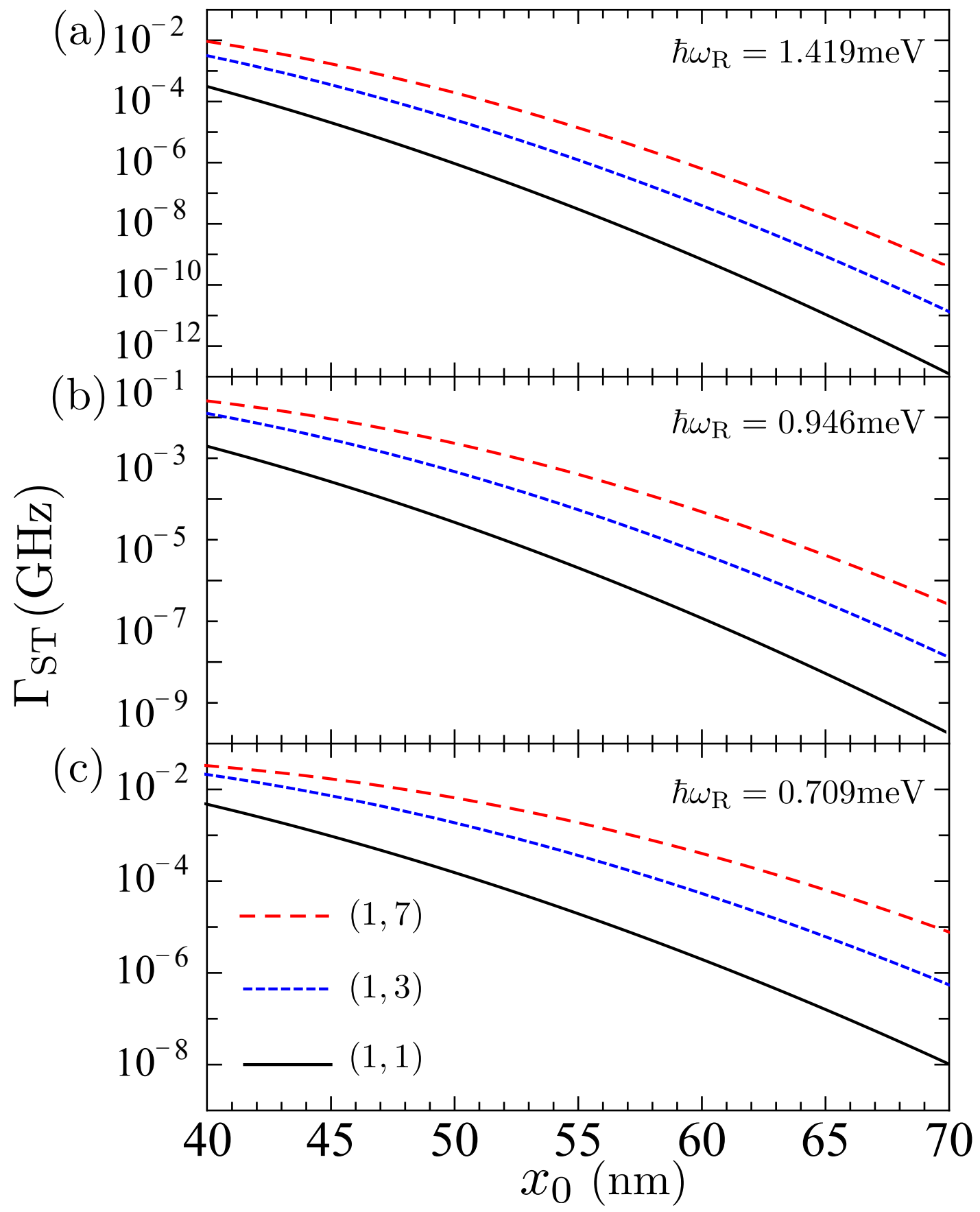

Fig. 2.2 The dephasing rate $\Gamma_{\mathrm{ST}}$ v.s. half dot distance $x_0$ in unbiased case for the three different electron configurations as indicated and the right dot confinement energy $\hbar\omega_{\mathrm{R}}$ being (a) 1.419 meV, (b) 0.946 meV and (c) 0.709 meV. The left dot confinement energy is fixed as $\hbar\omega_{\mathrm{L}} = 2.838$ meV.

Figure 2.2 shows the dephasing rate $\Gamma_{\rm ST}$ as a function of the half-dot distance $x_0$ for different confinement strengths $\hbar\omega_{\rm R}$, as indicated. The three values of the confinement strength in dot R ($\hbar\omega_{\rm R} = 1.419$ meV, 0.946 meV, and 0.709 meV) correspond to dot sizes of 28.076 nm, 33.981 nm, and 38.627 nm, respectively. Several features can be clearly observed from the figure:

1. The dephasing rate rapidly decreases with increasing $x_0$ in all cases. The results for $(1,1)$ are consistent with Ref. [115], and it is not surprising that the results for $(1,3)$ and $(1,7)$ follow a similar trend.

2. For a given confinement strength, the dephasing rate is greatest for $(1,7)$, as more electrons lead to larger contributions from $A_\phi$ (see Appendix 2.5.2), implying more channels for electron-phonon interaction. The dephasing rate is intermediate for $(1,3)$ and smallest for $(1,1)$.

3. When $x_0$ and $\hbar\omega_{\rm L}$ are fixed, the dephasing rate is greater when dot R is larger (smaller $\hbar\omega_{\rm R}$) and smaller when dot R is smaller (larger $\hbar\omega_{\rm R}$). This behavior is controlled by the integrals in Eqs. (2.70), (2.71), and (2.72) for the $(1,1)$, $(1,3)$, and $(1,7)$ cases, respectively. The left-hand side of Eqs. (2.70), (2.71), and (2.72) decreases with increasing $x_0$ or as the dots become smaller.

### 2.3.5 Biased Case and the Merit Figure

As the detuning $\Delta$ changes, the singlet states start to hybridize as described by Eq. (2.9) or Eq. (2.10). Fig. 2.3 shows the dephasing rate $\Gamma_{\rm ST}$ as a function of $\beta/\sqrt{1+\beta^2}$ for six different hybridized states (the normalization constant is omitted in the legend). For small $\beta$, the hybridization ratio $\beta/\sqrt{1+\beta^2} \approx \beta$ indicates the ratio of hybridization to states other than the $(1,1)$, $(1,3)$, and $(1,7)$ states considered.

Fig. 2.3(a) shows the range $0 \leq \beta/\sqrt{1+\beta^2} \leq 0.05$ while Fig. 2.3(b) shows the range $0.05 \leq \beta/\sqrt{1+\beta^2} \leq 0.10$. The dephasing rate $\Gamma_{\rm ST}$ increases monotonically with the hybridization ratio. The order of the results for states with mainly $(1,1)$, $(1,3)$, and $(1,7)$ character changes. Specifically, for $\beta/\sqrt{1+\beta^2} = 0$, the dephasing rate for the state $(1,1)$ is the smallest, consistent with the unbiased case. However, for $\beta/\sqrt{1+\beta^2} \gtrsim 0.02$, the dephasing rate for the state with mainly $(1,1)$ character becomes the largest, exceeding the rate for the state with mainly $(1,3)$ character by about 30

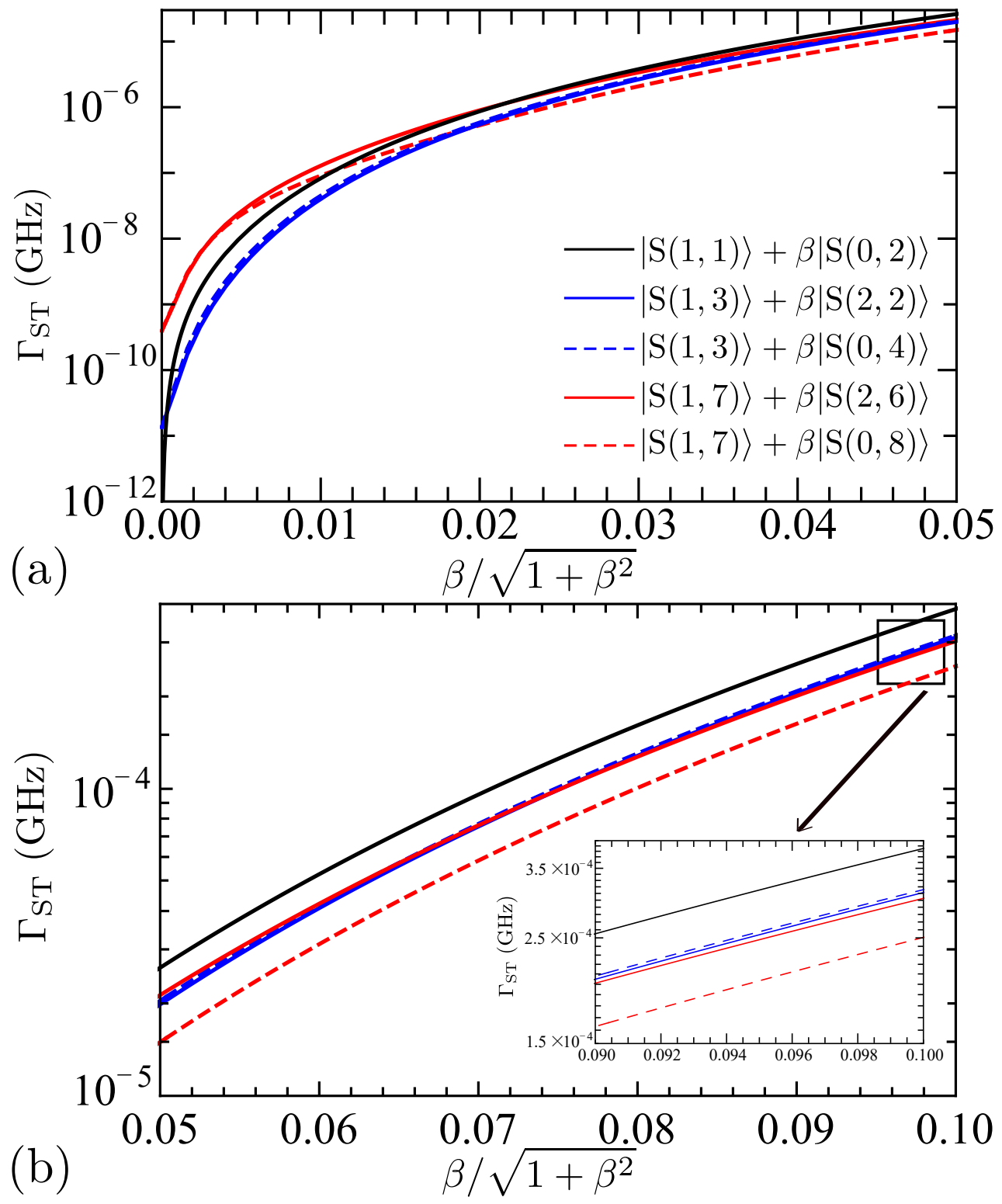

Fig. 2.3 The dephasing rate $\Gamma_{\mathrm{ST}}$ vs. $\beta/\sqrt{1+\beta^2}$ in the biased case for five different states as indicated (note that the normalization constant is omitted in the legend). (a) shows the range $0 \leq \beta/\sqrt{1+\beta^2} \leq 0.05$, and (b) the range $0.05 \leq \beta/\sqrt{1+\beta^2} \leq 0.1$, with an inset showing a zoomed-in view at the tail of the curves. Parameters: $x_0 = 70$ nm, $\hbar\omega_{\mathrm{L}} = 2.838$ meV, and $\hbar\omega_{\mathrm{R}} = 1.419$ meV.

The numerical results in Fig. 2.3(b) can be understood by inspecting the explicit forms of $\Gamma_{\mathrm{ST}}$ [Eq. (2.26)], $A_\phi$ [Eq. (2.73)], and the electron density of the fully occupied singlet state, $\langle S(0,2N)|\rho(\mathbf{q})|S(0,2N)\rangle$ [Eq. (2.75)]. Since $\Gamma_{\mathrm{ST}}$ depends on the magnitude of $|A_\phi|$ [Eq. (2.26)], at a fixed value of hybridization, $\beta/\sqrt{\beta^2+1}$, Eq. (2.73) suggests that the magnitudes of $A_\phi$ for different numbers of electrons depend on the magnitude of $\langle S(0,2N)|\rho(\mathbf{q})|S(0,2N)\rangle$. The reduction of dephasing for a larger number of electrons can be understood by examining the magnitudes of $|A_\phi|$ at the limit where $\beta/\sqrt{\beta^2+1} \approx 1$. In this limit, the additional nodes of the excited orbitals lead to extra terms with negative coefficients in the expression of $|A_\phi|$, resulting in a smaller magnitude of $|A_\phi|$. Physically, this can be interpreted as the reduction of dephasing due to the oscillations of the excited wavefunctions in position space, which interfere destructively with the fluctuations caused by the deformation potential. This analysis leads to the conclusion that the reduction in the dephasing rate with the number of elec-trons can be attributed to the larger electron density of the fully occupied singlet state, $\langle S(0,2N)|\rho(\mathbf{q})|S(0,2N)\rangle$. This argument can similarly be applied to the biased case where $|S(1,2N-1)\rangle$ hybridizes with $|S(2,2N-2)\rangle$ [Eq. (2.74) and Eq. (2.76)].

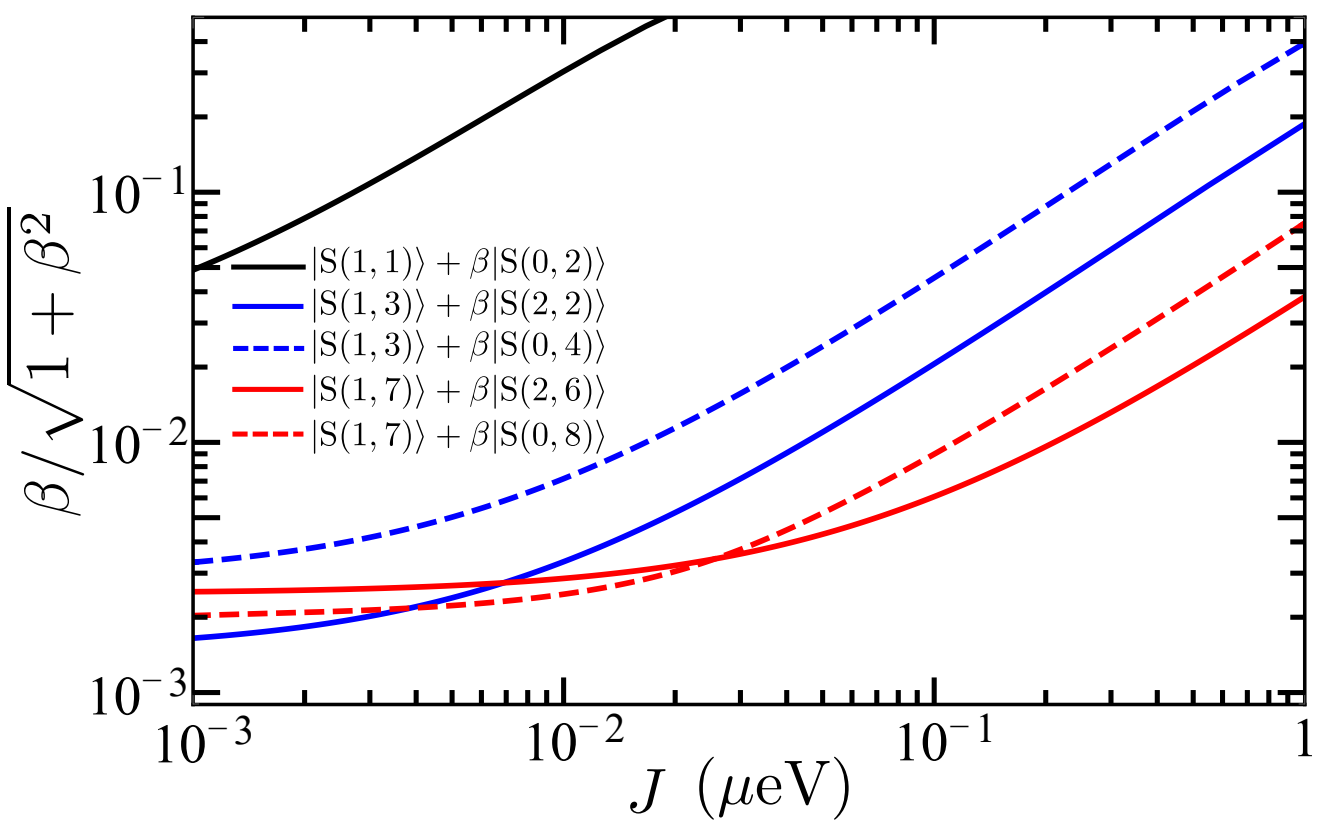

Fig. 2.4 $\beta/\sqrt{1+\beta^2}$ as a function of the exchange energy $J$ in the biased case for five different states as indicated. Parameters: $x_0 = 70$ nm, $\hbar\omega_\mathrm{L} = 2.838$ meV, $\hbar\omega_\mathrm{R} = 1.419$ meV.

Fig. 2.4 shows the hybridization ratio $\beta/\sqrt{1+\beta^2}$ versus the exchange interaction $J$, as calculated from Eq. (2.4). Generally, the more hybridized the singlet state, the larger the absolute value of detuning, and consequently, the greater the value of $J$. For the same value of $J$, the state with mainly $(1,1)$ character has a greater hybridization ratio [32].

To evaluate the performance of our system in practical situations, we define the merit $\mathcal{M} = J/\hbar\Gamma_\mathrm{ST}$ as the ratio between the exchange gate time given by $\hbar/J$ and the decay time given by $1/\Gamma_\mathrm{ST}$. The merit $\mathcal{M}$ assumes the negative role of electron-phonon interaction on the coherence of the logical states, as suggested by theoretical work [147] and experimental work [60] on the decoherence rate of a two-electron singlet-triplet qubit in a DQD device. This contrasts with other works showing that dissipation can enhance the stability of quantum systems [266, 97, 34, 33]. Specifically, Ref. [265] demonstrated that periodic driving can enhance the stability of a quantum metastable system. However, investigating the positive role of electron-phonon interaction in the coherence of a multielectron singlet-triplet qubit is beyond the scope of this work.

The merit $\mathcal{M}$ as a function of the exchange energy is shown in Fig. 2.5, which is the key result of this work. The non-monotonic behavior of the $\mathcal{M}$ vs. $J$ curves in Fig. 2.5 results from the varying rates of $\Gamma_\mathrm{ST}$ and $J$ as functions of $\beta/\sqrt{1+\beta^2}$. More importantly, the merit figure for states associated with $(1,3)$ and $(1,7)$ is greater than for those associated with $(1,1)$. This indicates that multi-electron quantum dots may offer advantages in reducing electron-phonon dephasing, which is the main result of this work. Fig. 2.5 is divided into two regions: $J < 0.04$ $\mu$eV (marked in yellow) and $J \geq 0.04$ $\mu$eV (marked in cyan). In the cyan region, the merit figures for states associated with $(1, 7)$ are greater than those for $(1, 3)$, while the merit figure for the state with $(1, 1)$ is the smallest. Given that practical operations of the qubit require the exchange interaction to be neither too small nor too large, in the regime of $J \geq 0.04$ $\mu$eV, having more electrons in the right dot implies a better merit figure, advantageous in experiments. This is the key finding of this work. We have verified that our conclusion holds for other experimentally relevant parameters, including the dot distance and dot sizes (confinement strength), with selective results shown in Appendix 2.5.3.

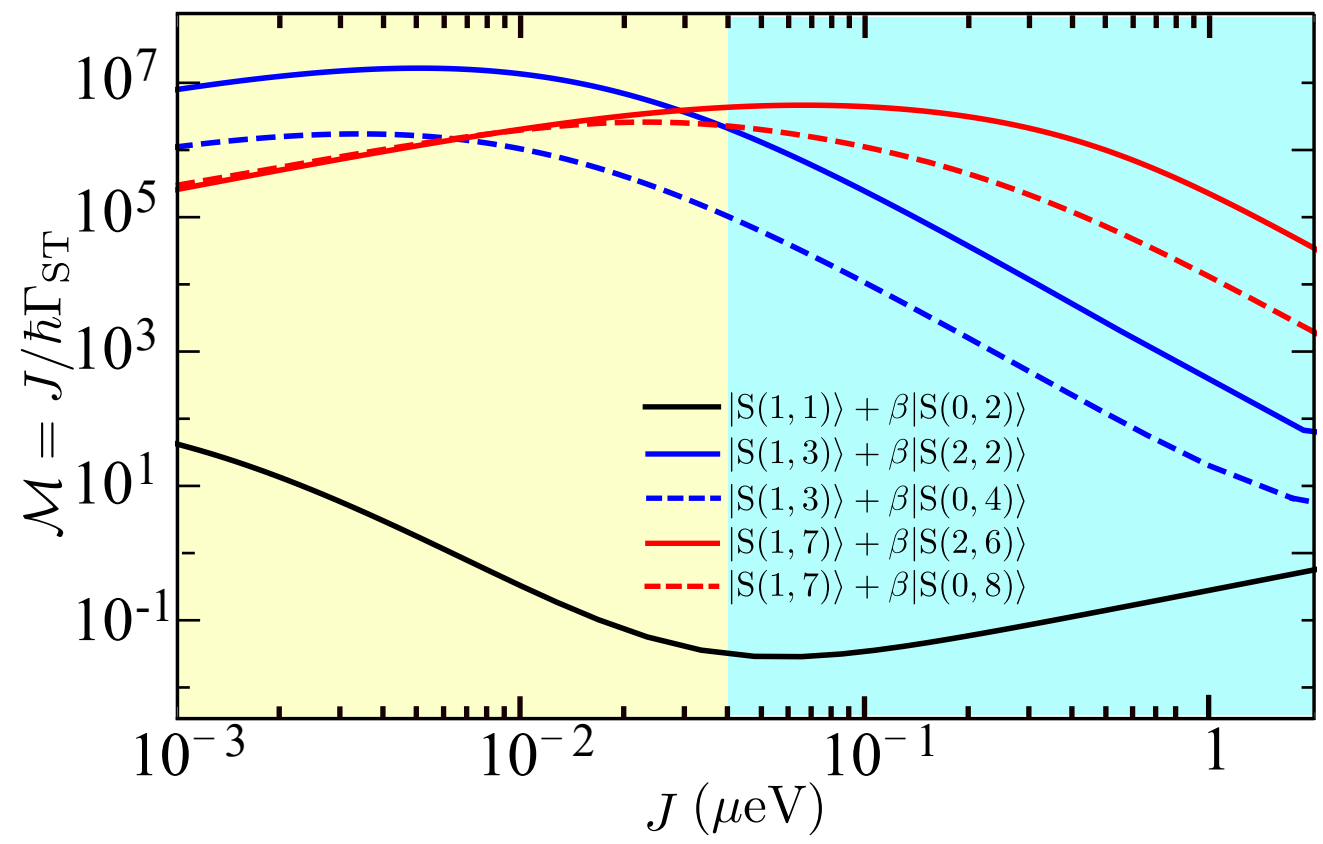

Fig. 2.5 The merit figure vs. exchange energy for five different states as indicated. Parameters: half-dot distance $x_0 = 70$ nm, $\hbar\omega_L = 2.838$ meV, and $\hbar\omega_R = 1.419$ meV. The yellow shaded area ($J < 0.04$ $\mu$eV) shows the regime where the merit figure for (1,3) and (1,7) is better than for (1,1), while the cyan shaded area ($J \geq 0.04$ $\mu$eV) shows the regime where the merit figure for (1,7) is the greatest, that for (1,3) is intermediate, and for (1,1) is the lowest.

This behavior is understandable from Fig. 2.3 and Fig. 2.4. Fig. 2.4 shows that at a fixed value of $\beta/\sqrt{1+\beta^2}$, $J$ is largest for states associated with $(1, 7)$, intermediate for $(1, 3)$, and smallest for $(1, 1)$, with appreciable differences. From Fig. 2.3, one sees that for the same value of $\beta/\sqrt{1+\beta^2}$, the values of $\Gamma_{ST}$ are close. Since $J$ is in the numerator of the merit figure, the merit figure follows the same trend as observed in Fig. 2.4.

Since the behavior of the merit figure follows the behavior of $J$, the critical value $J = 0.04$ $\mu$eV can be understood from the relationship between the number of electrons and the magnitude of $J$. For the regime in which $J > 0.04$ $\mu$eV and a larger number of electrons, at a larger value of hybridization $\beta/\sqrt{\beta^2+1}$, the wavefunctions of the excited orbitals are more extended in space, leading to enhanced overlap between wavefunctions in the left and right dots, resulting in larger exchange energy. Conversely, for the regime in which $J < 0.04$ $\mu$eV, at a small value of hybridization $\beta/\sqrt{\beta^2+1}$, the overlap occurs only at the tails of the valence orbitals, exhibiting similar forms for a four- and eight-electron singlet-triplet qubit, leading to a comparable magnitude of $J$ for both qubits.

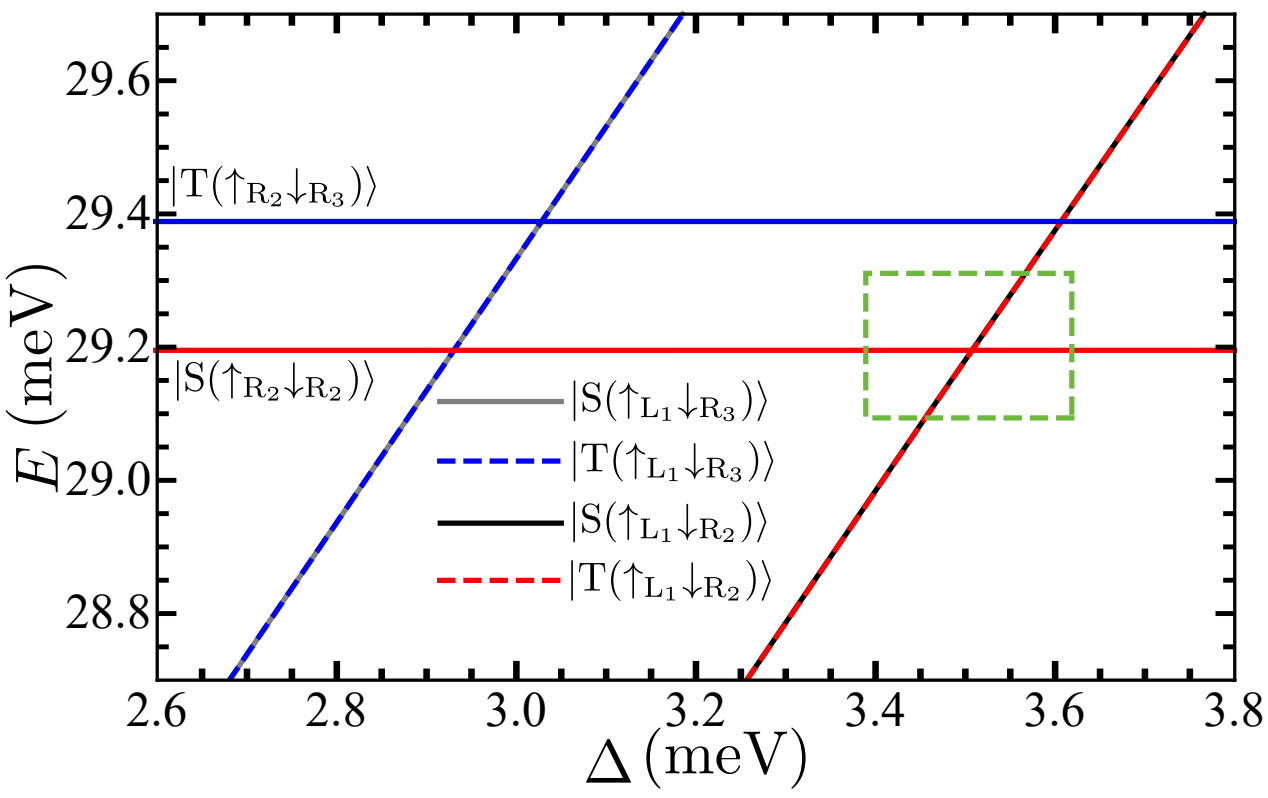

Fig. 2.6 Diagonal elements of DQD Hamiltonian occupied by four electrons as function of detuning, in bases of $\{|\mathrm{S}(\uparrow_{\mathrm{L}_1}\downarrow_{\mathrm{R}_2})\rangle\rangle$, $|\mathrm{T}(\uparrow_{\mathrm{L}_1}\downarrow_{\mathrm{R}_2})\rangle\rangle$, $|\mathrm{S}(\uparrow_{\mathrm{L}_1}\downarrow_{\mathrm{R}_3})\rangle\rangle$, $|\mathrm{T}(\uparrow_{\mathrm{L}_1}\downarrow_{\mathrm{R}_3})\rangle\rangle$, $|\mathrm{S}(\uparrow_{\mathrm{R}_2}\downarrow_{\mathrm{R}_3})\rangle\rangle$, $|\mathrm{T}(\uparrow_{\mathrm{R}_2}\downarrow_{\mathrm{R}_3})\rangle\rangle$, $|\mathrm{S}(\uparrow_{\mathrm{R}_2}\downarrow_{\mathrm{R}_2})\rangle\rangle\}$. $x_0 = 70$ nm, $\hbar\omega_{\mathrm{L}} = 2.838$ meV, $\hbar\omega_{\mathrm{R}} = 1.419$ meV, B=0.7 T.

## 2.4 Conclusion

In this work, we have calculated the dephasing rate, exchange energy, and merit figure of a multi-electron quantum-dot system with one electron in the left dot and 1, 3, or 7 electrons in the right dot. Our findings reveal that in the unbiased case, the dephasing rate generally increases with the number of electrons in the right dot. However, this trend does not necessarily hold in the biased case. Importantly, we have shown that in the experimentally relevant regime where $J \geq 0.04$ $\mu$eV, having more electrons in the right dot results in a better merit figure. These results suggest that multi-electron quantum dots may offer advantages in specific scenarios.

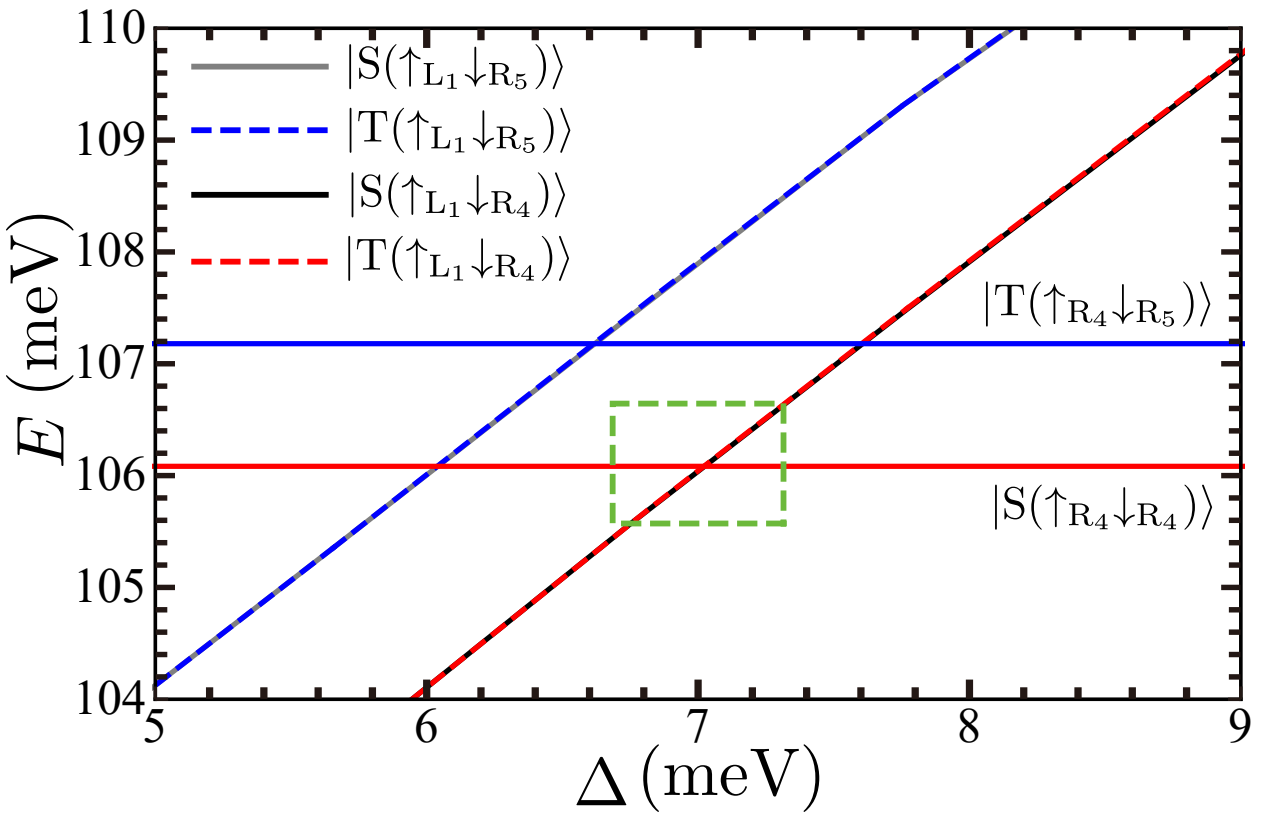

Fig. 2.7 Diagonal elements of DQD Hamiltonian occupied by eight electrons as function of detuning, in bases of $\{|\mathrm{S}(\uparrow_{\mathrm{L}_1}\downarrow_{\mathrm{R}_4})\rangle, |\mathrm{T}(\uparrow_{\mathrm{L}_1}\downarrow_{\mathrm{R}_4})\rangle, |\mathrm{S}(\uparrow_{\mathrm{L}_1}\downarrow_{\mathrm{R}_5})\rangle, |\mathrm{T}(\uparrow_{\mathrm{L}_1}\downarrow_{\mathrm{R}_5})\rangle, |\mathrm{S}(\uparrow_{\mathrm{R}_4}\downarrow_{\mathrm{R}_5})\rangle, |\mathrm{T}(\uparrow_{\mathrm{R}_4}\downarrow_{\mathrm{R}_5})\rangle, |\mathrm{S}(\uparrow_{\mathrm{R}_4}\downarrow_{\mathrm{R}_4})\rangle\}$. $x_0 = 70$ nm, $\hbar\omega_{\mathrm{L}} = 2.838$ meV, $\hbar\omega_{\mathrm{R}} = 1.419$ meV, B=0.7 T.

## 2.5 Appendix

### 2.5.1 System Hamiltonian and exchange energy

In $2N$-electron system, the effective Hamiltonian can be written in extended Hubbard model [125, 291]

$$\begin{aligned} H_e = &\sum_{j,\sigma} \varepsilon_{j,\sigma} c^\dagger_{j,\sigma} c_{j,\sigma} + \sum_{j<k,\sigma} (t_{j,k,\sigma} c^\dagger_{j,\sigma} c_{k,\sigma} + \text{H.c.}) \\ &+ \sum_j U_j n_{j\downarrow} n_{j\uparrow} + \sum_j U_{j,k} n_{j\sigma} n_{k\sigma'} \\ &+ \sum_{\sigma\sigma'} \sum_{j<k} U^e_{j,k} c^\dagger_{j,\sigma} c^\dagger_{k,\sigma'} c_{j,\sigma'} c_{k,\sigma}, \end{aligned} \tag{2.27}$$

where

$$U_j = \int \Psi^*_j(\mathbf{r}_1)\Psi^*_j(\mathbf{r}_2) C(\mathbf{r}_1, \mathbf{r}_2) \Psi_j(\mathbf{r}_1)\Psi_j(\mathbf{r}_2) \mathrm{d}\mathbf{r}^2, \tag{2.28}$$

$$U_{j,k} = \int \Psi^*_j(\mathbf{r}_1)\Psi^*_k(\mathbf{r}_2) C(\mathbf{r}_1, \mathbf{r}_2) \Psi_j(\mathbf{r}_1)\Psi_k(\mathbf{r}_2) \mathrm{d}\mathbf{r}^2, \tag{2.29}$$

$$U^e_{j,k} = \int \Psi^*_j(\mathbf{r}_1)\Psi^*_k(\mathbf{r}_2) C(\mathbf{r}_1, \mathbf{r}_2) \Psi_k(\mathbf{r}_1)\Psi_j(\mathbf{r}_2) \mathrm{d}\mathbf{r}^2, \tag{2.30}$$

$$t_{j,k} = \int \Psi^*_j(\mathbf{r}) \left[ \frac{\hbar^2}{2m^*} \nabla^2 + V(\mathbf{r}) \right] \Psi_k(\mathbf{r}) \mathrm{d}\mathbf{r}, \tag{2.31}$$

$$\varepsilon_j = \int \Psi_j^*(\mathbf{r}) \left[ \frac{\hbar^2}{2m^*} \nabla^2 + V(\mathbf{r}) \right] \Psi_j(\mathbf{r}) d\mathbf{r}, \tag{2.32}$$

$$C(\mathbf{r}_1, \mathbf{r}_2) = \frac{e^2}{\kappa \left| \mathbf{r}_1 - \mathbf{r}_2 \right|}. \tag{2.33}$$

here, $j$ and $k$ are orbital indices, for example, $\Psi_{\mathrm{R}_1}$, $\Psi_{\mathrm{L}_1}$, and $\Psi_{\mathrm{R}_2}$ represent F-D states on the orbits $\mathrm{R}_1$, $\mathrm{L}_1$, and $\mathrm{R}_2$ respectively. The symbols $\sigma$ and $\sigma^{'}$ denote spin states, which can be either up or down. $\varepsilon_{j,\sigma}$ is the onsite energy at orbital $j$, $t_{j,k,\sigma}$ represents the tunneling energy between the $j$th and $k$th orbitals, $U_j$ denotes the onsite Coulomb interaction in the $j$th orbital, and $U_{j,k}$ and $U^e_{j,k}$ represent the direct and exchange Coulomb interactions between the $j$th and $k$th orbitals, respectively.

We denote $2N$-electron Slater determinants as

$$|\mathrm{S}(\uparrow_{\mathrm{L}_1}\downarrow_{\mathrm{R}_m})\rangle = |\uparrow_{\mathrm{L}_1}\downarrow_{\mathrm{R}_m}\rangle + |\uparrow_{\mathrm{R}_m}\downarrow_{\mathrm{L}_1}\rangle, \tag{2.34}$$

$$|\mathrm{T}(\uparrow_{\mathrm{L}_1}\downarrow_{\mathrm{R}_m})\rangle = |\uparrow_{\mathrm{L}_1}\downarrow_{\mathrm{R}_m}\rangle - |\uparrow_{\mathrm{R}_m}\downarrow_{\mathrm{L}_1}\rangle, \tag{2.35}$$

$$|\mathrm{S}(\uparrow_{\mathrm{R}_N}\downarrow_{\mathrm{R}_m})\rangle = |\uparrow_{\mathrm{R}_N}\downarrow_{\mathrm{R}_m}\rangle + |\uparrow_{\mathrm{R}_m}\downarrow_{\mathrm{R}_N}\rangle, \tag{2.36}$$

$$|\mathrm{T}(\uparrow_{\mathrm{R}_N}\downarrow_{\mathrm{R}_m})\rangle = |\uparrow_{\mathrm{R}_N}\downarrow_{\mathrm{R}_m}\rangle - |\uparrow_{\mathrm{R}_m}\downarrow_{\mathrm{R}_N}\rangle, \tag{2.37}$$

where

$$|\uparrow_{\mathrm{L}_1}\downarrow_{\mathrm{R}_m}\rangle = |\uparrow_{\mathrm{L}_1}, \downarrow_{\mathrm{R}_m}, \uparrow_{\mathrm{R}_{N-1}}, \downarrow_{\mathrm{R}_{N-1}}, \ldots, \uparrow_{\mathrm{R}_1}, \downarrow_{\mathrm{R}_1}\rangle, \tag{2.38}$$

$$|\uparrow_{\mathrm{R}_m}\downarrow_{\mathrm{L}_1}\rangle = |\uparrow_{\mathrm{R}_m}, \downarrow_{\mathrm{L}_1}, \uparrow_{\mathrm{R}_{N-1}}, \downarrow_{\mathrm{R}_{N-1}}, \ldots, \uparrow_{\mathrm{R}_1}, \downarrow_{\mathrm{R}_1}\rangle, \tag{2.39}$$

$$|\uparrow_{\mathrm{R}_N}\downarrow_{\mathrm{R}_m}\rangle = |\uparrow_{\mathrm{R}_N}, \downarrow_{\mathrm{R}_m}, \uparrow_{\mathrm{R}_{N-1}}, \downarrow_{\mathrm{R}_{N-1}}, \ldots, \uparrow_{\mathrm{R}_1}, \downarrow_{\mathrm{R}_1}\rangle, \tag{2.40}$$

$$|\uparrow_{\mathrm{R}_m}\downarrow_{\mathrm{R}_N}\rangle = |\uparrow_{\mathrm{R}_m}, \downarrow_{\mathrm{R}_N}, \uparrow_{\mathrm{R}_{N-1}}, \downarrow_{\mathrm{R}_{N-1}}, \ldots, \uparrow_{\mathrm{R}_1}, \downarrow_{\mathrm{R}_1}\rangle, \tag{2.41}$$

are 2N-electron slater determinants, here $m \geqslant N$. Normalization coefficients are dropped for simplicity.

We can write effctive Hubbard Hamiltonian of four-electron in the bases: $\{|\mathrm{S}(\uparrow_{\mathrm{L}_1}\downarrow_{\mathrm{R}_2})\rangle, |\mathrm{T}(\uparrow_{\mathrm{L}_1}\downarrow_{\mathrm{R}_2})\rangle, |\mathrm{S}(\uparrow_{\mathrm{L}_1}\downarrow_{\mathrm{R}_3})\rangle, |\mathrm{T}(\uparrow_{\mathrm{L}_1}\downarrow_{\mathrm{R}_3})\rangle, |\mathrm{S}(\uparrow_{\mathrm{R}_2}\downarrow_{\mathrm{R}_3})\rangle, |\mathrm{T}(\uparrow_{\mathrm{R}_2}\downarrow_{\mathrm{R}_3})\rangle, |\mathrm{S}(\uparrow_{\mathrm{R}_2}\downarrow_{\mathrm{R}_2})\rangle\}$. With $2\varepsilon_{R_1}$ energy shift, diagonal elements of system Hamiltonian are:

$$\begin{aligned} \mathrm{diag}(|\mathrm{S}(\uparrow_{\mathrm{L}_1}\downarrow_{\mathrm{R}_2})\rangle) =& \varepsilon_{\mathrm{L}_1} + \varepsilon_{\mathrm{R}_2} + U_{\mathrm{R}_1} + 2U_{\mathrm{R}_1,\mathrm{L}_1} + 2U_{\mathrm{R}_1,\mathrm{R}_2} + U_{\mathrm{L}_1,\mathrm{R}_2} - U^e_{\mathrm{R}_1,\mathrm{L}_1} \\ & - U^e_{\mathrm{R}_1,\mathrm{R}_2} + U^e_{\mathrm{L}_1,\mathrm{R}_2}, \end{aligned} \tag{2.42}$$

$$\begin{aligned}\mathrm{diag}(|\mathrm{T}(\uparrow_{\mathrm{L}_1}\downarrow_{\mathrm{R}_2})\rangle\rangle) =&\varepsilon_{\mathrm{L}_1} + \varepsilon_{\mathrm{R}_2} + U_{\mathrm{R}_1} + 2U_{\mathrm{R}_1,\mathrm{L}_1} + 2U_{\mathrm{R}_1,\mathrm{R}_2} + U_{\mathrm{L}_1,\mathrm{R}_2} - U^e_{\mathrm{R}_1,\mathrm{L}_1} \\ &- U^e_{\mathrm{R}_1,\mathrm{R}_2} - U^e_{\mathrm{L}_1,\mathrm{R}_2},\end{aligned} \tag{2.43}$$

$$\begin{aligned}\mathrm{diag}(|\mathrm{S}(\uparrow_{\mathrm{L}_1}\downarrow_{\mathrm{R}_3})\rangle\rangle) =&\varepsilon_{\mathrm{L}_1} + \varepsilon_{\mathrm{R}_3} + U_{\mathrm{R}_1} + 2U_{\mathrm{R}_1,\mathrm{L}_1} + 2U_{\mathrm{R}_1,\mathrm{R}_3} + U_{\mathrm{L}_1,\mathrm{R}_3} - U^e_{\mathrm{R}_1,\mathrm{L}_1} \\ &- U^e_{\mathrm{R}_1,\mathrm{R}_3} + U^e_{\mathrm{L}_1,\mathrm{R}_3},\end{aligned} \tag{2.44}$$

$$\begin{aligned}\mathrm{diag}(|\mathrm{T}(\uparrow_{\mathrm{L}_1}\downarrow_{\mathrm{R}_3})\rangle\rangle) =&\varepsilon_{\mathrm{L}_1} + \varepsilon_{\mathrm{R}_3} + U_{\mathrm{R}_1} + 2U_{\mathrm{R}_1,\mathrm{L}_1} + 2U_{\mathrm{R}_1,\mathrm{R}_3} + U_{\mathrm{L}_1,\mathrm{R}_3} - U^e_{\mathrm{R}_1,\mathrm{L}_1} \\ &- U^e_{\mathrm{R}_1,\mathrm{R}_3} - U^e_{\mathrm{L}_1,\mathrm{R}_2},\end{aligned} \tag{2.45}$$

$$\begin{aligned}\mathrm{diag}(|\mathrm{S}(\uparrow_{\mathrm{R}_2}\downarrow_{\mathrm{R}_3})\rangle\rangle) =&\varepsilon_{\mathrm{R}_2} + \varepsilon_{\mathrm{R}_3} + U_{\mathrm{R}_1} + 2U_{\mathrm{R}_1,\mathrm{R}_2} + 2U_{\mathrm{R}_1,\mathrm{R}_3} + U_{\mathrm{R}_2,\mathrm{R}_3} - U^e_{\mathrm{R}_1,\mathrm{R}_2} \\ &- U^e_{\mathrm{R}_1,\mathrm{R}_3} + U^e_{\mathrm{R}_2,\mathrm{R}_3},\end{aligned} \tag{2.46}$$

$$\begin{aligned}\mathrm{diag}(|\mathrm{T}(\uparrow_{\mathrm{R}_2}\downarrow_{\mathrm{R}_3})\rangle\rangle) =&\varepsilon_{\mathrm{R}_2} + \varepsilon_{\mathrm{R}_3} + U_{\mathrm{R}_1} + 2U_{\mathrm{R}_1,\mathrm{R}_2} + 2U_{\mathrm{R}_1,\mathrm{R}_3} + U_{\mathrm{R}_2,\mathrm{R}_3} - U^e_{\mathrm{R}_1,\mathrm{R}_2} \\ &- U^e_{\mathrm{R}_1,\mathrm{R}_3} - U^e_{\mathrm{R}_2,\mathrm{R}_3},\end{aligned} \tag{2.47}$$

$$\mathrm{diag}(|\mathrm{S}(\uparrow_{\mathrm{R}_2}\downarrow_{\mathrm{R}_2})\rangle\rangle) =2\varepsilon_{\mathrm{R}_2} + U_{\mathrm{R}_1} + U_{\mathrm{R}_2} + 4U_{\mathrm{R}_1,\mathrm{R}_2} - 2U^e_{\mathrm{R}_1,\mathrm{R}_2}. \tag{2.48}$$

For eight-electron effctive Hubbard Hamiltonian can be written in the bases: $\{|\mathrm{S}(\uparrow_{\mathrm{L}_1}\downarrow_{\mathrm{R}_4}\rangle\rangle, |\mathrm{T}(\uparrow_{\mathrm{L}_1}\downarrow_{\mathrm{R}_4})\rangle, |\mathrm{S}(\uparrow_{\mathrm{L}_1}\downarrow_{\mathrm{R}_5})\rangle, |\mathrm{T}(\uparrow_{\mathrm{L}_1}\downarrow_{\mathrm{R}_5})\rangle, |\mathrm{S}(\uparrow_{\mathrm{R}_4}\downarrow_{\mathrm{R}_5})\rangle, |\mathrm{T}(\uparrow_{\mathrm{R}_4}\downarrow_{\mathrm{R}_5})\rangle, |\mathrm{S}(\uparrow_{\mathrm{R}_4}\downarrow_{\mathrm{R}_4})\rangle\}$, with energy shift of $2\varepsilon_{\mathrm{R}_1}+2\varepsilon_{\mathrm{R}_2}+2\varepsilon_{\mathrm{R}_3}$, diagonal elements are:

$$\begin{aligned}\mathrm{diag}(|\mathrm{S}(\uparrow_{\mathrm{L}_1}\downarrow_{\mathrm{R}_4})\rangle) =&\varepsilon_{\mathrm{L}_1}+\varepsilon_{\mathrm{R}_4}+U_{\mathrm{R}_1}+2U_{\mathrm{R}_1,\mathrm{L}_1}+4U_{\mathrm{R}_1,\mathrm{R}_2}+4U_{\mathrm{R}_1,\mathrm{R}_3}+2U_{\mathrm{R}_1,\mathrm{R}_4}\\&+2U_{\mathrm{L}_1,\mathrm{R}_2}+U_{\mathrm{L}_1,\mathrm{R}_4}+U_{\mathrm{R}_2}+4U_{\mathrm{R}_2,\mathrm{R}_3}+2U_{\mathrm{R}_2,\mathrm{R}_4}-U^e_{\mathrm{R}_1,\mathrm{R}_4}\\&+2U_{\mathrm{R}_3,\mathrm{R}_4}-U^e_{\mathrm{R}_1,\mathrm{L}_1}-2U^e_{\mathrm{R}_1,\mathrm{R}_3}-2U^e_{\mathrm{R}_1,\mathrm{R}_2}+U_{\mathrm{R}_3}-2U^e_{\mathrm{L}_1,\mathrm{R}_2}\\&-U^e_{\mathrm{L}_1,R3}-2U^e_{\mathrm{R}_2,\mathrm{R}_3}-U^e_{\mathrm{R}_2,\mathrm{R}_4}-U^e_{\mathrm{R}_3,R4}+U^e_{\mathrm{L}_1,\mathrm{R}_4},\end{aligned}\tag{2.49}$$

$$\begin{aligned}\mathrm{diag}(|\mathrm{T}(\uparrow_{\mathrm{L}_1}\downarrow_{\mathrm{R}_4})\rangle) =&\varepsilon_{\mathrm{L}_1}+\varepsilon_{\mathrm{R}_4}+U_{\mathrm{R}_1}+2U_{\mathrm{R}_1,\mathrm{L}_1}+4U_{\mathrm{R}_1,\mathrm{R}_2}+4U_{\mathrm{R}_1,\mathrm{R}_3}+2U_{\mathrm{R}_1,\mathrm{R}_4}\\&+2U_{\mathrm{L}_1,\mathrm{R}_2}+U_{\mathrm{L}_1,\mathrm{R}_4}+U_{\mathrm{R}_2}+4U_{\mathrm{R}_2,\mathrm{R}_3}+2U_{\mathrm{R}_2,\mathrm{R}_4}+2U_{\mathrm{R}_3,\mathrm{R}_4}\\&-U^e_{\mathrm{R}_1,\mathrm{L}_1}-2U^e_{\mathrm{R}_1,\mathrm{R}_3}-2U^e_{\mathrm{R}_1,\mathrm{R}_2}-U^e_{\mathrm{R}_1,\mathrm{R}_4}+U_{\mathrm{R}_3}-U^e_{\mathrm{L}_1,\mathrm{R}_3}\\&-2U^e_{\mathrm{L}_1,\mathrm{R}_2}-2U^e_{\mathrm{R}_2,\mathrm{R}_3}-U^e_{\mathrm{R}_2,\mathrm{R}_4}-U^e_{\mathrm{R}_3,R4}-U^e_{\mathrm{L}_1,\mathrm{R}_4},\end{aligned}\tag{2.50}$$

$$\begin{aligned}\mathrm{diag}(|\mathrm{S}(\uparrow_{\mathrm{L}_1}\downarrow_{\mathrm{R}_5})\rangle) =&\varepsilon_{\mathrm{L}_1}+\varepsilon_{\mathrm{R}_5}+U_{\mathrm{R}_1}+2U_{\mathrm{R}_1,\mathrm{L}_1}+4U_{\mathrm{R}_1,\mathrm{R}_2}+U_{\mathrm{R}_1,\mathrm{R}_3}+2U_{\mathrm{R}_1,\mathrm{R}_5}\\&+2U_{\mathrm{L}_1,\mathrm{R}_2}+2U_{\mathrm{L}_1,\mathrm{R}_3}+U_{\mathrm{L}_1,\mathrm{R}_5}+U_{\mathrm{R}_2}+4U_{\mathrm{R}_2,\mathrm{R}_3}+2U_{\mathrm{R}_2,\mathrm{R}_5}\\&+U_{\mathrm{R}_3}+2U_{\mathrm{R}_3,\mathrm{R}_5}-U^e_{\mathrm{R}_1,\mathrm{L}_1}-2U^e_{\mathrm{R}_1,\mathrm{R}_2}-2U^e_{\mathrm{R}_1,\mathrm{R}_3}-U^e_{\mathrm{R}_1,\mathrm{R}_5}\\&-U^e_{\mathrm{L}_1,\mathrm{R}_2}-2U^e_{\mathrm{L}_1,\mathrm{R}_3}-2U^e_{\mathrm{R}_2,\mathrm{R}_3}-U^e_{\mathrm{R}_2,\mathrm{R}_5}-U^e_{\mathrm{R}_3,\mathrm{R}_5}\\&+U^e_{\mathrm{L}_1,\mathrm{R}_5},\end{aligned}\tag{2.51}$$

$$\begin{aligned}\mathrm{diag}(|\mathrm{T}(\uparrow_{\mathrm{L}_1}\downarrow_{\mathrm{R}_5})\rangle) =&\varepsilon_{\mathrm{L}_1}+\varepsilon_{\mathrm{R}_5}+U_{\mathrm{R}_1}+2U_{\mathrm{R}_1,\mathrm{L}_1}+4U_{\mathrm{R}_1,\mathrm{R}_2}+U_{\mathrm{R}_1,\mathrm{R}_3}+2U_{\mathrm{R}_1,\mathrm{R}_5}\\&+2U_{\mathrm{L}_1,\mathrm{R}_2}+2U_{\mathrm{L}_1,\mathrm{R}_3}+U_{\mathrm{L}_1,\mathrm{R}_5}+U_{\mathrm{R}_2}+4U_{\mathrm{R}_2,\mathrm{R}_3}+2U_{\mathrm{R}_2,\mathrm{R}_5}\\&+U_{\mathrm{R}_3}+2U_{\mathrm{R}_3,\mathrm{R}_5}-U^e_{\mathrm{R}_1,\mathrm{L}_1}-2U^e_{\mathrm{R}_1,\mathrm{R}_2}-2U^e_{\mathrm{R}_1,\mathrm{R}_3}-U^e_{\mathrm{R}_1,\mathrm{R}_5}\\&-U^e_{\mathrm{L}_1,\mathrm{R}_2}-2U^e_{\mathrm{L}_1,\mathrm{R}_3}-2U^e_{\mathrm{R}_2,\mathrm{R}_3}-U^e_{\mathrm{R}_2,\mathrm{R}_5}-U^e_{\mathrm{R}_3,\mathrm{R}_5}\\&-U^e_{\mathrm{L}_1,\mathrm{R}_5},\end{aligned}\tag{2.52}$$

$$\begin{aligned}\mathrm{diag}(|\mathrm{S}(\uparrow_{\mathrm{R}_4}\downarrow_{\mathrm{R}_5})\rangle) =&\varepsilon_{\mathrm{R}_4} + \varepsilon_{\mathrm{R}_5} + U_{\mathrm{R}_1} + 4U_{\mathrm{R}_1,\mathrm{R}_3} - 2U^e_{\mathrm{R}_1,\mathrm{R}_3} + 2U_{\mathrm{R}_1,\mathrm{R}_4} - U^e_{\mathrm{R}_1,\mathrm{R}_4}\\ &+ 2U_{\mathrm{R}_1,\mathrm{R}_5} - U^e_{\mathrm{R}_1,\mathrm{R}_5} - 2U^e_{\mathrm{R}_1,\mathrm{R}_2} + 4U_{\mathrm{R}_1,\mathrm{R}_2} + U_{\mathrm{R}_2} + 4U_{\mathrm{R}_2,\mathrm{R}_3}\\ &+ 2U_{\mathrm{R}_2,\mathrm{R}_4} - U^e_{\mathrm{R}_2,\mathrm{R}_4} + 2U_{\mathrm{R}_2,\mathrm{R}_5} - U^e_{\mathrm{R}_2,\mathrm{R}_5} - 2U^e_{\mathrm{R}_2,\mathrm{R}_3} + U_{\mathrm{R}_3}\\ &+ 2U_{\mathrm{R}_3,\mathrm{R}_4} - U^e_{\mathrm{R}_3,\mathrm{R}_4} + 2U_{\mathrm{R}_3,\mathrm{R}_5} - U^e_{\mathrm{R}_3,\mathrm{R}_5} + U_{\mathrm{R}_4,\mathrm{R}_5}\\ &+ U^e_{\mathrm{R}_4,\mathrm{R}_5},\end{aligned} \tag{2.53}$$

$$\begin{aligned}\mathrm{diag}(|\mathrm{T}(\uparrow_{\mathrm{R}_4}\downarrow_{\mathrm{R}_5})\rangle) =&\varepsilon_{\mathrm{R}_4} + \varepsilon_{\mathrm{R}_5} + U_{\mathrm{R}_1} + 4U_{\mathrm{R}_1,\mathrm{R}_3} - 2U^e_{\mathrm{R}_1,\mathrm{R}_3} + 2U_{\mathrm{R}_1,\mathrm{R}_4} - U^e_{\mathrm{R}_1,\mathrm{R}_4}\\ &+ 2U_{\mathrm{R}_1,\mathrm{R}_5} - U^e_{\mathrm{R}_1,\mathrm{R}_5} - 2U^e_{\mathrm{R}_1,\mathrm{R}_2} + 4U_{\mathrm{R}_1,\mathrm{R}_2} + U_{\mathrm{R}_2} + U_{\mathrm{R}_3}\\ &+ 4U_{\mathrm{R}_2,\mathrm{R}_3} + 2U_{\mathrm{R}_2,\mathrm{R}_4} - U^e_{\mathrm{R}_2,\mathrm{R}_4} + 2U_{\mathrm{R}_2,\mathrm{R}_5} - U^e_{\mathrm{R}_2,\mathrm{R}_5}\\ &- U^e_{\mathrm{R}_3,\mathrm{R}_4} + 2U_{\mathrm{R}_3,\mathrm{R}_4} - 2U^e_{\mathrm{R}_2,\mathrm{R}_3} + 2U_{\mathrm{R}_3,\mathrm{R}_5} - U^e_{\mathrm{R}_3,\mathrm{R}_5}\\ &+ U_{\mathrm{R}_4,\mathrm{R}_5} - U^e_{\mathrm{R}_4,\mathrm{R}_5},\end{aligned} \tag{2.54}$$

$$\begin{aligned}\mathrm{diag}(|\mathrm{S}(\uparrow_{\mathrm{R}_4}\downarrow_{\mathrm{R}_4})\rangle) =&\varepsilon_{\mathrm{R}_4} + U_{\mathrm{R}_1} + 4U_{\mathrm{R}_1,\mathrm{R}_3} - 2U^e_{\mathrm{R}_1,\mathrm{R}_3} + 4U_{\mathrm{R}_1,\mathrm{R}_4} - 2U^e_{\mathrm{R}_1,\mathrm{R}_4}\\ &- 2U^e_{\mathrm{R}_1,\mathrm{R}_2} + 4U_{\mathrm{R}_1,\mathrm{R}_2} + U_{\mathrm{R}_2} + 4U_{\mathrm{R}_2,\mathrm{R}_3} + 4U_{\mathrm{R}_2,\mathrm{R}_4}\\ &- 2U^e_{\mathrm{R}_2,\mathrm{R}_3} - 2U^e_{\mathrm{R}_2,\mathrm{R}_4} + U_{\mathrm{R}_3} + 4U_{\mathrm{R}_3,V} - 2U^e_{\mathrm{R}_3,\mathrm{R}_4} + U_{\mathrm{R}_4,\mathrm{R}_4},\end{aligned} \tag{2.55}$$

Figures 2.6 and 2.7 show the diagonal elements of double quantum dots occupied by four electrons and eight electrons as functions of detuning $\Delta$ at $B = 0.7$ T. The states $|\mathrm{S}(\uparrow_{\mathrm{R}_2}\downarrow_{\mathrm{R}_2})\rangle$ and $|\mathrm{S}(\uparrow_{\mathrm{R}_4}\downarrow_{\mathrm{R}_4})\rangle$ have the lowest energy in their respective systems. Therefore, we can write the effective Hubbard Hamiltonian in the basis of $\{|\mathrm{T}(1, 2N-1)\rangle, |\mathrm{S}(1, 2N-1)\rangle, |\mathrm{S}(0, 2N)\rangle\}$ and calculate the single qubit exchange energy using Eq. (2.4) in the region marked by the dashed green line square in Figures 2.6 and 2.7.

Since the unbiased case corresponds to the minimum hybridization between singlet states, we consider the region where $\beta/\sqrt{1+\beta^2} < 0.01$ as the unbiased case. For example, for $2N = 4$ ($2N = 8$), $\Delta \sim 3.3$meV ($\Delta \sim 6.6$meV) can be considered the unbiased case as it corresponds to $\beta/\sqrt{1+\beta^2} \approx 0.01$. Similarly, we can calculate the exchange energy in the basis $\{|\mathrm{T}(1, 2N-1)\rangle, |\mathrm{S}(1, 2N-1)\rangle, |\mathrm{S}(2, 2N-2)\rangle\}$, where $|\mathrm{S}(2, 2N-2)\rangle$ has the lowest energy in the system.

## 2.5.2 Expression of Charge Distribution Difference

In form of slater determinant that contains spatial wave functions and spin states, the singlet and triplet states of unbiased case can be written as

$$|\mathrm{S}\rangle = a(|\uparrow_{\mathrm{L}_1}\downarrow_{\mathrm{R}_N} \cdots \uparrow_{\mathrm{R}_1}\downarrow_{\mathrm{R}_1}\rangle + |\uparrow_{\mathrm{R}_N}\downarrow_{\mathrm{L}_1} \cdots \uparrow_{\mathrm{R}_1}\downarrow_{\mathrm{R}_1}\rangle), \tag{2.56}$$

and

$$|\mathrm{T}\rangle = b(|\uparrow_{\mathrm{L}_1}\downarrow_{\mathrm{R}_N} \cdots \uparrow_{\mathrm{R}_1}\downarrow_{\mathrm{R}_1}\rangle - |\uparrow_{\mathrm{R}_N}\downarrow_{\mathrm{L}_1} \cdots \uparrow_{\mathrm{R}_1}\downarrow_{\mathrm{R}_1}\rangle), \tag{2.57}$$

where $|\mathrm{S}\rangle$ and $|\mathrm{T}\rangle$ satisfy

$$\langle \mathrm{S}|\mathrm{S}\rangle = 1, \langle \mathrm{T}|\mathrm{T}\rangle = 1, \langle \mathrm{S}|\mathrm{T}\rangle = 0, \tag{2.58}$$

therefore we can obtain

$$a = \frac{1}{\sqrt{2(1+\mathcal{I}_{N,\mathrm{S}})}}, b = \frac{1}{\sqrt{2(1-\mathcal{I}_{N,\mathrm{T}})}}, \tag{2.59}$$

as indicated in Eq. (2.5) and Eq. (2.6). $\mathcal{I}_{N,\mathrm{S}}$ and $\mathcal{I}_{N,\mathrm{T}}$ are factors dependent on electron numbers to be calculated below.

We denote $I_i = \langle \mathrm{L}_1|\mathrm{R}_i\rangle = \langle \mathrm{R}_i|\mathrm{L}_1\rangle$, where $1 \leq i < N$. $\mathrm{L}_1$ and $\mathrm{R}_N$ are wave functions (without spin part) based on Fock-Darwin states.

For $N > 1$, by applying the Slater-Condon rules [250, 10], we have

$$\mathcal{I}_{N,\mathrm{S}} = I_N^2 - \sum_{i=1}^{N-1} I_i^2, \tag{2.60}$$

and

$$\mathcal{I}_{N,\mathrm{T}} = I_N^2 + \sum_{i=1}^{N-1} I_i^2. \tag{2.61}$$

For unbiased case, we can express $A_\phi$ as

$$\begin{aligned} A_\phi &= \frac{1}{2}[\langle \mathrm{T}(1,2N-1)|\rho(\mathbf{q})|\mathrm{T}(1,2N-1)\rangle \\ &\quad -\langle \mathrm{S}(1,2N-1)|\rho(\mathbf{q})|\mathrm{S}(1,2N-1)\rangle] \\ &= A_\phi(\mathbf{q}_{\|})f(q_z), \end{aligned} \tag{2.62}$$

here $A_\phi(\mathbf{q}_{\|})$ is obtained from the $x$ and $y$ components of orbital states. $f(q_z)$ is solely determined by the $z$-direction wave function, given by

$$f(q_z) = \frac{\sin(q_z a_z)}{q_z a_z} \frac{-\pi^2}{(q_z a_z)^2 - \pi^2}, \tag{2.63}$$

where $q_z$ is $z$-component lattice momentum, $a_z = 3 \times 10^{-9}$m is width of the infinite square well for acoustic phonons [115].

For $N = 1$, the expression of $A_\phi$ has been explicitly shown in [115]. Here, we give a general expression of $A_\phi$ for $N > 1$:

$$\langle \mathrm{S}(1, 2N-1)|\rho(\mathbf{q})|\mathrm{S}(1, 2N-1)\rangle = \frac{\varrho_+}{1 + \mathcal{J}_{N,\mathrm{S}}}, \tag{2.64}$$

$$\langle \mathrm{T}(1, 2N-1)|\rho(\mathbf{q})|\mathrm{T}(1, 2N-1)\rangle = \frac{\varrho_-}{1 - \mathcal{J}_{N,\mathrm{T}}}, \tag{2.65}$$

where

$$\begin{aligned}
\varrho_\pm =& \rho_{\mathrm{L}_1,\mathrm{L}_1} + \rho_{\mathrm{R}_N,\mathrm{R}_N} + 2\sum_{\mathrm{i}=1}^{N-1} \rho_{\mathrm{R_i},\mathrm{R_i}} \pm I_N(\rho_{\mathrm{L}_1,\mathrm{R}_N} + \rho_{\mathrm{R}_N,\mathrm{L}_1}) \\
& - \sum_{\mathrm{i}=1}^{N-1} \left[ I_i(\rho_{\mathrm{L}_1,\mathrm{R_i}} + \rho_{\mathrm{R_i},\mathrm{L}_1}) + I_{\mathrm{i}}^2(\rho_{\mathrm{R_i},\mathrm{R_i}} + \rho_{\mathrm{R}_N,\mathrm{R}_N}) \right] \\
& \mp \sum_{\mathrm{i}=1}^{N-1} (-2I_N^2 \rho_{\mathrm{R_i},\mathrm{R_i}} + I_{\mathrm{i}} I_N \rho_{\mathrm{R_i},\mathrm{R}_N} + I_N I_{\mathrm{i}} \rho_{\mathrm{R}_N,\mathrm{R_i}}) \\
& + \sum_{\mathrm{j}=1}^{N-1} \sum_{\mathrm{i}=1,\mathrm{i}\neq\mathrm{j}}^{N-1} (I_{\mathrm{i}} I_{\mathrm{j}} \rho_{\mathrm{R_i},\mathrm{R_j}} - 2I_{\mathrm{j}}^2 \rho_{\mathrm{R_i},\mathrm{R_i}}).
\end{aligned} \tag{2.66}$$

Here, $I_i = \langle \mathrm{L}_1|\mathrm{R_i}\rangle$, $\rho_{\mathrm{R}_i,\mathrm{R}_j} = \langle \mathrm{R_i}|\rho|\mathrm{R_j}\rangle$, and similarily, $\rho_{\mathrm{L}_1,\mathrm{R_i}} = \langle \mathrm{L}_1|\rho|\mathrm{R_i}\rangle$. We then have

$$A_\phi = \frac{2I_N \Box_1 (1 - \sum_{i=1}^{N-1} I_i^2) + 2I_N^2 \Box_2}{1 - I_N^4 - (2 - \sum_{i=1}^{N-1} I_i^2) \sum_{i=1}^{N-1} I_i^2}, \tag{2.67}$$

$$\Box_1 = -(\rho_{\mathrm{L}_1,\mathrm{R}_N} + \rho_{\mathrm{R}_N,\mathrm{L}_1}) + \sum_{\mathrm{i}=1}^{N-1} I_{\mathrm{i}}(\rho_{\mathrm{R}_N,\mathrm{R}_i} + \rho_{\mathrm{R_i},\mathrm{R}_N}), \tag{2.68}$$

and

$$
\begin{aligned}
\mathbb{I}_2 =& \rho_{\mathrm{L}_1,\mathrm{L}_1} + \rho_{\mathrm{R}_N,\mathrm{R}_N} + 2\sum_{i=1}^{N-1} I_i^2 \sum_{\mathrm{i}=1}^{N-1} \rho_{\mathrm{R_i},\mathrm{R_i}} \\
& - \sum_{\mathrm{i}=1}^{N-1} \left[ I_i(\rho_{\mathrm{L}_1,\mathrm{R}_i} + \rho_{\mathrm{R_i},\mathrm{L}_1}) + I_{\mathrm{i}}^2(\rho_{\mathrm{R_i},\mathrm{R_i}} + \rho_{\mathrm{R}_N,\mathrm{R}_N}) \right] \\
& + \sum_{\mathrm{j}=1}^{N-1} \sum_{\mathrm{i}=1,\mathrm{i}\neq \mathrm{j}}^{N-1} (I_{\mathrm{i}} I_{\mathrm{j}} \rho_{\mathrm{R_i},\mathrm{R_j}} - 2I_{\mathrm{j}}^2 \rho_{\mathrm{R_i},\mathrm{R_i}}).
\end{aligned}
\tag{2.69}
$$

Here, it is straightforward to show that, for i $<$ j, we have $I_{\mathrm{i}} \ll 1$, $I_{\mathrm{i}} \ll I_{\mathrm{j}}$, $\rho_{\mathrm{R_i},\mathrm{R_i}} < \rho_{\mathrm{R_j},\mathrm{R_j}}$, $\rho_{\mathrm{L}_1,\mathrm{R_i}} < \rho_{\mathrm{L}_1,\mathrm{R_j}}$ . For (1,1), (1,3) and (1,7), we have $N = 1, 2, 4$, therefore

$$
I_1 = 2e^{-2x_0^2/(l_L^2+l_R^2)} l_L l_R/(l_L^2 + l_R^2), \tag{2.70}
$$

$$
I_2 = 4x_0 e^{-2x_0^2/(l_L^2+l_R^2)} l_L l_R^2/(l_L^2 + l_R^2)^2, \tag{2.71}
$$

$$
I_4 = 4\sqrt{2} x_0^2 e^{-2x_0^2/(l_L^2+l_R^2)} l_L l_R^3/(l_L^2 + l_R^2)^3, \tag{2.72}
$$

where $l_L$ is left dot confinement length and $l_R$ is right dot cofinement length that can be calculated from their confinement strength. Therefore in Eq. (2.67) numerator, $I_N^2 \mathbb{I}_2 \ll I_N \mathbb{I}_1$. As $N$ increases, $\mathbb{I}_1$ and $I_N$ also increases, eventually lead to increases of $|A_\phi|^2$ and the dephasing rate. One can also find that due to $A_\phi \sim I_N$, therefore $|A_\phi|^2$ decreases as $x_0$ and $\hbar\omega_{\mathrm{R}}$ increase.

In biased case, the explicit expression of $A_\phi$ at $N = 1$ can also be found in [115]. For $N > 1$, there are two situations of biased case in our consideration. From Eq. (2.9), we have

$$
\begin{aligned}
A_\phi =& \frac{1}{2}[\langle \mathrm{T}(1, 2N-1)|\rho(\mathbf{q})|\mathrm{T}(1, 2N-1)\rangle - \langle \mathrm{S}_{\mathrm{mix}}^{(0,2N)}|\rho(\mathbf{q})|\mathrm{S}_{\mathrm{mix}}^{(0,2N)}\rangle] \\
=& \frac{1}{2}[\langle \mathrm{T}(1, 2N-1)|\rho(\mathbf{q})|\mathrm{T}(1, 2N-1)\rangle \\
& - [\langle \mathrm{S}(1, 2N-1)|\rho(\mathbf{q})|\mathrm{S}(1, 2N-1)\rangle \\
& + 2\beta\langle \mathrm{S}(1, 2N-1)|\rho(\mathbf{q})|\mathbf{S}(0, 2N)\rangle \\
& + \beta^2\langle \mathrm{S}(0, 2N)|\rho(\mathbf{q})|\mathrm{S}(0, 2N)\rangle]/(1+\beta^2)],
\end{aligned}
\tag{2.73}
$$

and from Eq. (2.10), we have

$$\begin{aligned}A_\phi =&\frac{1}{2}[\langle \mathrm{T}(1,2N-1)|\rho(\mathbf{q})|\mathrm{T}(1,2N-1)\rangle - \langle \mathrm{S}^{(2,2N-2)}_{\mathrm{mix}}|\rho(\mathbf{q})|\mathrm{S}^{(2,2N-2)}_{\mathrm{mix}}\rangle] \\ =&\frac{1}{2}[\langle \mathrm{T}(1,2N-1)|\rho(\mathbf{q})|\mathrm{T}(1,2N-1)\rangle \\ &-[\langle \mathrm{S}(1,2N-1)|\rho(\mathbf{q})|\mathrm{S}(1,2N-1)\rangle \\ &+2\beta\langle \mathrm{S}(1,2N-1)|\rho(\mathbf{q})|\mathbf{S}(2,2N-2)\rangle \\ &+\beta^2\langle \mathrm{S}(2,2N-2)|\rho(\mathbf{q})|\mathrm{S}(2,2N-2)\rangle]/(1+\beta^2)],\end{aligned} \tag{2.74}$$

where

$$\langle \mathrm{S}(0,2N)|\rho(\mathbf{q})|\mathrm{S}(0,2N)\rangle] = 2\sum_{i=1}^{N}\rho_{\mathrm{R_i},\mathrm{R_i}}, \tag{2.75}$$

$$\begin{aligned}\langle \mathrm{S}(0,2N)|\rho(\mathbf{q})|\mathrm{S}(1,2N-1)\rangle =&\frac{1}{\sqrt{2(1+\mathcal{I}_{N,\mathrm{S}})}}(4I_N\sum_{\mathrm{i}=1}^{N-1}\rho_{\mathrm{R_i},\mathrm{R_i}} - 2\sum_{\mathrm{i}=1}^{N-1}I_{\mathrm{i}}\rho_{\mathrm{R_i},\mathrm{R_i}} \\ &+2I_N\rho_{\mathrm{R}_N,\mathrm{R}_N}),\end{aligned} \tag{2.76}$$

$$\begin{aligned}\langle \mathrm{S}(2,2N-2)|\rho(\mathbf{q})|\mathrm{S}(2,2N-2)\rangle =&2\rho_{\mathrm{L_1},\mathrm{L_1}}(1-\sum_{\mathrm{i}=1}^{N-1}I_{\mathrm{i}}^2) \\ &+2\sum_{\mathrm{i}=1}^{N-1}[\rho_{\mathrm{R_i},\mathrm{R_i}}(1-\sum_{\mathrm{i}=1}^{N-1}I_{\mathrm{i}}^2) \\ &-I_{\mathrm{i}}(\rho_{\mathrm{L_1},\mathrm{R_i}}+\rho_{\mathrm{R_i},\mathrm{L_1}})] \\ &+\sum_{\mathrm{i,j}\neq\mathrm{i}}^{N-1}(2I_{\mathrm{i}}I_{\mathrm{j}}\rho_{\mathrm{R_i},\mathrm{R_j}} - 4I_{\mathrm{i}}^2\rho_{\mathrm{R_i},\mathrm{R_j}} \\ &+O(I_{\mathrm{i}}^m\rho_{\mathrm{R_i},\mathrm{R_j}})),\end{aligned} \tag{2.77}$$

$$
\begin{aligned}
\langle \mathrm{S}(2,2N-2)|\rho(\mathbf{q})|\mathrm{S}(1,2N-1)\rangle = & \frac{1}{\sqrt{2(1+\mathcal{J}_{N,\mathrm{S}})}}[4\rho_{\mathrm{L}_1,\mathrm{L}_1}I_N + 2\rho_{\mathrm{R}_N,\mathrm{L}_1} \\
& + 2\rho_{\mathrm{L}_1,\mathrm{R}_N} + 2\sum_{\mathrm{i}=1}^{N-1}(4I_N\rho_{\mathrm{R}_\mathrm{i},\mathrm{R}_\mathrm{i}} \\
& - I_\mathrm{i}(\rho_{\mathrm{R}_\mathrm{i},\mathrm{R}_N} + \rho_{\mathrm{R}_N,\mathrm{R}_\mathrm{i}}) \\
& - 2I_\mathrm{i}I_N(\rho_{\mathrm{R}_\mathrm{i},\mathrm{L}_1} + \rho_{\mathrm{L}_1,\mathrm{R}_\mathrm{i}}) \\
& - I_\mathrm{i}^2(\rho_{\mathrm{R}_N,\mathrm{L}_1} + \rho_{\mathrm{L}_1,\mathrm{R}_N}))].
\end{aligned} \tag{2.78}
$$

Here, $O(I_\mathrm{i}^m\rho_{\mathrm{R}_\mathrm{i},\mathrm{R}_\mathrm{j}})$, $m > 2$ are higher-order terms that can be ignored due to the fact $I_\mathrm{i} \ll 1$.

### 2.5.3 Merit Figures for Various Quantum Dot Parameters

In Fig. 2.8, we show the merit figures calculated for three sets of parameters. Fig. 2.8(a) depicts the case with a short half-dot distance $x_0 = 50$ nm and a relatively strong confinement strength in the right dot $\hbar\omega_\mathrm{R} = 1.419$ meV. Fig. 2.8(b) shows the case with an intermediate half-dot distance $x_0 = 70$ nm and a relatively weak confinement strength $\hbar\omega_\mathrm{R} = 0.946$ meV. In these cases, the barrier between the two dots is low, making the $(1,7)$ electron occupancy ill-defined; therefore, only results for $(1,3)$ and $(1,1)$ are shown. We observe that the merit figure associated with $(1,3)$ is clearly higher than that with $(1,1)$, consistent with the findings in the main text.

In Fig. 2.8(c), results for all three cases $(1,7)$, $(1,3)$, and $(1,1)$ are shown. Again, these results support the main finding that, within a certain $J$ range, the merit figure for $(1,7)$ is the highest, for $(1,3)$ is intermediate, and for $(1,1)$ is the lowest.

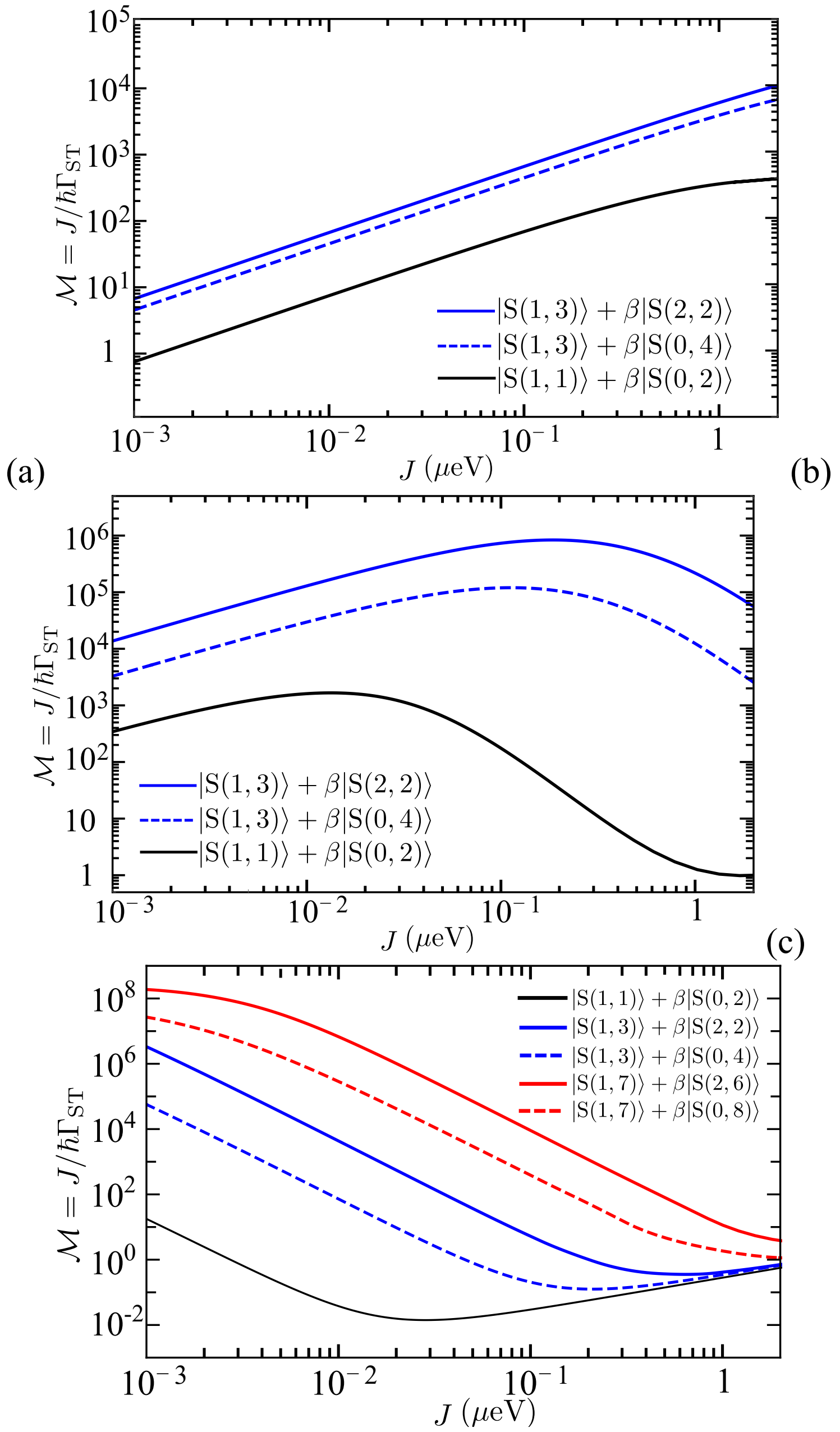

Fig. 2.8 The merit figures vs. the exchange interaction calculated for three different sets of parameters: (a) $x_0 = 50$ nm, $\hbar\omega_{\rm L} = 2.838$ meV, $\hbar\omega_{\rm R} = 1.419$ meV; (b) $x_0 = 70$ nm, $\hbar\omega_{\rm L} = 2.838$ meV, $\hbar\omega_{\rm R} = 0.946$ meV; (c) $x_0 = 80$ nm, $\hbar\omega_{\rm L} = 2.838$ meV, $\hbar\omega_{\rm R} = 1.419$ meV.

# Chapter 3

# Exploring entanglement spectrum and phase diagram in multi-electron quantum dot chains

## 3.1 Overview

We investigate the entanglement properties in semiconductor quantum dot systems modeled by extended Hubbard model, focusing on the impact of potential energy variations and electron interactions within a four-site quantum dot spin chain. Our study explores local and pairwise entanglement across configurations with electron counts $N = 4$ and $N = 6$, under different potential energy settings. By adjusting the potential energy in specific dots and examining the entanglement across various interaction regimes, we identify significant variations in the ground states of quantum dots. Our results reveal that local potential modifications lead to notable redistributions of electron configurations, significantly affecting the entanglement properties. These changes are depicted in phase diagrams that show entanglement dependencies on interaction strengths and potential energy adjustments, highlighting complex entanglement dynamics and phase transitions triggered by inter-dot interactions.

## 3.2 Background

Quantum entanglement plays a crucial role in various fields of quantum physics, including quantum communication and quantum information processing [22, 2]. In condensed matter physics, especially in many-body quantum systems, quantum entan-glement serves as a fundamental criterion for quantum phase transitions and many-body localization [4, 231, 127, 196]. Among various systems, semiconductor quantum dots have emerged as scalable, implementable, and precisely controllable [73, 91, 207, 314, 214, 76, 75, 234] platforms for simulating many-body systems of interest, in particular the Fermi-Hubbard physics [109, 268, 27, 54, 140, 280, 161]. The Fermi-Hubbard model provides a common framework for describing quantum dot systems in the regime of low temperatures and strong Coulomb interactions, finding extensive applications in the physical realization of quantum information processing [283, 51, 291, 284]. Consequently, a comprehensive understanding of quantum dots from the perspective of Fermi-Hubbard physics becomes imperative.

High-fidelity qubit gate operations [297, 35] and noise suppression schemes [175] commonly applied to conventional quantum dot systems, where each dot accommodates at most two electrons, traditionally rely on the monotonically increasing behavior of exchange energy as a function of detuning [167, 25, 12, 177, 181]. However, recent investigations [176, 173, 10, 242, 163, 209, 142, 112, 57, 38, 116, 72] have revealed the interesting properties of specific quantum dots capable of hosting more than two electrons, such as non-monotonic behavior of exchange energy with distinct sweet spots [38, 39], fast spin exchange dynamics [174], superexchange interactions between non-neighboring dots [56, 40, 211], and resilience to noise [8, 180, 39, 105]. These properties can be attributed to the influence of higher excited orbitals and can be effectively understood within the framework of the Full Configuration Interaction [220] and the extended Hubbard Model (EHM), which incorporates multiple energy levels.

The entanglement spectrum of the one-dimensional EHM in its ground state has been well-understood [79, 124, 6, 96]. Consequently, in the case of a half-filled system, the entanglement properties of a quantum dot spin chain can be effectively explained [1]. However, when there is a tilted potential energy difference among the dots, the mirror symmetry of the system is broken, which leads to the tunable entanglement values through the application of precise electron control using external electric fields [206]. These previous works have motivated us to investigate the entanglement spectrum of a quantum dot spin chain where each dot incorporates multiple energy levels. This exploration holds great potential for uncovering the rich physical properties of quantum dot systems.

In this study, we investigate the entanglement patterns of the ground states of multi-electron quantum dot systems using the EHM, which incorporates multiple orbitals within each dot. Our specific focus lies in characterizing the entanglement properties of one-site and two-site reduced density matrices. By computing and analyzing the entanglement spectrum for various system sizes, we uncover notable findings. Firstly, when there are no potential energy differences among the dots, the multi-electron quantum dot system can be accurately described by the EHM, either in a half-filled state or a non-half-filled state, depending on the total electron number. However, when a selected dot within the chain exhibits a potential energy difference relative to its neighboring dots, distinct system phases and phase boundaries emerge in the entanglement spectrum. These phases depend on the coupling strengths and potential energy difference values. The emergence of these phases indicates that the presence of a selected dot with a potential energy ladder profoundly impacts the electron configuration in its vicinity. This influence is more pronounced in small systems while limited in larger-size systems, due to the size effect.

This chapter is organized as follows. In Section 3.3, we present the EHM as a suitable framework for describing multi-electron quantum dot chain systems. Section 3.4 introduces the definition of one-site and two-site reduced density matrices and entanglement entropy for these systems. Our main results are presented in Section 3.5, starting with an examination of a system size of $L = 4$ and electron numbers $N = 4$ and $N = 6$. We analyze the entanglement spectrum properties with and without potential energy differences. Furthermore, we extend our analysis to larger system sizes as $L$ approaches infinity. Finally, we summarize our findings and provide concluding remarks in Section 3.6.

## 3.3 Extended Hubbard Model

We consider a Multiple-Quantum-Dot system (MQD) (schematically shown in Fig. 3.1), described by an EHM with short-range Coulomb interactions and tunneling restricted to nearest-neighbor sites within the same energy level and the nearest-neighbor energy level. The model can be described by the following Hamiltonian:

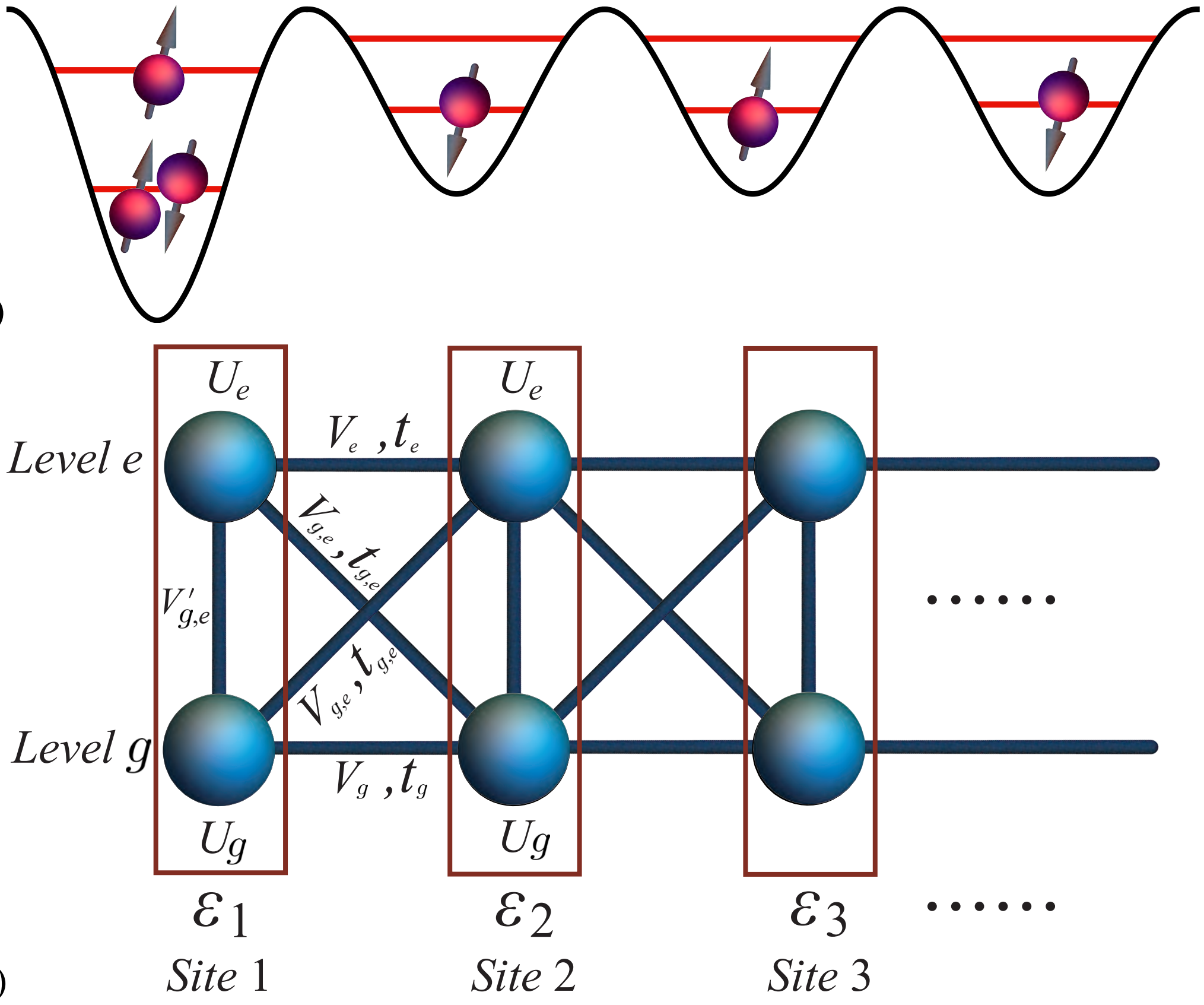

Fig. 3.1 (a) Schematic illustration of a $L = 4$ multi-electron quantum dot spin chain system hosting $N = 6$ electrons. (b) In two-level case, the equivalent asymmetric Hubbard ladder described by Hamiltonian (3.1). The box indicates for each site $i$, electrons on different energy levels have the same detuning energy $\varepsilon_i$. $g$ indicates the ground level and $e$ the first excited level.

$$\begin{aligned} H =& - \sum_{i,\nu,\overline{\nu},\sigma} (t_\nu c^\dagger_{i,\nu,\sigma} c_{i+1,\nu,\sigma} + t_{\nu,\overline{\nu}} c^\dagger_{i,\nu,\sigma} c_{i+1,\overline{\nu},\sigma} + \text{H.c.}) \\ &+ \sum_{i,\nu,\overline{\nu},\sigma} (V_\nu n_{i,\nu,\sigma} n_{i+1,\nu,\sigma'} + V_{\nu,\overline{\nu}} n_{i,\nu,\sigma} n_{i+1,\overline{\nu},\sigma'} \\ &+ V'_{\nu,\overline{\nu}} n_{i,\nu,\sigma} n_{i,\overline{\nu},\sigma'}) + \sum_{i,\nu} U_\nu n_{i,\nu\downarrow} n_{i,\nu\uparrow} \\ &+ \sum_{i,\sigma} \varepsilon_{i,\sigma} n_{i\sigma}, \end{aligned} \tag{3.1}$$

where $i$ indicates the quantum dot site, $\nu$ and $\overline{\nu}$ denotes different orbital level, which can be either ground orbital ($g$) or excited orbital ($e$), while $\sigma$ and $\sigma'$ refer to the spins that can be either up or down. $\varepsilon_{i,\sigma}$ is the potential energy at dot $i$, note that although in one quantum dot, electrons can occupy different orbitals, they share the same potential energy. $t_\nu$ is the tunneling energy between $i$th and $(i+1)$th site at $\nu$th orbital level, $t_{\nu,\overline{\nu}}$ is the tunneling energy between $i$th site at $\nu$th orbital level and $(i+1)$th site at $\overline{\nu}$th orbital level, i.e. $t_{g,e}$ or $t_{e,g}$. $U_\nu$ denotes the on-site Coulomb interaction in the $\nu$th orbital level, $V_\nu$ is the nearest direct Coulomb interaction between the $i$th and $(i+1)$th site at $\nu$th orbital, $V_{\nu,\overline{\nu}}$ is the nearest direct Coulomb interaction between the $i$th site at $\nu$th orbital and $(i+1)$th site at $\overline{\nu}$th orbital, and finally, $V'_{\nu,\overline{\nu}}$ is the nearest direct Coulomb interaction between the $\nu$th orbital and $\overline{\nu}$th orbital at $i$th site, i.e. $V_{g,e}$, $V_{e,g}$, $V'_{g,e}$, $V'_{e,g}$.

According to the Pauli exclusion principle, electrons have four occupation states $|v\rangle_{i,\nu}$=$|0\rangle_{i,\nu}$, $|\uparrow\rangle_{i,\nu}$, $|\downarrow\rangle_{i,\nu}$, $|\uparrow\downarrow\rangle_{i,\nu}$ in the $\nu$th orbital of the $i$th site. Thus, the dimension of the Hilbert space for an $L$-site MQD chain with $K$ orbitals per site is $4^{LK}$. The configuration basis states are $|v_1, v_2, ..., v_L\rangle = \prod_{i=1}^{L} |v_i\rangle_i$, where $|v_i\rangle_i = \prod_{\nu=1}^{K} |v\rangle_{i,\nu}$ represents the configuration basis for the i-th site. In this work, we numerically study MQD chains with $N$ and $N+2$ electrons in an $L = N$ sites systems, restricting our analysis to the ground and first excited orbital states ($\nu = g, e$) for each quantum dot.

## 3.4 Reduced Density Matrices and Entanglement

We first obtain the ground state (GS) $|\psi_{\text{GS}}\rangle$ of the system by diagonalizing the Hamiltonian. The GS can be expressed as a linear superposition of all possible electron

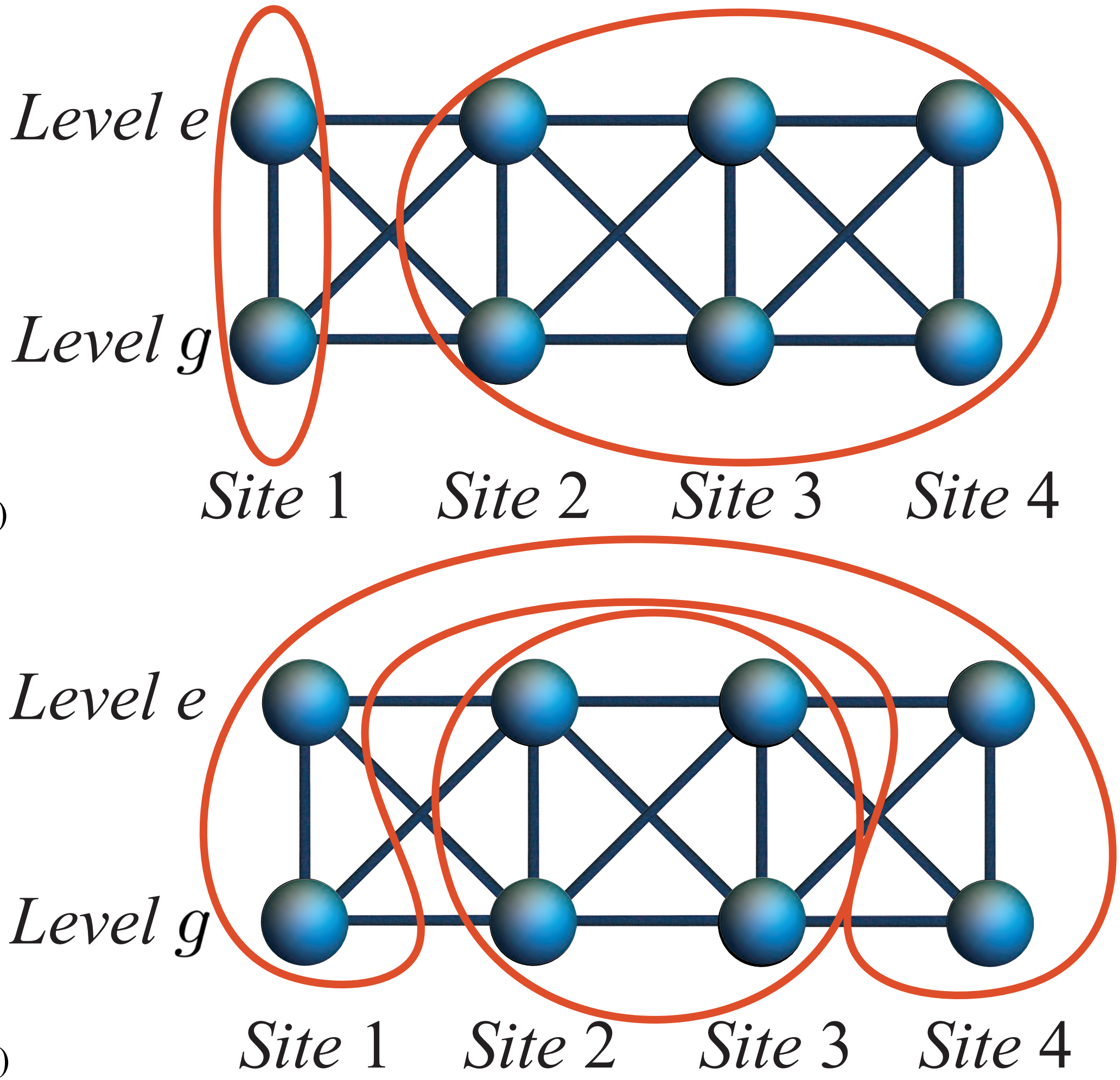

(a)

(b)

Fig. 3.2 Illustration of bipartite entanglement and quantum states in $L = 4$ two-level system (a) Local entanglement $E(\rho_1)$ and (b) Pairwise entanglement $E(\rho_{14}) = E(\rho_{23})$. The red circle indicates the selected partition. $g$ indicates the ground level and $e$ the first excited level.

configuration basis states $|\psi_m\rangle$ in the occupation number representation $|v_1, v_2...v_L\rangle$:

$$|\psi_{\mathrm{GS}}\rangle = \sum_m c_m|\psi_m\rangle, \tag{3.2}$$

where $c_m$ are the coefficients of the superposition.

The density matrix $\rho_{\mathrm{GS}}$ of the entire system can be expressed as a sum of the occupation probabilities $P_m$ of all electron configurations $|\psi_m\rangle$:

$$\rho_{\mathrm{GS}} = \sum_m P_m|\psi_m\rangle\langle\psi_m|. \tag{3.3}$$

To analyze entanglement, we divide the full system into subsystems A and B. The reduced density matrix $\rho_A$ for subsystem A is obtained by taking the partial trace of $\rho_{\mathrm{GS}}$:

$$\rho_A = \mathrm{Tr}_B\rho_{\mathrm{GS}}. \tag{3.4}$$

The von Neumann entropy $E(\rho_A)$ measures the entanglement between subsystem A and the remaining subsystem B, and is defined as:

$$E(\rho_A) = -\mathrm{Tr}(\rho_A \log_2 \rho_A). \tag{3.5}$$

### 3.4.1 Local Entanglement of Multi-electron Quantum Dot

In this paper, we focus on GaAs QD, since silicon QD is more complicated due to the valley splitting and phases [254]. In GaAs QD [10], in the two orbitals and within the parameters we considered, electrons prefer to doubly occupy ground states before filling the first excited states. Therefore the state space of a single site is spanned by nine bases: $\{|0,0\rangle, |\uparrow_g,0\rangle, |\downarrow_g,0\rangle, |\uparrow_g\downarrow_g,0\rangle, |\uparrow_g,\downarrow_e\rangle, |\downarrow_g,\uparrow_e\rangle, |\uparrow_g\downarrow_g,\uparrow_e\rangle, |\uparrow_g\downarrow_g,\downarrow_e\rangle, |\uparrow_g\downarrow_g,\uparrow_e\downarrow_e\rangle\}$. $0_g$ and $0_e$ represent cases with no electron occupying the ground and the first excited orbital, respectively. $\uparrow_g$, $\downarrow_g$, $\uparrow_e$, $\downarrow_e$ stand for an electron with spin up or down indicated as the arrow staying in the ground $(g)$ and the first excited orbital $(e)$ indicated in the subscript respectively.

The two-level one-site reduced density matrix for site $i$ can be written as

$$\rho_i = \mathrm{Tr}_i(\rho_{\mathrm{GS}}). \tag{3.6}$$

Expressing in terms of basis: $\{|0,0\rangle, |\uparrow_g,0\rangle, |\downarrow_g,0\rangle, |\uparrow_g\downarrow_g,0\rangle, |\uparrow_g,\downarrow_e\rangle, |\downarrow_g,\uparrow_e\rangle, |\uparrow_g\downarrow_g,\uparrow_e\rangle, |\uparrow_g\downarrow_g,\downarrow_e\rangle, |\uparrow_g\downarrow_g,\uparrow_e\downarrow_e\rangle\}$, $\rho_i$ can be written as a $9 \times 9$ matrix as follows:

$$\rho_i = \begin{pmatrix} v_{i,1} & & & & & & & & \\ & v_{i,2} & & & & & & & \\ & & v_{i,3} & & & & & & \\ & & & v_{i,4} & v_{i,a} & v_{i,b} & & & \\ & & & v_{i,a} & v_{i,5} & v_{i,c} & & & \\ & & & v_{i,b} & v_{i,c} & v_{i,6} & & & \\ & & & & & & v_{i,7} & & \\ & & & & & & & v_{i,8} & \\ & & & & & & & & v_{i,9} \end{pmatrix}.$$

Here, $v_{i,m}(m = 1, 2, ..., 9)$, $v_{i,a}$, $v_{i,b}$ and $v_{i,c}$ are determined by potential energy $\varepsilon$ of different dots and quantity $U$. In half-filled case, when there is no potential energy difference of all quantum dots, the local reduced density matrix $\rho_i$ can be simplified to one energy level case [96], with

$$v_{i,1} = 1 - v_{i,4} + v_{i,2} + v_{i,3}, \tag{3.7a}$$
$$v_{i,2} = \langle n_{i,g,\uparrow} \rangle - v_{i,4}, \tag{3.7b}$$
$$v_{i,3} = \langle n_{i,g,\downarrow} \rangle - v_{i,4}, \tag{3.7c}$$
$$v_{i,4} = \mathrm{Tr}(n_{i,g,\uparrow} n_{i,g,\downarrow} \rho_i) = \langle n_{g\uparrow} n_{g\downarrow} \rangle , \tag{3.7d}$$
$$v_{i,a} = v_{i,b} = v_{i,c} = 0, \tag{3.7e}$$
$$v_{i,5} = v_{i,6} = v_{i,7} = v_{i,8} = v_{i,9} = 0. \tag{3.7f}$$

When potential energy differences exist between quantum dots (in particular, in our work, only one site's potential energy is altered while the remaining sites have no potential energy difference), the contributions of $v_{i,5}$, $v_{i,6}$, $v_{i,7}$, $v_{i,8}$, $v_{i,9}$, $v_{i,a}$, $v_{i,b}$ and $v_{i,c}$ cannot be ignored. Therefore, the above expression of $v_{i,m}$ does not hold. However, we can still derive that $v_{i,2} = v_{i,3}$, $v_{i,5} = v_{i,6}$, and $v_{i,7} = v_{i,8}$. In particular, for GaAs, within the parameters we set (which will be explained in detail later), the basis of $|\uparrow_g, \downarrow_e\rangle$ and $|\downarrow_g, \uparrow_e\rangle$ are energetically unfavorable and therefore have no contribution, leading to $v_{i,5}, v_{i,6}, v_{i,a}, v_{i,b}, v_{i,c} \sim 0$ at any potential $V_i$. Thus, $\rho_i$ can be represented as a 7 × 7 diagonal matrix as:

$$\begin{aligned} \rho_i =& v_{i,1}|0_g, 0_e\rangle\langle 0_g, 0_e| + v_{i,2}|\uparrow_g, 0_e\rangle\langle\uparrow_g, 0_e| \\ &+ v_{i,3}|\downarrow_g, 0_e\rangle\langle\downarrow_g, 0_e| + v_{i,4}|\uparrow_g\downarrow_g, 0_e\rangle\langle\uparrow_g\downarrow_g, 0_e| \\ &+ v_{i,7}|\uparrow_g\downarrow_g, \uparrow_e\rangle\langle\uparrow_g\downarrow_g, \uparrow_e| + v_{i,8}|\uparrow_g\downarrow_g, \downarrow_e\rangle\langle\uparrow_g\downarrow_g, \downarrow_e| \\ &+ v_{i,9}|\uparrow_g\downarrow_g, \uparrow_e\downarrow_e\rangle\langle\uparrow_g\downarrow_g, \uparrow_e\downarrow_e|. \end{aligned} \tag{3.8}$$

For the $N = 4$ system, there are four distinct approaches to analyzing local bipartite entanglement: $E(\rho_1)$, $E(\rho_2)$, $E(\rho_3)$ and $E(\rho_4)$. An example of this can be seen in Fig. 3.2(a), which shows the local entanglement $E(\rho_1)$.

### 3.4.2 Pairwise Entanglement of Multi-electron Quantum Dot

Similarly, for site $i$ and site $j$, the two-site reduced density matrix can be written as

$$\rho_{ij} = \mathrm{Tr}_{ij}(\rho_{\mathrm{GS}}). \tag{3.9}$$

As depicted in Fig. 3.2(b). According to the nine bases considered for a single site in Sec. 3.4.1, the electrons in two sites with two orbitals have $9^2 = 81$ possible configurations. With respect to these bases, $\rho_{ij}$ can be described as an $81 \times 81$ matrix. Similar to one site case, where we dropped two energetically unfavorable bases $|\uparrow_g, \downarrow_e\rangle$ and $|\downarrow_g, \uparrow_e\rangle$, $\rho_{ij}$ can be described as a $49 \times 49$ matrix since electrons in two sites have $7^2 = 49$ occupation probabilities. There are three possible approaches to analyzing pairwise bipartite entanglement for the $N = 4$ system, : $E(\rho_{12})$ and $E(\rho_{34})$, $E(\rho_{13})$ and $E(\rho_{24})$, $E(\rho_{14})$ and $E(\rho_{23})$. Fig. 3.2(b) demonstrates one possible bipartite pairwise entanglement $E(\rho_{14})$ and $E(\rho_{23})$.

## 3.5 Results

In our GaAs quantum dots system setup, we have defined a set of parameters that can represent the properties of multi-electron dots [57, 38]. Accordingly, we set that the tunneling energy between the nearest sites is larger for lower orbitals, and is smaller for higher orbitals. This means that the tunneling between two ground orbitals is the greatest, followed by the tunneling between one ground orbital and one excited orbital, and finally, the tunneling between two excited orbitals, i.e., $t_e < t_{g,e} < t_g$. Similarly, within one single dot or between two nearest dots, the on-site Coulomb interaction energy and the nearest direct Coulomb interaction energy from higher orbitals are larger than those from lower energy levels, since the electron that occupies a higher orbital requires more energy, i.e., $U_g < V'_{g,e} < U_e$ and $V_g < V_{g,e} < V_e$. The numerical relation between $V_\nu$ and $U_\nu$ is referenced from [38, 226, 76, 191, 51], satisfying a strong repulsive on-site interaction regime in EHM [96], i.e., $V_\nu < U_\nu$ and $V_{g,e} < V'_{g,e}$. In one-dimensional EHM at half filling, the ratio between on-site Coulomb interaction $U_\nu$ and the nearest direct Coulomb interaction energy $V_\nu$ will lead to charge-density wave (CDW) order and spin-density wave (SDW) in the strong-coupling limit regime [96]. Specifically, for $U_g > 2V_g$, the ground state is a staggered charge-density-wave, and for $U_g < 2V_g$, the ground state is a staggered spin density wave. These spin order properties will also be apparent in our simulation results due to the chosen parameters, therefore our discussion will be split into two parts: $U_g > 2V_g$ and $U_g < 2V_g$. In this study, we have set our parameters as follows: $V_g = \alpha U_g$, $V_{g,e} = \alpha V'_{g,e}$, $Ve = \alpha U_e$, $V'_{g,e} = 1.5U_g$, $U_e = 2U_g$, $t_e = 0.3t_g$, $t_{g,e} = 0.6t_g$. Here according to the literature, the coupling strength ratio of $\alpha$ can be either set as 0.2 [226] or 0.7 [76, 191, 51], and $U = U_g/t_g$ is the main quantity parameter in the results.

### 3.5.1 Local Entanglement at Zero Potential Energy

Starting with an analysis of the local entanglement in the smallest system size ($L = 4$) for both electron number scenarios ($N = 4$ and $N = 6$), we first consider the case of $N = 4$ with $\alpha = 0.2$, as depicted in Fig. 3.3(a). It is apparent that for $N = 4$, the local entanglement at the end sites ($E(\rho_1) = E(\rho_L)$) is both equal and less than the local entanglement of the inner sites. This phenomenon arises from the preference of the end sites for single occupancy over the middle sites, particularly as the repulsive interaction increases [1]. With the increase in the repulsive interaction $U$ in the four-dot-four-electron system, specific configurations, such as $|\uparrow_g, \downarrow_g, \uparrow_g, \downarrow_g\rangle$, $|\downarrow_g, \uparrow_g, \downarrow_g, \uparrow_g\rangle$, $|\uparrow_g, \uparrow_g, \downarrow_g, \downarrow_g\rangle$, $|\downarrow_g, \downarrow_g, \uparrow_g, \downarrow_g\rangle$, $|\uparrow_g, \downarrow_g, \downarrow_g, \uparrow_g\rangle$, and $|\downarrow_g, \uparrow_g, \uparrow_g, \downarrow_g\rangle$, progressively dominate the ground state, as illustrated in Fig. 3.5(i).

For $N = 4$ and $\alpha = 0.7$, with $U_g < 2V_g$, akin to the behavior observed in charge density wave in large chain systems [96], electrons in a single dot tend to favor double occupancy over single occupancy. In a four-dot system, as $U$ increases, specific electron configurations such as $|\uparrow_g, \downarrow_g, 0, \uparrow_g\downarrow_g\rangle$, $|\downarrow_g, \uparrow_g, 0, \uparrow_g\downarrow_g\rangle$, $|\uparrow_g\downarrow_g, 0, \uparrow_g, \downarrow_g\rangle$, and $|\uparrow_g\downarrow_g, 0, \downarrow_g, \uparrow_g\rangle$ come to dominate the ground state configuration, as depicted in Fig. 3.6(i) (the above four states are all represented by $|\uparrow_g\downarrow_g, 0, \uparrow_g, \downarrow_g\rangle$ since they can be equally treated). This is related to the small size effect, since in such a system these configurations are most energetically favorable. Also, in Fig. 3.3(a), it is evident that $E(\rho_1) = E(\rho_4)$ and $E(\rho_2) = E(\rho_3)$, as all sites have an equal ratio of the four configurations of $|0\rangle$, $|\uparrow_g\rangle$, $|\downarrow_g\rangle$, and $|\uparrow_g\downarrow_g\rangle$. Specifically, $E(\rho_1)$ is almost equal to $E(\rho_2)$, with any differences being brought about by configuration states such as $|\uparrow_g\downarrow_g, 0, 0, \uparrow_g\downarrow_g\rangle$, illustrated in Fig. 3.6(h).

In the $L = 4$, $N = 6$, and $\alpha = 0.2$ system, entanglement is shown in Fig. 3.3(b). Due to the presence of two extra electrons (compared to the $N = 4$ case), the electron configurations of $|\uparrow_g\downarrow_g, \uparrow_g, \downarrow_g, \uparrow_g\downarrow_g\rangle$, $|\uparrow_g\downarrow_g, \uparrow_g, \uparrow_g\downarrow_g, \downarrow_g\rangle$, and $|\uparrow_g\downarrow_g, \uparrow_g\downarrow_g, \uparrow_g, \downarrow_g\rangle$ have the primary contribution to the system ground state $\psi_{\text{GS}}$, as shown in Fig. 3.7(i). For the ease of later discussion, we also introduce a notation describing the number of electrons in different sites. For example, $|\bullet\bullet, \bullet, \bullet, \bullet\bullet\rangle$, $|\bullet\bullet, \bullet, \bullet\bullet, \bullet\rangle$, and $|\bullet\bullet, \bullet\bullet, \bullet, \bullet\rangle$ represent the three aforementioned states occupancy respectively, where $\bullet$ or $\bullet\bullet$ represents a site occupied by one electron or two electrons respectively. We also use $\circ$ to express an empty site, so $|\bullet\bullet, \circ, \bullet\bullet, \bullet\bullet\rangle$ represents a case where site-1, site-3 and site-4 are doubly occupied while site-2 has no electron.

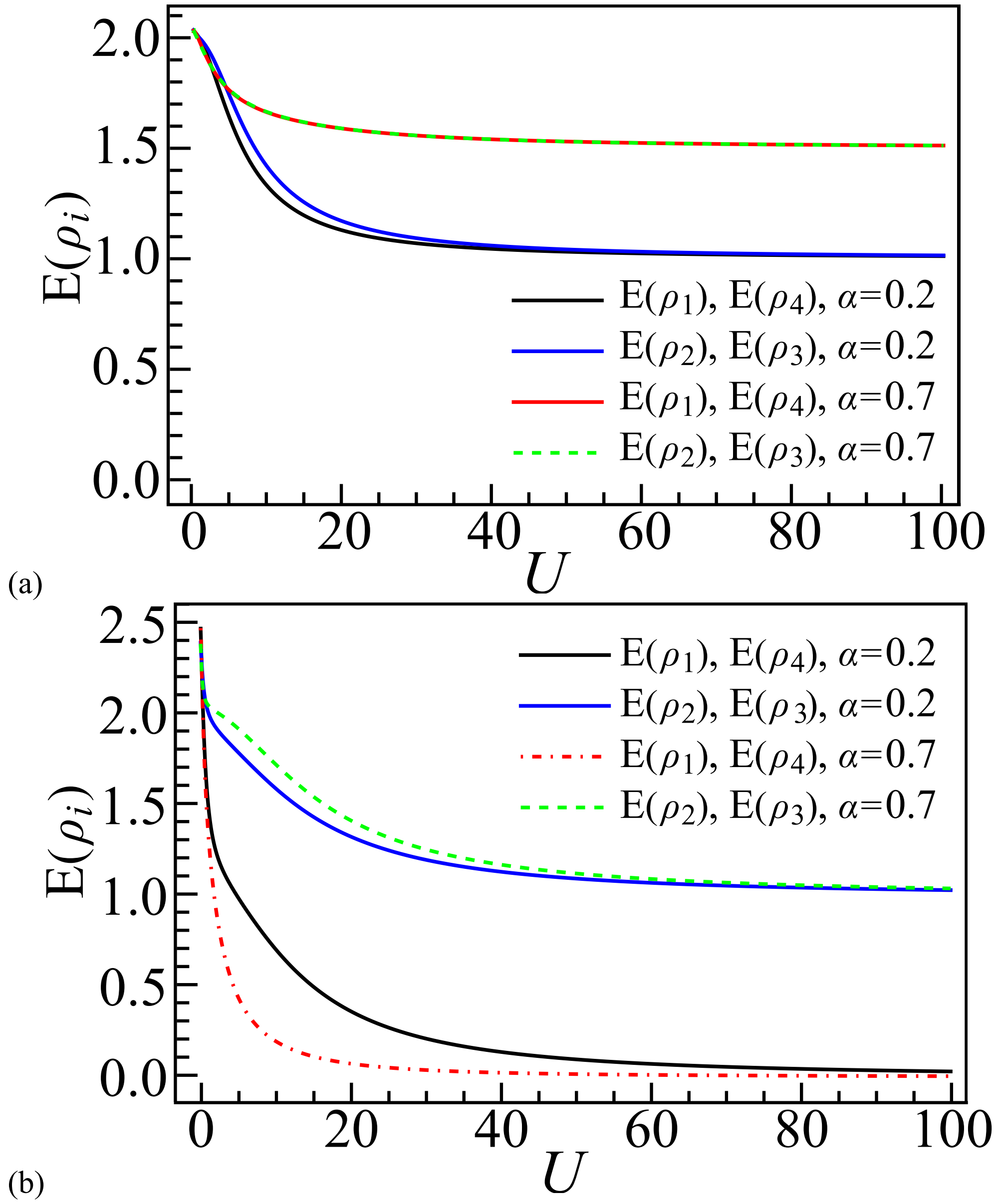

Fig. 3.3 Local entanglement $E(\rho_i)$ profiles for a four-site quantum dot system ($L = 4$) with coupling strengths $\alpha = 0.2$ or $\alpha = 0.7$, displayed as a function of interaction strength $U$. Panels (a) and (b) correspond to systems with four ($N = 4$) and six ($N = 6$) electrons, respectively, with zero detuning energy ($\varepsilon_i = 0$) at all sites. The entanglement measures $E(\rho_1)$ and $E(\rho_4)$ are equivalent, as are $E(\rho_2)$ and $E(\rho_3)$

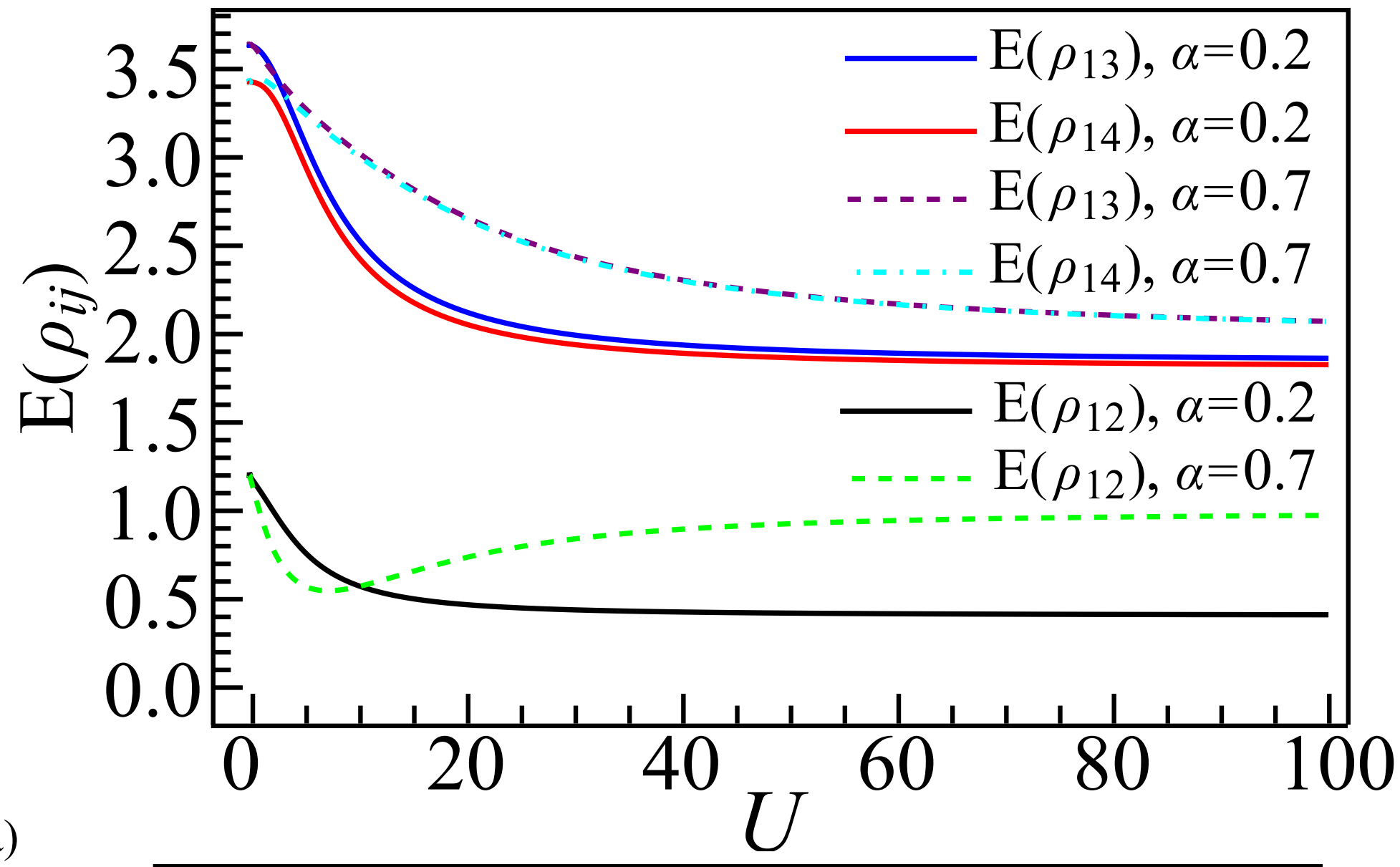

(a)

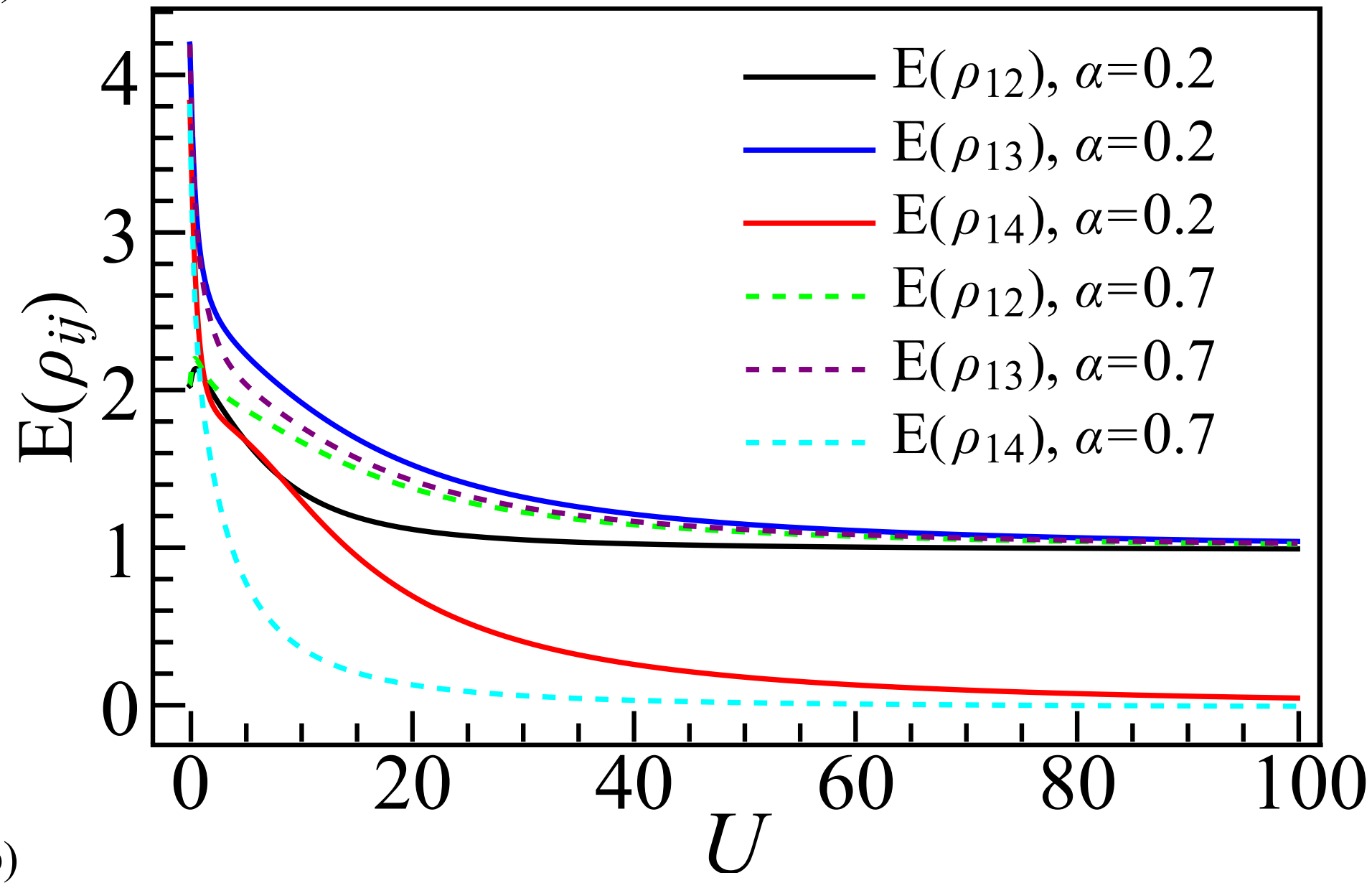

(b)

Fig. 3.4 These figures illustrate the pairwise entanglement metrics $E(\rho_{ij})$ for a four-site ($L = 4$) quantum dot array, analyzed under two coupling strength scenarios, $\alpha = 0.2$ and $\alpha = 0.7$. Displayed as functions of the interaction parameter $U$, panel (a) details configurations with four electrons ($N = 4$), and panel (b) with six electrons ($N = 6$), all with zero detuning energy at each site ($\varepsilon_i = 0$). The figures demonstrate equivalent entanglement values between dot pairs, specifically, $E(\rho_{12})$ with $E(\rho_{34})$, $E(\rho_{13})$ with $E(\rho_{24})$, and $E(\rho_{23})$ with $E(\rho_{14})$.

In the weak coupling regime, where $U \sim 0$, all electron configuration components have roughly the same proportion, thus $E(\rho_i)$ at $U \sim 0$ have similar values. As $U$ increases, the local entanglement of the end dots decreases more rapidly than that of the inner dots from the middle of the chain, and this rate of descent is even faster in the $N = 6$ case than the $N = 4$ case with $\alpha = 0.2$. This is due to the increasing dominance of the $|\bullet\bullet, \bullet, \bullet, \bullet\bullet\rangle$ configuration in the ground state, as depicted in Fig. 3.7(i). At $U \gg 1$, the inner dots tend to favor single occupancy, thereby resulting in similar values for $E(\rho_2)$ and $E(\rho_3)$ for both $N = 4$ and $N = 6$, while the end dots in the $N = 6$ case favor double occupancy, leading to a rapid decrease in the entanglement value.

For $N = 6$ and $\alpha = 0.7$, the system tends to favor double occupancy. Hence, the configurations $|\bullet\bullet, \bullet, \bullet, \bullet\bullet\rangle$, $|\bullet\bullet, \circ, \bullet\bullet, \bullet\bullet\rangle$ (also $|\bullet\bullet, \bullet\bullet, \circ, \bullet\bullet\rangle$) have a greater presence in the ground state compared to the $\alpha = 0.2$ case, as illustrated in Fig. 3.8(i). When compared to Fig. 3.7(i), the maximal probability of $|\bullet\bullet, \circ, \bullet\bullet, \bullet\bullet\rangle$ and $|\bullet\bullet, \bullet\bullet, \circ, \bullet\bullet\rangle$ in Fig. 3.8(i) has shifted toward smaller $U$. This indicates that all four sites in the $\alpha = 0.7$ setup prefer double occupancy over the $\alpha = 0.2$ case, leading to a faster decrease in $E(\rho_1)$ and $E(\rho_4)$, and a slower decrease in $E(\rho_2)$ and $E(\rho_3)$ compared to the $\alpha = 0.2$ scenario, since double occupancy contributes more to local entanglement.

### 3.5.2 Pairwise Entanglement at Zero Potential Energy

In $L = 4$ system with all quantum dots having equal potential energy ($\varepsilon_1 = \varepsilon_2 = \varepsilon_3 = \varepsilon_4 = 0$), mirror reflection symmetry ensures that the pairs of two-site reduced density matrices satisfy the relations $\rho_{12} = \rho_{34}$ and $\rho_{13} = \rho_{24}$. Additionally, due to the finite size effect inherent in the small system, it is observed that $\rho_{14} = \rho_{23}$, as illustrated in Figure 3.4.

For $N = 4$ and $\alpha = 0.2$, the entanglement results of $E(\rho_{12})$, $E(\rho_{13})$, and $E(\rho_{14})$ align well with the theoretical predictions for non-interacting systems ($\alpha = 0$), as eluci-dated in Ref. [1] and depicted in Figure 3.4(a). In the limit where $U \sim 0$, $E(\rho_{ij})$ has the same value for different $\alpha$ values since all Coulomb interactions are zero. Conversely, at $\alpha = 0.7$ with a positive $U$ value, the system demonstrates a preference for electron configurations such as $|\bullet\bullet, \circ, \bullet, \bullet\rangle$ and $|\bullet, \bullet, \circ, \bullet\bullet\rangle$. This preference equilibrates the entanglement levels $E(\rho_{13})$ and $E(\rho_{14})$ within the strong coupling regime, as illustrated in Figure 3.4(a). Within this regime, the probabilities for zero and single electron occupancy at sites 2 and 3 become comparable, as do the probabilities for single and double electron occupancy at sites 1 and 4, a phenomenon detailed in Figure 3.6(i). Concerning $E(\rho_{12})$, as depicted in the same figure, the diminished favorability of the state $|\bullet\bullet, \circ\rangle$ for the first and second sites leads to a reduction in the prevalence of the state $|\bullet\bullet, \circ, \circ, \bullet\bullet\rangle$ as $U$ increases. This reduction also leads to an increase in $E(\rho_{12})$ around $U \approx 7$, beyond which $E(\rho_{12})$ stabilizes to a constant value as $U$ continues to increase.

For $N = 6$ and $U = 0$, the uneven distribution of electrons leads to increased entanglement $E(\rho_{ij})$ compared to $N = 4$. This is particularly evident for $E(\rho_{12})$, as sites 1 and 2 are more likely to adopt the $|\bullet\bullet, \bullet\rangle$ configuration instead of the local half-filled state. Figures 3.7(i) and 3.8(i) illustrate that the electron arrangements $|\bullet\bullet, \bullet, \bullet, \bullet\bullet\rangle$, $|\bullet\bullet, \circ, \bullet\bullet, \bullet\bullet\rangle$, and $|\bullet, \bullet\bullet, \bullet, \bullet\bullet\rangle$ play a key role in determining the entanglement. For $\alpha = 0.7$, double occupancy is preferred, leading to a more rapid decline in configurations like $|\bullet, \bullet\bullet, \bullet, \bullet\bullet\rangle$ as $U$ increases, which in turn causes a quicker reduction in entanglement $E(\rho_{ij})$ compared to $\alpha = 0.2$. Regarding $E(\rho_{14})$, as $U$ moves into the strong coupling regime, $E(\rho_{14})$ approaches zero since sites 1 and 4 predominantly favor the $|\bullet\bullet\rangle$ configuration.

### 3.5.3 Entanglement Analysis for Non-zero Potential Energy with Four Electrons

Altering the potential energy of a specific quantum dot can significantly impact the entanglement behavior in the system, as demonstrated in Figures 3.5, 3.6, 3.7, and 3.8. For a particular quantum dot $i$, decreasing its potential energy causes electrons to congregate in this dot, which is reflected in the changes in the reduced density matrix elements: $v_{i,7}$, $v_{i,8}$, and $v_{i,9}$ increase, while $v_{i,1}$ to $v_{i,6}$ decrease.

In contrast, increasing the potential energy of dot $i$ leads to the dispersal of electrons to other dots, resulting in a decrease in all matrix elements of $\rho_i$ except for $v_{i,1}$, which corresponds to zero electron occupancy. In extreme cases, where the potential energy $\varepsilon_i$ undergoes significant changes, the electron configuration in this dot transi-tions to either $|0\rangle$ or $|\uparrow_g\downarrow_g, \uparrow_e\downarrow_e\rangle$, causing the local entanglement value to drop to zero, as shown in Figures 3.5(a), 3.6(a), 3.7(a), and 3.8(a). This phenomenon is particularly pronounced in the weakly coupling regime, where electrons have greater mobility. For instance, Figure 3.5(a) depicts the relationship between local entanglement $E(\rho_1)$, potential energy $\varepsilon_1$, and interaction strength $U$. In the weakly coupled regime ($U < 1$), as $\varepsilon_1$ deviates from zero, the value of $E(\rho_1)$ rapidly decreases from approximately 2 to 0. While in the strongly coupled regime ($U > 30$), electrons tend to remain separated in their respective quantum dots, adopting spin-wave-like configurations. Consequently, the local entanglement value approaches a limit of 1 as $U$ increases.

For coupling strength ratio set as $\alpha = 0.2$ and total electron number $N = 4$, we examine the system's favorable occupancy configurations to understand its entanglement diagram behavior. In the regime where the potential energy $\varepsilon_1$ is positive, an increase in $\varepsilon_1$ at a constant $U$ induces a transition in the main electron occupancy configuration components of the system's ground states from mostly $|\bullet, \bullet, \bullet, \bullet\rangle$ to the collection of $|\circ, \bullet\bullet, \circ, \bullet\bullet\rangle$, $|\circ, \bullet\bullet, \bullet, \bullet\rangle$, and $|\circ, \bullet\bullet, \bullet, \bullet\rangle$. Consequently, in the weakly coupled regime ($U < 1$), $E(\rho_1)$ undergoes a rapid decline, exhibiting distinct boundaries, while $E(\rho_2)$, $E(\rho_3)$, and $E(\rho_4)$ remain largely unchanged, as depicted in Figures 3.5(a)-(d). It is noteworthy that although the preferred electron occupancy configuration for dot 2 is $|\bullet\bullet\rangle$, the influence of other occupancy configurations like $|\bullet\rangle$ is significant, as shown in Figure 3.5(h), leading to a blurred boundary in $E(\rho_2)$.

In the strongly coupled regime ($U \gg 1$ and $\varepsilon_1 \ll U$), the system continues to favor the $|\bullet, \bullet, \bullet, \bullet\rangle$ occupancy configuration, where a substantial potential difference is required to alter the electron configurations. This transition is depicted in Figures 3.5(a), where an orange belt precedes the red entropy area at $\varepsilon_1 > 0$. It results from a rapid shift in preferred electron configurations, as shown in Figure 3.5(h). Adjacent to this belt, three regimes can be distinguished based on the coupling strength and the extent of potential energy influence: (1) the potential energy-influenced weak coupling regime, where $U \sim 1$ and $\varepsilon_1 \sim U$, allows electrons to be easily influenced by the potential energy difference between dots; (2) the potential energy-influenced strong coupling regime, representing the transition between weak and strong coupling regimes, where the potential energy can readily shift the system's favorable configurations; and (3) the strong coupling regime unaffected by potential energy, where $U \gg 1$ and the system remains largely unchanged by the relatively minor potential energy differences. These regimes are more distinguishable in the pairwise entanglement $E(\rho_{ij})$, as depicted in Figures 3.5(e)-(g). Near this belt (the potential energy-influenced strong coupling regime), the states $|0, \uparrow_g\downarrow_g\rangle$ are highly favored for the pair $\rho_{12}$, resulting in a low

entanglement value for $E(\rho_{12})$, while $|0, \uparrow_g\rangle$ and $|0, \downarrow_g\rangle$ are preferred for the pairs $\rho_{13}$ and $\rho_{14}$, leading to high entanglement values for $E(\rho_{13})$ and $E(\rho_{14})$.

In the regime where potential energy $\varepsilon_1 < 0$, multiple entanglement belts exist since one quantum dot can contain four electrons at most. In weakly coupled regimes, the decrease of potential energy $\varepsilon_1$ will quickly lead all electrons localized in site-1 since there are only four electrons in four quantum dots. More specifically, due to the size effect, the system is fully localized, and all entanglement values rapidly decline to zero, shown in Fig. 3.5(a)-(g). As the coupling strength $U$ increases, as shown in Figure 3.5(j), the favorable electron occupancy configurations of the ground states in the spin chain undergo a series of shifts: initially from $|\substack{\bullet\bullet\\\bullet\bullet}, \circ, \circ, \circ\rangle$ to $|\substack{\bullet\\\bullet\bullet}, \circ, \bullet, \circ\rangle$ and $|\substack{\bullet\\\bullet\bullet}, \circ, \circ, \bullet\rangle$, then to $|\bullet\bullet, \circ, \bullet, \bullet\rangle$, and eventually to $|\bullet, \bullet, \bullet, \bullet\rangle$. Here we use $\substack{\bullet\\\bullet\bullet}$ or $\substack{\bullet\bullet\\\bullet\bullet}$ represents a site occupied by three electron or four electrons respectively.

Firstly, in the intermediate phase where the preferred electron occupancy configurations are $|\substack{\bullet\\\bullet\bullet}, \circ, \bullet, \circ\rangle$ and $|\substack{\bullet\\\bullet\bullet}, \circ, \circ, \bullet\rangle$, three electrons tend to reside in the first dot, while the remaining electron occupies either the third or fourth dot. Consequently, $E(\rho_1)$, $E(\rho_3)$, and $E(\rho_4)$ exhibit higher local entanglement values, whereas $E(\rho_2)$ declines to a lower value, as depicted in Figures 3.5(a)-(d). This distribution demonstrates a transition in preferred states across the quantum dots from 1 to 4. Specifically, the value of $E(\rho_1)$ is associated with states indicative of three-electron occupancy $|\substack{\bullet\\\bullet\bullet}\rangle$, $E(\rho_2)$ corresponds to zero electron occupancy $|\circ\rangle$ (or $|0\rangle$), and $E(\rho_3)$ and $E(\rho_4)$ oscillate between one-electron occupancy $|\bullet\rangle$ and zero occupancy $|\circ\rangle$. Similarly, $E(\rho_{12})$ exhibits a high entanglement value, while $E(\rho_{13})$ and $E(\rho_{14})$ display even higher values.

Secondly, when the system's preferred occupancy configuration is $|\bullet\bullet, \circ, \bullet, \bullet\rangle$, the first dot favors double occupancy, and the second dot favors zero occupancy, while the third and fourth dots favor one-electron occupancy. As a result, $E(\rho_1)$, $E(\rho_2)$ and $E(\rho_{12})$ approach zero, while $E(\rho_3)$, $E(\rho_4)$, $E(\rho_{13})$ and $E(\rho_{14})$ become similar with high entanglement value as $U$ increases.

Lastly, in the region where the system favors the $|\bullet, \bullet, \bullet, \bullet\rangle$ occupancy configuration, all entanglement behaviors align with those in the $\varepsilon_1 \neq 0$ regime as the coupling strength becomes the dominant factor. Notably, the entanglement measures $E(\rho_3)$ and $E(\rho_4)$ exhibit smooth boundary transitions, indicating a preference for single-electron occupancy $|\bullet\rangle$ in both the third and fourth quantum dots at this boundary. Similarly to the $\varepsilon_1 = 0$ case, the system shows a preference for the configurations $|\uparrow_g, \downarrow_g, \uparrow_g, \downarrow_g\rangle$ and $|\downarrow_g, \uparrow_g, \downarrow_g, \uparrow_g\rangle$ over other spin state configurations, resulting in $E(\rho_{12}) < E(\rho_{13}) \sim E(\rho_{14})$.

When the coupling strength ratio is set to $\alpha = 0.7$, the system exhibits a preference for double occupancy over single occupancy, in line with the nature of EHM [79, 124, 6, 96]. This preference is maintained even when $\varepsilon_1 \neq 0$, as demonstrated in Figures 3.6(h) and 3.6(j). For $\varepsilon_1 > 0$, the favored electron occupancy configuration readily becomes $|\circ, \bullet\bullet, \circ, \bullet\bullet\rangle$ until $U \gg \varepsilon_1$, resulting in $E(\rho_i) \sim E(\rho_{ij}) \sim 0$ ($i$ for all sites from 1 to 4) when $U < \varepsilon_1$. Notably, in the weak coupling regime ($U \sim 1$), $E(\rho_1)$ equals zero, while $E(\rho_2)$, $E(\rho_3)$, $E(\rho_4)$, $E(\rho_{12})$, $E(\rho_{13})$ and $E(\rho_{14})$ experience a decrease in entanglement value, caused by the reduction of the electron occupancy configuration $|\circ, \bullet\bullet, \bullet, \bullet\rangle$, as shown in Figure 3.6(h).

For $\varepsilon_1 < 0$, the system similarly experiences three transitions, as illustrated in Figure 3.6(j). With increasing $U$, the electron occupancy in site-1 changes from 4 to 2, resulting in variations in the entanglement values across all sites. $E(\rho_1)$ remains nonzero only when the average electron number in this dot is 3, due to the presence of two favored configurations, either up or down in the excited state. $E(\rho_2)$ is predominantly zero, as this site is typically unoccupied by electrons, except along the boundary line where transitions between different system configurations render $E(\rho_2)$ nonzero. Regarding $E(\rho_3)$ and $E(\rho_4)$, their electron configurations tend to converge in the strong coupling regime, resulting in similar entanglement behaviors. For $E(\rho_{12})$, the occupancy in site-1 influences the behavior of $E(\rho_{12})$, making it similar to $E(\rho_1)$. For $E(\rho_{13})$ and $E(\rho_{14})$, their behavior in the regime where $U \gg \varepsilon_1$ is similar to $E(\rho_3)$ and $E(\rho_4)$, respectively. When $U \sim \varepsilon_1$, they also exhibit distinct features similar to $E(\rho_1)$.

### 3.5.4 Entanglement Analysis for Non-zero Potential Energy with Six Electrons

In contrast to the $N = 4$ case, the $N = 6$ system in a four-site lattice ($L = 4$) inherently exhibits an imbalance in electron configurations, necessitating the consideration of additional configurations.

In the strong coupling regime, where $U \gg \varepsilon_1$, Figures 3.7(h), 3.7(j), 3.8(h), and 3.8(j) demonstrate that for both $\varepsilon_1 > 0$ and $\varepsilon_1 < 0$, and for coupling ratios $\alpha = 0.2$ and $\alpha = 0.7$, the system's favored occupancy configuration is $|\bullet\bullet, \bullet, \bullet, \bullet\bullet\rangle$. This occupancy configuration leads to both $E(\rho_1)$ and $E(\rho_4)$ becoming zero, while $E(\rho_2)$ and $E(\rho_3)$ share the same entanglement value of approximately 1.2. When $U \sim \varepsilon_1$ and $\varepsilon_1 > 0$, the most favorable occupancy configuration for both $\alpha = 0.2$ and $\alpha = 0.7$ is $|\bullet, \bullet\bullet, \bullet, \bullet\bullet\rangle$, lead to $E(\rho_1) \sim E(\rho_3)$ and $E(\rho_2) \sim E(\rho_4)$. For $\varepsilon_1 < 0$, the most favorable occupancy configuration is $|\frac{\bullet}{\bullet\bullet}, \bullet, \bullet, \bullet\rangle$ for $\alpha = 0.2$, and for $\alpha = 0.7$, the configurations $|\frac{\bullet}{\bullet\bullet}, \circ, \bullet\bullet, \bullet\rangle$ and $|\frac{\bullet}{\bullet\bullet}, \circ, \bullet, \bullet\bullet\rangle$ are preferred. For $\alpha = 0.2$, the $|\bullet\rangle$ configuration of site-2 lead $E(\rho_2)$ and $E(\rho_{12})$ become non-zero, which is opposite for $\alpha = 0.7$ since site-2 favor $|\circ\rangle$ configuration. Therefore $E(\rho_2)$ becomes zero and $E(\rho_{12})$ behaves like $E(\rho_1)$.

In the weak coupling regime with $\varepsilon_1 > 0$, the system prefers specific electron configurations based on the coupling strength ratio $\alpha$. For $\alpha = 0.2$, the favored configurations are $|\circ, \bullet\bullet, \bullet\bullet, \bullet\bullet\rangle$ and $|\circ, \bullet\bullet, \bullet, \frac{\bullet}{\bullet\bullet}\rangle$, while for $\alpha = 0.7$, the preferences shift to $|\circ, \bullet\bullet, \bullet\bullet, \bullet\bullet\rangle$ and $|\circ, \frac{\bullet}{\bullet\bullet}, \circ, \frac{\bullet}{\bullet\bullet}\rangle$. As $U$ increases within this regime, a transition occurs: for $\alpha = 0.2$, the system changes towards occupancy $|\circ, \bullet\bullet, \bullet\bullet, \bullet\bullet\rangle$, causing all entanglement measures $E(\rho_i)$ and $E(\rho_{ij})$ to vanish. In contrast, for $\alpha = 0.7$, the system evolves towards the configuration $|\circ, \frac{\bullet}{\bullet\bullet}, \circ, \frac{\bullet}{\bullet\bullet}\rangle$, leading to a vanishing of $E(\rho_1)$, $E(\rho_3)$, and $E(\rho_{13})$, while $E(\rho_2)$, $E(\rho_4)$, $E(\rho_{12})$, and $E(\rho_{14})$ stabilize at a constant value.

In the weak coupling regime with $\varepsilon_1 < 0$, all four electrons are in the first dot. Both for $\alpha = 0.2$ and $\alpha = 0.7$, the system exhibits a preference for the configurations $|\frac{\bullet\bullet}{\bullet\bullet}, \circ, \bullet, \bullet\rangle$ and $|\frac{\bullet\bullet}{\bullet\bullet}, \circ, \bullet\bullet, \circ\rangle$. As a result, in these two coupling ratio settings, the entanglement measures $E(\rho_i)$ and $E(\rho_{ij})$ display similar patterns: $E(\rho_1)$ remains at zero, $E(\rho_2)$ and $E(\rho_{12})$ gently descend to zero, while $E(\rho_3)$, $E(\rho_4)$, $E(\rho_{13})$, and $E(\rho_{14})$ find equilibrium at a constant value. Notably, the values of $E(\rho_i)$ and $E(\rho_{ij})$ differ between $\alpha = 0.2$ and $\alpha = 0.7$, caused by different electron configuration ratio.

### 3.5.5 Boundaries of Entanglement Diagrams for Large System

In this section, we expand the entanglement diagram from a small, finite-size system to a larger spin chain quantum dot system. It is evident from the ground state of the finite-size system that advantageous electron configurations significantly influence the boundaries and values of the entanglement diagram. This analysis can be readily extended to larger systems by calculating the energy of the electron configuration obtained from the Hubbard model (see Eq. (3.1)). Since the system always favors the configuration with the lowest energy, which can be easily calculated and observed in small systems, we use this principle to infer the most favored configuration in larger systems.

For the case where $\alpha = 0.2$ and $N = L$, with $L$ denoting the length of the spin chain and indicating an average of one electron per quantum dot, the system exhibits a preference for single occupancy at each quantum dot, resulting in a spin density wave structure as described in previous studies [79, 124, 6, 96]. Figure 3.9(a) depicts the evolution of the dominant system configurations as the potential energy $\varepsilon_1$ transitions from positive to negative values, showcasing a sequence of dominant configurations across regimes **I** to **V**:

$$\textbf{I}: \quad |\circ, \bullet\bullet, \bullet, \bullet, \bullet, \bullet, \bullet, ...\bullet, \bullet, \bullet\rangle\,, \tag{3.10a}$$

$$\textbf{II}: \quad |\bullet, \bullet, \bullet, \bullet, \bullet, \bullet, \bullet, ...\bullet, \bullet, \bullet\rangle\,, \tag{3.10b}$$

$$\textbf{III}: \quad |\bullet\bullet, \circ, \bullet, \bullet, \bullet, \bullet, \bullet, ...\bullet, \bullet, \bullet\rangle\,, \tag{3.10c}$$

$$\textbf{IV}: \quad \left|\frac{\bullet}{\bullet\bullet}, \circ, \bullet, \circ, \bullet, \bullet, \bullet, ...\bullet, \bullet, \bullet\right\rangle\,, \tag{3.10d}$$

$$\textbf{V}: \quad \left|\frac{\bullet\bullet}{\bullet\bullet}, \circ, \bullet, \circ, \bullet, \circ, \bullet, ...\bullet, \bullet, \bullet\right\rangle\,. \tag{3.10e}$$

The energies associated with these configurations, as derived from the Hubbard model, are as follows: **I**: $U_g + (N-3)V_g$, **II**: $(N-1)V_g + \varepsilon_1$, **III**: $(N-3)V_g + U_g + 2\varepsilon_1$, **IV**: $(N-5)V_g + U_g + 2V'_{g,e} + 3\varepsilon_1$, **V**: $(N-7)V_g + U_g + U_e + 4V'_{g,e} + 4\varepsilon_1$. Consequently, the boundaries distinguishing these regions in Figure 3.9(a) can be calculated as follows:

$$\textbf{I-II}: \quad \varepsilon_1 = U_g - 2V_g, \tag{3.11a}$$

$$\textbf{II-III}: \quad \varepsilon_1 = -U_g + 2V_g, \tag{3.11b}$$

$$\textbf{III-IV}: \quad \varepsilon_1 = -2V'_{g,e} + 2V_g, \tag{3.11c}$$

$$\textbf{IV-V}: \quad \varepsilon_1 = -2V'_{g,e} - U_e + 3V_g. \tag{3.11d}$$

The boundaries between different regions mark the transitions between different electron occupancy configurations. In region **I**, the system exhibits a preference for the configuration $|\circ, \bullet\bullet, \bullet, \bullet, \bullet, \bullet, \bullet, ...\bullet, \bullet, \bullet\rangle$. This indicates that the first dot is unoccupied when $\varepsilon_1 > U_g - 2V_g$, and the extra electron from the first dot is likely to be found either in the second dot or at the last dot of the spin chain. This preference arises because the electron at these positions contributes only one $V_\nu$ interaction, while electrons in other positions contribute to $2V_\nu$ interactions. Similarly, in regions **III**, **IV**, and **V**, when the first dot accommodates more than one electron, the second dot tends to be unoccupied. This arrangement minimizes the Coulomb interaction between the first and second dots. Similarly, for the remaining dots, electrons tend to favor configurations where both neighboring dots are unoccupied and form spin density wave configurations, reducing the overall Coulomb interaction terms within the system.

Second, for $\alpha = 0.7$ with $N = L$, the system adopts a charge density wave structure [79, 124, 6, 96]. Figure 3.9(b) illustrates the progression of the dominant system configurations as the potential energy $\varepsilon_1$ shifts from positive to negative values, depicting a sequence of configurations that emerge in this transition. The configurations are:

$$\begin{aligned}
\textbf{I}: &\quad |\circ, \bullet\bullet, \circ, \bullet\bullet, \circ, \bullet\bullet, \circ, ..., \bullet\bullet, \circ, \bullet\bullet\rangle, && (3.12a)\\
\textbf{II}: &\quad |\bullet, \bullet, (\circ, \bullet\bullet, \circ, \bullet\bullet, \circ, ..., \bullet\bullet, \circ, \bullet\bullet)\rangle, && (3.12b)\\
\textbf{III}: &\quad |\circ, \bullet\bullet, \circ, \bullet\bullet, \circ, \bullet\bullet, \circ, ..., \bullet\bullet, \circ, \bullet\bullet\rangle, && (3.12c)\\
&\quad |\bullet\bullet, \circ, \bullet\bullet, \circ, \bullet\bullet, \circ, \bullet\bullet, ...\circ, \bullet\bullet, \circ\rangle, && (3.12d)\\
\textbf{IV}: &\quad |(\bullet\bullet, \circ, \bullet\bullet, \circ, \bullet\bullet, \circ, \bullet\bullet, ...\circ, \bullet\bullet, \circ)\rangle, && (3.12e)\\
\textbf{V}: &\quad \left|\frac{\bullet}{\bullet\bullet}, \circ, \bullet, \circ, (\bullet\bullet, \circ, \bullet\bullet, ...\circ, \bullet\bullet, \circ)\right\rangle, && (3.12f)\\
\textbf{VI}: &\quad \left|\frac{\bullet\bullet}{\bullet\bullet}, \circ, \bullet, \circ, \bullet, \circ, (\bullet\bullet, ...\circ, \bullet\bullet, \circ)\right\rangle. && (3.12g)
\end{aligned}$$

The energies corresponding to these configurations, as calculated from the Hubbard model, are as follows: **I**: $NU_g/2$, **II**: $(N-2)U_g/2+V_g+\varepsilon_1$, **III**: $NU_g/2$, **IV**: $NU_g/2+2\varepsilon_1$, **V**: $(N-4)U_g/2+U_g+V'_{g,e}+3\varepsilon_1$ **VI**: $(N-6)U_g/2+U_g+U_e+4V'_{g,e}+4\varepsilon_1$. Therefore, we can calculate the boundary functions of these regions in Fig. 3.9(b) as

$$\begin{aligned}
\textbf{I-II}: &\quad \varepsilon_1 = U_g - V_g, && (3.13a)\\
\textbf{II-III}: &\quad \varepsilon_1 = 0, && (3.13b)\\
\textbf{III-IV}: &\quad \varepsilon_1 = 0, && (3.13c)\\
\textbf{IV-V}: &\quad \varepsilon_1 = U_g - 2V'_{g,e}, && (3.13d)\\
\textbf{V-VI}: &\quad \varepsilon_1 = U_g - U_e - 2V'_{g,e}. && (3.13e)
\end{aligned}$$

In regions **I** and **IV**, the system adopts a global charge density wave structure, with the first dot being unoccupied and doubly occupied, respectively. Notably, in region **III**, the system exhibits a charge density wave pattern that arises from the superposition of two distinct configurations: (1) $|\circ, \bullet\bullet, \circ, \bullet\bullet, \circ, \bullet\bullet, \circ, ..., \bullet\bullet, \circ, \bullet\bullet\rangle$ and (2) $|\bullet\bullet, \circ, \bullet\bullet, \circ, \bullet\bullet, \circ, \bullet\bullet, ...\circ, \bullet\bullet, \circ\rangle$. In region **II**, both the first and second dots host a single electron. This arrangement minimizes the Coulomb interaction terms compared to alternative configurations. Specifically, the potential energy shift in the first dot and its interaction with the second dot yield a lower energy of $\varepsilon_1 + V_g$. In contrast, hosting two electrons in the second dot would result in a higher energy, given by $U_g$, thus making the single-electron configuration energetically favorable. For regions **V** and **VI**, apart from the first dot, the system prefers configurations where neighboring dots are unoccupied, maintaining the charge density wave structure throughout the rest of the system.

Third, in the case of $\alpha = 0.2$ with $N = L + 2$, the presence of two additional electrons raises the average electron count per dot above one. Consequently, only a portion of the system continues to exhibit a spin density wave structure. As illustrated in Figure 3.9(c), the dominant system configurations evolve as the potential energy $\varepsilon_1$ transitions from positive to negative values. The sequence of dominant configurations for regions **I** to **V** is as follows:

$$\begin{aligned}
\textbf{I}: &\quad |\circ, \bullet\bullet, \bullet, \bullet\bullet, \bullet, \bullet, \bullet, \ldots \bullet, \bullet, \bullet\bullet\rangle, && \text{(3.14a)}\\
\textbf{II}: &\quad |\bullet, \bullet\bullet, \bullet, \bullet, \bullet, \bullet, \bullet, \ldots \bullet, \bullet, \bullet\bullet\rangle, && \text{(3.14b)}\\
\textbf{III}: &\quad |\bullet\bullet, \bullet, \bullet, \bullet, \bullet, \bullet, \bullet, \ldots \bullet, \bullet, \bullet\bullet\rangle, && \text{(3.14c)}\\
\textbf{IV}: &\quad \left|\frac{\bullet}{\bullet\bullet}, \bullet, \bullet, \bullet, \bullet, \bullet, \bullet, \ldots \bullet, \bullet, \bullet\right\rangle, && \text{(3.14d)}\\
\textbf{V}: &\quad \left|\frac{\bullet\bullet}{\bullet\bullet}, \circ, \bullet, \bullet, \bullet, \bullet, \bullet, \ldots \bullet, \bullet, \bullet\right\rangle. && \text{(3.14e)}
\end{aligned}$$

The energy associated with each configuration in the Hubbard model obtained as follows: **I**: $3U_g + (N-6)V_g + 8V_g$, **II**: $\varepsilon_1 + 2U_g + (N-4)V_g + 6V_g$, **III**: $2\varepsilon_1 + (N-3)V_g + 2U_g + 4V_g$, **IV**: $3\varepsilon_1 + (N-2)V_g + 2V_g + V_{g,e} + 2V'_{g,e}$, **V**: $4\varepsilon_1 + (N-3)V_g + U_g + U_e + 4V'_{g,e}$. Accordingly, the boundary functions distinguishing these regions in Figure 3.9(c) are calculated as:

$$\begin{aligned}
\textbf{I-II}: &\quad \varepsilon_1 = U_g, && \text{(3.15a)}\\
\textbf{II-III}: &\quad \varepsilon_1 = V_g, && \text{(3.15b)}\\
\textbf{III-IV}: &\quad \varepsilon_1 = V_g + 2U_g - V_{g,e} - 2V'_{g,e}, && \text{(3.15c)}\\
\textbf{IV-V}: &\quad \varepsilon_1 = V_g + V_{g,e} - U_g - U_e - 2V'_{g,e}, && \text{(3.15d)}
\end{aligned}$$

In region **I**, the first dot is unoccupied, prompting the three additional electrons to distribute themselves along the chain to minimize Coulomb interactions: two electrons position themselves at the ends, while the third occupies a central position. This arrangement ensures minimal interaction with the electrons at the ends. Similarly, in regions **II** and **III**, the additional electrons also preferentially reside at the chain ends. Conversely, in regions **IV** and **V**, the extra electrons occupy the first dot, freeing up space along the rest of the chain for one electron per dot. Notably, in region **IV**, the electron in the second quantum dot remains localized rather than migrating to the third dot or further along the chain. This localization is evident when considering the configuration $\left|\frac{\bullet}{\bullet\bullet}, \bullet, \bullet, \bullet, \bullet, \bullet, \bullet, \ldots, \bullet, \bullet, \bullet\right\rangle$, where the electron in the second dot interacts with its adjacent electron with an energy of $3V_g + V_{g,e}$. Conversely, in the competitive configuration $\left|\frac{\bullet}{\bullet\bullet}, \circ, \bullet\bullet, \bullet, \bullet, \bullet, \bullet, \ldots, \bullet, \bullet, \bullet\right\rangle$, the electron in the third dot interacts with the fourth dot with an energy of $U_g + 2V_g$. This results in a higher total energy than the former configuration under the parameter setting $\alpha = 0.2$.

Last, in the case of $\alpha = 0.7$ with $N = L+2$, shown in Figure 3.9(d), the dominating system configuration as $\varepsilon_1$ changes from $\varepsilon_1 > 0$ to $\varepsilon_1 < 0$, the configurations will appear as the following sequences:

$$\textbf{I}: \quad |\circ, \bullet\bullet, \bullet, \bullet\bullet, \circ, \bullet\bullet, \circ, \ldots, \bullet\bullet, \bullet, \bullet\bullet\rangle, \tag{3.16a}$$

$$\textbf{II}: \quad |\bullet, \bullet\bullet, \bullet, \bullet\bullet, \circ, \bullet\bullet, \circ, \ldots, \bullet\bullet, \circ, \bullet\bullet\rangle, \tag{3.16b}$$

$$\textbf{III}: \quad |\bullet\bullet, \bullet, \bullet, \bullet\bullet, \circ, \bullet\bullet, \circ, \ldots, \bullet\bullet, \circ, \bullet\bullet\rangle, \tag{3.16c}$$

$$\textbf{IV}: \quad \left|\frac{\bullet}{\bullet\bullet}, \circ, \bullet, \bullet\bullet, \circ, \bullet\bullet, \circ, \ldots, \bullet\bullet, \circ, \bullet\bullet\right\rangle, \tag{3.16d}$$

$$\textbf{V}: \quad \left|\frac{\bullet\bullet}{\bullet\bullet}, \circ, \bullet\bullet, \circ, \bullet\bullet, \circ, \bullet\bullet, \circ, \ldots, \bullet\bullet, \circ\right\rangle. \tag{3.16e}$$

The energies corresponding to these configurations, as calculated from the Hubbard model, are as follows: **I**: $NU_g/2 + 8V_g$, **II**: $\varepsilon_1 + NU_g/2 + 6V_g$, **III**: $NU_g/2 + U_g + 8V_g + 2\varepsilon_1 + NU_g/2 + V_g + 4V_g + 2\varepsilon_1$, **IV**: $3\varepsilon_1 + NU_g/2 + 2V_g + 2V'_{g,e}$, **V**: $4\varepsilon_1 + NU_g/2 + U_e + 4V'_{g,e}$. Consequently, the boundaries distinguishing these regions in Figure 3.9(d) can be calculated as follows:

$$\textbf{I-II}: \quad \varepsilon_1 = 2V_g, \tag{3.17a}$$

$$\textbf{II-III}: \quad \varepsilon_1 = V_g, \tag{3.17b}$$

$$\textbf{III-IV}: \quad \varepsilon_1 = 3V_g - 2V'_{g,e}, \tag{3.17c}$$

$$\textbf{IV-V}: \quad \varepsilon_1 = 2V_g - U_e - 2V'_{g,e}, \tag{3.17d}$$

In regions **I** and **II**, the additional electrons—two in the former and one in the latter—have the flexibility to occupy any available sites along the spin chain. In region **III**, a distinctive arrangement emerges where two electrons specifically occupy sites 2 and 3. This localized occupation maintains a charge density wave structure throughout the remainder of the spin chain. The region **IV** exhibits a situation in which a single electron favors site 3, which is advantageous as it minimizes the Coulomb interaction, involving only a $2V_{\nu}$ contribution from the adjacent site 4, thereby optimizing the en-ergy configuration. Finally, the region **V** naturally evolves into a global charge density wave structure, where the electron distribution systematically alternates along the entire chain, reflecting a stable and energetically favorable arrangement. This structure highlights the intrinsic properties of the system under these specific conditions.

## 3.6 Conclusions

In this study, we systematically explored the entanglement properties of semiconductor quantum dots within a multi-site lattice, described by the EHM. Our investigations demonstrate that local and pairwise entanglement measures respond sensitively to interactions between Coulomb forces and tunneling effects, which are influenced by the system's electronic configurations and variations in external potential energies. Notably, the entanglement characteristics show distinct phase transitions influenced heavily by coupling strength ratios and variations in potential energy. We observed that varying the potential energy of a specific dot decisively alters ground state configurations and, consequently, entanglement measures, a phenomenon that is pronounced in both weak and strong coupling regimes. This indicates that potential energy modifications can effectively control entanglement in quantum dot systems.

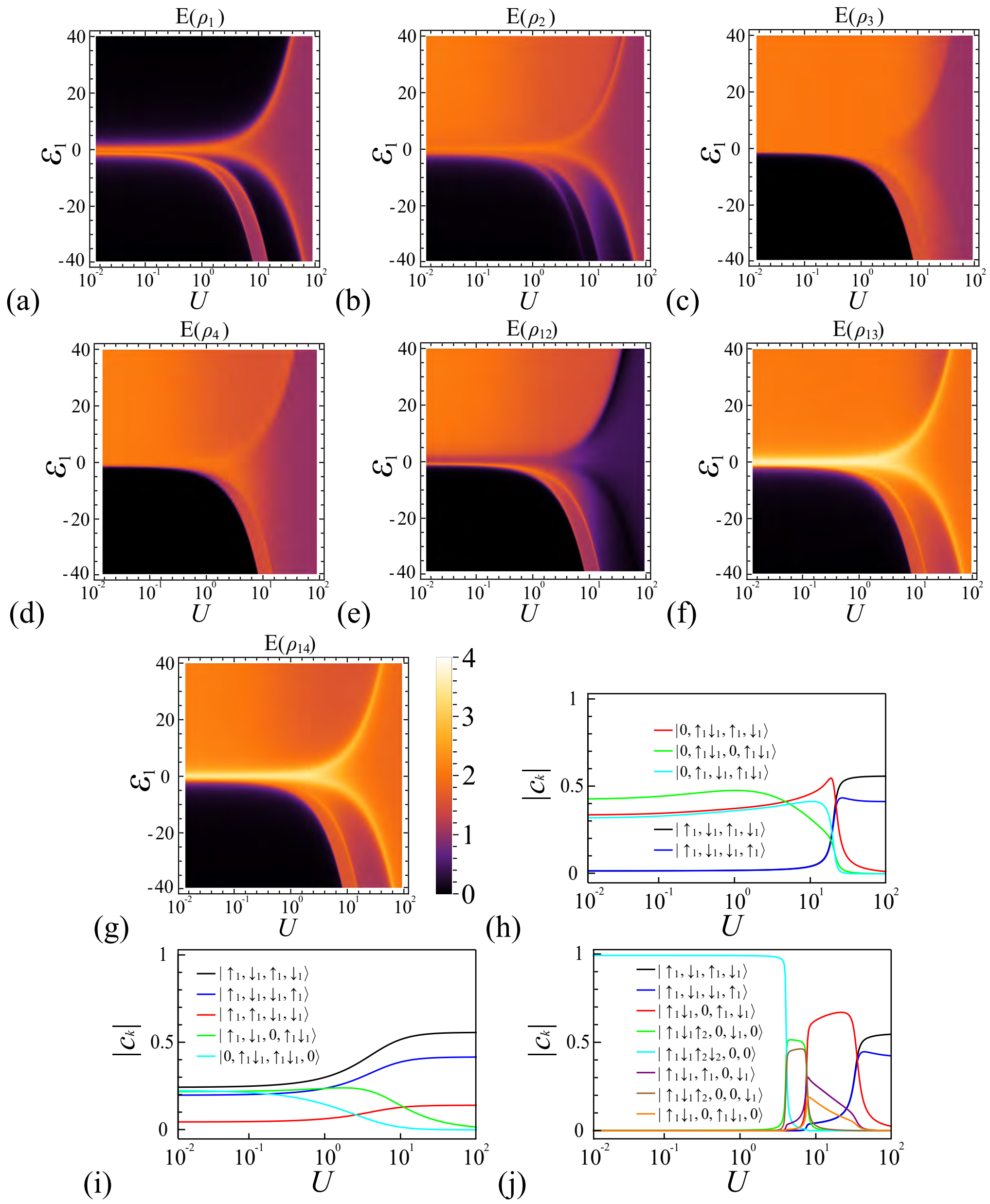

Fig. 3.5 Entanglement phase diagrams for a quantum dot system with four sites ($L = 4$) and four electrons ($N = 4$) under a coupling strength ratio of $\alpha = 0.2$. These diagrams are plotted as functions of the interaction strength $U$ and the potential energy $\varepsilon_1$. (a)-(d) local entanglement measures $E(\rho_1)$, $E(\rho_2)$, $E(\rho_3)$, and $E(\rho_4)$, respectively. (e)-(g) pairwise entanglement for dot pairs $E(\rho_{12})$, $E(\rho_{13})$, and $E(\rho_{14})$. (h)-(j) illustrate the proportions of selected advantageous electron configurations within the system's ground state, highlighting the influence of interaction parameters on system behavior, represent cases where $\varepsilon_1 = 20$, $\varepsilon_1 = 0$, and $\varepsilon_1 = -20$, respectively.

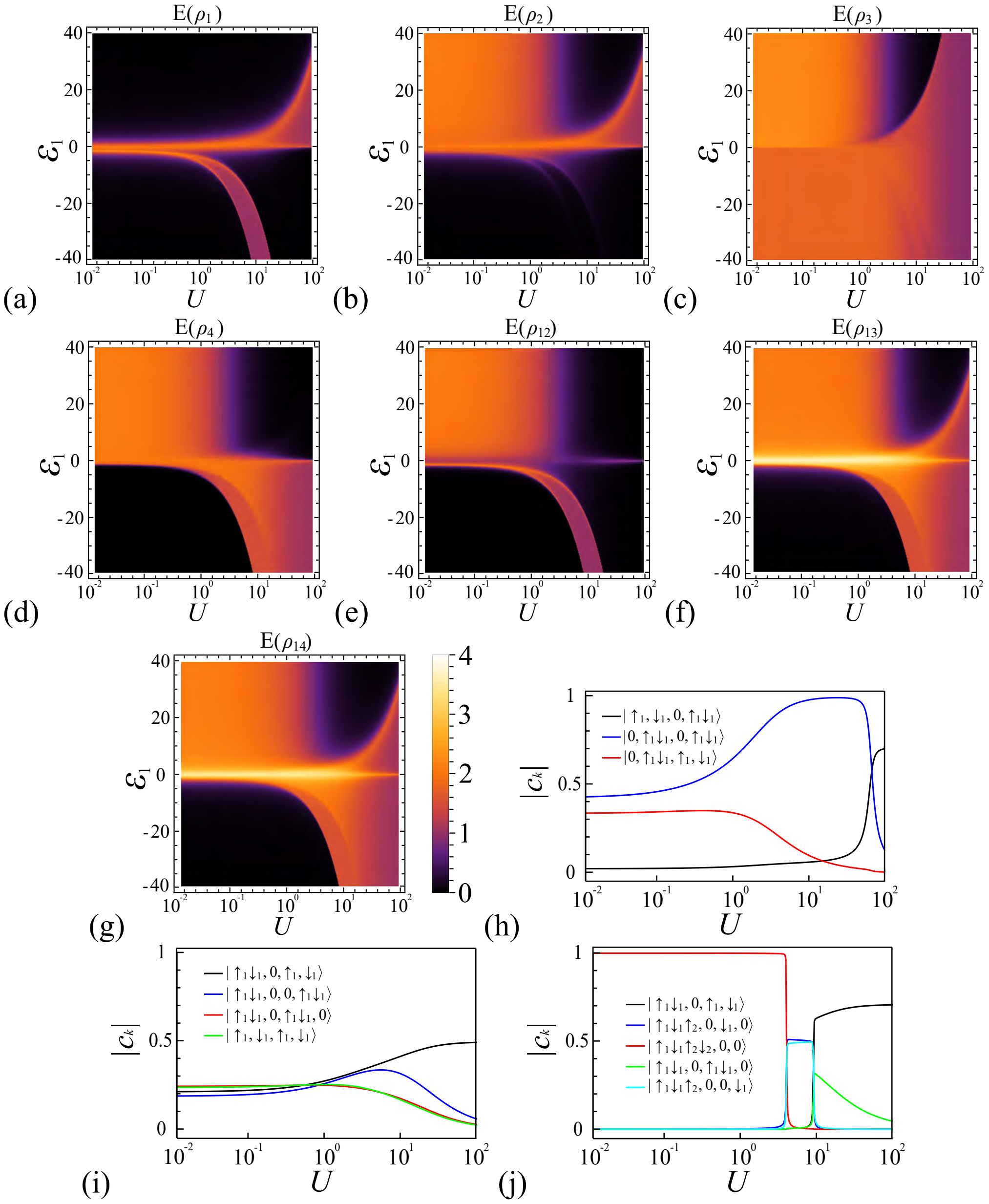

Fig. 3.6 Entanglement characteristics of a four-dot ($L = 4$), four-electron ($N = 4$) quantum dot system at a coupling ratio of $\alpha = 0.7$. Diagrams are plotted against interaction strength $U$ and potential energy $\varepsilon_1$. (a)-(d) local entanglement measures $E(\rho_1)$ to $E(\rho_4)$. (e)-(g) pairwise entanglement for dot pairs $E(\rho_{12})$, $E(\rho_{13})$, and $E(\rho_{14})$. The dominant electron configurations in the ground state corresponding to $\varepsilon_1 = 20$, $\varepsilon_1 = 0$, and $\varepsilon_1 = -20$ are represented by (h), (i), and (j) respectively.

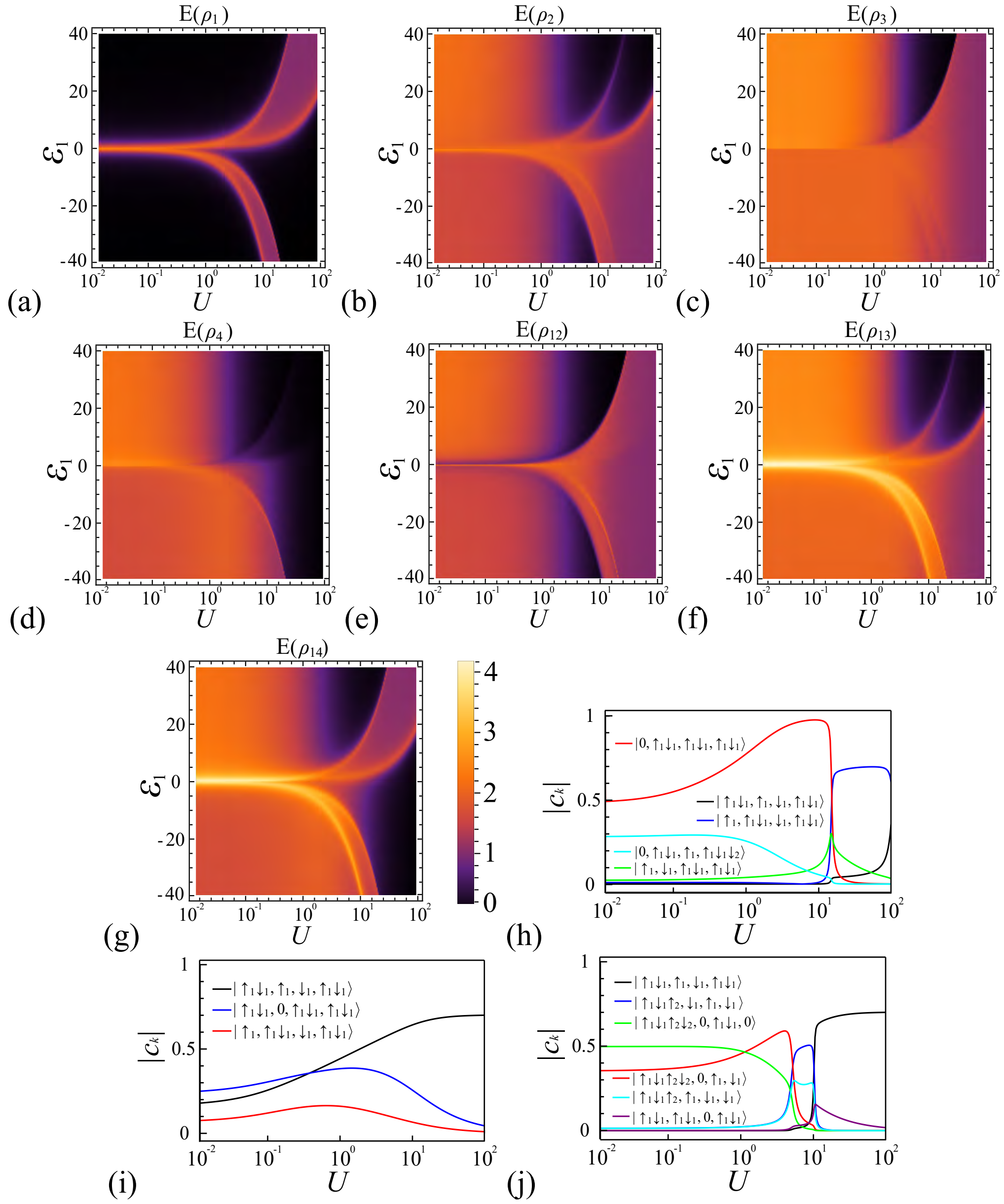

Fig. 3.7 Entanglement profiles for a four-site ($L = 4$), six-electron ($N = 6$) quantum dot system with a coupling strength of $\alpha = 0.2$. Charts are graphed according to interaction strength $U$ and potential energy $\varepsilon_1$. (a)-(d) depict local entanglement levels $E(\rho_1)$ through $E(\rho_4)$. (e)-(g) pairwise entanglement between dot pairs $E(\rho_{12})$, $E(\rho_{13})$, and $E(\rho_{14})$. (h)-(j) show the predominant electron configurations in the system's ground state, corresponding to scenarios where $\varepsilon_1 = 20$, $\varepsilon_1 = 0$, and $\varepsilon_1 = -20$, respectively.

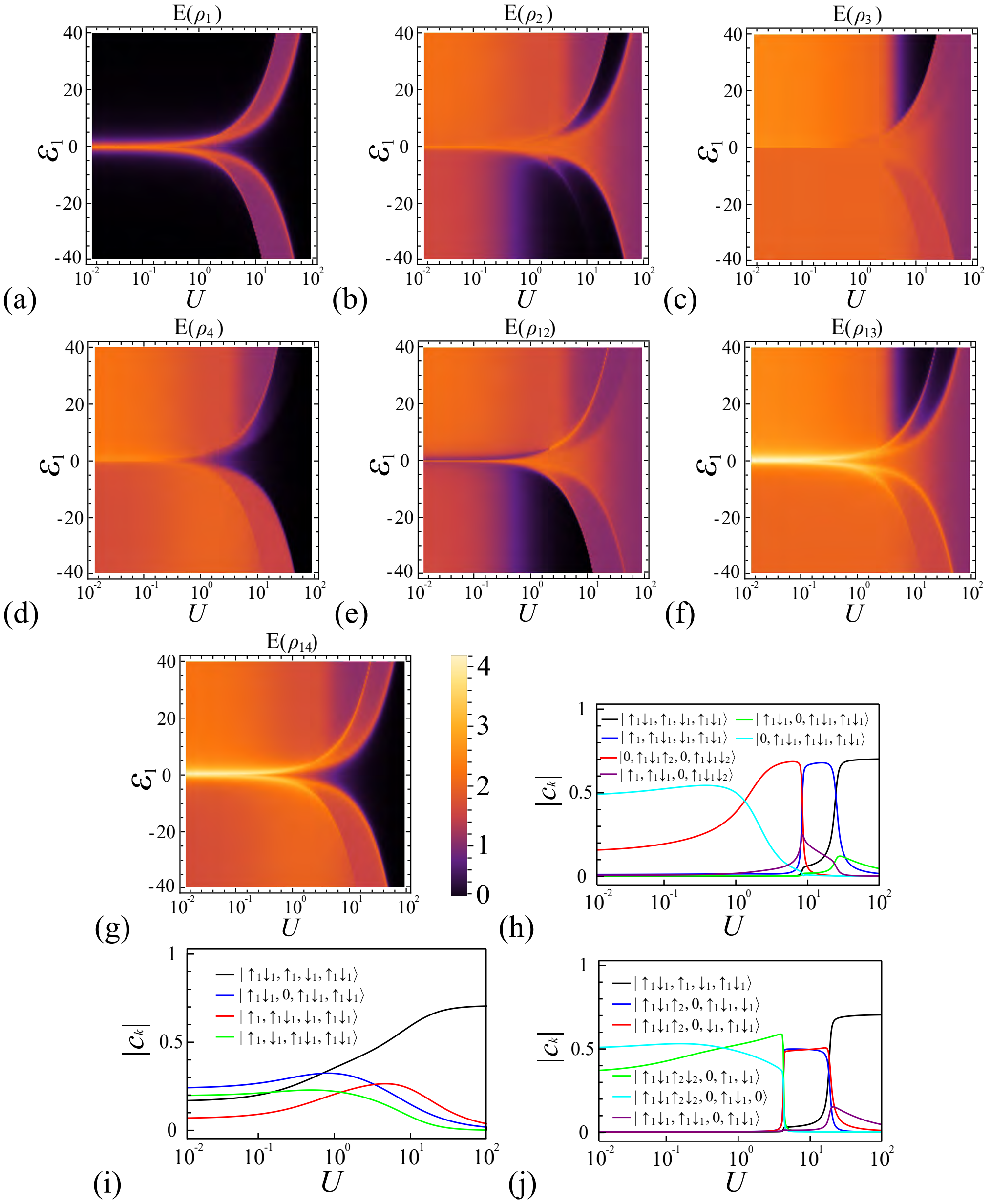

Fig. 3.8 Entanglement profiles for a four-site ($L = 4$), six-electron ($N = 6$) quantum dot system with a coupling strength of $\alpha = 0.7$. Charts are graphed according to interaction strength $U$ and potential energy $\varepsilon_1$. (a)-(d) depict local entanglement levels $E(\rho_1)$ through $E(\rho_4)$. (e)-(g) pairwise entanglement between dot pairs $E(\rho_{12})$, $E(\rho_{13})$, and $E(\rho_{14})$. (h)-(j) illustrate the dominant electron configurations in the system's ground state for $\varepsilon_1 = 20$, $\varepsilon_1 = 0$, and $\varepsilon_1 = -20$, respectively.

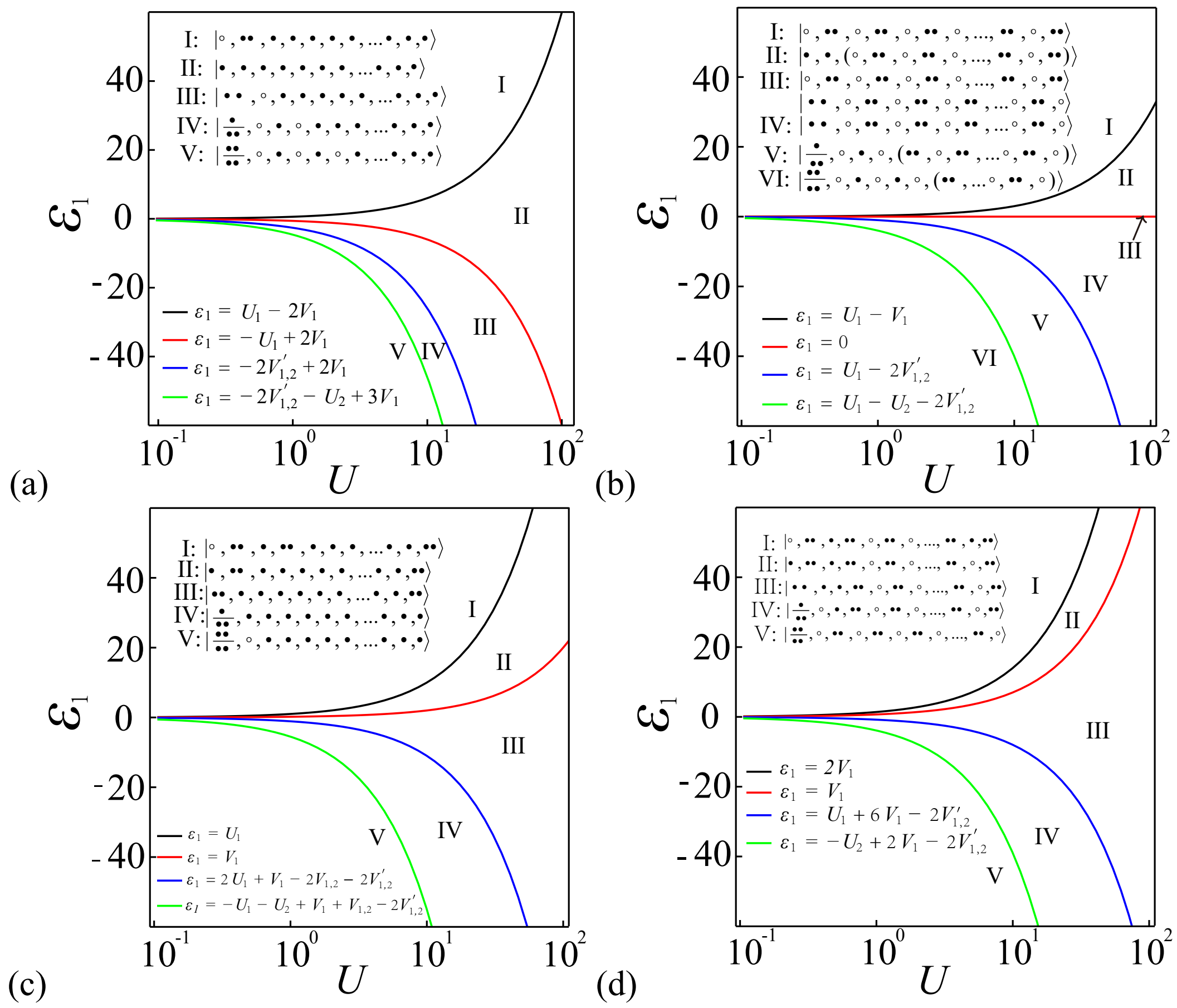

Fig. 3.9 The boundaries in the entanglement diagrams for large systems, which are derived by analyzing the energy of dominant configurations as determined by the EHM. For $N = L$, (a) $\alpha = 0.2$, (b) $\alpha = 0.7$. For $N = L + 2$, (c) $\alpha = 0.2$, (d) $\alpha = 0.7$.

# Chapter 4

# Constructing Three-Qubit Gate Pulse Sequences in Exchange-Only Spin System

## 4.1 Overview

In exchange-only (EO) qubits, single and two-qubit gates are implemented by precisely controlling exchange interactions within triple-quantum-dot (TQD) systems. Despite the progress in these foundational operations, the construction of large-scale EO qubits networks has remained largely unexplored. This study addresses several critical questions concerning the efficient construction of EO qubits in larger decoherence-free subspace(DFS) and the optimization of multi-qubit gate sequences. We present a practical methodology for EO qubits construction in a linear quantum dot spin chain system, where nine-qubit logical states form a DFS equivalent to three EO qubits. Leveraging quantum optimal control methods, we have derived optimized gate sequences, including a Toffoli gate sequence with 92 exchange pulses and 50 total time steps, significantly outperforming the conventional sequence by decomposition, which estimates requires 216 exchange pulses and 146 time steps. This optimized sequence enhances gate performance in the presence of noise and crosstalk. Furthermore, we explore the implementation of algorithms with reduced gate sequences. Our results demonstrate that this approach facilitates the practical realization of complex quantum algorithms on EO qubits, paving the way for scalable, fault-tolerant quantum computing. This research not only addresses the scalability challenges but also contributes to the robustness and efficiency of quantum operations in EO qubits systems.

## 4.2 Background

Silicon quantum dot-based spin quantum processors are emerging as a promising candidate platform due to their scalability, tunability, long-lived spin coherence, and advancements in high-fidelity qubit gate operations [264, 272, 224, 207, 24, 73, 76, 36, 250, 177, 268, 297, 35, 105, 308, 66]. Specifically, multiple spin-qubit configurations have been developed, including single-electron spin qubits, donor spin qubits, singlet-triplet spin qubits, EO spin qubits, hybrid qubits, and other alternative approaches [167, 188, 63, 226, 225, 113, 241, 232, 286]. High-fidelity gate operations are a critical aspect, and recent experiments have demonstrated universal control and fabrication of multi-quantum-dot arrays in linear, triangular, and square geometries [207, 272, 239, 3, 308, 20, 286].

Various methodologies have been investigated for achieving long-range spin coupling in quantum dot arrays, including charge transport, capacitive coupling, and spin-exchange interactions [62, 239, 77, 277]. Specifically, for EO qubits, both single-qubit and two-qubit gates can be implemented by precisely controlling the exchange interactions between adjacent spins in TQD systems [80, 302, 300, 182, 301, 270, 197, 219, 157, 111, 286].

Quantum gate implementation in EO qubits can be achieved through all-electronic control using partial swap operation sequences, which are susceptible to charge noise, crosstalk, and residual exchange interactions [117, 68, 5, 247, 121, 108]. To mitigate the influence of control pulse errors in EO qubits, numerous pulse correction schemes have been developed [74, 214, 175, 235, 49, 303, 248, 78]. Additionally, various sequence structure optimization techniques, both numerical and analytical, have been employed to identify optimal gate sequences [80, 302, 300, 182]. Utilizing genetic algorithms, Fong and Wandzura identified the most efficient exact CNOT gate sequence for linear geometries [80], which can be further analytically described using block-diagonal matrices in the effective spin particle representation [302, 300, 182]. Other optimal sequences for different quantum dot array geometries have also been discovered [233, 107]. Moreover, optimizations for the CNOT gate sequence, such as minimizing total gate operation time through reinforcement learning (RL) and reducing leakage errors, have been explored [270, 126].

Despite the significant advancements in the numerical and analytical processes for one- and two-qubit gates within the EO qubits system, the construction of large-scale quantum computing networks using EO qubits remains largely unexplored. Moreover, multi-qubit gates, such as $C^n$NOT gates, are crucial for quantum error correction and the execution of complex quantum algorithms. These considerations prompt several key questions: (1) How can EO qubits be efficiently integrated into larger DFS without compromising the generality of previously established gate sequences? (2) Is it possible to derive exact or approximate multi-qubit gate sequences based on current methodologies? (3) Are these sequences sufficiently optimized to enable the execution of complex quantum algorithms on EO qubits?

To address these questions, we first examined the construction of single-qubit gates and two-qubit gates in EO system. Then we demonstrate the effective construction of EO qubits in a linear quantum dot spin chain system, where nine-qubit logical states form a DFS equivalent to three EO qubits. In this framework, single-qubit and two-qubit gate sequences can be directly implemented in the three EO qubits systems without modification. Drawing inspiration from the Variational Quantum Algorithm (VQA), we combined the pulse-based brickwork ansatz with quantum optimal control methods, resulting in a Toffoli gate sequence consisting of 92 exchange pulses and 50 total time steps. This sequence is significantly shorter than the conventional Toffoli gate decomposition, achieving reductions of approximately 57% in exchange pulses and 66% in time steps compared to conventional methods, which require 216 exchange pulses and 146 time steps. This approach facilitates the search for the shortest exact multi-qubit gate sequences. Additionally, we compare the performance of these sequences under the influence of noise and residual exchange interactions; results show that the 92-exchange-pulse sequence is more robust. Finally, we derived several algorithm examples with shorter gate sequences.

This chapter is organized as follows: In Sections 4.3 and 4.4, we introduce the construction of single-qubit and two-qubit gates in the EO system. In Section 4.5, we present the angular momentum structure of the nine-spin system and its formation of a DFS, then examine the unitary operation structure of the two-qubit gate sequence under our angular momentum basis construction. Section 4.6 details the three-qubit gate sequence and structure based on quantum gate decomposition. In Section 4.7, we introduce the pulse-based brickwork ansatz and quantum optimal control(QOC) method used to obtain the 92-pulse Toffoli gate sequence. Then, in Section 4.8, we compare the performance of the Toffoli gate sequence obtained by gate decomposition and the 92-pulse Toffoli gate sequence under the influence of charge noise and crosstalk. Section 4.9 demonstrates that the unitary operator of a quantum algorithm can be expressed by a shorter sequence in the EO qubits system. Finally, in Section 4.10, we summarize our findings and offer concluding remarks.

## 4.3 Exchange-only Single-qubit Gates

The algebra for EO control of a triple-dot spin qubit is defined within the DFS of angular-momentum states, effectively resisting fluctuations in global magnetic fields[137, 7, 136, 63]. The total spin of the three electrons is denoted as $\mathbf{S} = \mathbf{S}_1 + \mathbf{S}_2 + \mathbf{S}_3$, where $\mathbf{S}_{12} = \mathbf{S}_1 + \mathbf{S}_2$ represents the total spin of the first two electrons, and $m$ is the projection of the total spin in an arbitrary direction, coupled to global magnetic fields. Utilizing the standard rules for angular momentum addition and Clebsch-Gordan coefficients, all eight spin states can be expressed as $|S_{12}, S; m\rangle$, where $S_{12}$ can be either 0 or 1, and $S$ can be either 1/2 or 3/2, here, we number them from $|A1\rangle$ to $|A8\rangle$:

$$
\begin{aligned}
|A1\rangle &= |S_{12} = 0, S = 1/2; m = 1/2\rangle = \frac{1}{\sqrt{2}}(|\uparrow\downarrow\uparrow\rangle - |\downarrow\uparrow\uparrow\rangle), \\
|A2\rangle &= |S_{12} = 0, S = 1/2; m = -1/2\rangle = \frac{1}{\sqrt{2}}(|\uparrow\downarrow\downarrow\rangle - |\downarrow\uparrow\downarrow\rangle), \\
|A3\rangle &= |S_{12} = 1, S = 1/2; m = 1/2\rangle = \frac{\sqrt{2}}{\sqrt{3}}|\uparrow\uparrow\downarrow\rangle - \frac{1}{\sqrt{6}}|\uparrow\downarrow\uparrow\rangle - \frac{1}{\sqrt{6}}|\downarrow\uparrow\uparrow\rangle, \\
|A4\rangle &= |S_{12} = 1, S = 1/2; m = -1/2\rangle = \frac{1}{\sqrt{6}}|\uparrow\downarrow\downarrow\rangle + \frac{1}{\sqrt{6}}|\downarrow\uparrow\downarrow\rangle - \frac{\sqrt{2}}{\sqrt{3}}|\downarrow\downarrow\uparrow\rangle, \\
|A5\rangle &= |S_{12} = 1, S = 3/2; m = 3/2\rangle = |\uparrow\uparrow\uparrow\rangle, \\
|A6\rangle &= |S_{12} = 1, S = 3/2; m = 1/2\rangle = \frac{1}{\sqrt{3}}(|\uparrow\uparrow\downarrow\rangle + |\uparrow\downarrow\uparrow\rangle + |\downarrow\uparrow\uparrow\rangle), \\
|A7\rangle &= |S_{12} = 1, S = 3/2; m = -1/2\rangle = \frac{1}{\sqrt{3}}(|\uparrow\downarrow\downarrow\rangle + |\downarrow\uparrow\downarrow\rangle + |\downarrow\downarrow\uparrow\rangle), \\
|A8\rangle &= |S_{12} = 1, S = 3/2; m = -3/2\rangle = |\downarrow\downarrow\downarrow\rangle.
\end{aligned} \tag{4.1}
$$

In these spin states, single qubit rotations can be achieved through a series of single exchange operations between neighboring spins[63], or by simultaneously activating multiple exchange interactions, such as those employed in the resonant exchange qubit[222, 179, 258, 223, 64], which maintains encoding by keeping intraqubit exchanges ”always on”[15, 235, 132]. This discussion focuses primarily on constructing gates based on serial single exchange operations.

Consider the exchange Hamiltonian:

$$H(t) = J_{12}(t)\mathbf{S_1} \cdot \mathbf{S_2} + J_{23}(t)\mathbf{S_2} \cdot \mathbf{S_3}, \tag{4.2}$$

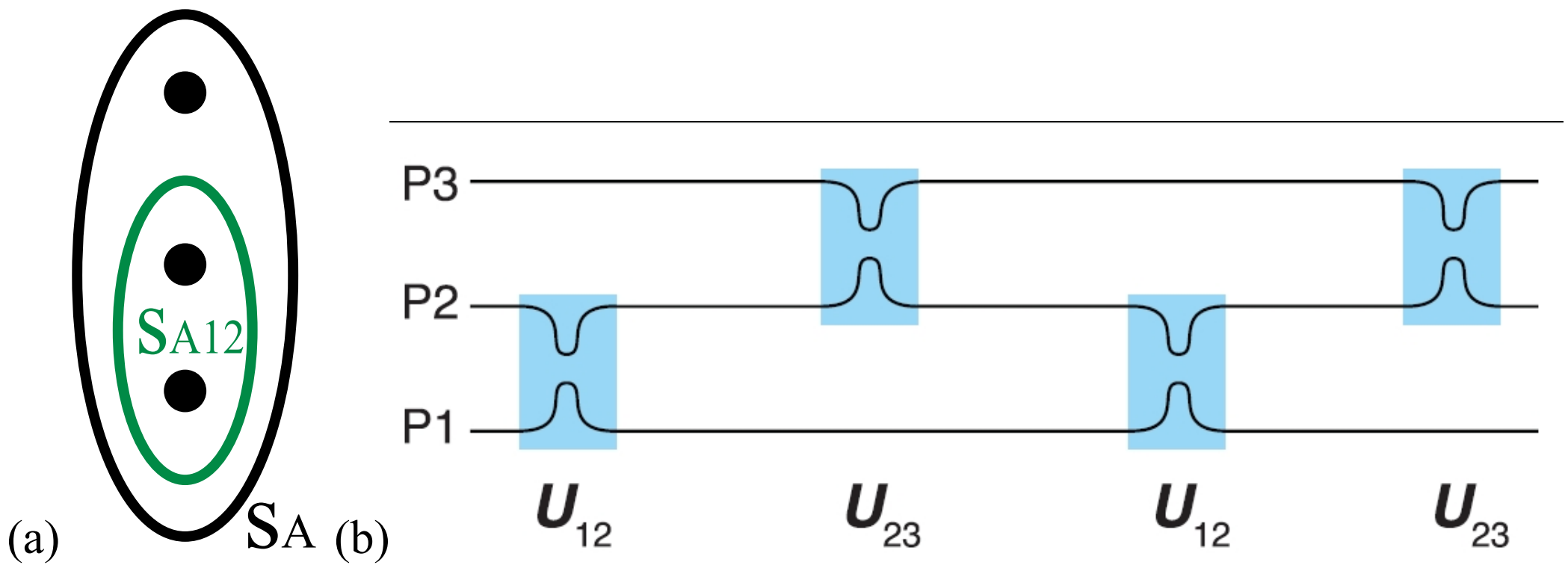

Fig. 4.1 (a)One DFS qubit in states $S_{A,1,2}$ with total spin $S_A = \frac{1}{2}$, represent by the outer oval, the inner oval represent the electrons of encoded spin state. (b)4-pulse brick structure for single qubit gate[286].

where the exchange is kept off except when pulsed, i.e., adiabatically switched on and off between pairs of spins. It is necessary to design pulse sequences that realize quantum gates on encoded qubits without leakage out of the encoded space. By pulsing this Hamiltonian for a duration $t$ with units of $\pi/J$, we obtain the time evolution operator $U_{ij}(t)$, where $ij$ represents the exchange between spin-$i$ and spin-$j$ is activated for time $t$. A sequence of four exchange pulses is sufficient to perform any single qubit gate rotation.

Figure 4.1(a) illustrates the encoding of three-qubit spins $(\bullet(\bullet\bullet)_a)_c$ (with $c = 1/2$ or $3/2$) considering the Hilbert space of the second and third electrons $(\bullet\bullet)_a$, spanned by the total spin states $a = 0$ and 1. Figure 4.1(b) shows the structure of the 4-pulse exchange operations. Using simple numerical search methods, we can determine the time values for the 4-pulse exchange time evolution $U(t_1, t_2, t_3, t_4) = U_{12}(t_1)U_{23}(t_2)U_{12}(t_3)U_{23}(t_4)$. Table 4.1 provides the pulse time parameters for single qubit gates used in the subsequent sections. Notably, literature provides a table of solutions for single-qubit Clifford operations[5], with an average exchange-pulse count of 2.7. A more complex matrix representation of the operation can be constructed with a five-pulse sequence[300] with specific durations.

## 4.4 Exchange-only Two-qubit Gates with Two DFS Qubits

### 4.4.1 Angular Momentum Structure of Six Spin DFS System

One commonly used method for implementing a two-qubit gate in an EO system involves an EO sequence among six spins arranged in specific spatial configurations.

| Gate | $U_{12}(t_1)$ | $U_{23}(t_2)$ | $U_{12}(t_3)$ | $U_{23}(t_4)$ |
|---|---|---|---|---|
| $H$ | 0.524218 | 1.09632 | 0.903682 | 0.475782 |
| $T$ | 1.08429 | 0.165715 | 1.08429 | 1.91571 |
| $T^\dagger$ | 0.915715 | 1.83429 | 0.915715 | 0.0842851 |
| $S$ | 1.17426 | 0.325739 | 1.17426 | 1.82574 |
| $S^\dagger$ | 0.825739 | 1.67426 | 0.825739 | 0.174261 |
| $X$ | 1.54944 | 1.19321 | 0.806788 | 1.45056 |
| $\sqrt{X}$ | 0.808687 | 0.447258 | 1.05274 | 0.191313 |
| $Z$ | 1.39183 | 0.608173 | 1.39183 | 1.60817 |

Table 4.1 Table of single-qubit gate sequence, adopted with a 4-pulse brick structure, obtained by the numerical search.

Employing a Genetic Algorithm, Fong and Wandzura [80] derived a sequential, gauge-independent CNOT gate sequence by encoding the system with a total angular momentum basis defined by six quantum numbers: $S_{A,B}$, $m_{A,B}$, $S_A$, $S_B$, $S_{A,1,2}$, and $S_{B,1,2}$. This sequence can also be analytically derived through specific constructions[302, 300, 301]. $S_{A,B}$ denotes the total spin of the system(DFS qubits A and B), while $m_{A,B}$ represents the total z-component of the system's spin. $S_A$ and $S_B$ correspond to the total spin of the DFS EO three-spin qubits A and B, respectively. $S_{A,1,2}$ and $S_{B,1,2}$ denote the spin of the two qubits that encode the logical information in DFS qubits A and B. The 64 basis states of the six-spin system, encoded based on these six quantum numbers, are detailed in Appendix 4.11.1. Within the total angular momentum basis, the exchange operator is represented as a block diagonal matrix consisting of one $5 \times 5$ spin-0 block, three identical $9 \times 9$ spin-1 blocks, five identical $5 \times 5$ spin-2 blocks, and a spin-3 block. The quantum numbers for the two DFS qubits are summarized in table 4.2, table 4.3 and table 4.4.

### 4.4.2 Two-qubit Gate Sequence and Matrix Representation

In this subsection, we adopt the notation introduced in the literature[302, 300, 301], where each spin in the six-dot array is represented by the symbol •, and groups of spins are enclosed in ovals labeled by their total spin. The three-spin qubit states are defined as $|a\rangle = (\bullet(\bullet\bullet)_a)_{1/2}$ with $a = 0$ or 1, while $|NC\rangle = (\bullet(\bullet\bullet)_a)_{3/2}$ are non-computational states.

| Spin-0 | 1 | 2 | 3 | 4 | 5 |
|---|---|---|---|---|---|
| $S_{A,B}$ | 0 | 0 | 0 | 0 | 0 |
| $m_{A,B}$ | 0 | 0 | 0 | 0 | 0 |
| $S_A$ | $\frac{1}{2}$ | $\frac{1}{2}$ | $\frac{1}{2}$ | $\frac{1}{2}$ | $\frac{3}{2}$ |
| $S_B$ | $\frac{1}{2}$ | $\frac{1}{2}$ | $\frac{1}{2}$ | $\frac{1}{2}$ | $\frac{3}{2}$ |
| $S_{A,1,2}$ | 0 | 0 | 1 | 1 | 1 |
| $S_{B,1,2}$ | 0 | 1 | 0 | 1 | 1 |

Table 4.2 Quantum numbers for two DFS qubits in the total spin-0 subspace: $S_{A,B}$ (total spin of six physical qubits), $m_{A,B}$ (total spin-z), $S_A$ and $S_B$ (total spins of DFS qubits A and B), $S_{A,1,2}$ and $S_{B,1,2}$ (spins of first two qubits of DFS qubits A and B). The top row indicates basis vector indices in the total angular momentum basis. Basis vectors 1–4 are valid encoded states; basis vector 5 is a leaked state.

| Spin-1 | 6 | 7 | 8 | 9 | 10 | 11 | 12 | 13 | 14 |
|---|---|---|---|---|---|---|---|---|---|
| $S_{A,B}$ | 1 | 1 | 1 | 1 | 1 | 1 | 1 | 1 | 1 |
| $m_{A,B}$ | 1 | 1 | 1 | 1 | 1 | 1 | 1 | 1 | 1 |
| $S_A$ | $\frac{1}{2}$ | $\frac{1}{2}$ | $\frac{1}{2}$ | $\frac{1}{2}$ | $\frac{1}{2}$ | $\frac{1}{2}$ | $\frac{3}{2}$ | $\frac{3}{2}$ | $\frac{3}{2}$ |
| $S_B$ | $\frac{1}{2}$ | $\frac{1}{2}$ | $\frac{1}{2}$ | $\frac{1}{2}$ | $\frac{3}{2}$ | $\frac{3}{2}$ | $\frac{1}{2}$ | $\frac{1}{2}$ | $\frac{3}{2}$ |
| $S_{A,1,2}$ | 0 | 0 | 1 | 1 | 0 | 1 | 1 | 1 | 1 |
| $S_{B,1,2}$ | 0 | 1 | 0 | 1 | 1 | 1 | 0 | 1 | 1 |

Table 4.3 Quantum numbers for two DFS qubits in the total spin-1, basis vectors 6–9 are valid encoded states, basis vectors 10 and 11 are unleaked in DFS qubit A, 12 and 13 are unleaked in DFS qubit B.

| Spin-2 | 15 | 16 | 17 | 18 | 19 |
|---|---|---|---|---|---|
| $S_{A,B}$ | 2 | 2 | 2 | 2 | 2 |
| $m_{A,B}$ | 2 | 2 | 2 | 2 | 2 |
| $S_A$ | $\frac{1}{2}$ | $\frac{1}{2}$ | $\frac{3}{2}$ | $\frac{3}{2}$ | $\frac{3}{2}$ |
| $S_B$ | $\frac{3}{2}$ | $\frac{3}{2}$ | $\frac{1}{2}$ | $\frac{1}{2}$ | $\frac{3}{2}$ |
| $S_{A,1,2}$ | 0 | 1 | 1 | 1 | 1 |
| $S_{B,1,2}$ | 1 | 1 | 0 | 1 | 1 |

Table 4.4 Quantum numbers for two DFS qubits in the total spin-2, basis vectors 15 and 16 are unleaked in DFS qubit A, 17 and 18 are unleaked in DFS qubit B.

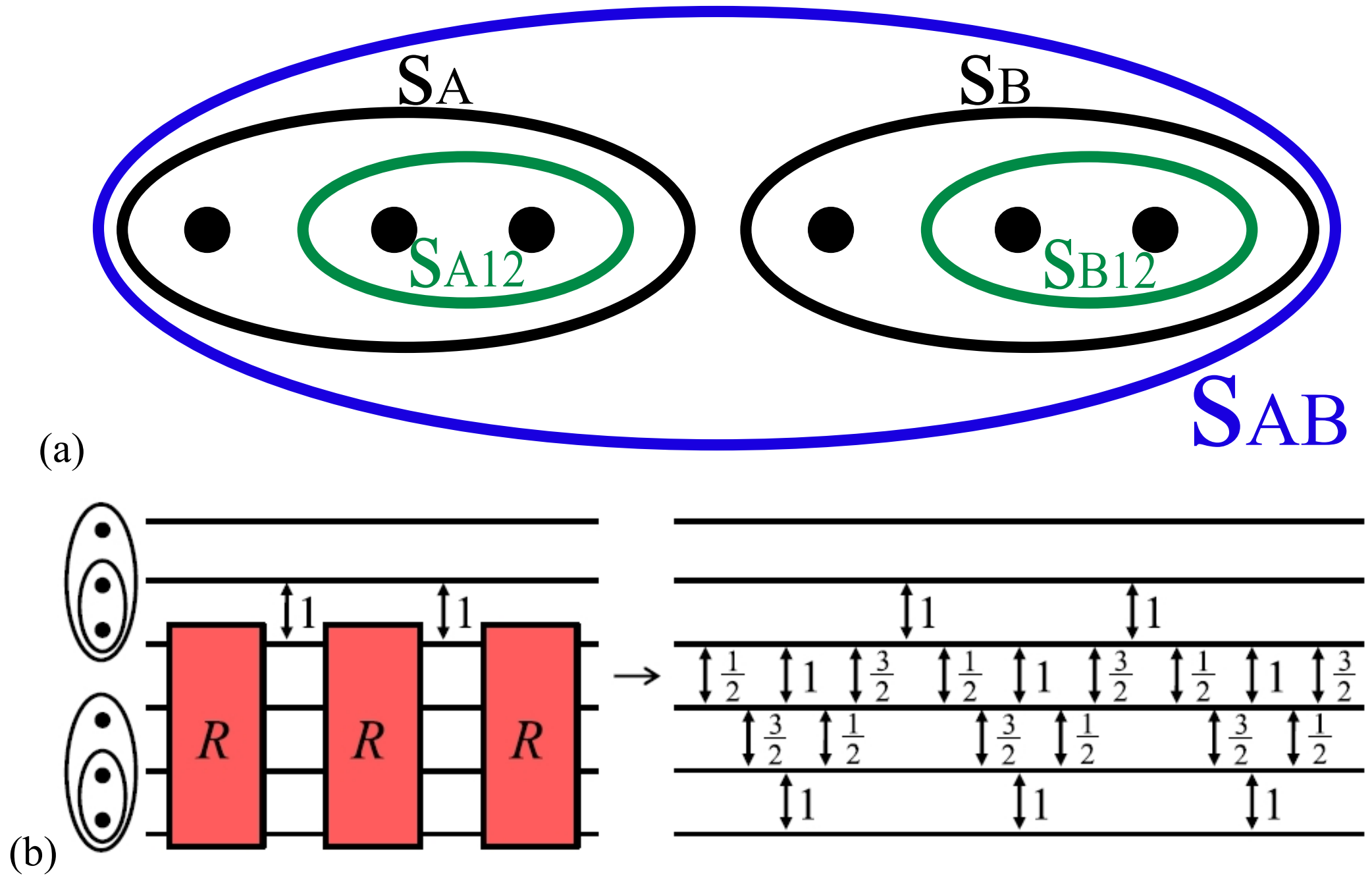

Fig. 4.2 (a) Two three-spin EO qubits in states $S_{A,1,2}$ and $S_{B,1,2}$, with total spin $S_A = S_B = \frac{1}{2}$, $S_{AB} = 0, 1$. (b) CNOT gate sequence in [300], with four local rotations omitted.

Under this notation, a basis change is required to transition from the standard three spin-1/2 qubit basis $|a\rangle = (\bullet(\bullet\bullet)_a)_{1/2}$ to $|a\rangle = ((\bullet\bullet)_{a'}\bullet)_{1/2}$ with $a' = 0$ or $1$. Fong and Wandzura[80] identified a 22-pulse CNOT gate sequence for a pair of three-spin qubits with state labels $a$ and $b$ in the bases $((\bullet(\bullet\bullet)_a)_{1/2}((\bullet\bullet)\bullet_b)_{1/2})$, which is also equivalent to the sequence derived by Daniel Zeuch and N. E. Bonesteel[301] in the bases $((\bullet(\bullet\bullet)_{a'})_{1/2}(\bullet(\bullet\bullet)_{b'})_{1/2})$. Additional two-qubit gate sequences have been discovered, including leakage-controlled, low gate time, and approximate gate sequences[270, 126].

Expressing the exchange unitaries of the CNOT gate sequence[300], the sequence is shown in Figure 4.2, in the total angular momentum basis and computing their product yields the following matrix for the spin-0 subspace:

$$e^{i\theta_1}\begin{pmatrix} 1 & 0 & 0 & 0 & 0 \\ 0 & 1 & 0 & 0 & 0 \\ 0 & 0 & 0 & 1 & 0 \\ 0 & 0 & 1 & 0 & 0 \\ 0 & 0 & 0 & 0 & e^{i\theta_0} \end{pmatrix},$$

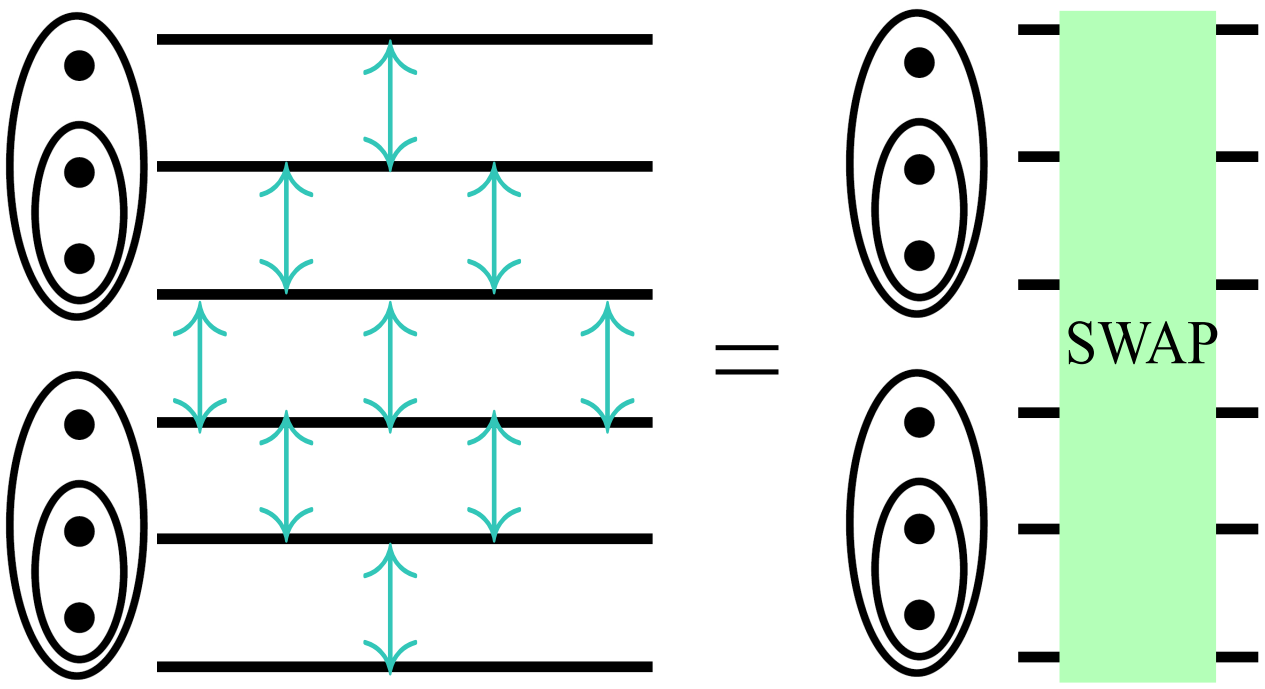

Fig. 4.3 SWAP gate sequence with nine pulses.

and for the spin-1 subspace:

$$e^{i\theta_1}\begin{pmatrix} 1 & 0 & 0 & 0 & 0 & 0 & 0 & 0 & 0 \\ 0 & 1 & 0 & 0 & 0 & 0 & 0 & 0 & 0 \\ 0 & 0 & 0 & 1 & 0 & 0 & 0 & 0 & 0 \\ 0 & 0 & 1 & 0 & 0 & 0 & 0 & 0 & 0 \\ 0 & 0 & 0 & 0 & -\frac{11}{16} & -\frac{5\sqrt{3}}{16} & 0 & 0 & -\frac{\sqrt{15}}{8} \\ 0 & 0 & 0 & 0 & -\frac{5\sqrt{3}}{16} & -\frac{1}{16} & 0 & 0 & \frac{3\sqrt{5}}{8} \\ 0 & 0 & 0 & 0 & 0 & 0 & 0 & 1 & 0 \\ 0 & 0 & 0 & 0 & 0 & 0 & 1 & 0 & 0 \\ 0 & 0 & 0 & 0 & -\frac{\sqrt{15}}{8} & \frac{3\sqrt{5}}{8} & 0 & 0 & -\frac{1}{4} \end{pmatrix}.$$

Weinstein et al. propose a 15 $\pi$-pulse SWAP gate that changes the qubit indices from $((\bullet_1\bullet_2\bullet_3)(\bullet_4\bullet_5\bullet_6))$ to $((\bullet_6\bullet_5\bullet_4)(\bullet_3\bullet_2\bullet_1))$[286]. Here, shown in Figure 4.3, we employ a 9 $\pi$-pulse SWAP sequence with the matrix sequence:

$$U_{\text{swap}}^{9-\text{pulse}} = U_{34}(1)U_{23}(1)U_{12}(1)U_{45}(1)U_{34}(1)U_{23}(1)U_{56}(1)U_{45}(1)U_{34}(1). \quad (4.3)$$

This sequence change the qubit indices from $((\bullet_1\bullet_2\bullet_3)(\bullet_4\bullet_5\bullet_6))$ to $((\bullet_4\bullet_5\bullet_6)(\bullet_1\bullet_2\bullet_3))$, resulting in the following matrix for the spin-0 subspace:

$$e^{i\theta_2}\begin{pmatrix} 1 & 0 & 0 & 0 & 0 \\ 0 & 0 & 1 & 0 & 0 \\ 0 & 1 & 0 & 0 & 0 \\ 0 & 0 & 0 & 1 & 0 \\ 0 & 0 & 0 & 0 & 1 \end{pmatrix},$$

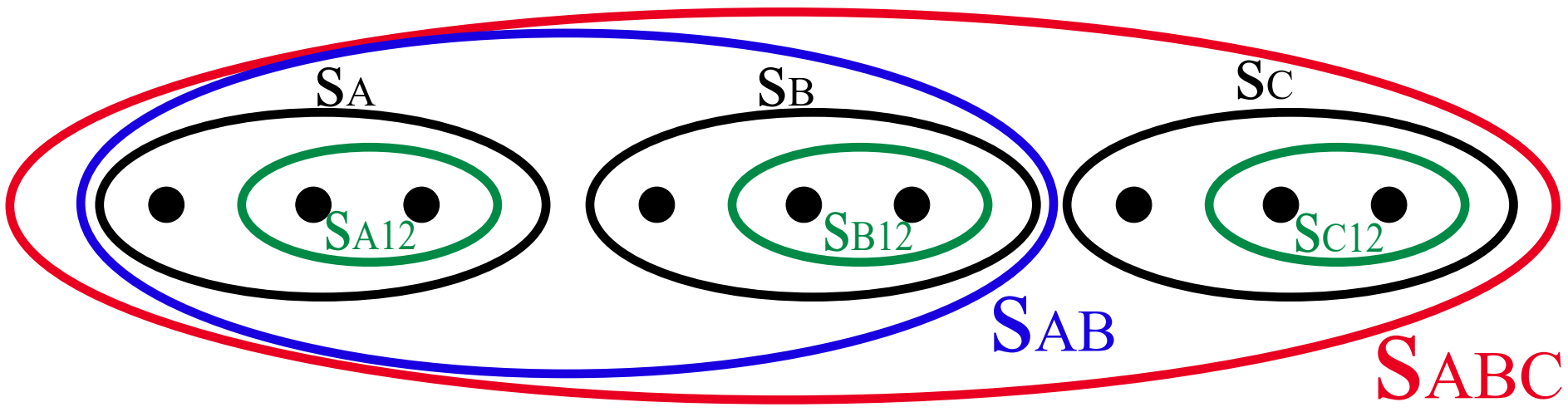

Fig. 4.4 In an intuitive way, three three-spin EO qubits in states $S_{A,1,2}$, $S_{B,1,2}$ and $S_{C,1,2}$, with effective total spin $S_A = S_B = S_C = \frac{1}{2}$, $S_{AB} = 0, 1$ and $S_{ABC} = \frac{1}{2}, \frac{3}{2}$.

and for the spin-1 subspace:

$$e^{i\theta_3}\begin{pmatrix} 1 & 0 & 0 & 0 & 0 & 0 & 0 & 0 & 0 \\ 0 & 0 & 1 & 0 & 0 & 0 & 0 & 0 & 0 \\ 0 & 1 & 0 & 0 & 0 & 0 & 0 & 0 & 0 \\ 0 & 0 & 0 & 1 & 0 & 0 & 0 & 0 & 0 \\ 0 & 0 & 0 & 0 & 0 & 0 & 1 & 0 & 0 \\ 0 & 0 & 0 & 0 & 0 & 0 & 0 & 1 & 0 \\ 0 & 0 & 0 & 0 & 1 & 0 & 0 & 0 & 0 \\ 0 & 0 & 0 & 0 & 0 & 1 & 0 & 0 & 0 \\ 0 & 0 & 0 & 0 & 0 & 0 & 0 & 0 & 1 \end{pmatrix}.$$

For the CNOT gate, the spin-0 and spin-1 subspaces share the same global phase $e^{i\theta_1}$. In contrast, for the SWAP gate, the spin-0 and spin-1 subspaces have different global phases $e^{i\theta_2}$ and $e^{i\theta_3}$, respectively.

## 4.5 Angular Momentum Structure of DFS in Nine Spin System

### 4.5.1 Bases

To implement a three-qubit gate within the EO qubits system, a minimum of three EO qubits, which correspond to nine physical qubits, is required, as shown in Figure 4.4. For encoding these three EO qubits, the quantum numbers of the nine-spin system are employed to maintain the block diagonal structure and ensure the applicability of previously identified sequences in the new basis. Consequently, both the total spin $S$ and its z-component $m$ are crucial. This approach parallels the treatment of two EO qubits as described by Fong and Wandzura[80], where the quantum numbers for the total spin and the spins of each DFS qubit $A$, $B$, and $C$ are essential; specifically, $S_A$, $S_B$, $S_C$, $S_{A,1,2}$, $S_{B,1,2}$, and $S_{C,1,2}$. For the entire system, $S_{A,B,C}$ (total spin of nine physical qubits) and $m_{A,B,C}$ (total spin-z) are required to describe the Hilbert space. Furthermore, the quantum number $S_{A,B}$, which represents the total spin of the first two EO qubits $A$ and $B$, plays a crucial role as an intermediate spin system. It is essential for preserving the angular momentum structure established by Fong and Wandzura[80] within the $S_{A,B,C}$ blocks that describe the entire system.

| Spin-1/2 | 1 | 2 | 3 | 4 | 5 | 6 | 7 | 8 | 9 | 10 |
|---|---|---|---|---|---|---|---|---|---|---|
| $S_{A,B,C}$ | $\frac{1}{2}$ | $\frac{1}{2}$ | $\frac{1}{2}$ | $\frac{1}{2}$ | $\frac{1}{2}$ | $\frac{1}{2}$ | $\frac{1}{2}$ | $\frac{1}{2}$ | $\frac{1}{2}$ | $\frac{1}{2}$ |
| $m_{A,B,C}$ | $\frac{1}{2}$ | $\frac{1}{2}$ | $\frac{1}{2}$ | $\frac{1}{2}$ | $\frac{1}{2}$ | $\frac{1}{2}$ | $\frac{1}{2}$ | $\frac{1}{2}$ | $\frac{1}{2}$ | $\frac{1}{2}$ |
| $S_{A,B}$ | 0 | 0 | 0 | 0 | 0 | 0 | 0 | 0 | 0 | 0 |
| $S_A$ | $\frac{1}{2}$ | $\frac{1}{2}$ | $\frac{1}{2}$ | $\frac{1}{2}$ | $\frac{1}{2}$ | $\frac{1}{2}$ | $\frac{1}{2}$ | $\frac{1}{2}$ | $\frac{3}{2}$ | $\frac{3}{2}$ |
| $S_B$ | $\frac{1}{2}$ | $\frac{1}{2}$ | $\frac{1}{2}$ | $\frac{1}{2}$ | $\frac{1}{2}$ | $\frac{1}{2}$ | $\frac{1}{2}$ | $\frac{1}{2}$ | $\frac{3}{2}$ | $\frac{3}{2}$ |
| $S_C$ | $\frac{1}{2}$ | $\frac{1}{2}$ | $\frac{1}{2}$ | $\frac{1}{2}$ | $\frac{1}{2}$ | $\frac{1}{2}$ | $\frac{1}{2}$ | $\frac{1}{2}$ | $\frac{1}{2}$ | $\frac{1}{2}$ |
| $S_{A,1,2}$ | 0 | 0 | 0 | 0 | 1 | 1 | 1 | 1 | 1 | 1 |
| $S_{B,1,2}$ | 0 | 0 | 1 | 1 | 0 | 0 | 1 | 1 | 1 | 1 |
| $S_{C,1,2}$ | 0 | 1 | 0 | 1 | 0 | 1 | 0 | 1 | 0 | 1 |

Table 4.5 Quantum numbers for three DFS qubits in the total spin-1/2, basis vectors 1–8 are valid encoded states, basis vectors 9 and 10 are unleaked in DFS qubit C, but leaked in DFS qubit A and DFS qubit B.

| Spin-1/2 | 11 | 12 | 13 | 14 | 15 | 16 | 17 | 18 | 19 | 20 | 21 | 22 | 23 | 24 | 25 | 26 | 27 | 28 |
|---|---|---|---|---|---|---|---|---|---|---|---|---|---|---|---|---|---|---|
| $S_{A,B,C}$ | $\frac{1}{2}$ | $\frac{1}{2}$ | $\frac{1}{2}$ | $\frac{1}{2}$ | $\frac{1}{2}$ | $\frac{1}{2}$ | $\frac{1}{2}$ | $\frac{1}{2}$ | $\frac{1}{2}$ | $\frac{1}{2}$ | $\frac{1}{2}$ | $\frac{1}{2}$ | $\frac{1}{2}$ | $\frac{1}{2}$ | $\frac{1}{2}$ | $\frac{1}{2}$ | $\frac{1}{2}$ | $\frac{1}{2}$ |
| $m_{A,B,C}$ | $\frac{1}{2}$ | $\frac{1}{2}$ | $\frac{1}{2}$ | $\frac{1}{2}$ | $\frac{1}{2}$ | $\frac{1}{2}$ | $\frac{1}{2}$ | $\frac{1}{2}$ | $\frac{1}{2}$ | $\frac{1}{2}$ | $\frac{1}{2}$ | $\frac{1}{2}$ | $\frac{1}{2}$ | $\frac{1}{2}$ | $\frac{1}{2}$ | $\frac{1}{2}$ | $\frac{1}{2}$ | $\frac{1}{2}$ |
| $S_{A,B}$ | 1 | 1 | 1 | 1 | 1 | 1 | 1 | 1 | 1 | 1 | 1 | 1 | 1 | 1 | 1 | 1 | 1 | 1 |
| $S_A$ | $\frac{1}{2}$ | $\frac{1}{2}$ | $\frac{1}{2}$ | $\frac{1}{2}$ | $\frac{1}{2}$ | $\frac{1}{2}$ | $\frac{1}{2}$ | $\frac{1}{2}$ | $\frac{1}{2}$ | $\frac{1}{2}$ | $\frac{1}{2}$ | $\frac{1}{2}$ | $\frac{3}{2}$ | $\frac{3}{2}$ | $\frac{3}{2}$ | $\frac{3}{2}$ | $\frac{3}{2}$ | $\frac{3}{2}$ |
| $S_B$ | $\frac{1}{2}$ | $\frac{1}{2}$ | $\frac{1}{2}$ | $\frac{1}{2}$ | $\frac{1}{2}$ | $\frac{1}{2}$ | $\frac{1}{2}$ | $\frac{1}{2}$ | $\frac{3}{2}$ | $\frac{3}{2}$ | $\frac{3}{2}$ | $\frac{3}{2}$ | $\frac{1}{2}$ | $\frac{1}{2}$ | $\frac{1}{2}$ | $\frac{1}{2}$ | $\frac{3}{2}$ | $\frac{3}{2}$ |
| $S_C$ | $\frac{1}{2}$ | $\frac{1}{2}$ | $\frac{1}{2}$ | $\frac{1}{2}$ | $\frac{1}{2}$ | $\frac{1}{2}$ | $\frac{1}{2}$ | $\frac{1}{2}$ | $\frac{1}{2}$ | $\frac{1}{2}$ | $\frac{1}{2}$ | $\frac{1}{2}$ | $\frac{1}{2}$ | $\frac{1}{2}$ | $\frac{1}{2}$ | $\frac{1}{2}$ | $\frac{1}{2}$ | $\frac{1}{2}$ |
| $S_{A,1,2}$ | 0 | 0 | 0 | 0 | 1 | 1 | 1 | 1 | 0 | 0 | 1 | 1 | 1 | 1 | 1 | 1 | 1 | 1 |
| $S_{B,1,2}$ | 0 | 0 | 1 | 1 | 0 | 0 | 1 | 1 | 1 | 1 | 1 | 1 | 0 | 0 | 1 | 1 | 1 | 1 |
| $S_{C,1,2}$ | 0 | 1 | 0 | 1 | 0 | 1 | 0 | 1 | 0 | 1 | 0 | 1 | 0 | 1 | 0 | 1 | 0 | 1 |

Table 4.6 Quantum numbers for three DFS qubits in the total spin-1/2, basis vectors 11–18 are valid encoded states, basis vectors 19-28 are unleaked in DFS qubit C, while leaked in DFS qubit A or DFS qubit B.

| Spin-1/2 | 29 | 30 | 31 | 32 | 33 | 34 | 35 | 36 | 37 | 38 | 39 | 40 | 41 | 42 |
|---|---|---|---|---|---|---|---|---|---|---|---|---|---|---|
| $S_{A,B,C}$ | $\frac{1}{2}$ | $\frac{1}{2}$ | $\frac{1}{2}$ | $\frac{1}{2}$ | $\frac{1}{2}$ | $\frac{1}{2}$ | $\frac{1}{2}$ | $\frac{1}{2}$ | $\frac{1}{2}$ | $\frac{1}{2}$ | $\frac{1}{2}$ | $\frac{1}{2}$ | $\frac{1}{2}$ | $\frac{1}{2}$ |
| $m_{A,B,C}$ | $\frac{1}{2}$ | $\frac{1}{2}$ | $\frac{1}{2}$ | $\frac{1}{2}$ | $\frac{1}{2}$ | $\frac{1}{2}$ | $\frac{1}{2}$ | $\frac{1}{2}$ | $\frac{1}{2}$ | $\frac{1}{2}$ | $\frac{1}{2}$ | $\frac{1}{2}$ | $\frac{1}{2}$ | $\frac{1}{2}$ |
| $S_{A,B}$ | 1 | 1 | 1 | 1 | 1 | 1 | 1 | 1 | 1 | 2 | 2 | 2 | 2 | 2 |
| $S_A$ | $\frac{1}{2}$ | $\frac{1}{2}$ | $\frac{1}{2}$ | $\frac{1}{2}$ | $\frac{1}{2}$ | $\frac{1}{2}$ | $\frac{3}{2}$ | $\frac{3}{2}$ | $\frac{3}{2}$ | $\frac{1}{2}$ | $\frac{1}{2}$ | $\frac{3}{2}$ | $\frac{3}{2}$ | $\frac{3}{2}$ |
| $S_B$ | $\frac{1}{2}$ | $\frac{1}{2}$ | $\frac{1}{2}$ | $\frac{1}{2}$ | $\frac{3}{2}$ | $\frac{3}{2}$ | $\frac{1}{2}$ | $\frac{1}{2}$ | $\frac{3}{2}$ | $\frac{3}{2}$ | $\frac{3}{2}$ | $\frac{1}{2}$ | $\frac{1}{2}$ | $\frac{3}{2}$ |
| $S_C$ | $\frac{3}{2}$ | $\frac{3}{2}$ | $\frac{3}{2}$ | $\frac{3}{2}$ | $\frac{3}{2}$ | $\frac{3}{2}$ | $\frac{3}{2}$ | $\frac{3}{2}$ | $\frac{3}{2}$ | $\frac{3}{2}$ | $\frac{3}{2}$ | $\frac{3}{2}$ | $\frac{3}{2}$ | $\frac{3}{2}$ |
| $S_{A,1,2}$ | 0 | 0 | 1 | 1 | 0 | 1 | 1 | 1 | 1 | 0 | 1 | 1 | 1 | 1 |
| $S_{B,1,2}$ | 0 | 1 | 0 | 1 | 1 | 1 | 0 | 1 | 1 | 1 | 1 | 0 | 1 | 1 |
| $S_{C,1,2}$ | 1 | 1 | 1 | 1 | 1 | 1 | 1 | 1 | 1 | 1 | 1 | 1 | 1 | 1 |

Table 4.7 Quantum numbers for three DFS qubits in the total spin-1/2, basis vectors 29–42 are leaked in DFS qubit C.

| Spin-3/2 | 43 | 44 | 45 | 46 | 47 | 48 | 49 | 50 | 51 | 52 | 53 | 54 | 55 | 56 | 57 | 58 | 59 | 60 |
|---|---|---|---|---|---|---|---|---|---|---|---|---|---|---|---|---|---|---|
| $S_{A,B,C}$ | $\frac{3}{2}$ | $\frac{3}{2}$ | $\frac{3}{2}$ | $\frac{3}{2}$ | $\frac{3}{2}$ | $\frac{3}{2}$ | $\frac{3}{2}$ | $\frac{3}{2}$ | $\frac{3}{2}$ | $\frac{3}{2}$ | $\frac{3}{2}$ | $\frac{3}{2}$ | $\frac{3}{2}$ | $\frac{3}{2}$ | $\frac{3}{2}$ | $\frac{3}{2}$ | $\frac{3}{2}$ | $\frac{3}{2}$ |
| $m_{A,B,C}$ | $\frac{3}{2}$ | $\frac{3}{2}$ | $\frac{3}{2}$ | $\frac{3}{2}$ | $\frac{3}{2}$ | $\frac{3}{2}$ | $\frac{3}{2}$ | $\frac{3}{2}$ | $\frac{3}{2}$ | $\frac{3}{2}$ | $\frac{3}{2}$ | $\frac{3}{2}$ | $\frac{3}{2}$ | $\frac{3}{2}$ | $\frac{3}{2}$ | $\frac{3}{2}$ | $\frac{3}{2}$ | $\frac{3}{2}$ |
| $S_{A,B}$ | 1 | 1 | 1 | 1 | 1 | 1 | 1 | 1 | 1 | 1 | 1 | 1 | 1 | 1 | 1 | 1 | 1 | 1 |
| $S_A$ | $\frac{1}{2}$ | $\frac{1}{2}$ | $\frac{1}{2}$ | $\frac{1}{2}$ | $\frac{1}{2}$ | $\frac{1}{2}$ | $\frac{1}{2}$ | $\frac{1}{2}$ | $\frac{1}{2}$ | $\frac{1}{2}$ | $\frac{1}{2}$ | $\frac{1}{2}$ | $\frac{3}{2}$ | $\frac{3}{2}$ | $\frac{3}{2}$ | $\frac{3}{2}$ | $\frac{3}{2}$ | $\frac{3}{2}$ |
| $S_B$ | $\frac{1}{2}$ | $\frac{1}{2}$ | $\frac{1}{2}$ | $\frac{1}{2}$ | $\frac{1}{2}$ | $\frac{1}{2}$ | $\frac{1}{2}$ | $\frac{1}{2}$ | $\frac{3}{2}$ | $\frac{3}{2}$ | $\frac{3}{2}$ | $\frac{3}{2}$ | $\frac{1}{2}$ | $\frac{1}{2}$ | $\frac{1}{2}$ | $\frac{1}{2}$ | $\frac{3}{2}$ | $\frac{3}{2}$ |
| $S_C$ | $\frac{1}{2}$ | $\frac{1}{2}$ | $\frac{1}{2}$ | $\frac{1}{2}$ | $\frac{1}{2}$ | $\frac{1}{2}$ | $\frac{1}{2}$ | $\frac{1}{2}$ | $\frac{1}{2}$ | $\frac{1}{2}$ | $\frac{1}{2}$ | $\frac{1}{2}$ | $\frac{1}{2}$ | $\frac{1}{2}$ | $\frac{1}{2}$ | $\frac{1}{2}$ | $\frac{1}{2}$ | $\frac{1}{2}$ |
| $S_{A,1,2}$ | 0 | 0 | 0 | 0 | 1 | 1 | 1 | 1 | 0 | 0 | 1 | 1 | 1 | 1 | 1 | 1 | 1 | 1 |
| $S_{B,1,2}$ | 0 | 0 | 1 | 1 | 0 | 0 | 1 | 1 | 1 | 1 | 1 | 1 | 0 | 0 | 1 | 1 | 1 | 1 |
| $S_{C,1,2}$ | 0 | 1 | 0 | 1 | 0 | 1 | 0 | 1 | 0 | 1 | 0 | 1 | 0 | 1 | 0 | 1 | 0 | 1 |

Table 4.8 Quantum numbers for three DFS qubits in the total spin-3/2, basis vectors 43–50 are valid encoded states, basis vectors 43-60 are unleaked in DFS qubit C, while leaked in DFS qubit A or DFS qubit B. These bases have corresponding quantum numbers to total spin-1/2 basis vectors 11-28.

Using these nine quantum numbers, the structure of the Hamiltonian $512 \times 512$ matrix, written in the basis $|S_{A,B,C}, m_{A,B,C}, S_{A,B}, S_A, S_B, S_C, S_{A,1,2}, S_{B,1,2}, S_{C,1,2}\rangle$, is block diagonalized:

| Spin-3/2 | 61 | 62 | 63 | 64 | 65 | 66 | 67 | 68 | 69 | 70 |
|---|---|---|---|---|---|---|---|---|---|---|
| $S_{A,B,C}$ | $\frac{3}{2}$ | $\frac{3}{2}$ | $\frac{3}{2}$ | $\frac{3}{2}$ | $\frac{3}{2}$ | $\frac{3}{2}$ | $\frac{3}{2}$ | $\frac{3}{2}$ | $\frac{3}{2}$ | $\frac{3}{2}$ |
| $m_{A,B,C}$ | $\frac{3}{2}$ | $\frac{3}{2}$ | $\frac{3}{2}$ | $\frac{3}{2}$ | $\frac{3}{2}$ | $\frac{3}{2}$ | $\frac{3}{2}$ | $\frac{3}{2}$ | $\frac{3}{2}$ | $\frac{3}{2}$ |
| $S_{A,B}$ | 2 | 2 | 2 | 2 | 2 | 2 | 2 | 2 | 2 | 2 |
| $S_A$ | $\frac{1}{2}$ | $\frac{1}{2}$ | $\frac{1}{2}$ | $\frac{1}{2}$ | $\frac{3}{2}$ | $\frac{3}{2}$ | $\frac{3}{2}$ | $\frac{3}{2}$ | $\frac{3}{2}$ | $\frac{3}{2}$ |
| $S_B$ | $\frac{3}{2}$ | $\frac{3}{2}$ | $\frac{3}{2}$ | $\frac{3}{2}$ | $\frac{1}{2}$ | $\frac{1}{2}$ | $\frac{1}{2}$ | $\frac{1}{2}$ | $\frac{3}{2}$ | $\frac{3}{2}$ |
| $S_C$ | $\frac{1}{2}$ | $\frac{1}{2}$ | $\frac{1}{2}$ | $\frac{1}{2}$ | $\frac{1}{2}$ | $\frac{1}{2}$ | $\frac{1}{2}$ | $\frac{1}{2}$ | $\frac{1}{2}$ | $\frac{1}{2}$ |
| $S_{A,1,2}$ | 0 | 0 | 1 | 1 | 1 | 1 | 1 | 1 | 1 | 1 |
| $S_{B,1,2}$ | 1 | 1 | 1 | 1 | 0 | 0 | 1 | 1 | 1 | 1 |
| $S_{C,1,2}$ | 0 | 1 | 0 | 1 | 0 | 1 | 0 | 1 | 0 | 1 |

Table 4.9 Quantum numbers for three DFS qubits in the total spin-3/2, basis vectors 61–70 are leaked states.

$$H_{9spin} = \left(\begin{array}{c|c|c|c} S_{A,B,C} = 1/2 & \mathbf{0} & \mathbf{0} & \cdots \\ \hline \mathbf{0} & S_{A,B,C} = 3/2 & \mathbf{0} & \cdots \\ \hline \mathbf{0} & \mathbf{0} & S_{A,B,C} = 5/2 & \cdots \\ \hline \vdots & \vdots & \vdots & \ddots \end{array}\right).$$

Specifically, the bases for $S_{A,B,C} = 1/2$ and $S_{A,B,C} = 3/2$ are detailed in Appendix 4.11.2. The quantum numbers $S_{A,B,C} = 1/2$ for the three DFS qubits are summarized in table 4.5, table 4.6 and table 4.7. The quantum numbers $S_{A,B,C} = 3/2$ for the three DFS qubits are summarized in table 4.8, table 4.9 and table 4.10.

Any effective DFS qubit unitary operation should encompass two $8 \times 8$ three-DFS unleaked blocks within the $S_{A,B,C} = 1/2$ $42 \times 42$ block and one $8 \times 8$ three-DFS unleaked block within the $S_{A,B,C} = 3/2$ $48 \times 48$ block:

$$U^{\text{Operation}}_{S_{A,B,C}=1/2,3/2} = \left(\begin{array}{c|c|c|c|c|c} U^{\text{unleak}}_{8\times8} & \mathbf{0} & \mathbf{0} & \mathbf{0} & \mathbf{0} & \mathbf{0} \\ \hline \mathbf{0} & U^{\text{leak}}_{2\times2} & \mathbf{0} & U^{\text{leak}}_{2\times24} & \mathbf{0} & \mathbf{0} \\ \hline \mathbf{0} & \mathbf{0} & U^{\text{unleak}}_{8\times8} & \mathbf{0} & \mathbf{0} & \mathbf{0} \\ \hline \mathbf{0} & U^{\text{leak}}_{24\times2} & \mathbf{0} & U^{\text{leak}}_{24\times24} & \mathbf{0} & \mathbf{0} \\ \hline \mathbf{0} & \mathbf{0} & \mathbf{0} & \mathbf{0} & U^{\text{unleak}}_{8\times8} & \mathbf{0} \\ \hline \mathbf{0} & \mathbf{0} & \mathbf{0} & \mathbf{0} & \mathbf{0} & U^{\text{leak}}_{40\times40} \end{array}\right).$$

| Spin-3/2 | 71 | 72 | 73 | 74 | 75 | 76 | 77 | 78 | 79 | 80 |
|---|---|---|---|---|---|---|---|---|---|---|
| $S_{A,B,C}$ | $\frac{3}{2}$ | $\frac{3}{2}$ | $\frac{3}{2}$ | $\frac{3}{2}$ | $\frac{3}{2}$ | $\frac{3}{2}$ | $\frac{3}{2}$ | $\frac{3}{2}$ | $\frac{3}{2}$ | $\frac{3}{2}$ |
| $m_{A,B,C}$ | $\frac{3}{2}$ | $\frac{3}{2}$ | $\frac{3}{2}$ | $\frac{3}{2}$ | $\frac{3}{2}$ | $\frac{3}{2}$ | $\frac{3}{2}$ | $\frac{3}{2}$ | $\frac{3}{2}$ | $\frac{3}{2}$ |
| $S_{A,B}$ | 0 | 0 | 0 | 0 | 0 | 1 | 1 | 1 | 1 | 1 |
| $S_A$ | $\frac{1}{2}$ | $\frac{1}{2}$ | $\frac{1}{2}$ | $\frac{1}{2}$ | $\frac{3}{2}$ | $\frac{1}{2}$ | $\frac{1}{2}$ | $\frac{1}{2}$ | $\frac{1}{2}$ | $\frac{1}{2}$ |
| $S_B$ | $\frac{1}{2}$ | $\frac{1}{2}$ | $\frac{1}{2}$ | $\frac{1}{2}$ | $\frac{3}{2}$ | $\frac{1}{2}$ | $\frac{1}{2}$ | $\frac{1}{2}$ | $\frac{1}{2}$ | $\frac{3}{2}$ |
| $S_C$ | $\frac{3}{2}$ | $\frac{3}{2}$ | $\frac{3}{2}$ | $\frac{3}{2}$ | $\frac{3}{2}$ | $\frac{3}{2}$ | $\frac{3}{2}$ | $\frac{3}{2}$ | $\frac{3}{2}$ | $\frac{3}{2}$ |
| $S_{A,1,2}$ | 0 | 0 | 1 | 1 | 1 | 0 | 0 | 1 | 1 | 0 |
| $S_{B,1,2}$ | 0 | 1 | 0 | 1 | 1 | 0 | 1 | 0 | 1 | 1 |
| $S_{C,1,2}$ | 1 | 1 | 1 | 1 | 1 | 1 | 1 | 1 | 1 | 1 |

| Spin-3/2 | 81 | 82 | 83 | 84 | 85 | 86 | 87 | 88 | 89 | 90 |
|---|---|---|---|---|---|---|---|---|---|---|
| $S_{A,B,C}$ | $\frac{3}{2}$ | $\frac{3}{2}$ | $\frac{3}{2}$ | $\frac{3}{2}$ | $\frac{3}{2}$ | $\frac{3}{2}$ | $\frac{3}{2}$ | $\frac{3}{2}$ | $\frac{3}{2}$ | $\frac{3}{2}$ |
| $m_{A,B,C}$ | $\frac{3}{2}$ | $\frac{3}{2}$ | $\frac{3}{2}$ | $\frac{3}{2}$ | $\frac{3}{2}$ | $\frac{3}{2}$ | $\frac{3}{2}$ | $\frac{3}{2}$ | $\frac{3}{2}$ | $\frac{3}{2}$ |
| $S_{A,B}$ | 1 | 1 | 1 | 1 | 2 | 2 | 2 | 2 | 2 | 3 |
| $S_A$ | $\frac{1}{2}$ | $\frac{3}{2}$ | $\frac{3}{2}$ | $\frac{3}{2}$ | $\frac{1}{2}$ | $\frac{1}{2}$ | $\frac{3}{2}$ | $\frac{3}{2}$ | $\frac{3}{2}$ | $\frac{3}{2}$ |
| $S_B$ | $\frac{3}{2}$ | $\frac{1}{2}$ | $\frac{1}{2}$ | $\frac{3}{2}$ | $\frac{3}{2}$ | $\frac{3}{2}$ | $\frac{1}{2}$ | $\frac{1}{2}$ | $\frac{3}{2}$ | $\frac{3}{2}$ |
| $S_C$ | $\frac{3}{2}$ | $\frac{3}{2}$ | $\frac{3}{2}$ | $\frac{3}{2}$ | $\frac{3}{2}$ | $\frac{3}{2}$ | $\frac{3}{2}$ | $\frac{3}{2}$ | $\frac{3}{2}$ | $\frac{3}{2}$ |
| $S_{A,1,2}$ | 1 | 1 | 1 | 1 | 0 | 1 | 1 | 1 | 1 | 1 |
| $S_{B,1,2}$ | 1 | 0 | 1 | 1 | 1 | 1 | 1 | 0 | 1 | 1 |
| $S_{C,1,2}$ | 1 | 1 | 1 | 1 | 1 | 1 | 1 | 1 | 1 | 1 |

Table 4.10 Quantum numbers for three DFS qubits in the total spin-3/2, basis vectors 71–90 are leaked states.

Here, any effective DFS three-qubit operations are encoded in $U^{\text{unleak}}_{8\times 8}$.

### 4.5.2 Linear arrangement

We have Hamiltonian and operation:

$$H_{ij}(t) = J_{ij}(t)\mathbf{S_i} \cdot \mathbf{S_j}, \tag{4.4}$$

$$U_{ij}(t) = exp(-iH_{ij}(t)t_c). \tag{4.5}$$

Follow the spin arrangement and sequence obtained by Daniel Zeuch and N.E. Bonesteel, in this work, we consider the linear arrangement and let the three DFS qubits are connected end to end and face the same direction $((\bullet(\bullet\bullet)_a)_{1/2}(\bullet(\bullet\bullet)_b)_{1/2}(\bullet(\bullet\bullet)_c)_{1/2})$.

Therefore in $S_{A,B,C} = 1/2$ 42 × 42 block and $S_{A,B,C} = 3/2$ 48 × 48 block, we have the matrix representation of the Hamiltonian $H_{ij}^{90\times 90}$ below:

$$\begin{aligned}H_{12}^{90\times 90} = \text{diag}(&A_1 \otimes I_4, \mathbf{0}_2, A_1 \otimes I_4, A_1 \otimes I_2, \mathbf{0}_6, A_1 \otimes I_2, A_1, \mathbf{0}_3, A_1, \mathbf{0}_3,\\ &A_1 \otimes I_4, A_1 \otimes I_2, \mathbf{0}_6, A_1 \otimes I_2, \mathbf{0}_6, A_1 \otimes I_2, \mathbf{0}_1, A_1 \otimes I_2, \mathbf{0}_3, A_1, \mathbf{0}_4)\end{aligned} \tag{4.6}$$

$$\begin{aligned}H_{23}^{90\times 90} = \text{diag}(&-I_4, \mathbf{0}_6, -I_4, \mathbf{0}_4, -I_2, \mathbf{0}_8, -I_2, \mathbf{0}_2, -I_1, \mathbf{0}_4,\\ &-I_1, \mathbf{0}_4, -I_4, \mathbf{0}_4, -I_2, \mathbf{0}_8, -I_2, \mathbf{0}_8, -I_2, \mathbf{0}_3, -I_2, \mathbf{0}_3,\\ &-I_2, \mathbf{0}_2, -I_1, \mathbf{0}_4, -I_1, \mathbf{0}_5)\end{aligned} \tag{4.7}$$

$$H_{34}^{90\times 90} = \text{diag}(A_2 \otimes I_2, A_3 \otimes I_2, A_3, A_4, A_3 \otimes I_2, A_4 \otimes I_2, A_2, A_3, A_4, \mathbf{0}_1) \tag{4.8}$$

$$\begin{aligned}H_{45}^{90\times 90} = \text{diag}(&I_2 \otimes (A_1 \otimes I_2), \mathbf{0}_2, I_2 \otimes (A_1 \otimes I_2), \mathbf{0}_4, A_1 \otimes I_2,\\ &\mathbf{0}_2, I_2 \otimes A_1, \mathbf{0}_2, A_1, \mathbf{0}_3, A_1, \mathbf{0}_1, I_2 \otimes (A_1 \otimes I_2),\\ &\mathbf{0}_4, A_1 \otimes I_2, \mathbf{0}_6, A_1 \otimes I_2, \mathbf{0}_2, I_2 \otimes A_1, \mathbf{0}_1, I_2 \otimes A_1,\\ &\mathbf{0}_2, A_1, \mathbf{0}_3, A_1, \mathbf{0}_2)\end{aligned} \tag{4.9}$$

$$\begin{aligned}H_{56}^{90\times 90} = \text{diag}(&-I_2, \mathbf{0}_2, -I_2, \mathbf{0}_4, -I_2, \mathbf{0}_2, -I_2, \mathbf{0}_6, -I_2, \mathbf{0}_4,\\ &-I_1, \mathbf{0}_1, -I_1, \mathbf{0}_3, -I_1, \mathbf{0}_4, -I_1, \mathbf{0}_2, -I_2, \mathbf{0}_2, -I_2, \mathbf{0}_6,\\ &-I_2, \mathbf{0}_8, -I_2, \mathbf{0}_4, -I_1, \mathbf{0}_1, -I_1, \mathbf{0}_2, -I_1, \mathbf{0}_1, -I_1, \mathbf{0}_3,\\ &-I_1, \mathbf{0}_4, -I_1, \mathbf{0}_3)\end{aligned} \tag{4.10}$$

$$H_{67}^{90\times 90} = \text{diag}(A_5, A_6) \tag{4.11}$$

$$H_{78}^{90\times 90} = \text{diag}(I_{14} \otimes A_1, \mathbf{0}_{14}, I_{14} \otimes A_1, \mathbf{0}_{20}) \tag{4.12}$$

$$H_{89}^{90\times 90} = \text{diag}(I_{14} \otimes A_7, \mathbf{0}_{14}, I_{14} \otimes A_7, \mathbf{0}_{20}) \tag{4.13}$$

$$A_1=\begin{pmatrix} -\frac{1}{4} & -\frac{\sqrt{3}}{4} \\ -\frac{\sqrt{3}}{4} & -\frac{3}{4} \end{pmatrix}, A_2 = \begin{pmatrix} -\frac{1}{4} & 0 & -\frac{\sqrt{3}}{4} & 0 & 0 \\ 0 & -\frac{1}{4} & 0 & \frac{1}{4\sqrt{3}} & -\frac{1}{\sqrt{6}} \\ -\frac{\sqrt{3}}{4} & 0 & -\frac{3}{4} & 0 & 0 \\ 0 & \frac{1}{4\sqrt{3}} & 0 & -\frac{1}{12} & \frac{1}{3\sqrt{2}} \\ 0 & -\frac{1}{\sqrt{6}} & 0 & \frac{1}{3\sqrt{2}} & -\frac{2}{3} \end{pmatrix},$$

$$A_3 = \begin{pmatrix} -\frac{1}{4} & 0 & \frac{1}{4\sqrt{3}} & 0 & 0 & 0 & \frac{1}{\sqrt{6}} & 0 & 0 \\ 0 & -\frac{1}{4} & 0 & -\frac{1}{12\sqrt{3}} & 0 & \frac{\sqrt{2/3}}{3} & 0 & -\frac{1}{3\sqrt{6}} & -\frac{\sqrt{5/6}}{3} \\ \frac{1}{4\sqrt{3}} & 0 & -\frac{1}{12} & 0 & 0 & 0 & -\frac{1}{3\sqrt{2}} & 0 & 0 \\ 0 & -\frac{1}{12\sqrt{3}} & 0 & -\frac{11}{36} & \frac{\sqrt{2/3}}{3} & \frac{2\sqrt{2}}{9} & 0 & \frac{1}{9\sqrt{2}} & \frac{\sqrt{5/2}}{9} \\ 0 & 0 & 0 & \frac{\sqrt{2/3}}{3} & -\frac{1}{4} & -\frac{5}{12\sqrt{3}} & 0 & -\frac{1}{6\sqrt{3}} & -\frac{\sqrt{5/3}}{6} \\ 0 & \frac{\sqrt{2/3}}{3} & 0 & \frac{2\sqrt{2}}{9} & -\frac{5}{12\sqrt{3}} & -\frac{19}{36} & 0 & \frac{1}{18} & \frac{\sqrt{5}}{18} \\ \frac{1}{\sqrt{6}} & 0 & -\frac{1}{3\sqrt{2}} & 0 & 0 & 0 & -\frac{2}{3} & 0 & 0 \\ 0 & -\frac{1}{3\sqrt{6}} & 0 & \frac{1}{9\sqrt{2}} & -\frac{1}{6\sqrt{3}} & \frac{1}{18} & 0 & -\frac{1}{9} & -\frac{\sqrt{5}}{9} \\ 0 & -\frac{\sqrt{5/6}}{3} & 0 & \frac{\sqrt{5/2}}{9} & -\frac{\sqrt{5/3}}{6} & \frac{\sqrt{5}}{18} & 0 & -\frac{\sqrt{5}}{9} & -\frac{5}{9} \end{pmatrix},$$

$$A_4 = \begin{pmatrix} -\frac{1}{4} & \frac{1}{4\sqrt{3}} & 0 & \frac{1}{2\sqrt{3}} & \frac{1}{2\sqrt{3}} \\ \frac{1}{4\sqrt{3}} & -\frac{1}{12} & 0 & -\frac{1}{6} & -\frac{1}{6} \\ 0 & 0 & 0 & 0 & 0 \\ \frac{1}{2\sqrt{3}} & -\frac{1}{6} & 0 & -\frac{1}{3} & -\frac{1}{3} \\ \frac{1}{2\sqrt{3}} & -\frac{1}{6} & 0 & -\frac{1}{3} & -\frac{1}{3} \end{pmatrix}, A_7=\begin{pmatrix} -1 & 0 \\ 0 & 0 \end{pmatrix},$$ $I_n$ is $n$-dimensional identity matrix, $\mathbf{0}_n$ is $n$-dimensional zero matrix. Here, $A_5$ is 42×42 block, and $A_6$ is 48×48 block is not given due to its size.

### 4.5.3 CNOT Gate Structure

We have CNOT operation[300] between the first DFS qubit A and the second DFS qubit B(A is control qubit) shown in Figure 4.2(a), and the sequence can be expressed as:

$$U^{\mathrm{CNOT_{AB}}} = U_{45}(p1)U_{56}(p2)R_{\mathrm{AB}}U_{23}(1)R_{\mathrm{AB}}U_{23}(1)R_{\mathrm{AB}}U_{56}(2-p2)U_{45}(2-p1), \tag{4.14}$$

where $R_{\mathrm{AB}} = U_{34}(0.5)U_{45}(1.5)U_{34}(0.5)U_{56}(1)U_{45}(0.5)U_{34}(1.5)$, $p1$=ArcCos$(2\sqrt{3}/3-1)/\pi - 1$, $p2$=ArcSin$(2\sqrt{3}/3-1)/\pi$.

For $S_{A,B,C} = 1/2$ block, we have

$$U_{42\times 42}^{\mathrm{CNOT_{AB}}} = \left(\begin{array}{c|c|c|c|c} U_{8\times 8}^{\mathrm{CNOT_{AB}}} & \mathbf{0} & \mathbf{0} & \mathbf{0} & \mathbf{0} \\ \hline \mathbf{0} & -I_2 & \mathbf{0} & \mathbf{0} & \mathbf{0} \\ \hline \mathbf{0} & \mathbf{0} & U_{8\times 8}^{\mathrm{CNOT_{AB}}} & \mathbf{0} & \mathbf{0} \\ \hline \mathbf{0} & \mathbf{0} & \mathbf{0} & U_{10\times 10}^{\mathrm{CNOT_{AB}}} & \mathbf{0} \\ \hline \mathbf{0} & \mathbf{0} & \mathbf{0} & \mathbf{0} & U_{14\times 14}^{\mathrm{CNOT_{AB}}} \end{array}\right),$$

where

$$U_{8\times 8}^{\mathrm{CNOT_{AB}}} = \begin{pmatrix} 1 & 0 & 0 & 0 & 0 & 0 & 0 & 0 \\ 0 & 1 & 0 & 0 & 0 & 0 & 0 & 0 \\ 0 & 0 & 1 & 0 & 0 & 0 & 0 & 0 \\ 0 & 0 & 0 & 1 & 0 & 0 & 0 & 0 \\ 0 & 0 & 0 & 0 & 0 & 0 & 1 & 0 \\ 0 & 0 & 0 & 0 & 0 & 0 & 0 & 1 \\ 0 & 0 & 0 & 0 & 1 & 0 & 0 & 0 \\ 0 & 0 & 0 & 0 & 0 & 1 & 0 & 0 \end{pmatrix},$$

$$U_{10\times 10}^{\mathrm{CNOT_{AB}}} = \begin{pmatrix} -\frac{11}{16} & 0 & \frac{5\sqrt{3}}{16} & 0 & 0 & 0 & 0 & 0 & \frac{\sqrt{15}}{8} & 0 \\ 0 & -\frac{11}{16} & 0 & \frac{5\sqrt{3}}{16} & 0 & 0 & 0 & 0 & 0 & \frac{\sqrt{15}}{8} \\ \frac{5\sqrt{3}}{16} & 0 & -\frac{1}{16} & 0 & 0 & 0 & 0 & 0 & \frac{3\sqrt{5}}{8} & 0 \\ 0 & \frac{5\sqrt{3}}{16} & 0 & -\frac{1}{16} & 0 & 0 & 0 & 0 & 0 & \frac{3\sqrt{5}}{8} \\ 0 & 0 & 0 & 0 & 0 & 0 & 1 & 0 & 0 & 0 \\ 0 & 0 & 0 & 0 & 0 & 0 & 0 & 1 & 0 & 0 \\ 0 & 0 & 0 & 0 & 1 & 0 & 0 & 0 & 0 & 0 \\ 0 & 0 & 0 & 0 & 0 & 1 & 0 & 0 & 0 & 0 \\ \frac{\sqrt{15}}{8} & 0 & \frac{3\sqrt{5}}{8} & 0 & 0 & 0 & 0 & 0 & -\frac{1}{4} & 0 \\ 0 & \frac{\sqrt{15}}{8} & 0 & \frac{3\sqrt{5}}{8} & 0 & 0 & 0 & 0 & 0 & -\frac{1}{4} \end{pmatrix},$$

$$
U^{\mathrm{CNOT_{AB}}}_{14\times 14} = \left(\begin{array}{cccc|ccccc|ccccc}
1 & 0 & 0 & 0 & 0 & 0 & 0 & 0 & 0 & 0 & 0 & 0 & 0 & 0 \\
0 & 1 & 0 & 0 & 0 & 0 & 0 & 0 & 0 & 0 & 0 & 0 & 0 & 0 \\
0 & 0 & 0 & 1 & 0 & 0 & 0 & 0 & 0 & 0 & 0 & 0 & 0 & 0 \\
0 & 0 & 1 & 0 & 0 & 0 & 0 & 0 & 0 & 0 & 0 & 0 & 0 & 0 \\
\hline
0 & 0 & 0 & 0 & -\frac{11}{16} & \frac{5\sqrt{3}}{16} & 0 & 0 & \frac{\sqrt{15}}{8} & 0 & 0 & 0 & 0 & 0 \\
0 & 0 & 0 & 0 & \frac{5\sqrt{3}}{16} & -\frac{1}{16} & 0 & 0 & \frac{3\sqrt{5}}{8} & 0 & 0 & 0 & 0 & 0 \\
0 & 0 & 0 & 0 & 0 & 0 & 0 & 1 & 0 & 0 & 0 & 0 & 0 & 0 \\
0 & 0 & 0 & 0 & 0 & 0 & 1 & 0 & 0 & 0 & 0 & 0 & 0 & 0 \\
0 & 0 & 0 & 0 & \frac{\sqrt{15}}{8} & \frac{3\sqrt{5}}{8} & 0 & 0 & -\frac{1}{4} & 0 & 0 & 0 & 0 & 0 \\
\hline
0 & 0 & 0 & 0 & 0 & 0 & 0 & 0 & 0 & -\frac{11}{16} & -\frac{3\sqrt{3}}{16} & 0 & 0 & -\frac{3\sqrt{3}}{8} \\
0 & 0 & 0 & 0 & 0 & 0 & 0 & 0 & 0 & -\frac{3\sqrt{3}}{16} & \frac{15}{16} & 0 & 0 & -\frac{1}{8} \\
0 & 0 & 0 & 0 & 0 & 0 & 0 & 0 & 0 & 0 & 0 & 0 & 1 & 0 \\
0 & 0 & 0 & 0 & 0 & 0 & 0 & 0 & 0 & 0 & 0 & 1 & 0 & 0 \\
0 & 0 & 0 & 0 & 0 & 0 & 0 & 0 & 0 & -\frac{3\sqrt{3}}{8} & -\frac{1}{8} & 0 & 0 & \frac{3}{4}
\end{array}\right).
$$

The effective CNOT gate between DFS qubit A and DFS qubit B is $U^{\mathrm{CNOT_{AB}}}_{8\times 8}$. The first $U^{\mathrm{CNOT_{AB}}}_{8\times 8}$ block and $(-I_2)$ formed the spin-1/2 and $S_{A,B} = 0$ subspace, $I_2$ is two-dimensional identity matrix. The second $U^{\mathrm{CNOT_{AB}}}_{8\times 8}$ block, with $U^{\mathrm{CNOT_{AB}}}_{10\times 10}$ and $U^{\mathrm{CNOT_{AB}}}_{14\times 14}$ form the spin-1/2 and $S_{A,B} = 1$ subspace.

For the CNOT operation between the second DFS B qubit and the third DFS qubit C (where the second qubit is the control qubit), by equivalently shifting the previous sequence of DFS qubits A and B to DFS qubits B and C, we have:

$$
U^{\mathrm{CNOT_{BC}}} = U_{78}(p1)U_{89}(p2)R_{\mathrm{BC}}U_{56}(1)R_{\mathrm{BC}}U_{56}(1)R_{\mathrm{BC}}U_{89}(2-p2)U_{78}(2-p1), \tag{4.15}
$$

where $R_{\mathrm{BC}} = U_{67}(0.5)U_{78}(1.5)U_{67}(0.5)U_{89}(1)U_{78}(0.5)U_{67}(1.5), p1$=ArcCos$(2\sqrt{3}/3-1)/\pi - 1, p2$=ArcSin$(2\sqrt{3}/3-1)/\pi$.

For $S_{A,B,C} = 1/2$ block, we have

$$U_{42\times 42}^{\text{CNOT}_{\text{BC}}} = \left(\begin{array}{c|c|c|c|c} U_{8\times 8}^{\text{CNOT}_{\text{BC}}} & \mathbf{0} & \mathbf{0} & \mathbf{0} & \mathbf{0} \\ \hline \mathbf{0} & X & \mathbf{0} & \mathbf{0} & \mathbf{0} \\ \hline \mathbf{0} & \mathbf{0} & U_{8\times 8}^{\text{CNOT}_{\text{BC}}} & \mathbf{0} & \mathbf{0} \\ \hline \mathbf{0} & \mathbf{0} & \mathbf{0} & U_{10\times 10}^{\text{CNOT}_{\text{BC}}} & \mathbf{0} \\ \hline \mathbf{0} & \mathbf{0} & \mathbf{0} & \mathbf{0} & U_{14\times 14}^{\text{CNOT}_{\text{BC}}} \end{array}\right),$$

where $X$ is pauli-x matrix, and

$$U_{8\times 8}^{\text{CNOT}_{\text{BC}}} = \begin{pmatrix} 1 & 0 & 0 & 0 & 0 & 0 & 0 & 0 \\ 0 & 1 & 0 & 0 & 0 & 0 & 0 & 0 \\ 0 & 0 & 0 & 1 & 0 & 0 & 0 & 0 \\ 0 & 0 & 1 & 0 & 0 & 0 & 0 & 0 \\ 0 & 0 & 0 & 0 & 1 & 0 & 0 & 0 \\ 0 & 0 & 0 & 0 & 0 & 1 & 0 & 0 \\ 0 & 0 & 0 & 0 & 0 & 0 & 0 & 1 \\ 0 & 0 & 0 & 0 & 0 & 0 & 1 & 0 \end{pmatrix},$$

$$U_{10\times 10}^{\text{CNOT}_{\text{BC}}} = \begin{pmatrix} 0 & 1 & 0 & 0 & 0 & 0 & 0 & 0 & 0 & 0 \\ 1 & 0 & 0 & 0 & 0 & 0 & 0 & 0 & 0 & 0 \\ 0 & 0 & 0 & 1 & 0 & 0 & 0 & 0 & 0 & 0 \\ 0 & 0 & 1 & 0 & 0 & 0 & 0 & 0 & 0 & 0 \\ 0 & 0 & 0 & 0 & 1 & 0 & 0 & 0 & 0 & 0 \\ 0 & 0 & 0 & 0 & 0 & 1 & 0 & 0 & 0 & 0 \\ 0 & 0 & 0 & 0 & 0 & 0 & 0 & 1 & 0 & 0 \\ 0 & 0 & 0 & 0 & 0 & 0 & 1 & 0 & 0 & 0 \\ 0 & 0 & 0 & 0 & 0 & 0 & 0 & 0 & 0 & 1 \\ 0 & 0 & 0 & 0 & 0 & 0 & 0 & 0 & 1 & 0 \end{pmatrix},$$

### 4.5.4 SWAP Gate Structure

The SWAP gate sequence between the first DFS qubit and the second DFS qubit is

$$U^{\text{SWAP}_{\text{AB}}} = U_{34}(1)U_{23}(1)U_{12}(1)U_{45}(1)U_{34}(1)U_{23}(1)U_{56}(1)U_{45}(1)U_{34}(1), \quad (4.16)$$

$$U^{\mathrm{CNOT_{BC}}}_{14\times14} = \begin{pmatrix}
\frac{-11}{16} & \frac{5\sqrt{3}}{16} & 0 & 0 & \frac{5\sqrt{3}}{16\sqrt{2}} & 0 & 0 & 0 & 0 & \frac{-3\sqrt{5}}{16\sqrt{2}} & 0 & 0 & 0 & 0 \\
\frac{5\sqrt{3}}{16} & \frac{-1}{16} & 0 & 0 & \frac{15}{16\sqrt{2}} & 0 & 0 & 0 & 0 & \frac{-3\sqrt{15}}{16\sqrt{2}} & 0 & 0 & 0 & 0 \\
0 & 0 & \frac{-11}{16} & \frac{5\sqrt{3}}{16} & 0 & \frac{5\sqrt{3}}{16\sqrt{2}} & 0 & 0 & 0 & 0 & \frac{-3\sqrt{5}}{16\sqrt{2}} & 0 & 0 & 0 \\
0 & 0 & \frac{5\sqrt{3}}{16} & \frac{-1}{16} & 0 & \frac{15}{16\sqrt{2}} & 0 & 0 & 0 & 0 & \frac{-3\sqrt{15}}{16\sqrt{2}} & 0 & 0 & 0 \\
\frac{5\sqrt{3}}{16\sqrt{2}} & \frac{15}{16\sqrt{2}} & 0 & 0 & \frac{-17}{32} & 0 & 0 & 0 & 0 & \frac{-3\sqrt{15}}{32} & 0 & 0 & 0 & 0 \\
0 & 0 & \frac{5\sqrt{3}}{16\sqrt{2}} & \frac{15}{16\sqrt{2}} & 0 & \frac{-17}{32} & 0 & 0 & 0 & 0 & \frac{-3\sqrt{15}}{32} & 0 & 0 & 0 \\
0 & 0 & 0 & 0 & 0 & 0 & \frac{-11}{16} & \frac{-5\sqrt{3}}{32} & \frac{-5\sqrt{15}}{32} & 0 & 0 & 0 & \frac{-3\sqrt{5}}{32} & \frac{-3\sqrt{5}}{32} \\
0 & 0 & 0 & 0 & 0 & 0 & \frac{-5\sqrt{3}}{32} & \frac{7}{8} & \frac{-3\sqrt{5}}{32} & 0 & 0 & \frac{-3\sqrt{5}}{32} & \frac{\sqrt{15}}{16} & \frac{\sqrt{15}}{32} \\
0 & 0 & 0 & 0 & 0 & 0 & \frac{-5\sqrt{15}}{32} & \frac{-3\sqrt{5}}{32} & \frac{1}{2} & 0 & 0 & \frac{3}{32} & \frac{7\sqrt{3}}{32} & \frac{\sqrt{3}}{4} \\
\frac{-3\sqrt{5}}{16\sqrt{2}} & \frac{-3\sqrt{15}}{16\sqrt{2}} & 0 & 0 & \frac{-3\sqrt{15}}{32} & 0 & 0 & 0 & 0 & \frac{-23}{32} & 0 & 0 & 0 & 0 \\
0 & 0 & \frac{-3\sqrt{5}}{16\sqrt{2}} & \frac{-3\sqrt{15}}{16\sqrt{2}} & 0 & \frac{-3\sqrt{15}}{32} & 0 & 0 & 0 & 0 & \frac{-23}{32} & 0 & 0 & 0 \\
0 & 0 & 0 & 0 & 0 & 0 & 0 & \frac{-3\sqrt{5}}{32} & \frac{3}{32} & 0 & 0 & \frac{-11}{16} & \frac{9\sqrt{3}}{32} & \frac{-9\sqrt{3}}{32} \\
0 & 0 & 0 & 0 & 0 & 0 & \frac{-3\sqrt{5}}{32} & \frac{\sqrt{15}}{16} & \frac{7\sqrt{3}}{32} & 0 & 0 & \frac{9\sqrt{3}}{32} & 0 & \frac{-23}{32} \\
0 & 0 & 0 & 0 & 0 & 0 & \frac{-3\sqrt{5}}{32} & \frac{\sqrt{15}}{32} & \frac{\sqrt{3}}{4} & 0 & 0 & \frac{-9\sqrt{3}}{32} & \frac{-23}{32} & 0
\end{pmatrix}.$$

For $S_{A,B,C} = 1/2$ block, we have $U_{42\times 42}^{\mathrm{SWAP_{AB}}} =$

$$\left(\begin{array}{c|c|c|c|c|c|c|c|c|c}
-U_{8\times8}^{\mathrm{SWAP_{AB}}} & \mathbf{0} & \mathbf{0} & \mathbf{0} & \mathbf{0} & \mathbf{0} & \mathbf{0} & \mathbf{0} & \mathbf{0} & \mathbf{0} \\ \hline
\mathbf{0} & -I_2 & \mathbf{0} & \mathbf{0} & \mathbf{0} & \mathbf{0} & \mathbf{0} & \mathbf{0} & \mathbf{0} & \mathbf{0} \\ \hline
\mathbf{0} & \mathbf{0} & U_{8\times8}^{\mathrm{SWAP_{AB}}} & \mathbf{0} & \mathbf{0} & \mathbf{0} & \mathbf{0} & \mathbf{0} & \mathbf{0} & \mathbf{0} \\ \hline
\mathbf{0} & \mathbf{0} & \mathbf{0} & U_{8\times8}^{\mathrm{S_{AB}}} & \mathbf{0} & \mathbf{0} & \mathbf{0} & \mathbf{0} & \mathbf{0} & \mathbf{0} \\ \hline
\mathbf{0} & \mathbf{0} & \mathbf{0} & \mathbf{0} & I_2 & \mathbf{0} & \mathbf{0} & \mathbf{0} & \mathbf{0} & \mathbf{0} \\ \hline
\mathbf{0} & \mathbf{0} & \mathbf{0} & \mathbf{0} & \mathbf{0} & U_{4\times4}^{\mathrm{Swap}} & \mathbf{0} & \mathbf{0} & \mathbf{0} & \mathbf{0} \\ \hline
\mathbf{0} & \mathbf{0} & \mathbf{0} & \mathbf{0} & \mathbf{0} & \mathbf{0} & U_{4\times4}^{\mathrm{S_{AB}}} & \mathbf{0} & \mathbf{0} & \mathbf{0} \\ \hline
\mathbf{0} & \mathbf{0} & \mathbf{0} & \mathbf{0} & \mathbf{0} & \mathbf{0} & \mathbf{0} & 1 & \mathbf{0} & \mathbf{0} \\ \hline
\mathbf{0} & \mathbf{0} & \mathbf{0} & \mathbf{0} & \mathbf{0} & \mathbf{0} & \mathbf{0} & \mathbf{0} & U_{4\times4}^{\mathrm{S_{AB}}} & \mathbf{0} \\ \hline
\mathbf{0} & \mathbf{0} & \mathbf{0} & \mathbf{0} & \mathbf{0} & \mathbf{0} & \mathbf{0} & \mathbf{0} & \mathbf{0} & -1
\end{array}\right),$$

$$U_{8\times8}^{\mathrm{SWAP_{AB}}} = \begin{pmatrix}
1 & 0 & 0 & 0 & 0 & 0 & 0 & 0 \\
0 & 1 & 0 & 0 & 0 & 0 & 0 & 0 \\
0 & 0 & 0 & 0 & 1 & 0 & 0 & 0 \\
0 & 0 & 0 & 0 & 0 & 1 & 0 & 0 \\
0 & 0 & 1 & 0 & 0 & 0 & 0 & 0 \\
0 & 0 & 0 & 1 & 0 & 0 & 0 & 0 \\
0 & 0 & 0 & 0 & 0 & 0 & 1 & 0 \\
0 & 0 & 0 & 0 & 0 & 0 & 0 & 1
\end{pmatrix}, U_{8\times8}^{\mathrm{S_{AB}}} = \begin{pmatrix}
0 & 0 & 0 & 0 & 1 & 0 & 0 & 0 \\
0 & 0 & 0 & 0 & 0 & 1 & 0 & 0 \\
0 & 0 & 0 & 0 & 0 & 0 & 1 & 0 \\
0 & 0 & 0 & 0 & 0 & 0 & 0 & 1 \\
1 & 0 & 0 & 0 & 0 & 0 & 0 & 0 \\
0 & 1 & 0 & 0 & 0 & 0 & 0 & 0 \\
0 & 0 & 1 & 0 & 0 & 0 & 0 & 0 \\
0 & 0 & 0 & 1 & 0 & 0 & 0 & 0
\end{pmatrix},$$

$$U_{4\times4}^{\mathrm{Swap}} = \begin{pmatrix}
1 & 0 & 0 & 0 \\
0 & 0 & 1 & 0 \\
0 & 1 & 0 & 0 \\
0 & 0 & 0 & 1
\end{pmatrix}, U_{4\times4}^{\mathrm{S_{AB}}} = \begin{pmatrix}
0 & 0 & 1 & 0 \\
0 & 0 & 0 & 1 \\
1 & 0 & 0 & 0 \\
0 & 1 & 0 & 0
\end{pmatrix}.$$

And the SWAP gate sequence between the second DFS qubit and the third DFS qubit is

$$U^{\mathrm{SWAP_{BC}}} = U_{67}(1)U_{56}(1)U_{45}(1)U_{78}(1)U_{67}(1)U_{56}(1)U_{89}(1)U_{78}(1)U_{67}(1). \quad (4.17)$$

For $S_{A,B,C} = 1/2$ block, we have $U^{\mathrm{SWAP_{BC}}}_{42\times 42} =$

$$\left(\begin{array}{c|c|c|c|c|c|c|c}
U^{\mathrm{SWAP_{BC}}}_{8\times 8} & \mathbf{0} & \sqrt{3}U^{\mathrm{SWAP_{BC}}}_{8\times 8} & \mathbf{0} & \mathbf{0} & \mathbf{0} & \mathbf{0} & \mathbf{0} \\ \hline
\mathbf{0} & \mathbf{0} & \mathbf{0} & \mathbf{0} & \mathbf{0} & \mathbf{0} & \mathbf{0} & U^{\mathrm{SWAP_{BC}}}_{2\times 10} \\ \hline
\sqrt{3}U^{\mathrm{SWAP_{BC}}}_{8\times 8} & \mathbf{0} & -U^{\mathrm{SWAP_{BC}}}_{8\times 8} & \mathbf{0} & \mathbf{0} & \mathbf{0} & \mathbf{0} & \mathbf{0} \\ \hline
\mathbf{0} & \mathbf{0} & \mathbf{0} & \mathbf{0} & \mathbf{0} & \mathbf{0} & I_4 & \mathbf{0} \\ \hline
\mathbf{0} & \mathbf{0} & \mathbf{0} & \mathbf{0} & U^{\mathrm{Swap}}_{4\times 4} & \mathbf{0} & \mathbf{0} & \mathbf{0} \\ \hline
\mathbf{0} & \mathbf{0} & \mathbf{0} & \mathbf{0} & \mathbf{0} & \mathbf{0} & \mathbf{0} & U^{\mathrm{SWAP'_{BC}}}_{2\times 10} \\ \hline
\mathbf{0} & \mathbf{0} & \mathbf{0} & I_4 & \mathbf{0} & \mathbf{0} & \mathbf{0} & \mathbf{0} \\ \hline
\mathbf{0} & (U^{\mathrm{SWAP_{BC}}}_{2\times 10})^{\top} & \mathbf{0} & \mathbf{0} & \mathbf{0} & (U^{\mathrm{SWAP'_{BC}}}_{2\times 10})^{\top} & \mathbf{0} & U^{\mathrm{SWAP_{BC}}}_{10\times 10}
\end{array}\right),$$

$$\text{where } U^{\mathrm{SWAP_{BC}}}_{8\times 8} = \begin{pmatrix}
\frac{1}{2} & 0 & 0 & 0 & 0 & 0 & 0 & 0 \\
0 & \frac{1}{2} & 0 & 0 & 0 & 0 & 0 & 0 \\
0 & 0 & 0 & \frac{1}{2} & 0 & 0 & 0 & 0 \\
0 & 0 & \frac{1}{2} & 0 & 0 & 0 & 0 & 0 \\
0 & 0 & 0 & 0 & \frac{1}{2} & 0 & 0 & 0 \\
0 & 0 & 0 & 0 & 0 & \frac{1}{2} & 0 & 0 \\
0 & 0 & 0 & 0 & 0 & 0 & 0 & \frac{1}{2} \\
0 & 0 & 0 & 0 & 0 & 0 & \frac{1}{2} & 0
\end{pmatrix},$$

$$U^{\mathrm{SWAP_{BC}}}_{2\times 10} = \begin{pmatrix}
0 & 0 & \frac{\sqrt{\frac{3}{2}}}{2} & 0 & 0 & 0 & 0 & \frac{\sqrt{\frac{5}{2}}}{2} & 0 & 0 \\
0 & 0 & 0 & \frac{\sqrt{\frac{3}{2}}}{2} & 0 & 0 & 0 & 0 & \frac{\sqrt{\frac{5}{2}}}{2} & 0
\end{pmatrix},$$

$$U^{\mathrm{SWAP'_{BC}}}_{2\times 10} = \begin{pmatrix}
0 & 0 & \frac{\sqrt{\frac{5}{2}}}{2} & 0 & 0 & 0 & 0 & -\frac{\sqrt{\frac{3}{2}}}{2} & 0 & 0 \\
0 & 0 & 0 & \frac{\sqrt{\frac{5}{2}}}{2} & 0 & 0 & 0 & 0 & -\frac{\sqrt{\frac{3}{2}}}{2} & 0
\end{pmatrix},$$

$$U^{\mathrm{SWAP_{BC}}}_{10\times 10} = \begin{pmatrix}
\frac{1}{4} & 0 & 0 & 0 & 0 & -\frac{\sqrt{15}}{4} & 0 & 0 & 0 & 0 \\
0 & \frac{1}{4} & 0 & 0 & 0 & 0 & -\frac{\sqrt{15}}{4} & 0 & 0 & 0 \\
0 & 0 & 0 & 0 & 0 & 0 & 0 & 0 & 0 & 0 \\
0 & 0 & 0 & 0 & 0 & 0 & 0 & 0 & 0 & 0 \\
0 & 0 & 0 & 0 & -\frac{1}{2} & 0 & 0 & 0 & 0 & -\frac{\sqrt{3}}{2} \\
-\frac{\sqrt{15}}{4} & 0 & 0 & 0 & 0 & -\frac{1}{4} & 0 & 0 & 0 & 0 \\
0 & -\frac{\sqrt{15}}{4} & 0 & 0 & 0 & 0 & -\frac{1}{4} & 0 & 0 & 0 \\
0 & 0 & 0 & 0 & 0 & 0 & 0 & 0 & 0 & 0 \\
0 & 0 & 0 & 0 & 0 & 0 & 0 & 0 & 0 & 0 \\
0 & 0 & 0 & 0 & -\frac{\sqrt{3}}{2} & 0 & 0 & 0 & 0 & \frac{1}{2}
\end{pmatrix}.$$

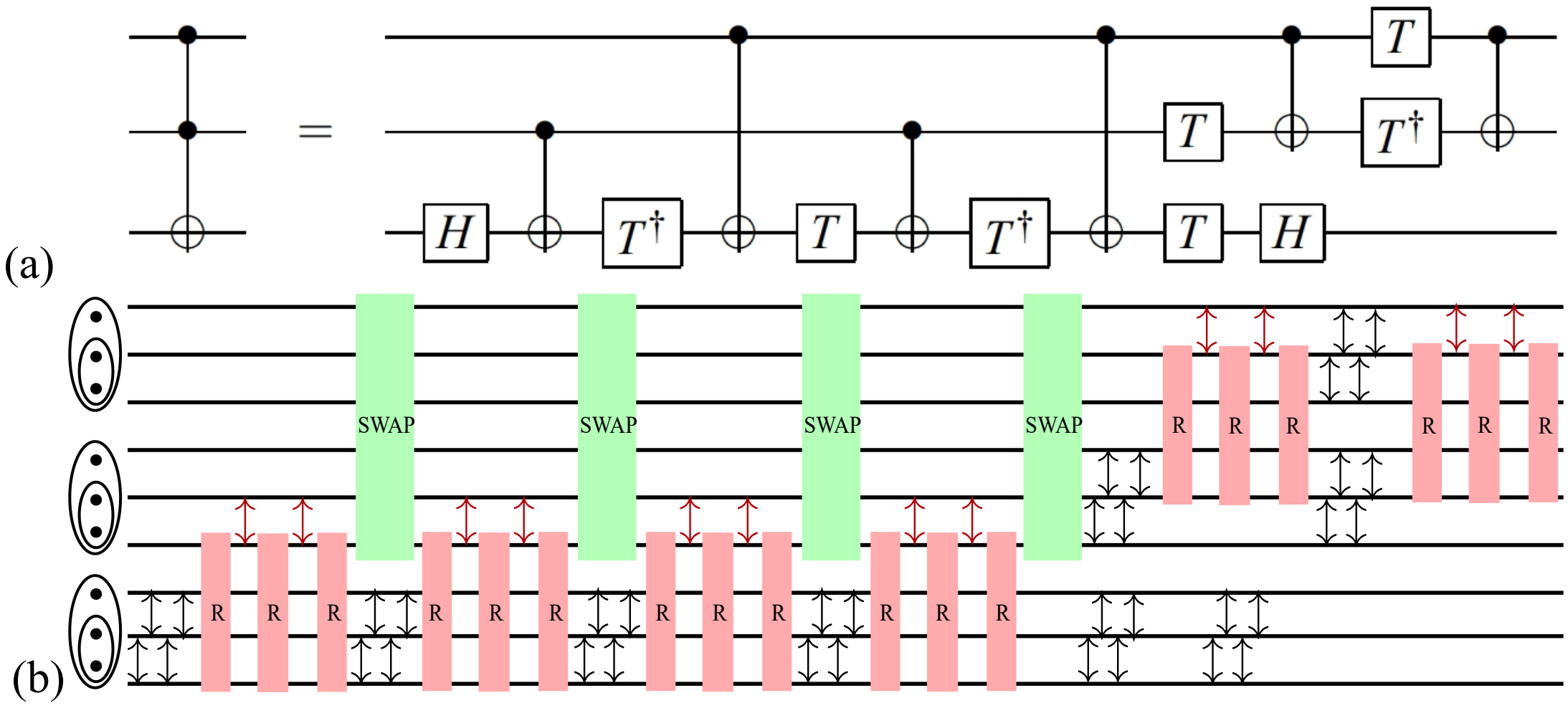

Fig. 4.5 (a) The Toffoli gate (CCNOT) can be decomposed into 6 CNOT gates and 9 single-qubit gates. This common decomposition uses controlled Clifford operators, breaking down the Toffoli into simpler quantum gates like Hadamard (H), T, and $T^\dagger$ gates, along with CNOTs. Respectively. (b) The EO qubits Toffoli gate sequence structure by decomposition, with 216 pulses and 146 time steps.

## 4.6 Three Qubit Gate Sequences and Structures Based on Gate Decomposition

### 4.6.1 Toffoli Gate

A Toffoli-class three-qubit gate has been successfully implemented on a silicon TQD quantum device, demonstrating quantum error correction (QEC) and mitigating one-qubit phase-flip errors [253, 271]. The standard Toffoli gate can be synthesized using single-qubit gates and CNOT gates [192, 46, 9], as illustrated in Figure 4.5(a). The most efficient decomposition requires 6 CNOT gates and 9 single-qubit gates. Notably, the type of single-qubit gate in Toffoli gate decomposition depends on the order of the CNOT gates selected.

Although the T gate can be costly in standard quantum circuits, in EO (electron spin) qubits, any single-qubit gate requires only four or less time-ordered pulses, making the cost of all single-qubit gates equivalent. The pulse cost for the CNOT gate is 22 (including 4 local-rotation pulses for DFS qubit B) for the Fong-Wandzura arrangement $((\bullet(\bullet\bullet)_a)1/2((\bullet\bullet)_b\bullet)1/2)$ [80]. For the arrangement $((\bullet(\bullet\bullet)_a)1/2(\bullet(\bullet\bullet)_b)1/2)$, the CNOT pulse cost is 24 (including 4 local-rotation pulses for DFS qubit B) [300]. The SWAP gate, crucial for reordering qubits, has a cost of 9 pulses, as shown in Equation 4.16.

Considering a linear geometry of a 9-spin system with only nearest-neighbor exchanges, the connectivity of the equivalent three-qubit quantum circuit is inherently limited. The arrangement is $((\bullet(\bullet\bullet)_a)_{1/2}(\bullet(\bullet\bullet)_b)_{1/2}(\bullet(\bullet\bullet)_c)_{1/2})$. Entangling gates can only be implemented between adjacent qubits. Thus, implementing a CNOT gate between distant qubits necessitates the use of SWAP gates, which can be expressed as $\mathrm{CNOT(A,C)} = \mathrm{SWAP(B,C)} \cdot \mathrm{CNOT(A,B)} \cdot \mathrm{SWAP(B,C)}$. Consequently, the pulse cost for a Toffoli gate in a linear EO qubits system is estimates 216 pulses, which includes $2\times4$ pulses for Hadamard gates, $7\times4$ pulses for T gates, $6\times24$ pulses for CNOT gates, and $4\times9$ pulses for SWAP gates. Additionally, for the total time steps, each single-qubit gate requires 4 time steps, each CNOT gate requires 19 time steps(with two extra local rotation time steps), and each SWAP gate requires 5 time steps. Therefore, the decomposed Toffoli gate sequence requires a total of 146 time steps, as shown in Figure 4.5(b).

The Toffoli sequence in a nine-spin EO system has the matrix block diagonalized structure, for $S_{A,B,C} = 1/2$ block, we have,

$$U^{\mathrm{CCNOT}}_{42\times42} = \left(\begin{array}{c|c|c|c|c} U^{\mathrm{CCNOT}}_{8\times8} & \mathbf{0} & \mathbf{0} & \mathbf{0} & \mathbf{0} \\ \hline \mathbf{0} & -iX & \mathbf{0} & \mathbf{0} & \mathbf{0} \\ \hline \mathbf{0} & \mathbf{0} & U^{\mathrm{CCNOT}}_{8\times8} & \mathbf{0} & \mathbf{0} \\ \hline \mathbf{0} & \mathbf{0} & \mathbf{0} & U^{\mathrm{CCNOT}}_{10\times10} & \mathbf{0} \\ \hline \mathbf{0} & \mathbf{0} & \mathbf{0} & \mathbf{0} & U^{\mathrm{CCNOT}}_{14\times14} \end{array}\right).$$

And for $S_{A,B,C} = 3/2$ block, we have:

$$U^{\mathrm{CCNOT}}_{48\times48} = \left(\begin{array}{c|c|c|c} U^{\mathrm{CCNOT}}_{8\times8} & \mathbf{0} & \mathbf{0} & \mathbf{0} \\ \hline \mathbf{0} & U^{\mathrm{CCNOT}}_{10\times10} & \mathbf{0} & \mathbf{0} \\ \hline \mathbf{0} & \mathbf{0} & U^{\mathrm{CCNOT}'}_{10\times10} & \mathbf{0} \\ \hline \mathbf{0} & \mathbf{0} & \mathbf{0} & U^{\mathrm{CCNOT}}_{20\times20} \end{array}\right).$$

In these two blocks, we have $C_1 = -\pi/4$, $C_2 = (\sqrt{2}-1-i)$, $C_3 = (e^{3i\pi/4}-i)$, and

$$U_{8\times 8}^{\mathrm{CCNOT}} = \begin{pmatrix} 1 & 0 & 0 & 0 & 0 & 0 & 0 & 0 \\ 0 & 1 & 0 & 0 & 0 & 0 & 0 & 0 \\ 0 & 0 & 1 & 0 & 0 & 0 & 0 & 0 \\ 0 & 0 & 0 & 1 & 0 & 0 & 0 & 0 \\ 0 & 0 & 0 & 0 & 1 & 0 & 0 & 0 \\ 0 & 0 & 0 & 0 & 0 & 1 & 0 & 0 \\ 0 & 0 & 0 & 0 & 0 & 0 & 0 & 1 \\ 0 & 0 & 0 & 0 & 0 & 0 & 1 & 0 \end{pmatrix},$$

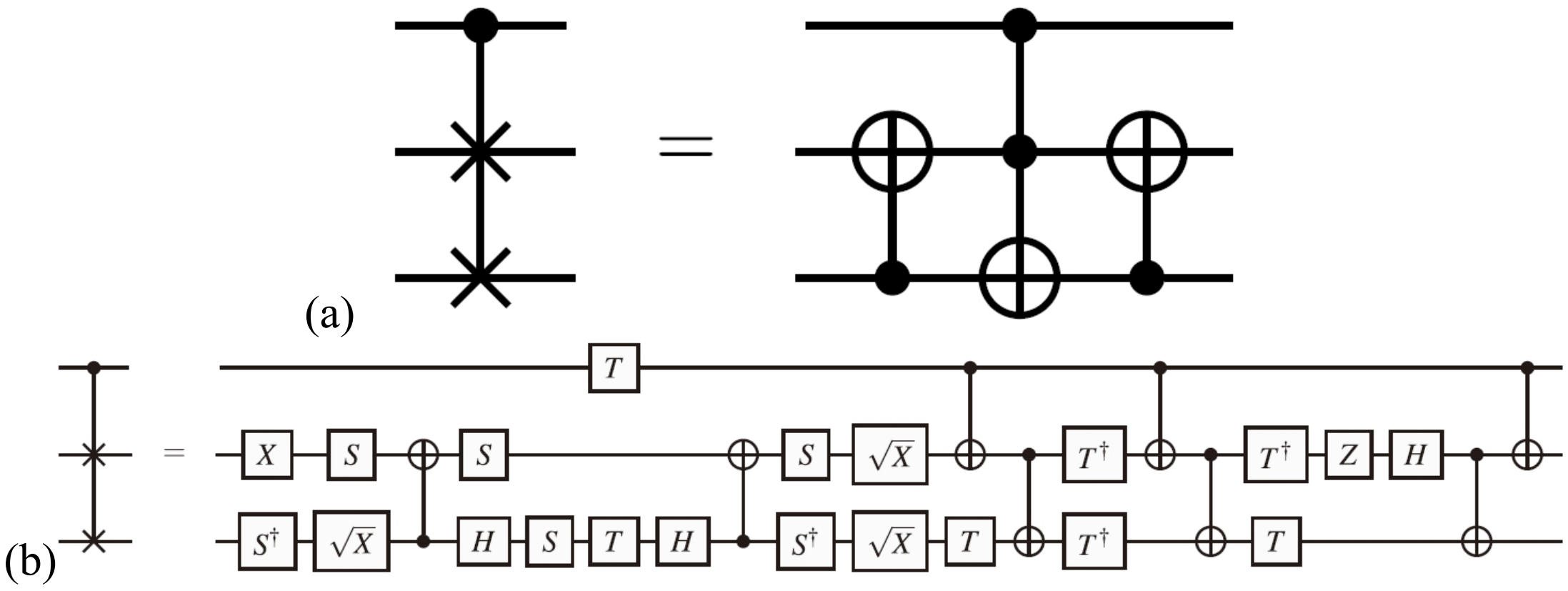

Fig. 4.6 The Controlled-SWAP (CSWAP) gate, also known as the Fredkin gate, can be decomposed into simpler quantum gates. (a) A common decomposition of the CSWAP gate consists of two CNOT gates and one Toffoli gate. (b)CSWAP gate circuit simplified with the Toffoli gate further decomposed into 6 CNOT gates and 9 single qubit gates.

$U_{20\times 20}^{\mathrm{CCNOT}}$ is not given due to it's size.

### 4.6.2 CSWAP Gate

The CSWAP gate can be decomposed into one Toffoli gate with two CNOT gate, or with 6 CNOT gates, 2 inverse CNOT gates, and 21 single-qubit gates, shown in Figure 4.6. For $S_{A,B,C} = 1/2$ block, the block itself is not block diagnoalied:

$$U_{42\times 42}^{\mathrm{CSWAP}} = \left(\begin{array}{c|c|c|c} U_{8\times 8}^{\mathrm{CSWAP}} & \mathbf{0} & \mathbf{0} & \mathbf{0} \\ \hline \mathbf{0} & U_{2\times 2}^{\mathrm{CSWAP}} & \mathbf{0} & U_{2\times 24}^{\mathrm{CSWAP}} \\ \hline \mathbf{0} & \mathbf{0} & U_{8\times 8}^{\mathrm{CSWAP}} & \mathbf{0} \\ \hline \mathbf{0} & U_{24\times 2}^{\mathrm{CSWAP}} & \mathbf{0} & U_{24\times 24}^{\mathrm{CSWAP}} \end{array}\right),$$

and for $S_{A,B,C} = 3/2$ block, it contain two block:

$$
U^{\text{CCNOT}}_{10\times 10} = \begin{pmatrix}
\frac{181+75e^{i\pi/4}}{256} & 0 & \frac{5\sqrt{3}C_2}{256\sqrt{2}} & 0 & 0 & 0 & 0 & 0 & \frac{15\sqrt{15}C_2}{128\sqrt{2}} & 0 \\
0 & \frac{181+75e^{i\pi/4}}{256} & 0 & \frac{5\sqrt{3}C_2}{256\sqrt{2}} & 0 & 0 & 0 & 0 & 0 & \frac{15\sqrt{15}C_2}{128\sqrt{2}} \\
0 & \frac{5\sqrt{3}C_3}{256} & 0 & \frac{-e^{3i\pi/4}-255i}{256} & 0 & 0 & 0 & 0 & 0 & \frac{3\sqrt{5}C_3}{128} \\
\frac{5\sqrt{3}C_3}{256} & 0 & \frac{-e^{3i\pi/4}-255i}{256} & 0 & 0 & 0 & 0 & 0 & \frac{3\sqrt{5}C_3}{128} & 0 \\
0 & 0 & 0 & 0 & e^{iC_1} & 0 & 0 & 0 & 0 & 0 \\
0 & 0 & 0 & 0 & 0 & e^{iC_1} & 0 & 0 & 0 & 0 \\
0 & 0 & 0 & 0 & 0 & 0 & 0 & e^{iC_1} & 0 & 0 \\
0 & 0 & 0 & 0 & 0 & 0 & e^{iC_1} & 0 & 0 & 0 \\
0 & \frac{-15\sqrt{15}C_3}{128} & 0 & \frac{3\sqrt{5}C_3}{128} & 0 & 0 & 0 & 0 & 0 & \frac{45e^{3i\pi/4}-19i}{64} \\
\frac{-15\sqrt{15}C_3}{128} & 0 & \frac{3\sqrt{5}C_3}{128} & 0 & 0 & 0 & 0 & 0 & \frac{45e^{3i\pi/4}-19i}{64} & 0
\end{pmatrix},
$$

$$U_{10\times 10}^{\mathrm{CCNOT}'} = \begin{pmatrix} \frac{229+27e^{i\pi/4}}{256} & 0 & \frac{45\sqrt{3}C_2}{256\sqrt{2}} & 0 & 0 & 0 & 0 & 0 & \frac{-3\sqrt{3}C_2}{128\sqrt{2}} & 0 \\ 0 & \frac{229+27e^{i\pi/4}}{256} & 0 & \frac{45\sqrt{3}C_2}{256\sqrt{2}} & 0 & 0 & 0 & 0 & 0 & \frac{-3\sqrt{3}C_2}{128\sqrt{2}} \\ 0 & \frac{45\sqrt{3}C_3}{256} & 0 & \frac{-225e^{3i\pi/4}-31i}{256} & 0 & 0 & 0 & 0 & 0 & \frac{15m}{128} \\ \frac{45\sqrt{3}m}{256} & 0 & \frac{-225e^{3i\pi/4}-31i}{256} & 0 & 0 & 0 & 0 & 0 & \frac{15(e^{3i\pi/4}-i)}{128} & 0 \\ 0 & 0 & 0 & 0 & e^{iC_1} & 0 & 0 & 0 & 0 & 0 \\ 0 & 0 & 0 & 0 & 0 & e^{iC_1} & 0 & 0 & 0 & 0 \\ 0 & 0 & 0 & 0 & 0 & 0 & 0 & e^{iC_1} & 0 & 0 \\ 0 & 0 & 0 & 0 & 0 & 0 & e^{iC_1} & 0 & 0 & 0 \\ 0 & \frac{-3\sqrt{3}C_3}{128} & 0 & \frac{15C_3}{128} & 0 & 0 & 0 & 0 & 0 & \frac{-e^{3i\pi/4}-63i}{64} \\ \frac{-3\sqrt{3}C_3}{128} & 0 & \frac{15C_3}{128} & 0 & 0 & 0 & 0 & 0 & \frac{-e^{3i\pi/4}-63i}{64} & 0 \end{pmatrix},$$

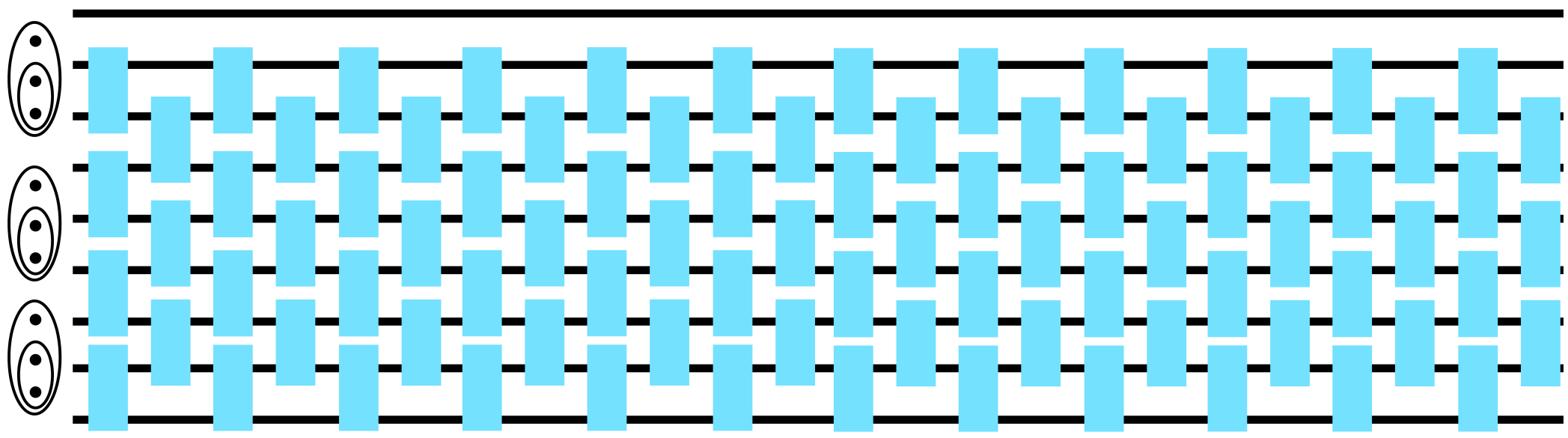

Fig. 4.7 A pulse-based brickwork ansatz with variable time steps. Each step uses 3-4 exchange pulses, alternating between odd and even qubit pairs. Odd steps: $J_{34}$, $J_{56}$, $J_{78}$. Even steps: $J_{23}$, $J_{45}$, $J_{67}$, $J_{89}$

$$U^{\text{CSWAP}}_{48\times48} = \left(\begin{array}{c|c} U^{\text{CSWAP}}_{8\times8} & \mathbf{0} \\ \hline \mathbf{0} & U^{\text{CSWAP}}_{40\times40} \end{array}\right),$$

where

$$U^{\text{CSWAP}}_{8\times8} = \begin{pmatrix} 1 & 0 & 0 & 0 & 0 & 0 & 0 & 0 \\ 0 & 1 & 0 & 0 & 0 & 0 & 0 & 0 \\ 0 & 0 & 1 & 0 & 0 & 0 & 0 & 0 \\ 0 & 0 & 0 & 1 & 0 & 0 & 0 & 0 \\ 0 & 0 & 0 & 0 & 1 & 0 & 0 & 0 \\ 0 & 0 & 0 & 0 & 0 & 0 & 1 & 0 \\ 0 & 0 & 0 & 0 & 0 & 1 & 0 & 0 \\ 0 & 0 & 0 & 0 & 0 & 0 & 0 & 1 \end{pmatrix}$$, and other matrix are leaked states block.

## 4.7 Obtain Shorter Sequences

### 4.7.1 Pulse-based Brickwork Ansatz

Variational quantum algorithms (VQAs) and quantum optimal control are two rapidly evolving fields in quantum computing that have shown significant interconnections in recent research. VQAs leverage classical optimization to adjust quantum circuit parameters, while QOC focuses on finding optimal ways to manipulate quantum systems. Recent studies have explored how insights from QOC can enhance VQA performance, as demonstrated in the work by Magann et al.[171]. De Keijzer et al.[53] introduced pulse-based VQAs that directly optimize control pulses using QOC principles, potentially offering advantages over standard gate-based approaches. Di Paolo et al. (2020)[43] proposed QOC-inspired ansatz for VQAs, incorporating symmetry-breaking unitaries to improve algorithm efficiency. The fundamental similarities between QOC problems and VQAs were highlighted by Li and Wang (2023)[165], suggesting potential cross-pollination of techniques. Furthermore, Koch et al. (2022)[145] emphasized the relationship and overlapping research areas between QOC and VQAs in a strategic report on quantum technologies. These studies collectively demonstrate a growing recognition of the synergies between VQAs and QOC, pointing towards a promising direction for advancing both fields through collaborative research and shared insights.

Building on this understanding, we took inspiration from one of the most commonly used block-layered ansatz in VQA and the commutation relation between qubits. We employed a symmetry brick structure ansatz to determine three-qubit gate sequences in our encoded spin system[159, 98, 304]. Similar to the analytic equivalent CNOT gate sequence in the $((\bullet(\bullet\bullet)a)1/2(\bullet(\bullet\bullet)b)1/2)$ encoded structure, we hypothesize that the CCNOT gate in our encoding system $((\bullet(\bullet\bullet)a)1/2(\bullet(\bullet\bullet)b)1/2(\bullet(\bullet\bullet)c)1/2)$ also does not require the exchange pulse $J_{12}$, as shown in Figure 4.7. This assumption aims to reduce the total number of exchange pulses.

Consequently, we established the pulse-based brickwork ansatz with adjustable time steps. In this ansatz, each time step comprises 3 or 4 commutated exchange pulses, as shown in Figure 4.7. The odd-numbered time steps contain exchange pulses $J_{34}$, $J_{56}$, and $J_{78}$, while the even-numbered time steps contain exchange pulses $J_{23}$, $J_{45}$, $J_{67}$, and $J_{89}$. It is worth noting that although the pulse-based brickwork ansatz maximizes the use of each time step, it is not the only possible ansatz. Other approaches, such as the staircase ansatz and N-local circuits ansatz, are also feasible and remain to be explored.

### 4.7.2 Krotov Method

Following the framework of brickwork ansatz, we define the piecewise system Hamiltonian as follows:

$$H(t) = \begin{cases} H_1(t), & \text{for odd time steps,} \\ H_2(t), & \text{for even time steps.} \end{cases} \tag{4.18}$$

Here, the Hamiltonians $H_1(t)$ and $H_2(t)$ are given by:

$$H_1(t) = J_{34}(t)\mathbf{S_3} \cdot \mathbf{S_4} + J_{56}(t)\mathbf{S_5} \cdot \mathbf{S_6} + J_{78}(t)\mathbf{S_7} \cdot \mathbf{S_8}, \tag{4.19}$$

$$H_2(t) = J_{23}(t)\mathbf{S_2} \cdot \mathbf{S_3} + J_{45}(t)\mathbf{S_4} \cdot \mathbf{S_5} + J_{67}(t)\mathbf{S_6} \cdot \mathbf{S_7} + J_{89}(t)\mathbf{S_8} \cdot \mathbf{S_9}. \tag{4.20}$$

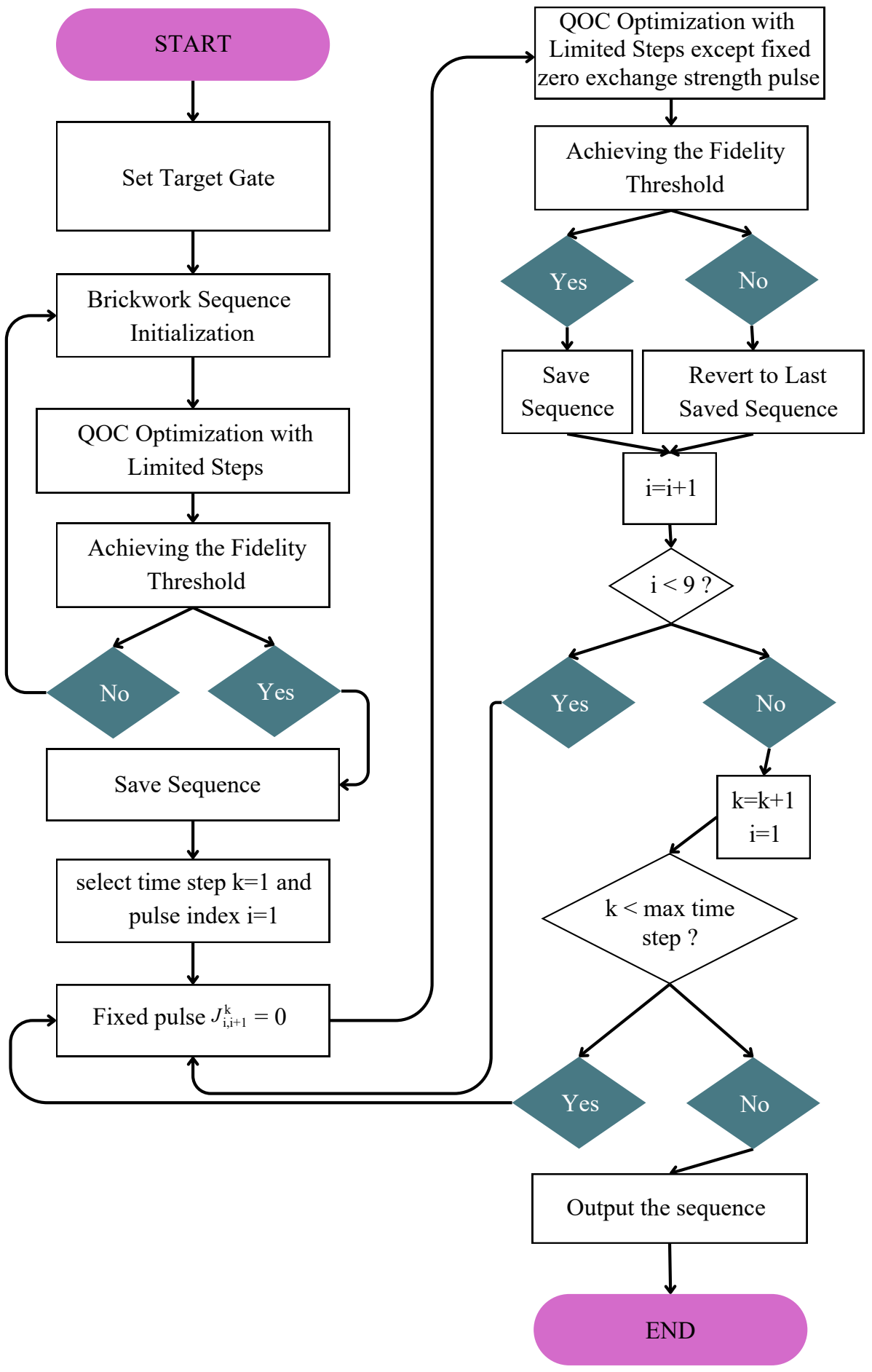

(a)

Fig. 4.8 The flowchart for finding a new sequence consists of two parts: the first part involves obtaining an initial sequence that meets the fidelity threshold within a limited number of optimization steps, and the second part focuses on shortening the resulting sequence.

By transforming the Hamiltonian into the angular momentum bases (as detailed in Appendix 4.11.2), we convert it into a quantum control problem. The Hamiltonian $\hat{H}(t)$ now depends on seven piecewise time-dependent control fields: $J_{23}(t)$, $J_{34}(t)$, $J_{45}(t)$, $J_{56}(t)$, $J_{67}(t)$, $J_{78}(t)$, and $J_{89}(t)$. According to Krotov's method, the optimization functional is:

$$J[\{|\phi_k^{(i)}(t)\rangle\}, \{J_{l,l+1}^{(i)}(t)\}] = J_T(\{|\phi_k^{(i)}(t)\rangle\}) + \sum_{l,l+1} \int_0^T g_a(J_{l,l+1}^{(i)}(t))dt$$
$$+ \int_0^T g_b(\{\phi_k^{(i)}(t)\})dt, \tag{4.21}$$

where $\{|\phi_k^{(i)}(t)\rangle\}$ are the time-evolved initial states $\{|\phi_k\rangle\}$ under the controls $\{J_{l,l+1}^{(i)}(t)\}$ of the i-th iteration. For a three-qubit gate, $\{|\phi_k\rangle\}$ represents all its logical three-qubit basis states: $|000\rangle$, $|001\rangle$, $|010\rangle$, $|011\rangle$, $|100\rangle$, $|101\rangle$, $|110\rangle$, and $|111\rangle$. To obtain the exchange gate sequence for a three-qubit gate, the functional of the quantum gate $U_{gate}$ must satisfy:

$$J_{T,re} = 1 - \frac{1}{N}\mathrm{Re}\left[\sum_{k=1}^{N} \tau_k\right], \tag{4.22}$$

with $\tau_k = \langle\phi_k^{\text{target}}|\phi_k(T)\rangle$, $|\phi_k^{\text{target}}\rangle = U_{gate}|\phi_k\rangle$, and $N$ being the dimension of the logical subspace.

After setting the initial guess $J_{l,l+1}^{(0)}(t)$, the optimized field $J_{l,l+1}^{(i)}(t)$ in iteration $i$ is updated as follows:

$$J_{l,l+1}^{(i)}(t) = J_{l,l+1}^{(i-1)}(t) + \Delta J_{l,l+1}^{(i)}(t), \tag{4.23}$$

where

$$\Delta J_{l,l+1}^{(i)}(t) = \frac{S_l(t)}{\lambda_{a,l}}\mathrm{Im}\left[\sum_{k=1}^{N}\left\langle \chi_k^{i-1}(t)\right|\left(\left.\frac{\partial \hat{H}}{\partial J_{l,l+1}(t)}\right|_{(i)}\right)\left|\phi_k^{i}(t)\right\rangle\right], \tag{4.24}$$

with the forward propagation equation of motion for $|\phi_k^{(i)}(t)\rangle$ given by:

$$\frac{\partial}{\partial t}|\phi_k^{(i)}(t)\rangle = -\frac{\mathrm{i}}{\hbar}\hat{H}^{(i)}|\phi_k^{(i)}(t)\rangle. \tag{4.25}$$

The co-states $|\chi_k^{i-1}(t)\rangle$ are propagated backward in time under the guess controls of iteration $i-1$, satisfying:

$$\frac{\partial}{\partial t}|\chi_k^{i-1}(t)\rangle = -\frac{\mathrm{i}}{\hbar}\hat{H}^{\dagger(i-1)}|\chi_k^{i-1}(t)\rangle + \left.\frac{\partial g_b}{\partial\langle\phi_k|}\right|_{(i-1)}, \tag{4.26}$$

with the boundary condition:

$$|\chi_k^{i-1}(T)\rangle = -\frac{\partial J_T}{\partial\langle\phi_k(T)|}\bigg|_{(i-1)}. \tag{4.27}$$

This detailed formulation and iterative optimization approach enable the effective design and control of quantum gates, ensuring high fidelity in quantum operations. Figure 4.8 illustrates the algorithm employed to derive the final sequence. Utilizing quantum optimal control methods, we first generate an initial sequence that meets the desired fidelity threshold for the target unitary matrix within a specified total time step. This initial sequence is refined through a process of gradually deleting unnecessary pulses by fixing the strength of certain exchange pulses to zero and optimizing the remaining sequence. This refinement excludes any pulses that have previously been fixed at zero strength. The process is iterated until no further non-zero exchange strength pulses can be removed within the constraints of the optimization steps, indicating that the search has been completed.

The pseudocode of the sequence searching process is provided in Appendix 4.11.3. It is noteworthy that while the Krotov method has been utilized for this numerical sequence search due to its guaranteed monotonic convergence[186, 90, 81], alternative quantum optimal control(QOC) techniques such as Gradient Ascent Pulse Engineering(GRAPE) [118, 309] or Reinforcement Learning(RL) [270, 126, 309] can also be effectively employed to these kinds of questions, which remains to be explored in the future.

## 4.8 Sequence Comparison

### 4.8.1 CCNOT Gate Sequence by QOC

Using the Krotov method and setting the total time steps to 55 (192 pulses), we obtained the raw sequence data for the Toffoli gate, as detailed in Appendix 4.11.4. This raw sequence includes two $8 \times 8$ three-DFS qubit unleaked blocks in the $S_{A,B,C} = 1/2$ $42 \times 42$ block and one $8 \times 8$ three-DFS qubit unleaked block in the $S_{A,B,C} = 3/2$ $48 \times 48$ block. Following this, we removed unnecessary pulses and further optimized the pulse parameters, resulting in a refined 92-pulse sequence with 50 time steps, as shown in Table 4.13.

The Toffoli 92-pulse sequence in a nine-spin EO system also has the matrix block diagonalized structure, for $S_{A,B,C} = 1/2$ block, we have,

$$U^{\text{CCNOT}}_{42\times 42} = \left(\begin{array}{c|c|c|c|c} U^{\text{CCNOT}}_{8\times 8} & \mathbf{0} & \mathbf{0} & \mathbf{0} & \mathbf{0} \\ \hline \mathbf{0} & X & \mathbf{0} & \mathbf{0} & \mathbf{0} \\ \hline \mathbf{0} & \mathbf{0} & U^{\text{CCNOT}}_{8\times 8} & \mathbf{0} & \mathbf{0} \\ \hline \mathbf{0} & \mathbf{0} & \mathbf{0} & U^{\text{Leak1}}_{10\times 10} & \mathbf{0} \\ \hline \mathbf{0} & \mathbf{0} & \mathbf{0} & \mathbf{0} & U^{\text{Leak}}_{14\times 14} \end{array}\right).$$

And for $S_{A,B,C} = 3/2$ block, we have:

$$U^{\text{CCNOT}}_{48\times 48} = \left(\begin{array}{c|c|c|c} U^{\text{CCNOT}}_{8\times 8} & \mathbf{0} & \mathbf{0} & \mathbf{0} \\ \hline \mathbf{0} & U^{\text{Leak1}}_{10\times 10} & \mathbf{0} & \mathbf{0} \\ \hline \mathbf{0} & \mathbf{0} & U^{\text{Leak2}}_{10\times 10} & \mathbf{0} \\ \hline \mathbf{0} & \mathbf{0} & \mathbf{0} & U^{\text{Leak}}_{20\times 20} \end{array}\right).$$

Here $U^{\text{CCNOT}}_{8\times 8}$ is CCNOT gate matrix, $X$ is pauli-x matrix. $U^{\text{Leak1}}_{10\times 10}$, $U^{\text{Leak2}}_{10\times 10}$, $U^{\text{Leak3}}_{10\times 10}$, $U^{\text{Leak}}_{14\times 14}$, $U^{\text{Leak}}_{20\times 20}$ are leak blocks, each block have a global phase corresponds to the decomposition sequence blocks($U^{\text{CCNOT}}_{10\times 10}$, $U^{\text{CCNOT}'}_{10\times 10}$, $U^{\text{CCNOT}}_{14\times 14}$, $U^{\text{CCNOT}}_{20\times 20}$). For example, we have

$$U^{\text{Leak1}}_{10\times 10} = \left(\begin{array}{c|c|c} U^{\text{Leak1}}_{4\times 4} & \mathbf{0} & U^{\text{Leak1}}_{4\times 2} \\ \hline \mathbf{0} & U^{\text{CNOT}}_{4\times 4} & \mathbf{0} \\ \hline U^{\text{Leak1}}_{2\times 4} & \mathbf{0} & U^{\text{Leak1}}_{2\times 2} \end{array}\right).$$

where

$$U^{\text{Leak1}}_{4\times 4} = \begin{pmatrix} 0.817e^{2.283i} & 0 & 0.429e^{-0.733i} & 0 \\ 0 & 0.817e^{2.283i} & 0 & 0.429e^{-0.733i} \\ 0.429e^{1.722i} & 0 & 0.454e^{1.849i} & \frac{4}{9} \\ 0 & 0.429e^{1.722i} & \frac{4}{9} & 0.454e^{1.849i} \end{pmatrix},$$

$$U^{\text{Leak1}}_{4\times 2} = \begin{pmatrix} 0.383e^{-0.731i} & 0 \\ 0 & 0.383e^{-0.731i} \\ 0.406e^{1.849i} & -0.496904 \\ -0.496904 & 0.406e^{1.849i} \end{pmatrix},$$

$$U^{\text{Leak1}}_{2\times 4} = \begin{pmatrix} 0.383e^{1.722i} & 0 & 0.406e^{1.849i} & -0.496904 \\ 0 & 0.383e^{1.722i} & -0.496904 & 0.406e^{1.849i} \end{pmatrix},$$

$$U^{\text{CNOT}}_{4\times 4} = \begin{pmatrix} 1 & 0 & 0 & 0 \\ 0 & 1 & 0 & 0 \\ 0 & 0 & 0 & 1 \\ 0 & 0 & 1 & 0 \end{pmatrix}, U^{\text{Leak1}}_{2\times 2} = \begin{pmatrix} 0.363e^{1.849i} & \frac{5}{9} \\ \frac{5}{9} & 0.363e^{1.849i} \end{pmatrix}.$$

This sequence can be manually divided into five parts based on its operation target on different qubits, as depicted in Figure 4.9(a): 1. $U^1_{A,B}$, a unitary operation on DFS qubits A and B from time step 1 to 15, consisting of 24 pulses. 2. $U^2_{A,B,C}$, a unitary operation on DFS qubits A, B, and C from time step 16 to 27, consisting of 29 pulses. 3. $U^3_{B,C}$, a unitary operation on DFS qubits B and C from time step 28 to 38, consisting of 18 pulses. 4. $U^4_{A,B,C}$, a unitary operation on DFS qubits A, B, and C from time step 39 to 40, consisting of 6 pulses. 5. $U^5_{A,B}$, a unitary operation on DFS qubits A and B from time step 41 to 50, consisting of 15 pulses.

It is worth noting that this partitioning is not unique, and different time steps can also be obtained for these unitary operations.

Alternatively, based on observation and applying the commutation rule, this sequence can be conveniently divided into three segments, as illustrated in Figure 4.9(b). In this 92-pulse sequence of the Toffoli gate, the yellow and blue segments represent two-qubit operations on DFS qubits A and B, while the red segment represents a two-qubit operation on DFS qubits B and C. Interestingly, similar to the sequence discovered previously[80, 300],these three two-qubit operations do not require exchange pulses between the first and second spins of the first DFS qubit. Specifically, the yellow and blue segments do not require $J_{12}$, and the red segment does not require $J_{45}$.

### 4.8.2 Noise Performance Comparison

In this section, we analyze the impact of common noise models on the fidelity of qubit gates in semiconductor quantum dot devices, focusing specifically on charge noise and pulse crosstalk. Charge noise and crosstalk are both modeled as control-dependent and quasi-static, with charge noise represented as $\delta J_{i,i+1} = \alpha_{i,i+1} J_{i,i+1}$. This model accounts for a shift in control strength, $J_{i,i+1} \rightarrow J_{i,i+1} + \delta J_{i,i+1}$. Crosstalk effects from neighboring qubits are modeled as $J_{i,i+1}\mathbf{S_i}\cdot\mathbf{S_{i+1}} \rightarrow J_{i,i+1}\mathbf{S_i}\cdot\mathbf{S_{i+1}} + \delta J_{i,i+1}\mathbf{S_{i+1}}\cdot\mathbf{S_{i+2}} + \delta J_{i,i+1}\mathbf{S_{i-1}}\cdot\mathbf{S_i}$. The fidelity of the quantum gates is calculated using the formula $F = (d + \left|\text{Tr}[U^\dagger_{\text{ideal}} U_{\text{actual}}]\right|^2)/d(d+1)$, where $U_{\text{ideal}}$ represents the target Toffoli gate and $U_{\text{actual}}$ represents the 8x8 three-DFS qubit unleaked block under noise.

Our results, illustrated in Figure 4.10, highlight the performance of different sequences under charge noise and pulse crosstalk. The black and red lines represent the performance of the Toffoli gate decomposition sequence. Due to the exact nature of the single and CNOT gate sequence solutions, the Toffoli gate decomposition sequence shows an increase in infidelity with increasing noise strength $\delta J/J$. Conversely, the numerical 92-pulse sequence, optimized to an infidelity of $\sim 10^{-11}$, exhibits a con-stant infidelity as $\delta J/J$ approaches zero, with the slopes of the blue and green curves decreasing to zero. Comparing the performance under charge noise, the infidelity of the decomposition sequence (red curve) is consistently higher than that of the 92-pulse sequence (green curve) before reaching our optimization limit. This trend also holds true in the presence of crosstalk, as shown by the black and blue curves, where the 92-pulse sequence outperforms the decomposition sequence.

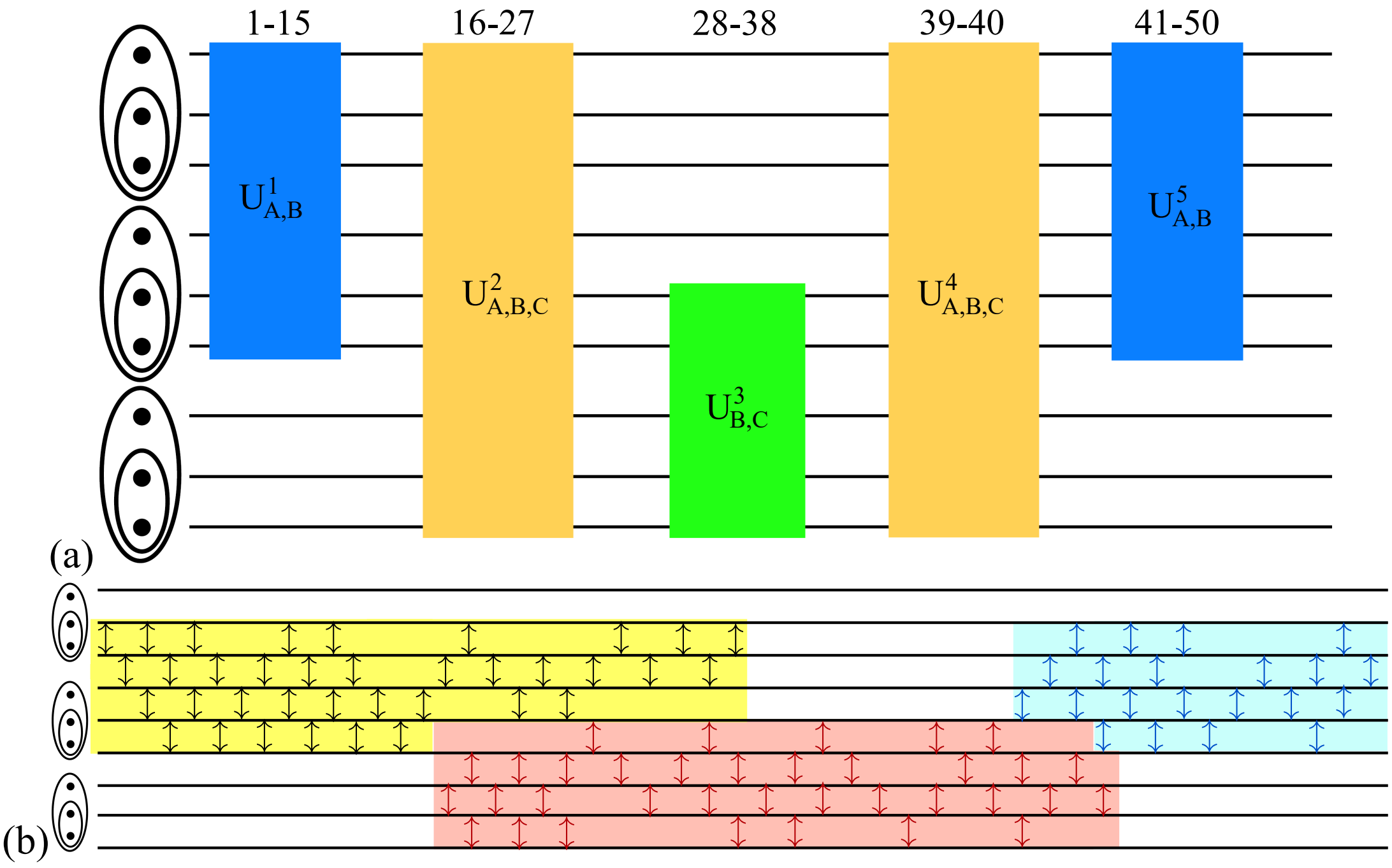

Fig. 4.9 The 92-pulse Toffoli gate sequence is structured in two ways: (a) It is divided into five parts based on target qubits, involving different combinations of qubits A, B, and C, with specified pulse counts for each part. (b) It is segmented into three sections based on operation types - two yellow/blue segments for two-qubit operations on DFS qubits A and B (without $J_{12}$), and a red segment for two-qubit operations on DFS qubits B and C (without $J_{45}$). This structure resembles previously discovered sequences.

In conclusion, the newly optimized 92-pulse sequence demonstrates enhanced robustness against both charge noise and crosstalk compared to the traditional Toffoli gate decomposition sequence. Additionally, the 92-pulse sequence requires fewer total pulses and time steps, making it a more efficient and reliable choice for quantum gate implementation in semiconductor quantum dot devices. This significant improvement in performance under various noise conditions highlights the potential of the 92-pulse sequence for more robust quantum computing applications.

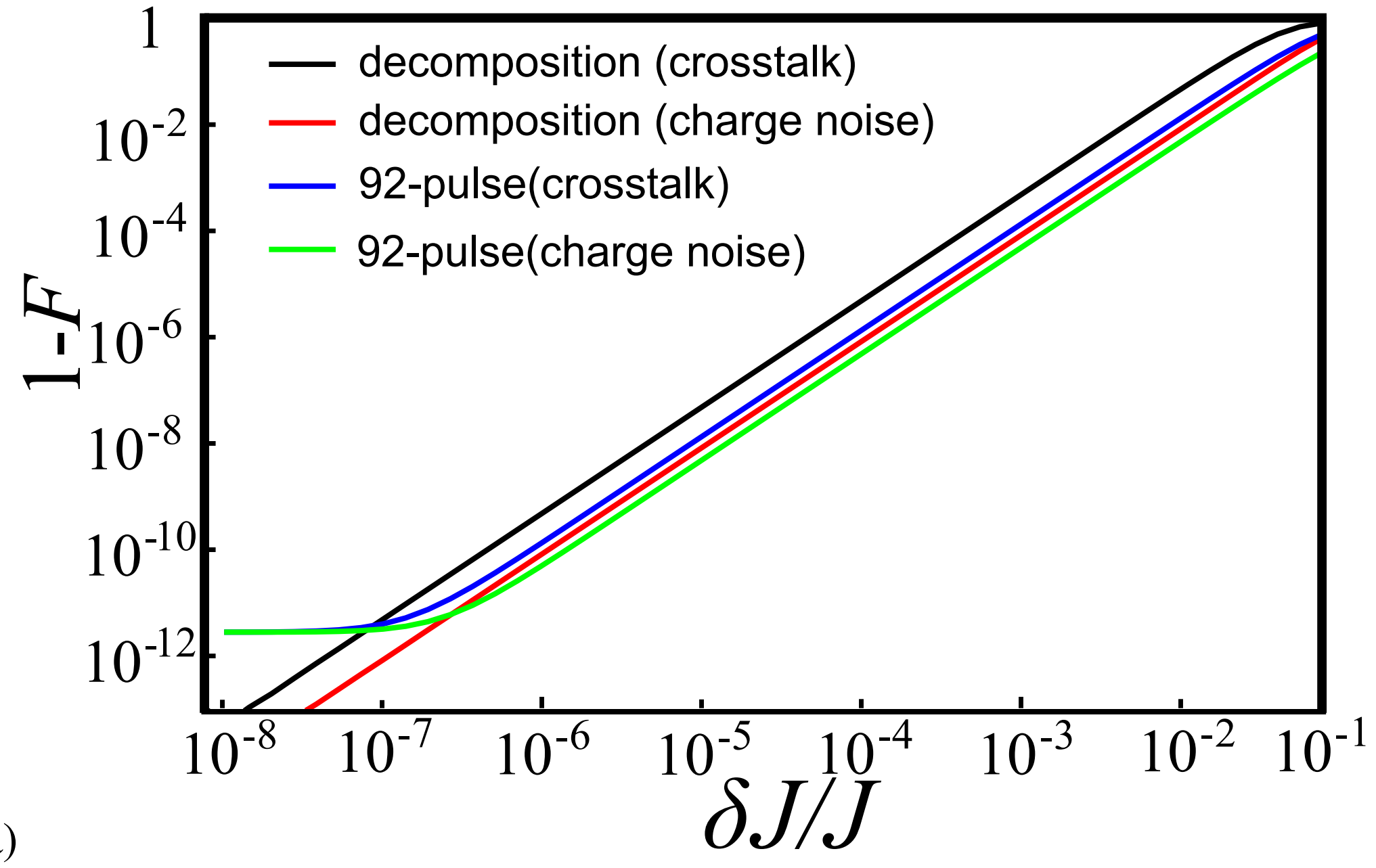

(a)

Fig. 4.10 Infidelity as a function of $\delta J/J$. The green(blue) curve represents the newly optimized 92-pulse sequence performance under charge noise(crosstalk), and the red(black) represents the Toffoli gate sequence under decomposition performance under charge noise(crosstalk).

## 4.9 Algorithm on EO Qubits

Quantum algorithms leverage the principles of quantum mechanics to solve certain computational problems more efficiently than classical algorithms. These algorithms exploit quantum phenomena such as superposition, entanglement, and quantum parallelism [192, 16]. Notable quantum algorithms include Shor's algorithm for factoring integers [237], Grover's algorithm for unstructured search [94], and the Quantum Fourier Transform (QFT), which is a key component in many quantum algorithms [70].

Quantum circuits are the computational model used to implement quantum algorithms. A quantum circuit consists of quantum bits (qubits) and quantum gates that manipulate these qubits [59]. One of the fundamental quantum states used in quantum computing is the Greenberger-Horne-Zeilinger (GHZ) state, which is a maximally entangled state involving multiple qubits [92, 213]. GHZ states are crucial for various quantum information tasks, including quantum error correction [31, 244], quantum cryptography, and quantum metrology [89]. Recent experiments have demonstrated the generation and verification of GHZ states on quantum devices with up to 27 qubits, highlighting the progress in near-term quantum computers [240, 93]. These states are

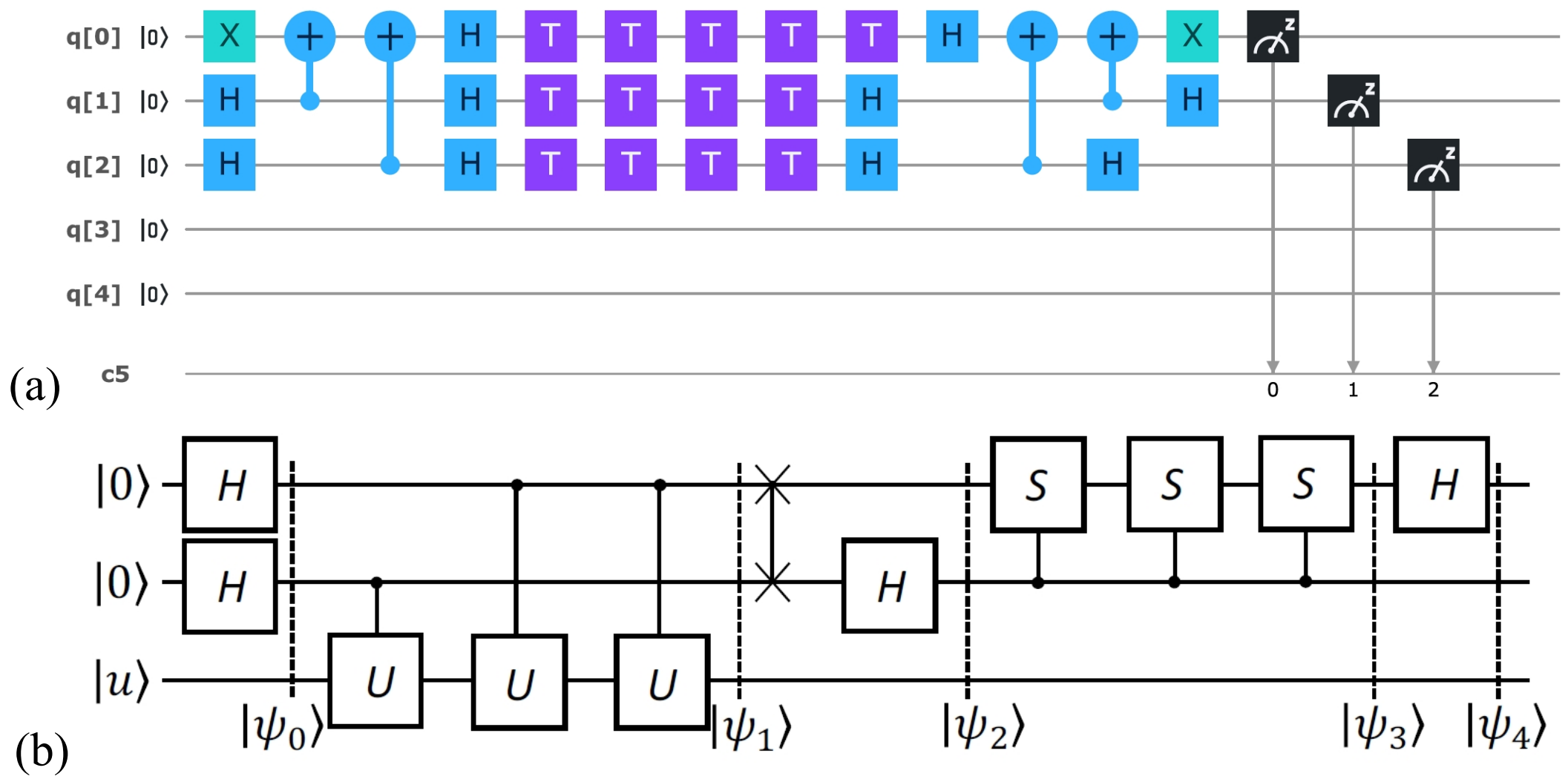

Fig. 4.11 (a)Three-qubit GHZ state $|-\rangle = (|000\rangle - |111\rangle)/\sqrt{2}$ circuit in [128]. (b)Quantum phase estimation(QPE) circuit in [58].

also used to benchmark the performance of quantum devices and to study the effects of quantum decoherence and error mitigation techniques [184, 260].

The GHZ state can be created with the quantum circuit in various ways. For instance, we consider the creation of three-qubit GHZ state $|-\rangle = (|000\rangle - |111\rangle)/\sqrt{2}$[128], shown in Figure 4.11(a). In this quantum circuit, 2 X-gate, 10 H-gate, 13 T-gate, and 4 CNOT gates have been applied. In EO qubits system, this circuit requires more than 200 pulses and 60 time steps to separately implement these single-qubit gates and two-qubit gates. We can easliy get the matrix representation of this quantum circuit:

$$U^{\mathrm{GHZ}} = \begin{pmatrix} \frac{\sqrt{2}-1-i}{2\sqrt{2}} & 0 & 0 & 0 & \frac{\sqrt{2}+1+i}{2\sqrt{2}} & 0 & 0 & 0 \\ 0 & \frac{-\sqrt{2}+1+i}{2\sqrt{2}} & 0 & 0 & 0 & \frac{-\sqrt{2}-1-i}{2\sqrt{2}} & 0 & 0 \\ 0 & 0 & \frac{-\sqrt{2}+1+i}{2\sqrt{2}} & 0 & 0 & 0 & \frac{-\sqrt{2}-1-i}{2\sqrt{2}} & 0 \\ 0 & 0 & 0 & \frac{\sqrt{2}-1-i}{2\sqrt{2}} & 0 & 0 & 0 & \frac{\sqrt{2}+1+i}{2\sqrt{2}} \\ \frac{\sqrt{2}+1+i}{2\sqrt{2}} & 0 & 0 & 0 & \frac{\sqrt{2}-1-i}{2\sqrt{2}} & 0 & 0 & 0 \\ 0 & \frac{-\sqrt{2}-1-i}{2\sqrt{2}} & 0 & 0 & 0 & \frac{-\sqrt{2}+1+i}{2\sqrt{2}} & 0 & 0 \\ 0 & 0 & \frac{-\sqrt{2}-1-i}{2\sqrt{2}} & 0 & 0 & 0 & \frac{-\sqrt{2}+1+i}{2\sqrt{2}} & 0 \\ 0 & 0 & 0 & \frac{\sqrt{2}+1+i}{2\sqrt{2}} & 0 & 0 & 0 & \frac{\sqrt{2}-1-i}{2\sqrt{2}} \end{pmatrix}.$$

It is easy to find that this matrix has a unique symmetry. Surprisingly, we found a 9-pulse sequence for this unitary matrix operation, shown in Figure 4.12. This short sequence contains two 4-pulse operations on DFS qubits A and C and a single swap pulse inside DFS qubit B; this proves that finding a shorter sequence for unitary matrix operations in specific algorithms in the EO qubits system is possible. This shorter sequence avoids multiple qubit gate implementation, requires less gate time, and is less affected by noise.

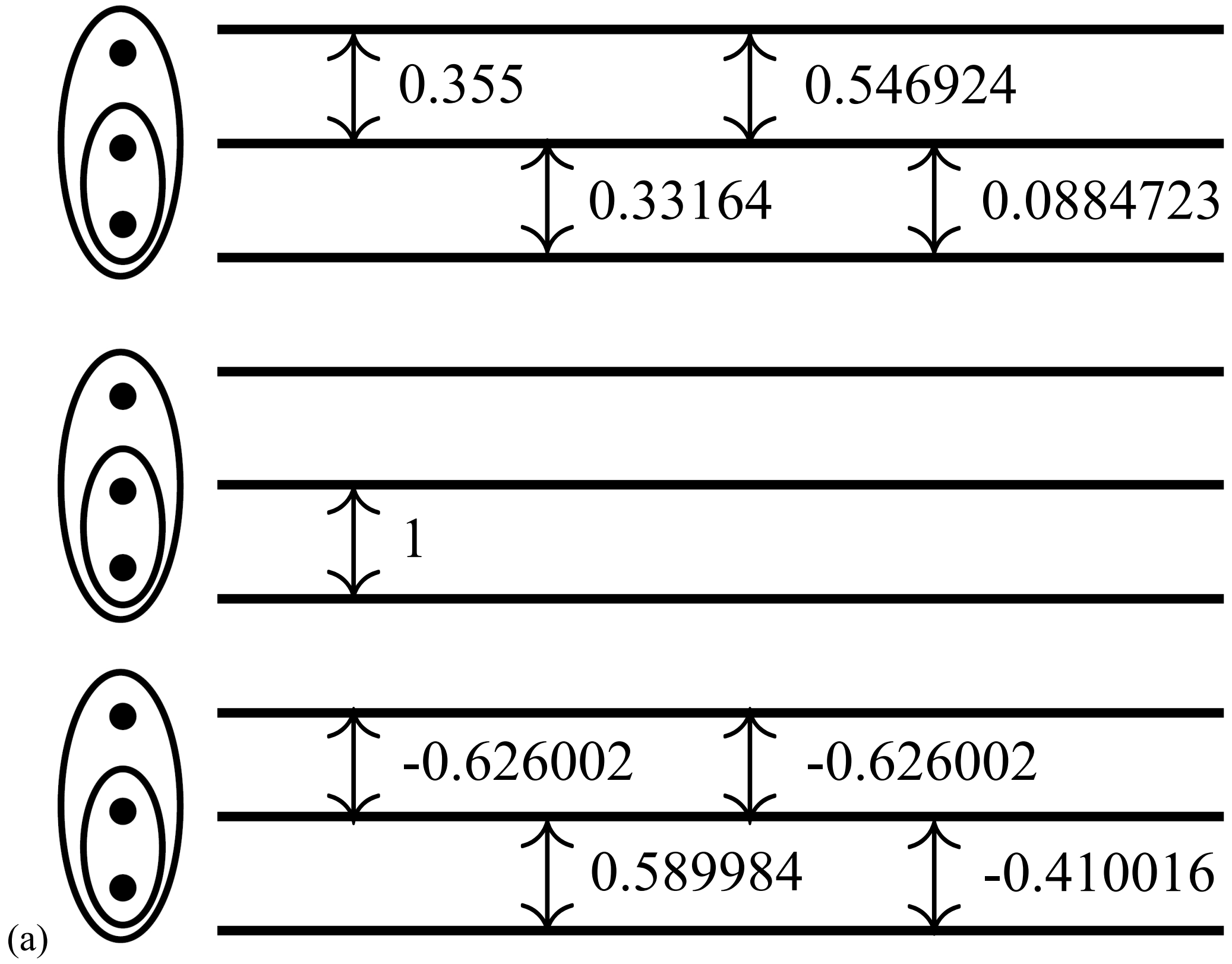

Fig. 4.12 The 9-pulse sequence for the unitary matrix operation $U^{\mathrm{GHZ}}$ in DFS qubit system.

Another example is quantum phase estimation(QPE) circuit[58] shown in Figure 4.11(b), for this circuit, its equivalent matrix representation is:

$$U^{\mathrm{QPE}} = \begin{pmatrix} \frac{1}{2} & -\frac{i}{2} & \frac{1}{2} & \frac{i}{2} & 0 & 0 & 0 & 0 \\ \frac{i}{2} & \frac{1}{2} & -\frac{i}{2} & \frac{1}{2} & 0 & 0 & 0 & 0 \\ 0 & 0 & 0 & 0 & \frac{1}{2} & -\frac{1}{2} & \frac{1}{2} & \frac{1}{2} \\ 0 & 0 & 0 & 0 & \frac{1}{2} & \frac{1}{2} & -\frac{1}{2} & \frac{1}{2} \\ \frac{1}{2} & \frac{i}{2} & \frac{1}{2} & -\frac{i}{2} & 0 & 0 & 0 & 0 \\ -\frac{i}{2} & \frac{1}{2} & \frac{i}{2} & \frac{1}{2} & 0 & 0 & 0 & 0 \\ 0 & 0 & 0 & 0 & \frac{1}{2} & \frac{1}{2} & \frac{1}{2} & -\frac{1}{2} \\ 0 & 0 & 0 & 0 & -\frac{1}{2} & \frac{1}{2} & \frac{1}{2} & \frac{1}{2} \end{pmatrix}.$$

For this unitary matrix, it lacks of symmetry, therefore, in EO qubits system, it require more exchange sequence to form such operation, we found a 160-pulse sequence and optimazed to $1 - 10^{-5}$.

## 4.10 Conclusions

In this work, we have addressed the efficient construction and optimization of exchange-only (EO) qubits within a decoherence-free subspace (DFS) framework, focusing on the development and refinement of multi-qubit gate sequences. Our study introduced a practical methodology for the construction of EO qubits in a linear quantum dot spin chain system, demonstrating the formation of a nine-qubit DFS equivalent to three EO qubits. Leveraging well-developed quantum optimal control techniques, we derived optimized gate sequences that significantly enhance the performance of EO qubits.

We presented a Toffoli gate sequence comprising 92 exchange pulses and 50 total time steps, a substantial improvement over the conventional decomposition requiring 216 exchange pulses and 146 time steps. This optimization not only reduces the operational complexity but also improves gate fidelity in the presence of noise and pulse errors. Furthermore, we explored the implementation of various two-qubit and three-qubit algorithms, showcasing the potential for reduced gate sequences and enhanced algorithmic efficiency.

Our results highlight the practical realization of complex quantum algorithms on EO qubits, paving the way for scalable and fault-tolerant quantum computing. The methodologies and optimizations presented in this study contribute to the robustness and efficiency of quantum operations in EO qubits systems, addressing critical scalability challenges.

In conclusion, the integration of EO qubits into larger DFS structures, coupled with precise control and optimization techniques, represents a significant advancement in the field of quantum computing. This research lays the groundwork for future explorations into the scalability and practical implementation of EO qubits, ultimately contributing to the development of robust, high-fidelity quantum computing architectures.

## 4.11 Appendix

### 4.11.1 Six Spin Bases Table

The six-spin system's 64 basis states can be encoded using six quantum numbers, represented as $|S_{A,B}, m_{A,B}, S_A, S_B, S_{A,1,2}, S_{B,1,2}\rangle$. Here, $S_{A,B}$ is the total spin of the entire system (DFS qubits A and B), $m_{A,B}$ is the total z-component of spin for the entire system, $S_{A,B}$ is the total spin of DFS qubits A and B, $S_A$, $S_B$ are the total spins of

individual DFS qubits A and B, and $S_{A,1,2}$, $S_{B,1,2}$ represent the spins of the two qubits encoding logical information in each DFS qubit. This encoding scheme provides a complete description of the six-spin system's quantum state, with basis states labeled from $|B1\rangle$ to $|B64\rangle$, for $S_{A,B} = 0$ :

$$|B1\rangle = |0,0,\frac{1}{2},\frac{1}{2},0,0\rangle = \frac{1}{\sqrt{2}}(|A1\rangle|A2\rangle - |A2\rangle|A1\rangle), \tag{4.28}$$

$$|B2\rangle = |0,0,\frac{1}{2},\frac{1}{2},0,1\rangle = \frac{1}{\sqrt{2}}(|A1\rangle|A4\rangle - |A2\rangle|A3\rangle), \tag{4.29}$$

$$|B3\rangle = |0,0,\frac{1}{2},\frac{1}{2},1,0\rangle = \frac{1}{\sqrt{2}}(|A3\rangle|A2\rangle - |A4\rangle|A1\rangle), \tag{4.30}$$

$$|B4\rangle = |0,0,\frac{1}{2},\frac{1}{2},1,1\rangle = \frac{1}{\sqrt{2}}(|A3\rangle|A4\rangle - |A4\rangle|A3\rangle), \tag{4.31}$$

$$|B5\rangle = |0,0,\frac{3}{2},\frac{3}{2},1,1\rangle = \frac{1}{2}(|A5\rangle|A8\rangle - |A6\rangle|A7\rangle + |A7\rangle|A6\rangle - |A8\rangle|A5\rangle). \tag{4.32}$$

For $S_{A,B} = 1$ :

$$|B6\rangle = |1,-1,\frac{1}{2},\frac{1}{2},0,0\rangle = |A2\rangle|A2\rangle, \tag{4.33}$$

$$|B7\rangle = |1,-1,\frac{1}{2},\frac{1}{2},0,1\rangle = |A2\rangle|A4\rangle, \tag{4.34}$$

$$|B8\rangle = |1,-1,\frac{1}{2},\frac{1}{2},1,0\rangle = |A4\rangle|A2\rangle, \tag{4.35}$$

$$|B9\rangle = |1,-1,\frac{1}{2},\frac{1}{2},1,1\rangle = |A4\rangle|A4\rangle, \tag{4.36}$$

$$|B10\rangle = |1,-1,\frac{1}{2},\frac{3}{2},0,1\rangle = \frac{1}{2}(|A2\rangle|A7\rangle - \sqrt{3}|A1\rangle|A8\rangle), \tag{4.37}$$

$$|B11\rangle = |1,-1,\frac{1}{2},\frac{3}{2},1,1\rangle = \frac{1}{2}(|A4\rangle|A7\rangle - \sqrt{3}|A3\rangle|A8\rangle), \tag{4.38}$$

$$|B12\rangle = |1,-1,\frac{3}{2},\frac{1}{2},1,0\rangle = \frac{1}{2}(|A7\rangle|A2\rangle - \sqrt{3}|A8\rangle|A1\rangle), \tag{4.39}$$

$$|B13\rangle = |1,-1,\frac{3}{2},\frac{1}{2},1,1\rangle = \frac{1}{2}(|A7\rangle|A4\rangle - \sqrt{3}|A8\rangle|A3\rangle), \tag{4.40}$$

$$|B14\rangle = |1,-1,\frac{3}{2},\frac{3}{2},1,1\rangle = \frac{1}{\sqrt{10}}(\sqrt{3}|A6\rangle|A8\rangle - 2|A7\rangle|A7\rangle + \sqrt{3}|A8\rangle|A6\rangle), \tag{4.41}$$

$$|B15\rangle = |1,0,\frac{1}{2},\frac{1}{2},0,0\rangle = \frac{1}{\sqrt{2}}(|A1\rangle|A2\rangle + |A2\rangle|A1\rangle), \tag{4.42}$$

$$|B16\rangle = |1,0,\frac{1}{2},\frac{1}{2},0,1\rangle = \frac{1}{\sqrt{2}}(|A1\rangle|A4\rangle + |A2\rangle|A3\rangle), \tag{4.43}$$

$$|B17\rangle = |1,0,\frac{1}{2},\frac{1}{2},1,0\rangle = \frac{1}{\sqrt{2}}(|A3\rangle|A2\rangle + |A4\rangle|A1\rangle), \tag{4.44}$$

$$|B18\rangle = |1,0,\frac{1}{2},\frac{1}{2},1,1\rangle = \frac{1}{\sqrt{2}}(|A3\rangle|A4\rangle + |A4\rangle|A3\rangle), \tag{4.45}$$

$$|B19\rangle = |1,0,\frac{1}{2},\frac{1}{2},1,1\rangle = \frac{1}{\sqrt{2}}(|A2\rangle|A6\rangle - |A1\rangle|A7\rangle), \tag{4.46}$$

$$|B20\rangle = |1,0,\frac{1}{2},\frac{1}{2},1,1\rangle = \frac{1}{\sqrt{2}}(|A4\rangle|A6\rangle - |A3\rangle|A7\rangle), \tag{4.47}$$

$$|B21\rangle = |1,0,\frac{1}{2},\frac{1}{2},1,1\rangle = \frac{1}{\sqrt{2}}(|A6\rangle|A2\rangle - |A7\rangle|A1\rangle), \tag{4.48}$$

$$|B22\rangle = |1,0,\frac{1}{2},\frac{1}{2},1,1\rangle = \frac{1}{\sqrt{2}}(|A6\rangle|A4\rangle - |A7\rangle|A3\rangle), \tag{4.49}$$

$$|B23\rangle = |1,0,\frac{3}{2},\frac{3}{2},1,1\rangle = \frac{1}{\sqrt{20}}(3|A5\rangle|A8\rangle - |A6\rangle|A7\rangle - |A7\rangle|A6\rangle + 3|A8\rangle|A5\rangle), \tag{4.50}$$

$$|B24\rangle = |1,1,\frac{1}{2},\frac{1}{2},0,0\rangle = |A1\rangle|A1\rangle, \tag{4.51}$$

$$|B25\rangle = |1,1,\frac{1}{2},\frac{1}{2},0,1\rangle = |A1\rangle|A3\rangle, \tag{4.52}$$

$$|B26\rangle = |1,1,\frac{1}{2},\frac{1}{2},1,0\rangle = |A3\rangle|A1\rangle, \tag{4.53}$$

$$|B27\rangle = |1,1,\frac{1}{2},\frac{1}{2},1,1\rangle = |A3\rangle|A3\rangle, \tag{4.54}$$

$$|B28\rangle = |1,1,\frac{1}{2},\frac{3}{2},0,1\rangle = \frac{1}{2}(\sqrt{3}|A2\rangle|A5\rangle - |A1\rangle|A6\rangle), \tag{4.55}$$

$$|B29\rangle = |1,1,\frac{1}{2},\frac{3}{2},1,1\rangle = \frac{1}{2}(\sqrt{3}|A4\rangle|A5\rangle - |A3\rangle|A6\rangle), \tag{4.56}$$

$$|B30\rangle = |1,1,\frac{3}{2},\frac{1}{2},1,0\rangle = \frac{1}{2}(\sqrt{3}|A5\rangle|A2\rangle - |A6\rangle|A1\rangle), \tag{4.57}$$

$$|B31\rangle = |1,1,\frac{3}{2},\frac{1}{2},1,1\rangle = \frac{1}{2}(\sqrt{3}|A5\rangle|A4\rangle - |A6\rangle|A3\rangle), \tag{4.58}$$

$$|B32\rangle = |1,1,\frac{3}{2},\frac{3}{2},1,1\rangle = \frac{1}{\sqrt{10}}(\sqrt{3}|A5\rangle|A7\rangle - 2|A6\rangle|A6\rangle + \sqrt{3}|A7\rangle|A5\rangle). \tag{4.59}$$

For $S_{A,B} = 2$ :

$$|B33\rangle = |2, -2, \frac{1}{2}, \frac{3}{2}, 0, 1\rangle = |A2\rangle|A8\rangle, \tag{4.60}$$

$$|B34\rangle = |2, -2, \frac{1}{2}, \frac{3}{2}, 1, 1\rangle = |A4\rangle|A8\rangle, \tag{4.61}$$

$$|B35\rangle = |2, -2, \frac{3}{2}, \frac{1}{2}, 1, 0\rangle = |A8\rangle|A2\rangle, \tag{4.62}$$

$$|B36\rangle = |2, -2, \frac{3}{2}, \frac{1}{2}, 1, 1\rangle = |A8\rangle|A4\rangle, \tag{4.63}$$

$$|B37\rangle = |2, -2, \frac{3}{2}, \frac{3}{2}, 1, 1\rangle = \frac{1}{\sqrt{2}}(|A7\rangle|A8\rangle - |A8\rangle|A7\rangle), \tag{4.64}$$

$$|B38\rangle = |2, -1, \frac{1}{2}, \frac{3}{2}, 0, 1\rangle = \frac{1}{2}(\sqrt{3}|A2\rangle|A7\rangle + |A1\rangle|A8\rangle), \tag{4.65}$$

$$|B39\rangle = |2, -1, \frac{1}{2}, \frac{3}{2}, 1, 1\rangle = \frac{1}{2}(\sqrt{3}|A4\rangle|A7\rangle + |A3\rangle|A8\rangle), \tag{4.66}$$

$$|B40\rangle = |2, -1, \frac{3}{2}, \frac{1}{2}, 1, 0\rangle = \frac{1}{2}(\sqrt{3}|A7\rangle|A2\rangle + |A8\rangle|A1\rangle), \tag{4.67}$$

$$|B41\rangle = |2, -1, \frac{3}{2}, \frac{1}{2}, 1, 1\rangle = \frac{1}{2}(\sqrt{3}|A7\rangle|A4\rangle + |A8\rangle|A3\rangle), \tag{4.68}$$

$$|B42\rangle = |2, -1, \frac{3}{2}, \frac{3}{2}, 1, 1\rangle = \frac{1}{\sqrt{2}}(|A6\rangle|A8\rangle - |A8\rangle|A6\rangle), \tag{4.69}$$

$$|B43\rangle = |2, 0, \frac{1}{2}, \frac{3}{2}, 0, 1\rangle = \frac{1}{\sqrt{2}}(|A2\rangle|A6\rangle + |A1\rangle|A7\rangle), \tag{4.70}$$

$$|B44\rangle = |2, 0, \frac{1}{2}, \frac{3}{2}, 1, 1\rangle = \frac{1}{\sqrt{2}}(|A4\rangle|A6\rangle + |A3\rangle|A7\rangle), \tag{4.71}$$

$$|B45\rangle = |2, 0, \frac{3}{2}, \frac{1}{2}, 1, 0\rangle = \frac{1}{\sqrt{2}}(|A6\rangle|A2\rangle + |A7\rangle|A1\rangle), \tag{4.72}$$

$$|B46\rangle = |2, 0, \frac{3}{2}, \frac{1}{2}, 1, 1\rangle = \frac{1}{\sqrt{2}}(|A6\rangle|A4\rangle + |A7\rangle|A3\rangle), \tag{4.73}$$

$$|B47\rangle = |2, 0, \frac{3}{2}, \frac{3}{2}, 1, 1\rangle = \frac{1}{2}(|A5\rangle|A8\rangle + |A6\rangle|A7\rangle - |A7\rangle|A6\rangle - |A8\rangle|A5\rangle), \tag{4.74}$$

$$|B48\rangle = |2, 1, \frac{1}{2}, \frac{3}{2}, 0, 1\rangle = \frac{1}{2}(|A2\rangle|A5\rangle + \sqrt{3}|A1\rangle|A6\rangle), \tag{4.75}$$

$$|B49\rangle = |2, 1, \frac{1}{2}, \frac{3}{2}, 1, 1\rangle = \frac{1}{2}(|A4\rangle|A5\rangle + \sqrt{3}|A3\rangle|A6\rangle), \tag{4.76}$$

$$|B50\rangle = |2, 1, \frac{3}{2}, \frac{1}{2}, 1, 0\rangle = \frac{1}{2}(|A5\rangle|A2\rangle + \sqrt{3}|A6\rangle|A1\rangle), \tag{4.77}$$

$$|B51\rangle = |2,1,\frac{3}{2},\frac{1}{2},1,1\rangle = \frac{1}{2}(|A5\rangle|A4\rangle + \sqrt{3}|A6\rangle|A3\rangle), \tag{4.78}$$

$$|B52\rangle = |2,1,\frac{3}{2},\frac{3}{2},1,1\rangle = \frac{1}{\sqrt{2}}(|A5\rangle|A7\rangle - |A7\rangle|A5\rangle), \tag{4.79}$$

$$|B53\rangle = |2,2,\frac{1}{2},\frac{3}{2},0,1\rangle = |A1\rangle|A5\rangle, \tag{4.80}$$

$$|B54\rangle = |2,2,\frac{1}{2},\frac{3}{2},1,1\rangle = |A3\rangle|A5\rangle, \tag{4.81}$$

$$|B55\rangle = |2,2,\frac{3}{2},\frac{1}{2},1,0\rangle = |A5\rangle|A1\rangle, \tag{4.82}$$

$$|B56\rangle = |2,2,\frac{3}{2},\frac{1}{2},1,1\rangle = |A5\rangle|A3\rangle, \tag{4.83}$$

$$|B57\rangle = |2,2,\frac{3}{2},\frac{3}{2},1,1\rangle = \frac{1}{\sqrt{2}}(|A5\rangle|A6\rangle - |A6\rangle|A5\rangle). \tag{4.84}$$

For $S_{A,B} = 3$ :

$$|B58\rangle = |3,-3,\frac{3}{2},\frac{3}{2},1,1\rangle = |A8\rangle|A8\rangle, \tag{4.85}$$

$$|B59\rangle = |3,-2,\frac{3}{2},\frac{3}{2},1,1\rangle = \frac{1}{\sqrt{2}}(|A7\rangle|A8\rangle + |A8\rangle|A7\rangle), \tag{4.86}$$

$$|B60\rangle = |3,-1,\frac{3}{2},\frac{3}{2},1,1\rangle = \frac{1}{\sqrt{5}}(|A6\rangle|A8\rangle + \sqrt{3}|A7\rangle|A7\rangle + |A8\rangle|A6\rangle), \tag{4.87}$$

$$|B61\rangle = |3,0,\frac{3}{2},\frac{3}{2},1,1\rangle = \frac{1}{\sqrt{20}} = (|A5\rangle|A8\rangle + 3|A6\rangle|A7\rangle + 3|A7\rangle|A6\rangle + |A8\rangle|A5\rangle), \tag{4.88}$$

$$|B62\rangle = |3,1,\frac{3}{2},\frac{3}{2},1,1\rangle = \frac{1}{\sqrt{5}}(|A5\rangle|A7\rangle + \sqrt{3}|A6\rangle|A6\rangle + |A7\rangle|A5\rangle), \tag{4.89}$$

$$|B63\rangle = |3,2,\frac{3}{2},\frac{3}{2},1,1\rangle = \frac{1}{\sqrt{2}}(|A5\rangle|A6\rangle + |A6\rangle|A5\rangle), \tag{4.90}$$

$$|B64\rangle = |3,3,\frac{3}{2},\frac{3}{2},1,1\rangle = |A5\rangle|A5\rangle. \tag{4.91}$$

### 4.11.2 Nine Spin Bases Table

The nine-spin system basis states can be encoded using nine quantum numbers, represented as $|S_{A,B,C}, m_{A,B,C}, S_{A,B}, S_A, S_B, S_C, S_{A,1,2}, S_{B,1,2}, S_{C,1,2}\rangle$. Here, $S_{A,B,C}$ is the total spin of the entire system (DFS qubits A, B, and C), $m_{A,B,C}$ is the total z-component of spin for the entire system, $S_{A,B}$ is the total spin of DFS qubits A and

B, $S_A$, $S_B$, $S_C$ are the total spins of individual DFS qubits A, B, and C, and $S_{A,1,2}$, $S_{B,1,2}$, $S_{C,1,2}$ represent the spins of the two qubits encoding logical information in each DFS qubit. This encoding provides a complete description of the nine-spin system's quantum state. The basis states are divided into two blocks based on $S_{A,B,C}$: a 42 × 42 block for $S_{A,B,C} = 1/2$ (labeled $|C1\rangle$ to $|C42\rangle$) and a 48 × 48 block for $S_{A,B,C} = 3/2$ (labeled $|C43\rangle$ to $|C90\rangle$).

For $S_{A,B,C} = \frac{1}{2}$ :

$$|C1\rangle = |\frac{1}{2}, \frac{1}{2}, 0, \frac{1}{2}, \frac{1}{2}, \frac{1}{2}, 0, 0, 0\rangle = |B1\rangle|A1\rangle, \tag{4.92}$$

$$|C2\rangle = |\frac{1}{2}, \frac{1}{2}, 0, \frac{1}{2}, \frac{1}{2}, \frac{1}{2}, 0, 0, 1\rangle = |B1\rangle|A3\rangle, \tag{4.93}$$

$$|C3\rangle = |\frac{1}{2}, \frac{1}{2}, 0, \frac{1}{2}, \frac{1}{2}, \frac{1}{2}, 0, 1, 0\rangle = |B2\rangle|A1\rangle, \tag{4.94}$$

$$|C4\rangle = |\frac{1}{2}, \frac{1}{2}, 0, \frac{1}{2}, \frac{1}{2}, \frac{1}{2}, 0, 1, 1\rangle = |B2\rangle|A3\rangle, \tag{4.95}$$

$$|C5\rangle = |\frac{1}{2}, \frac{1}{2}, 0, \frac{1}{2}, \frac{1}{2}, \frac{1}{2}, 1, 0, 0\rangle = |B3\rangle|A1\rangle, \tag{4.96}$$

$$|C6\rangle = |\frac{1}{2}, \frac{1}{2}, 0, \frac{1}{2}, \frac{1}{2}, \frac{1}{2}, 1, 0, 1\rangle = |B3\rangle|A3\rangle, \tag{4.97}$$

$$|C7\rangle = |\frac{1}{2}, \frac{1}{2}, 0, \frac{1}{2}, \frac{1}{2}, \frac{1}{2}, 1, 1, 0\rangle = |B4\rangle|A1\rangle, \tag{4.98}$$

$$|C8\rangle = |\frac{1}{2}, \frac{1}{2}, 0, \frac{1}{2}, \frac{1}{2}, \frac{1}{2}, 1, 1, 1\rangle = |B4\rangle|A3\rangle, \tag{4.99}$$

$$|C9\rangle = |\frac{1}{2}, \frac{1}{2}, 0, \frac{3}{2}, \frac{3}{2}, \frac{1}{2}, 1, 1, 0\rangle = |B5\rangle|A1\rangle, \tag{4.100}$$

$$|C10\rangle = |\frac{1}{2}, \frac{1}{2}, 0, \frac{3}{2}, \frac{3}{2}, \frac{1}{2}, 1, 1, 1\rangle = |B5\rangle|A1\rangle, \tag{4.101}$$

$$|C11\rangle = |\frac{1}{2}, \frac{1}{2}, 1, \frac{1}{2}, \frac{1}{2}, \frac{1}{2}, 0, 0, 0\rangle = \sqrt{\frac{2}{3}}|B24\rangle|A2\rangle - \sqrt{\frac{1}{3}}|B15\rangle|A1\rangle, \tag{4.102}$$

$$|C12\rangle = |\frac{1}{2}, \frac{1}{2}, 1, \frac{1}{2}, \frac{1}{2}, \frac{1}{2}, 0, 0, 1\rangle = \sqrt{\frac{2}{3}}|B24\rangle|A4\rangle - \sqrt{\frac{1}{3}}|B15\rangle|A3\rangle, \tag{4.103}$$

$$|C13\rangle = |\frac{1}{2}, \frac{1}{2}, 1, \frac{1}{2}, \frac{1}{2}, \frac{1}{2}, 0, 1, 0\rangle = \sqrt{\frac{2}{3}}|B25\rangle|A2\rangle - \sqrt{\frac{1}{3}}|B16\rangle|A1\rangle, \tag{4.104}$$

$$|C14\rangle = |\frac{1}{2},\frac{1}{2},1,\frac{1}{2},\frac{1}{2},\frac{1}{2},0,1,1\rangle = \sqrt{\frac{2}{3}}|B25\rangle|A4\rangle - \sqrt{\frac{1}{3}}|B16\rangle|A3\rangle, \quad (4.105)$$

$$|C15\rangle = |\frac{1}{2},\frac{1}{2},1,\frac{1}{2},\frac{1}{2},\frac{1}{2},1,0,0\rangle = \sqrt{\frac{2}{3}}|B26\rangle|A2\rangle - \sqrt{\frac{1}{3}}|B17\rangle|A1\rangle, \quad (4.106)$$

$$|C16\rangle = |\frac{1}{2},\frac{1}{2},1,\frac{1}{2},\frac{1}{2},\frac{1}{2},1,0,1\rangle = \sqrt{\frac{2}{3}}|B26\rangle|A4\rangle - \sqrt{\frac{1}{3}}|B17\rangle|A3\rangle, \quad (4.107)$$

$$|C17\rangle = |\frac{1}{2},\frac{1}{2},1,\frac{1}{2},\frac{1}{2},\frac{1}{2},1,1,0\rangle = \sqrt{\frac{2}{3}}|B27\rangle|A2\rangle - \sqrt{\frac{1}{3}}|B18\rangle|A1\rangle, \quad (4.108)$$

$$|C18\rangle = |\frac{1}{2},\frac{1}{2},1,\frac{1}{2},\frac{1}{2},\frac{1}{2},1,1,1\rangle = \sqrt{\frac{2}{3}}|B27\rangle|A4\rangle - \sqrt{\frac{1}{3}}|B18\rangle|A1\rangle, \quad (4.109)$$

$$|C19\rangle = |\frac{1}{2},\frac{1}{2},1,\frac{1}{2},\frac{3}{2},\frac{1}{2},0,1,0\rangle = \sqrt{\frac{2}{3}}|B28\rangle|A2\rangle - \sqrt{\frac{1}{3}}|B19\rangle|A1\rangle, \quad (4.110)$$

$$|C20\rangle = |\frac{1}{2},\frac{1}{2},1,\frac{1}{2},\frac{3}{2},\frac{1}{2},0,1,1\rangle = \sqrt{\frac{2}{3}}|B28\rangle|A4\rangle - \sqrt{\frac{1}{3}}|B19\rangle|A3\rangle, \quad (4.111)$$

$$|C21\rangle = |\frac{1}{2},\frac{1}{2},1,\frac{1}{2},\frac{3}{2},\frac{1}{2},1,1,0\rangle = \sqrt{\frac{2}{3}}|B29\rangle|A2\rangle - \sqrt{\frac{1}{3}}|B20\rangle|A1\rangle, \quad (4.112)$$

$$|C22\rangle = |\frac{1}{2},\frac{1}{2},1,\frac{1}{2},\frac{3}{2},\frac{1}{2},1,1,1\rangle = \sqrt{\frac{2}{3}}|B29\rangle|A4\rangle - \sqrt{\frac{1}{3}}|B20\rangle|A3\rangle, \quad (4.113)$$

$$|C23\rangle = |\frac{1}{2},\frac{1}{2},1,\frac{3}{2},\frac{1}{2},\frac{1}{2},1,0,0\rangle = \sqrt{\frac{2}{3}}|B30\rangle|A2\rangle - \sqrt{\frac{1}{3}}|B21\rangle|A1\rangle, \quad (4.114)$$

$$|C24\rangle = |\frac{1}{2},\frac{1}{2},1,\frac{3}{2},\frac{1}{2},\frac{1}{2},1,0,1\rangle = \sqrt{\frac{2}{3}}|B30\rangle|A4\rangle - \sqrt{\frac{1}{3}}|B21\rangle|A3\rangle, \quad (4.115)$$

$$|C25\rangle = |\frac{1}{2},\frac{1}{2},1,\frac{3}{2},\frac{1}{2},\frac{1}{2},1,1,0\rangle = \sqrt{\frac{2}{3}}|B31\rangle|A2\rangle - \sqrt{\frac{1}{3}}|B22\rangle|A1\rangle, \quad (4.116)$$

$$|C26\rangle = |\frac{1}{2},\frac{1}{2},1,\frac{3}{2},\frac{1}{2},\frac{1}{2},1,1,1\rangle = \sqrt{\frac{2}{3}}|B31\rangle|A4\rangle - \sqrt{\frac{1}{3}}|B22\rangle|A3\rangle, \quad (4.117)$$

$$|C27\rangle = |\frac{1}{2},\frac{1}{2},1,\frac{3}{2},\frac{3}{2},\frac{1}{2},1,1,0\rangle = \sqrt{\frac{2}{3}}|B32\rangle|A2\rangle - \sqrt{\frac{1}{3}}|B23\rangle|A1\rangle, \quad (4.118)$$

$$|C28\rangle = |\frac{1}{2},\frac{1}{2},1,\frac{3}{2},\frac{3}{2},\frac{1}{2},1,1,1\rangle = \sqrt{\frac{2}{3}}|B32\rangle|A4\rangle - \sqrt{\frac{1}{3}}|B23\rangle|A3\rangle, \quad (4.119)$$

$$|C29\rangle = |\frac{1}{2},\frac{1}{2},1,\frac{1}{2},\frac{1}{2},\frac{3}{2},0,0,1\rangle = \sqrt{\frac{1}{2}}|B06\rangle|A5\rangle - \sqrt{\frac{1}{3}}|B15\rangle|A6\rangle + \sqrt{\frac{1}{6}}|B24\rangle|A7\rangle, \tag{4.120}$$

$$|C30\rangle = |\frac{1}{2},\frac{1}{2},1,\frac{1}{2},\frac{1}{2},\frac{3}{2},0,1,1\rangle = \sqrt{\frac{1}{2}}|B07\rangle|A5\rangle - \sqrt{\frac{1}{3}}|B16\rangle|A6\rangle + \sqrt{\frac{1}{6}}|B25\rangle|A7\rangle, \tag{4.121}$$

$$|C31\rangle = |\frac{1}{2},\frac{1}{2},1,\frac{1}{2},\frac{1}{2},\frac{3}{2},1,0,1\rangle = \sqrt{\frac{1}{2}}|B08\rangle|A5\rangle - \sqrt{\frac{1}{3}}|B17\rangle|A6\rangle + \sqrt{\frac{1}{6}}|B26\rangle|A7\rangle, \tag{4.122}$$

$$|C32\rangle = |\frac{1}{2},\frac{1}{2},1,\frac{1}{2},\frac{1}{2},\frac{3}{2},1,1,1\rangle = \sqrt{\frac{1}{2}}|B09\rangle|A5\rangle - \sqrt{\frac{1}{3}}|B18\rangle|A6\rangle + \sqrt{\frac{1}{6}}|B27\rangle|A7\rangle, \tag{4.123}$$

$$|C33\rangle = |\frac{1}{2},\frac{1}{2},1,\frac{1}{2},\frac{3}{2},\frac{3}{2},0,1,1\rangle = \sqrt{\frac{1}{2}}|B10\rangle|A5\rangle - \sqrt{\frac{1}{3}}|B19\rangle|A6\rangle + \sqrt{\frac{1}{6}}|B28\rangle|A7\rangle, \tag{4.124}$$

$$|C34\rangle = |\frac{1}{2},\frac{1}{2},1,\frac{1}{2},\frac{3}{2},\frac{3}{2},1,1,1\rangle = \sqrt{\frac{1}{2}}|B11\rangle|A5\rangle - \sqrt{\frac{1}{3}}|B20\rangle|A6\rangle + \sqrt{\frac{1}{6}}|B29\rangle|A7\rangle, \tag{4.125}$$

$$|C35\rangle = |\frac{1}{2},\frac{1}{2},1,\frac{3}{2},\frac{1}{2},\frac{3}{2},1,0,1\rangle = \sqrt{\frac{1}{2}}|B12\rangle|A5\rangle - \sqrt{\frac{1}{3}}|B21\rangle|A6\rangle + \sqrt{\frac{1}{6}}|B30\rangle|A7\rangle, \tag{4.126}$$

$$|C36\rangle = |\frac{1}{2},\frac{1}{2},1,\frac{3}{2},\frac{1}{2},\frac{3}{2},1,1,1\rangle = \sqrt{\frac{1}{2}}|B13\rangle|A5\rangle - \sqrt{\frac{1}{3}}|B22\rangle|A6\rangle + \sqrt{\frac{1}{6}}|B31\rangle|A7\rangle, \tag{4.127}$$

$$|C37\rangle = |\frac{1}{2},\frac{1}{2},1,\frac{3}{2},\frac{3}{2},\frac{3}{2},1,1,1\rangle = \sqrt{\frac{1}{2}}|B14\rangle|A5\rangle - \sqrt{\frac{1}{3}}|B23\rangle|A6\rangle + \sqrt{\frac{1}{6}}|B32\rangle|A7\rangle, \tag{4.128}$$

$$\begin{aligned}|C38\rangle =&|\frac{1}{2},\frac{1}{2},2,\frac{1}{2},\frac{3}{2},\frac{3}{2},0,1,1\rangle \\ =& -\sqrt{\frac{1}{10}}|B38\rangle|A5\rangle + \sqrt{\frac{1}{5}}|B43\rangle|A6\rangle - \sqrt{\frac{3}{10}}|B48\rangle|A7\rangle + \sqrt{\frac{2}{5}}|B53\rangle|A8\rangle,\end{aligned} \tag{4.129}$$

$$\begin{aligned}|C39\rangle =&|\frac{1}{2},\frac{1}{2},2,\frac{1}{2},\frac{3}{2},\frac{3}{2},1,1,1\rangle \\ =& -\sqrt{\frac{1}{10}}|B39\rangle|A5\rangle + \sqrt{\frac{1}{5}}|B44\rangle|A6\rangle - \sqrt{\frac{3}{10}}|B49\rangle|A7\rangle + \sqrt{\frac{2}{5}}|B54\rangle|A8\rangle,\end{aligned} \tag{4.130}$$

$$\begin{aligned}|C40\rangle =&|\frac{1}{2},\frac{1}{2},2,\frac{3}{2},\frac{1}{2},\frac{3}{2},1,0,1\rangle \\ =&-\sqrt{\frac{1}{10}}|B40\rangle|A5\rangle+\sqrt{\frac{1}{5}}|B45\rangle|A6\rangle-\sqrt{\frac{3}{10}}|B50\rangle|A7\rangle+\sqrt{\frac{2}{5}}|B55\rangle|A8\rangle,\end{aligned} \tag{4.131}$$

$$\begin{aligned}|C41\rangle =&|\frac{1}{2},\frac{1}{2},2,\frac{3}{2},\frac{1}{2},\frac{3}{2},1,1,1\rangle \\ =&-\sqrt{\frac{1}{10}}|B41\rangle|A5\rangle+\sqrt{\frac{1}{5}}|B46\rangle|A6\rangle-\sqrt{\frac{3}{10}}|B51\rangle|A7\rangle+\sqrt{\frac{2}{5}}|B56\rangle|A8\rangle,\end{aligned} \tag{4.132}$$

$$\begin{aligned}|C42\rangle =&|\frac{1}{2},\frac{1}{2},2,\frac{3}{2},\frac{3}{2},\frac{3}{2},1,1,1\rangle \\ =&-\sqrt{\frac{1}{10}}|B42\rangle|A5\rangle+\sqrt{\frac{1}{5}}|B47\rangle|A6\rangle-\sqrt{\frac{3}{10}}|B52\rangle|A7\rangle+\sqrt{\frac{2}{5}}|B57\rangle|A8\rangle,\end{aligned} \tag{4.133}$$

For $S_{A,B,C} = \frac{3}{2}$ :

$$|C43\rangle = |\frac{3}{2},\frac{3}{2},1,\frac{1}{2},\frac{1}{2},\frac{1}{2},0,0,0\rangle = |B24\rangle|A1\rangle, \tag{4.134}$$

$$|C44\rangle = |\frac{3}{2},\frac{3}{2},1,\frac{1}{2},\frac{1}{2},\frac{1}{2},0,0,1\rangle = |B24\rangle|A3\rangle, \tag{4.135}$$

$$|C45\rangle = |\frac{3}{2},\frac{3}{2},1,\frac{1}{2},\frac{1}{2},\frac{1}{2},0,1,0\rangle = |B25\rangle|A1\rangle, \tag{4.136}$$

$$|C46\rangle = |\frac{3}{2},\frac{3}{2},1,\frac{1}{2},\frac{1}{2},\frac{1}{2},0,1,1\rangle = |B25\rangle|A3\rangle, \tag{4.137}$$

$$|C47\rangle = |\frac{3}{2},\frac{3}{2},1,\frac{1}{2},\frac{1}{2},\frac{1}{2},1,0,0\rangle = |B26\rangle|A1\rangle, \tag{4.138}$$

$$|C48\rangle = |\frac{3}{2},\frac{3}{2},1,\frac{1}{2},\frac{1}{2},\frac{1}{2},1,0,1\rangle = |B26\rangle|A3\rangle, \tag{4.139}$$

$$|C49\rangle = |\frac{3}{2},\frac{3}{2},1,\frac{1}{2},\frac{1}{2},\frac{1}{2},1,1,0\rangle = |B27\rangle|A1\rangle, \tag{4.140}$$

$$|C50\rangle = |\frac{3}{2},\frac{3}{2},1,\frac{1}{2},\frac{1}{2},\frac{1}{2},1,1,1\rangle = |B27\rangle|A3\rangle, \tag{4.141}$$

$$|C51\rangle = |\frac{3}{2}, \frac{3}{2}, 1, \frac{1}{2}, \frac{3}{2}, \frac{1}{2}, 0, 1, 0\rangle = |B28\rangle|A1\rangle, \tag{4.142}$$

$$|C52\rangle = |\frac{3}{2}, \frac{3}{2}, 1, \frac{1}{2}, \frac{3}{2}, \frac{1}{2}, 0, 1, 1\rangle = |B28\rangle|A3\rangle, \tag{4.143}$$

$$|C53\rangle = |\frac{3}{2}, \frac{3}{2}, 1, \frac{1}{2}, \frac{3}{2}, \frac{1}{2}, 1, 1, 0\rangle = |B29\rangle|A1\rangle, \tag{4.144}$$

$$|C54\rangle = |\frac{3}{2}, \frac{3}{2}, 1, \frac{1}{2}, \frac{3}{2}, \frac{1}{2}, 1, 1, 1\rangle = |B29\rangle|A3\rangle, \tag{4.145}$$

$$|C55\rangle = |\frac{3}{2}, \frac{3}{2}, 1, \frac{3}{2}, \frac{1}{2}, \frac{1}{2}, 1, 0, 0\rangle = |B30\rangle|A1\rangle, \tag{4.146}$$

$$|C56\rangle = |\frac{3}{2}, \frac{3}{2}, 1, \frac{3}{2}, \frac{1}{2}, \frac{1}{2}, 1, 0, 1\rangle = |B30\rangle|A3\rangle, \tag{4.147}$$

$$|C57\rangle = |\frac{3}{2}, \frac{3}{2}, 1, \frac{3}{2}, \frac{1}{2}, \frac{1}{2}, 1, 1, 0\rangle = |B31\rangle|A1\rangle, \tag{4.148}$$

$$|C58\rangle = |\frac{3}{2}, \frac{3}{2}, 1, \frac{3}{2}, \frac{1}{2}, \frac{1}{2}, 1, 1, 1\rangle = |B31\rangle|A3\rangle, \tag{4.149}$$

$$|C59\rangle = |\frac{3}{2}, \frac{3}{2}, 1, \frac{3}{2}, \frac{3}{2}, \frac{1}{2}, 1, 1, 0\rangle = |B32\rangle|A1\rangle, \tag{4.150}$$

$$|C60\rangle = |\frac{3}{2}, \frac{3}{2}, 1, \frac{3}{2}, \frac{3}{2}, \frac{1}{2}, 1, 1, 1\rangle = |B32\rangle|A3\rangle, \tag{4.151}$$

$$|C61\rangle = |\frac{3}{2}, \frac{3}{2}, 2, \frac{1}{2}, \frac{3}{2}, \frac{1}{2}, 0, 1, 0\rangle = \sqrt{\frac{4}{5}}|B53\rangle|A2\rangle - \sqrt{\frac{1}{5}}|B48\rangle|A1\rangle, \tag{4.152}$$

$$|C62\rangle = |\frac{3}{2}, \frac{3}{2}, 2, \frac{1}{2}, \frac{3}{2}, \frac{1}{2}, 0, 1, 1\rangle = \sqrt{\frac{4}{5}}|B53\rangle|A4\rangle - \sqrt{\frac{1}{5}}|B48\rangle|A3\rangle, \tag{4.153}$$

$$|C63\rangle = |\frac{3}{2}, \frac{3}{2}, 2, \frac{1}{2}, \frac{3}{2}, \frac{1}{2}, 1, 1, 0\rangle = \sqrt{\frac{4}{5}}|B54\rangle|A2\rangle - \sqrt{\frac{1}{5}}|B49\rangle|A1\rangle, \tag{4.154}$$

$$|C64\rangle = |\frac{3}{2}, \frac{3}{2}, 2, \frac{1}{2}, \frac{3}{2}, \frac{1}{2}, 1, 1, 1\rangle = \sqrt{\frac{4}{5}}|B54\rangle|A4\rangle - \sqrt{\frac{1}{5}}|B49\rangle|A3\rangle, \tag{4.155}$$

$$|C65\rangle = |\frac{3}{2}, \frac{3}{2}, 2, \frac{3}{2}, \frac{1}{2}, \frac{1}{2}, 1, 0, 0\rangle = \sqrt{\frac{4}{5}}|B55\rangle|A2\rangle - \sqrt{\frac{1}{5}}|B50\rangle|A1\rangle, \tag{4.156}$$

$$|C66\rangle = |\frac{3}{2}, \frac{3}{2}, 2, \frac{3}{2}, \frac{1}{2}, \frac{1}{2}, 1, 0, 1\rangle = \sqrt{\frac{4}{5}}|B55\rangle|A4\rangle - \sqrt{\frac{1}{5}}|B50\rangle|A3\rangle, \tag{4.157}$$

$$|C67\rangle = |\frac{3}{2}, \frac{3}{2}, 2, \frac{3}{2}, \frac{1}{2}, \frac{1}{2}, 1, 1, 0\rangle = \sqrt{\frac{4}{5}}|B56\rangle|A2\rangle - \sqrt{\frac{1}{5}}|B51\rangle|A1\rangle, \tag{4.158}$$

$$|C68\rangle = |\frac{3}{2},\frac{3}{2},2,\frac{3}{2},\frac{1}{2},\frac{1}{2},1,1,1\rangle = \sqrt{\frac{4}{5}}|B56\rangle|A4\rangle - \sqrt{\frac{1}{5}}|B51\rangle|A3\rangle, \quad (4.159)$$

$$|C69\rangle = |\frac{3}{2},\frac{3}{2},2,\frac{3}{2},\frac{3}{2},\frac{1}{2},1,1,0\rangle = \sqrt{\frac{4}{5}}|B57\rangle|A2\rangle - \sqrt{\frac{1}{5}}|B52\rangle|A1\rangle, \quad (4.160)$$

$$|C70\rangle = |\frac{3}{2},\frac{3}{2},2,\frac{3}{2},\frac{3}{2},\frac{1}{2},1,1,1\rangle = \sqrt{\frac{4}{5}}|B57\rangle|A4\rangle - \sqrt{\frac{1}{5}}|B52\rangle|A3\rangle, \quad (4.161)$$

$$|C71\rangle = |\frac{3}{2},\frac{3}{2},0,\frac{1}{2},\frac{1}{2},\frac{3}{2},0,0,1\rangle = |B01\rangle|A5\rangle, \quad (4.162)$$

$$|C72\rangle = |\frac{3}{2},\frac{3}{2},0,\frac{1}{2},\frac{1}{2},\frac{3}{2},0,1,1\rangle = |B02\rangle|A5\rangle, \quad (4.163)$$

$$|C73\rangle = |\frac{3}{2},\frac{3}{2},0,\frac{1}{2},\frac{1}{2},\frac{3}{2},1,0,1\rangle = |B03\rangle|A5\rangle, \quad (4.164)$$

$$|C74\rangle = |\frac{3}{2},\frac{3}{2},0,\frac{1}{2},\frac{1}{2},\frac{3}{2},1,1,1\rangle = |B04\rangle|A5\rangle, \quad (4.165)$$

$$|C75\rangle = |\frac{3}{2},\frac{3}{2},0,\frac{3}{2},\frac{3}{2},\frac{3}{2},1,1,1\rangle = |B05\rangle|A5\rangle, \quad (4.166)$$

$$|C76\rangle = |\frac{3}{2},\frac{3}{2},0,\frac{1}{2},\frac{1}{2},\frac{3}{2},0,0,1\rangle = \sqrt{\frac{3}{5}}|B15\rangle|A5\rangle - \sqrt{\frac{2}{5}}|B24\rangle|A6\rangle, \quad (4.167)$$

$$|C77\rangle = |\frac{3}{2},\frac{3}{2},0,\frac{1}{2},\frac{1}{2},\frac{3}{2},0,1,1\rangle = \sqrt{\frac{3}{5}}|B16\rangle|A5\rangle - \sqrt{\frac{2}{5}}|B25\rangle|A6\rangle, \quad (4.168)$$

$$|C78\rangle = |\frac{3}{2},\frac{3}{2},0,\frac{1}{2},\frac{1}{2},\frac{3}{2},1,0,1\rangle = \sqrt{\frac{3}{5}}|B17\rangle|A5\rangle - \sqrt{\frac{2}{5}}|B26\rangle|A6\rangle, \quad (4.169)$$

$$|C79\rangle = |\frac{3}{2},\frac{3}{2},0,\frac{1}{2},\frac{1}{2},\frac{3}{2},1,1,1\rangle = \sqrt{\frac{3}{5}}|B18\rangle|A5\rangle - \sqrt{\frac{2}{5}}|B27\rangle|A6\rangle, \quad (4.170)$$

$$|C80\rangle = |\frac{3}{2},\frac{3}{2},0,\frac{1}{2},\frac{3}{2},\frac{3}{2},0,1,1\rangle = \sqrt{\frac{3}{5}}|B19\rangle|A5\rangle - \sqrt{\frac{2}{5}}|B28\rangle|A6\rangle, \quad (4.171)$$

$$|C81\rangle = |\frac{3}{2},\frac{3}{2},0,\frac{1}{2},\frac{3}{2},\frac{3}{2},1,1,1\rangle = \sqrt{\frac{3}{5}}|B20\rangle|A5\rangle - \sqrt{\frac{2}{5}}|B29\rangle|A6\rangle, \quad (4.172)$$

$$|C82\rangle = |\frac{3}{2},\frac{3}{2},0,\frac{3}{2},\frac{1}{2},\frac{3}{2},1,0,1\rangle = \sqrt{\frac{3}{5}}|B21\rangle|A5\rangle - \sqrt{\frac{2}{5}}|B30\rangle|A6\rangle, \quad (4.173)$$

$$|C83\rangle = |\frac{3}{2},\frac{3}{2},0,\frac{3}{2},\frac{1}{2},\frac{3}{2},1,1,1\rangle = \sqrt{\frac{3}{5}}|B22\rangle|A5\rangle - \sqrt{\frac{2}{5}}|B31\rangle|A6\rangle, \quad (4.174)$$

$$|C84\rangle = |\frac{3}{2},\frac{3}{2},0,\frac{3}{2},\frac{3}{2},\frac{3}{2},1,1,1\rangle = \sqrt{\frac{3}{5}}|B23\rangle|A5\rangle - \sqrt{\frac{2}{5}}|B32\rangle|A6\rangle, \quad (4.175)$$

$$|C85\rangle = |\frac{3}{2},\frac{3}{2},0,\frac{1}{2},\frac{3}{2},\frac{3}{2},0,1,1\rangle = \sqrt{\frac{1}{5}}|B43\rangle|A5\rangle - \sqrt{\frac{2}{5}}|B48\rangle|A6\rangle + \sqrt{\frac{2}{5}}|B53\rangle|A7\rangle, \quad (4.176)$$

$$|C86\rangle = |\frac{3}{2},\frac{3}{2},0,\frac{1}{2},\frac{3}{2},\frac{3}{2},1,1,1\rangle = \sqrt{\frac{1}{5}}|B44\rangle|A5\rangle - \sqrt{\frac{2}{5}}|B49\rangle|A6\rangle + \sqrt{\frac{2}{5}}|B54\rangle|A7\rangle, \quad (4.177)$$

$$|C87\rangle = |\frac{3}{2},\frac{3}{2},0,\frac{3}{2},\frac{1}{2},\frac{3}{2},1,0,1\rangle = \sqrt{\frac{1}{5}}|B45\rangle|A5\rangle - \sqrt{\frac{2}{5}}|B50\rangle|A6\rangle + \sqrt{\frac{2}{5}}|B55\rangle|A7\rangle, \quad (4.178)$$

$$|C88\rangle = |\frac{3}{2},\frac{3}{2},0,\frac{3}{2},\frac{1}{2},\frac{3}{2},1,1,1\rangle = \sqrt{\frac{1}{5}}|B46\rangle|A5\rangle - \sqrt{\frac{2}{5}}|B51\rangle|A6\rangle + \sqrt{\frac{2}{5}}|B56\rangle|A7\rangle, \quad (4.179)$$

$$|C89\rangle = |\frac{3}{2},\frac{3}{2},0,\frac{3}{2},\frac{3}{2},\frac{3}{2},1,1,1\rangle = \sqrt{\frac{1}{5}}|B47\rangle|A5\rangle - \sqrt{\frac{2}{5}}|B52\rangle|A6\rangle + \sqrt{\frac{2}{5}}|B57\rangle|A7\rangle, \quad (4.180)$$

$$\begin{aligned}|C90\rangle =& |\frac{3}{2},\frac{3}{2},3,\frac{3}{2},\frac{3}{2},\frac{3}{2},1,1,1\rangle \\ =& -\sqrt{\frac{1}{35}}|B61\rangle|A5\rangle + \sqrt{\frac{4}{35}}|B62\rangle|A6\rangle + \sqrt{\frac{10}{35}}|B63\rangle|A7\rangle + \sqrt{\frac{20}{35}}|B64\rangle|A8\rangle.\end{aligned} \quad (4.181)$$

### 4.11.3 Searching Algorithm Pseudocode

### 4.11.4 Exchange Sequence Data

**Algorithm 1** Krotov's method with pulse deletion for quantum control

**Initialize:**
$L \leftarrow$ initial number of time segments ▷ Set initial number of segments
$J^L_{i,i+1} \leftarrow$ initial swap pulse sequence ▷ Initial guess for swap pulses
$U_{\text{gate}} \leftarrow$ target quantum gate
infidelity_threshold $\leftarrow$ desired infidelity threshold
max_iterations $\leftarrow$ maximum Krotov iterations
**repeat**
 **Optimize Swap Pulses:**
 $J^L_{i,i+1} \leftarrow \text{Krotov_optimization}(J^L_{i,i+1}, U_{\text{gate}})$
 infidelity $\leftarrow \text{calculate_infidelity}(J^L_{i,i+1}, U_{\text{gate}})$
**until** infidelity $<$ infidelity_threshold
**Main Loop:**
deleted_pulse $\leftarrow \emptyset$ ▷ Empty set of deleted pulses
**for** $k = 1, ..., L$ **do**
 **for** $i = 1, ..., n-1$ **do**
  **if** $\{k, i\} \notin$ deleted_pulse **then**
   temp_J $\leftarrow \left\{J^n_{i,i+1}\right\}$
   $J^k_{i,i+1} \leftarrow 0$ ▷ Delete this swap pulse
   $\left\{J^n_{i,i+1}\right\} \leftarrow \text{Krotov_optimization}(\left\{J^n_{i,i+1}\right\} \mid_{\substack{J^m_{l,l+1}=0 \\ \{m,l\}\in\text{deleted_pulse}}}, U_{\text{gate}}, \text{max_iterations})$
    **with** $J^m_{l,l+1}$ **fixed at** 0 **if** $\{m, l\} \in$ deleted_pulse ▷ Do not update
deleted pulses
   $\bar{F} \leftarrow \text{calculate_infidelity}(\left\{J^n_{i,i+1}\right\}, U_{\text{gate}})$
   **if** $\bar{F} > \bar{F}_{threshold}$ **then**
    $\left\{J^n_{i,i+1}\right\} \leftarrow$ temp_J ▷ Restore sequence if infidelity threshold not
met
   **end if**
  **end if**
 **end for**
 **if** max_fidelity $> 0$ **then**
  deleted_pulse $\leftarrow$ deleted_pulse $\cup \{(k, i) \mid J^k_{i,i+1} = 0\}$
 **end if**
**end for**

| Step | $J_{2,3}$ | $J_{3,4}$ | $J_{4,5}$ | $J_{5,6}$ | $J_{6,7}$ | $J_{7,8}$ | $J_{8,9}$ |
|---|---|---|---|---|---|---|---|
| 1 | 0 | 0.171280 | 0 | 0.057610 | 0 | -0.113705 | 0 |
| 2 | -0.101342 | 0 | 0.177459 | 0 | 0 | 0 | 0.009875 |
| 3 | 0 | 0.328343 | 0 | 0.363308 | 0 | -0.075663 | 0 |
| 4 | 0.630905 | 0 | -0.609766 | 0 | 0 | 0 | 0.036130 |
| 5 | 0 | 0.391954 | 0 | 0.355324 | 0 | -0.012761 | 0 |
| 6 | -0.119032 | 0 | -0.362184 | 0 | 0 | 0 | 0.015486 |
| 7 | 0 | 0.267867 | 0 | -0.540480 | 0 | -0.015869 | 0 |
| 8 | -0.934082 | 0 | -0.046239 | 0 | 0 | 0 | -0.445718 |
| 9 | 0 | -0.821827 | 0 | 0.005940 | 0 | 0.775597 | 0 |
| 10 | -0.004373 | 0 | -0.547811 | 0 | 0 | 0 | -0.163575 |
| 11 | 0 | 0.668425 | 0 | 0.260181 | 0 | 0.004333 | 0 |
| 12 | 0.396335 | 0 | -0.817440 | 0 | 0 | 0 | -0.013776 |
| 13 | 0 | -0.441627 | 0 | -0.646837 | 0 | -0.048896 | 0 |
| 14 | 0.287595 | 0 | -0.434438 | 0 | 0 | 0 | -0.496229 |
| 15 | 0 | 0.952645 | 0 | 0.791044 | 0 | -0.880058 | 0 |
| 16 | 0.011442 | 0 | -0.544495 | 0 | 0 | 0 | -0.346432 |
| 17 | 0 | -0.506475 | 0 | -0.048677 | 0 | -0.081481 | 0 |
| 18 | -0.092802 | 0 | -0.476621 | 0 | 0 | 0 | -0.149473 |
| 19 | 0 | -0.542894 | 0 | -0.294417 | 0 | -0.666237 | 0 |
| 20 | 0.323522 | 0 | 0.129986 | 0 | 0.690990 | 0 | 0.102330 |
| 21 | 0 | -0.485848 | 0 | 0 | 0 | -0.567432 | 0 |
| 22 | -0.056718 | 0 | 0.487515 | 0 | -0.750763 | 0 | 0.597217 |
| 23 | 0 | -0.700015 | 0 | 0 | 0 | -0.897555 | 0 |
| 24 | 0.761959 | 0 | 0.797867 | 0 | -1.145150 | 0 | 0.503777 |
| 25 | 0 | 0.541105 | 0 | -1.383010 | 0 | 0.421025 | 0 |
| 26 | -0.094056 | 0 | 0 | 0 | -0.912389 | 0 | 0.391495 |
| 27 | 0 | -0.368176 | 0 | -1.434450 | 0 | -1.027200 | 0 |
| 28 | 0.858183 | 0 | 0 | 0 | 1.018070 | 0 | -0.627970 |

Table 4.11 Fully optimized 55-time step raw sequence table, detailing the steps from time step 1 to time step 28.

| Step | $J_{2,3}$ | $J_{3,4}$ | $J_{4,5}$ | $J_{5,6}$ | $J_{6,7}$ | $J_{7,8}$ | $J_{8,9}$ |
|---|---|---|---|---|---|---|---|
| 29 | 0 | 0.042789 | 0 | 1.001480 | 0 | 0.632431 | 0 |
| 30 | -0.024318 | 0 | 0 | 0 | 0.993828 | 0 | -0.671796 |
| 31 | 0 | -0.016895 | 0 | 0.937422 | 0 | -0.139507 | 0 |
| 32 | -0.564398 | 0 | 0 | 0 | 1.312100 | 0 | 0.140317 |
| 33 | 0 | -0.745372 | 0 | 0.503377 | 0 | -0.438745 | 0 |
| 34 | 0.013151 | 0 | 0 | 0 | 1.360290 | 0 | -0.521566 |
| 35 | 0 | -0.026124 | 0 | 0.693291 | 0 | 0.831985 | 0 |
| 36 | -0.382984 | 0 | 0 | 0 | -1.954380 | 0 | 0.395606 |
| 37 | 0 | -0.331504 | 0 | -0.470088 | 0 | -0.711540 | 0 |
| 38 | -0.343993 | 0 | 0 | 0 | -1.684190 | 0 | -0.250346 |
| 39 | 0 | -0.718306 | 0 | 0.709398 | 0 | 0.572356 | 0 |
| 40 | 0.595562 | 0 | 0 | 0 | 0.785922 | 0 | 0.598140 |
| 41 | 0 | 0.027358 | 0 | -0.132902 | 0 | -0.306381 | 0 |
| 42 | 0.320719 | 0 | -0.654041 | 0 | 0.513170 | 0 | 0.301276 |
| 43 | 0 | 0.391232 | 0 | -0.319153 | 0 | 0.862290 | 0 |
| 44 | 0.962250 | 0 | -0.133952 | 0 | 0.000011 | 0 | -0.104524 |
| 45 | 0 | -0.101004 | 0 | -0.277882 | 0 | 0.066079 | 0 |
| 46 | 0.443214 | 0 | -0.572575 | 0 | 0 | 0 | -0.037455 |
| 47 | 0 | 0.578033 | 0 | -0.378393 | 0 | -0.018994 | 0 |
| 48 | 0.857235 | 0 | 0.327071 | 0 | 0 | 0 | -0.230227 |
| 49 | 0 | -0.275722 | 0 | -1.084370 | 0 | -0.100310 | 0 |
| 50 | -0.171715 | 0 | -1.122270 | 0 | 0 | 0 | 0.051434 |
| 51 | 0 | -0.638390 | 0 | 0.033564 | 0 | -0.088825 | 0 |
| 52 | 0.011886 | 0 | -0.261533 | 0 | 0 | 0 | 0.537031 |
| 53 | 0 | 0.393078 | 0 | -0.241973 | 0 | 0.236710 | 0 |
| 54 | -0.424537 | 0 | 1.315600 | 0 | 0 | 0 | 0.428487 |
| 55 | 0 | -0.576510 | 0 | -0.132951 | 0 | -0.034890 | 0 |

Table 4.12 Fully optimized 55-time step raw sequence table, detailing the steps from time step 28 to time step 55.

| Step | $J_{2,3}$ | $J_{3,4}$ | $J_{4,5}$ | $J_{5,6}$ | $J_{6,7}$ | $J_{7,8}$ | $J_{8,9}$ |
|---|---|---|---|---|---|---|---|
| 1 | -0.361356 | 0 | 0 | 0 | 0 | 0 | 0 |
| 2 | 0 | 0.46694 | 0 | 0 | 0 | 0 | 0 |
| 3 | 0.474996 | 0 | -0.6808 | 0 | 0 | 0 | 0 |
| 4 | 0 | 0.463992 | 0 | 0.683111 | 0 | 0 | 0 |
| 5 | -0.921766 | 0 | -0.352727 | 0 | 0 | 0 | 0 |
| 6 | 0 | -0.647173 | 0 | -1.29074 | 0 | 0 | 0 |
| 7 | 0 | 0 | -0.783947 | 0 | 0 | 0 | 0 |
| 8 | 0 | 0.289479 | 0 | 0.463229 | 0 | 0 | 0 |
| 9 | 0.23616 | 0 | -1.03236 | 0 | 0 | 0 | 0 |
| 10 | 0 | -0.517465 | 0 | -0.341 | 0 | 0 | 0 |
| 11 | 0.445978 | 0 | -0.591699 | 0 | 0 | 0 | 0 |
| 12 | 0 | 0.757532 | 0 | 1.29282 | 0 | 0 | 0 |
| 13 | 0 | 0 | -0.580782 | 0 | 0 | 0 | 0 |
| 14 | 0 | 0 | 0 | -0.355991 | 0 | 0 | 0 |
| 15 | 0 | 0 | -0.547022 | 0 | 0 | 0 | 0 |
| 16 | 0 | -0.76173 | 0 | 0 | 0 | -0.401742 | 0 |
| 17 | 0.400511 | 0 | 0 | 0 | 0.561134 | 0 | -0.985982 |
| 18 | 0 | -0.490934 | 0 | 0 | 0 | -0.49032 | 0 |
| 19 | 0 | 0 | 0.613639 | 0 | -0.240236 | 0 | 0.782226 |
| 20 | 0 | -0.562063 | 0 | 0 | 0 | -0.574828 | 0 |
| 21 | 0 | 0 | 0.708219 | 0 | -1.47277 | 0 | -0.0251349 |
| 22 | 0 | 0.448977 | 0 | -0.754719 | 0 | 0 | 0 |
| 23 | 0.780178 | 0 | 0 | 0 | -0.409208 | 0 | 0 |
| 24 | 0 | -0.279016 | 0 | 0 | 0 | -1.07902 | 0 |
| 25 | -0.585912 | 0 | 0 | 0 | -0.313122 | 0 | 0 |
| 26 | 0 | -1.07795 | 0 | 0.812474 | 0 | 0.370976 | 0 |
| 27 | 0.768569 | 0 | 0 | 0 | 0.20101 | 0 | -1.44753 |
| 28 | 0 | 0 | 0 | 0 | 0 | -0.571905 | 0 |
| 29 | 0 | 0 | 0 | 0 | 1.31895 | 0 | 0.349844 |
| 30 | 0 | 0 | 0 | 0.724231 | 0 | -0.647533 | 0 |
| 31 | 0 | 0 | 0 | 0 | 1.8308 | 0 | 0 |
| 32 | 0 | 0 | 0 | 0 | 0 | 1.34169 | 0 |
| 33 | 0 | 0 | 0 | 0 | 0 | 0 | 0.418598 |
| 34 | 0 | 0 | 0 | 0.481629 | 0 | -0.807275 | 0 |
| 35 | 0 | 0 | 0 | 0 | -1.59216 | 0 | 0 |
| 36 | 0 | 0 | 0 | 0.650856 | 0 | -0.0284115 | 0 |
| 37 | 0 | 0 | -0.692437 | 0 | 1.29184 | 0 | 0.45061 |
| 38 | 0 | 0.51052 | 0 | 0 | 0 | -0.529812 | 0 |
| 39 | 0.610313 | 0 | -0.0895998 | 0 | 0.577097 | 0 | 0 |
| 40 | 0 | -0.538592 | 0 | -0.649904 | 0 | 1.46932 | 0 |
| 41 | 0.6718 | 0 | -0.648449 | 0 | 0 | 0 | 0 |
| 42 | 0 | 0.586782 | 0 | -0.353683 | 0 | 0 | 0 |
| 43 | 0.669013 | 0 | 0.378554 | 0 | 0 | 0 | 0 |
| 44 | 0 | 0 | 0 | -1.18546 | 0 | 0 | 0 |
| 45 | 0 | 0 | -1.44203 | 0 | 0 | 0 | 0 |
| 46 | 0 | -1.01647 | 0 | 0 | 0 | 0 | 0 |
| 47 | 0 | 0 | -0.617632 | 0 | 0 | 0 | 0 |
| 48 | 0 | 0.823515 | 0 | -0.376096 | 0 | 0 | 0 |
| 49 | -0.475713 | 0 | 1.31782 | 0 | 0 | 0 | 0 |
| 50 | 0 | -0.556544 | 0 | 0 | 0 | 0 | 0 |

Table 4.13 Fully optimized 92-pulse, 50-time step sequence table, detailing the steps from time step 1 to time step 50.

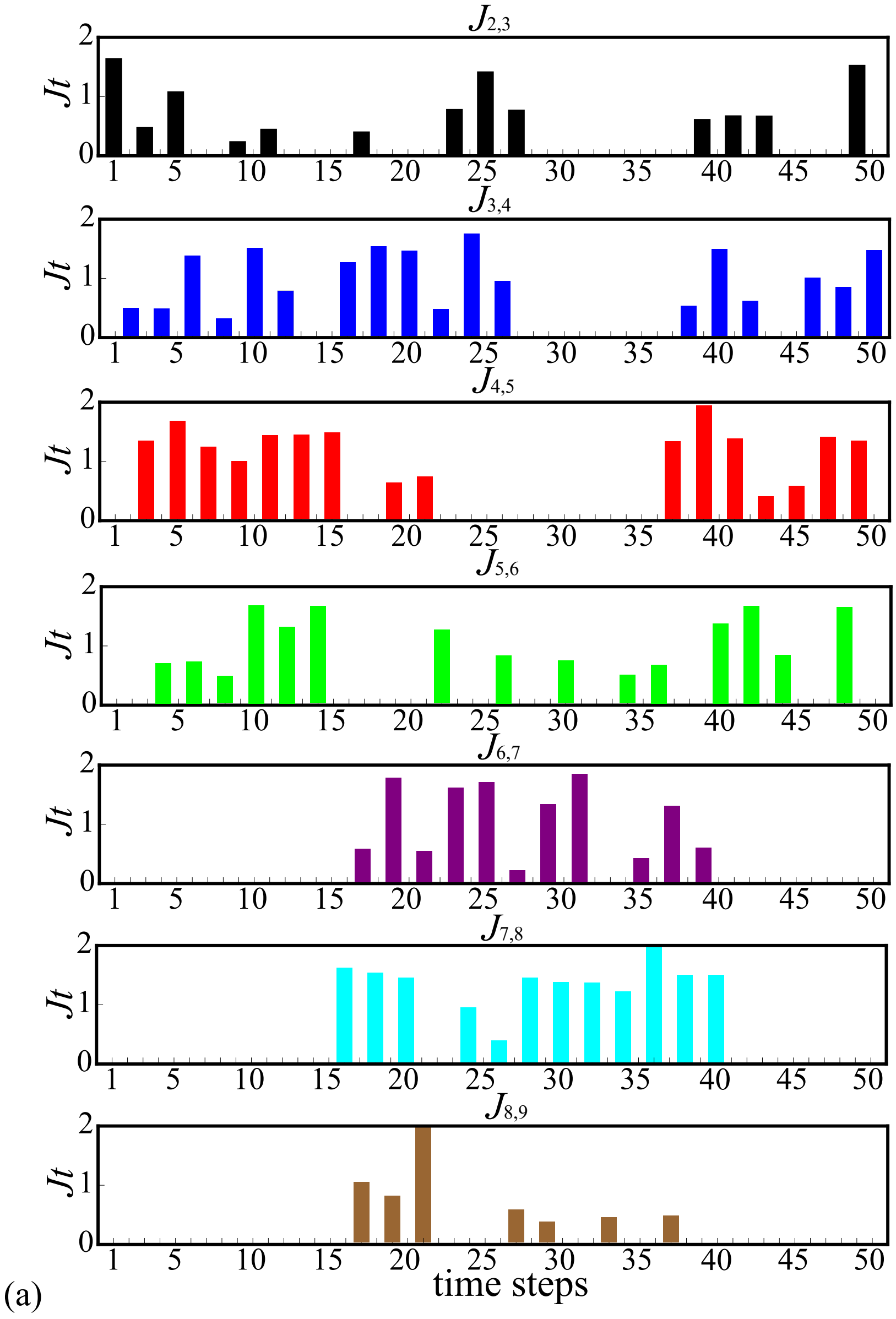

Fig. 4.13 Exchange pulse profiles for 92-pulse sequence, ordered in time step.

# Chapter 5

# Summary and Outlook

## 5.1 Summary

In this thesis, I investigate various physical effects and control methods in quantum dot chain systems. Specifically, I focus on the behavior of electron-phonon coupling in multi-electron quantum dots, the entanglement phase diagram in multi-electron quantum dot chains as the dot potential varies, and the construction of three-qubit gates and the search for exchange sequences in a nine-quantum-dot, nine-electron system.

Chapter 1 introduces the fundamental concepts of the quantum dot chain system, focusing on the extended Hubbard model, the double quantum dot system, and electron-phonon coupling. Chapter 2 delves into electron-phonon coupling in a multielectron double quantum dot system, examining the dephasing rates in unbiased and biased cases. I show that multielectron quantum dots may offer advantages under certain conditions. In Chapter 3, I investigate entanglement entropy in a multielectron quantum dot spin chain system using the extended Hubbard model, showing that local and pairwise entanglement is influenced by Coulomb interactions, tunneling strengths, and potential energy variations, which can significantly impact ground state configurations and entanglement entropy. In Chapter 4, I construct the decoherence-free subspace in a nine-spin quantum dot chain and explore operation sequences in the quantum dot spin chain system based on the Heisenberg model, describing the nine-spin system within a nine-quantum-dot arrangement. I creatively used the Krotov method of quantum optimal control and pulse-based brickwork ansatz, and individually identified a more efficient pulse-level operation sequence for an exchange-only quantum dot spin chain system, potentially enhancing the development of concise quantum algorithm representations.

## 5.2 Future work on theory on electron-phonon spin dephasing in GaAs multi-electron double quantum dots

In Chapter 2, we explore the electron-phonon interaction within a double quantum dot system, where one dot hosts multiple electrons while the other contains a single electron. The system's magnetic field is tuned such that the ground and excited states are primarily singlet and triplet states. However, recent experiments have demonstrated that spin states with high spin angular momentum can form under relatively high magnetic fields [169, 142].

In [169], experiments reveal that at a magnetic field of $B = 2.6\,\mathrm{T}$, a new signal indicating the transition to quintet states is observed. Further increasing the magnetic field to $B = 4.4\,\mathrm{T}$ results in the formation of a six-electron septet state. Additionally, [142] demonstrates in a gate-defined GaAs/AlGaAs single quantum dot, using spin filtering with quantum Hall edge states coupled to the dot, that the relaxation rates of high-spin states can be measured. The findings indicate that the relaxation rates of high-spin states, such as the three-electron quartet state and the four-electron quintet state, are significantly faster than those of low-spin states.

These results suggest that under high magnetic fields, the ground and excited states of the system can be significantly altered. The physical mechanisms underlying the coupling between electron higher spin states in double quantum dots and other factors, such as phonons, remain inadequately explored in the literature.

## 5.3 Future work on exploring entanglement spectrum and phase diagram in multi-electron quantum dot chains

In Chapter 3, we investigate the entanglement behavior in a four-quantum-dot spin chain, considering systems with either four or six electrons. Our study reveals a variety of behaviors as we adjust the strength ratio between the on-site Coulomb interaction and the nearest-neighbor Coulomb interactions. Notably, recent literature reveals diverse dynamics associated with the valley phase in silicon quantum dot spin chains [254, 26]. Consequently, more extensive research on entanglement with valley phase and quantum simulation in silicon quantum dot spin chains is essential for future studies.

Additionally, advanced computational methods, such as the quantum tensor network approach, have potential shortcuts for calculating the dynamic behavior in quantum dot spin chains [259, 200]. These findings suggest promising directions for future research on the behavior of quantum dot spin chains with multi-electron occupancy or electrons in higher orbitals.

## 5.4 Future work on constructing Three-Qubit Gate Pulse Sequences in Exchange-Only Spin System

Recent experimental progress [286] has demonstrated that implementing a two-qubit gate sequence using only voltage pulses is feasible, marking a significant advancement in the realization of quantum dot spin qubits. The literature [233, 126, 3] has explored the potential geometries of six-spin quantum dot systems, indicating that alternative sequences can be implemented with specific quantum dot array configurations. However, for larger quantum dot arrays, the complexity of geometric arrangements for exchange-only qubits, such as non-linear three-qubit gates, has not been thoroughly examined.

Furthermore, the local geometry of a single exchange-only (EO) qubit can influence the length and efficiency of the two-qubit gate sequence, raising similar considerations for 2D arrays. Figure 5.1 illustrates five possible nine-spin geometries that remain to be discussed, which could offer shorter sequences and improved performance under noise conditions. Notably, each geometry also has more possible pulse-based brickwork ansatz. For larger arrays of quantum dots, this results in a complex tiling problem that requires detailed investigation. Moreover, and more importantly, quantum algorithms based on gate synthesis can greatly increase the number of pulses due to traditional quantum gate decomposition, leading to an exponential rise in pulses number. Therefore, in EO qubits, more concise algorithmic pulse implementations should be sought for specific geometric configurations.

Figure 5.1(a) shows three exchange-only (EO) qubits arranged in parallel. Each EO qubit can be translated spatially, with one to three possible exchange pulses with adjacent EO qubits. Figure 5.1(b) illustrates three EO qubits, with one EO qubit parallel to the adjacent two EO qubits. The two EO qubits on the right are connected head-to-tail, and the leftmost EO qubit can have one or two possible exchange pulses with the upper right EO qubit and one possible exchange pulse with the lower right EO qubit.

Figure 5.1(c) displays three triangular EO qubits connected at their vertices, with each vertex having one possible exchange pulse with the other two EO qubits, repre-

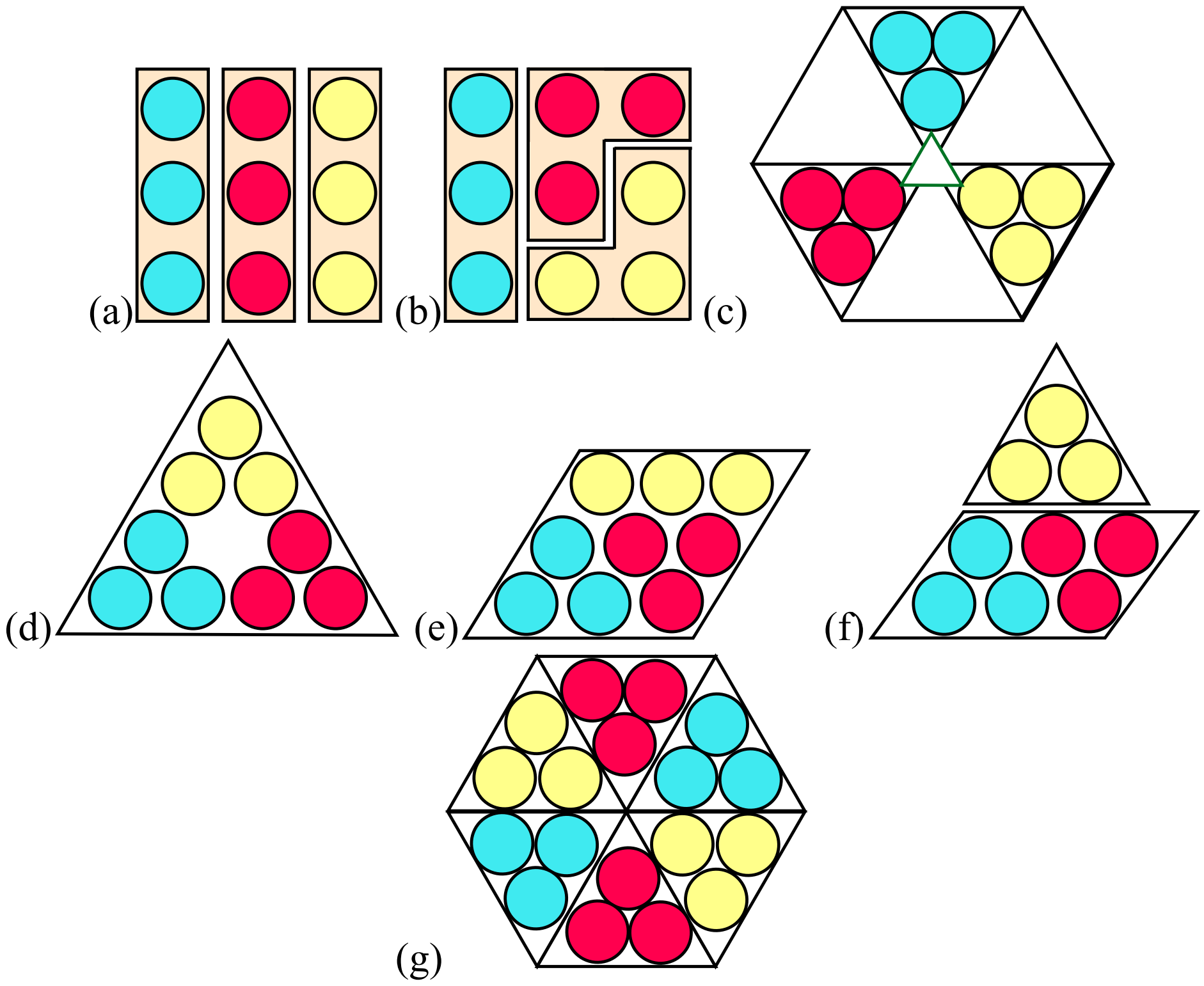

Fig. 5.1 Three DFS qubits are noted by different colors, and each DFS qubit contains three spins. (a)-(b)Two examples of square geometry for three EO qubits. (c)-(g)Five examples of triangle geometry for three or more EO qubits.

sented by the green triangle in the center. Figure 5.1(d) shows an alternative arrangement of three triangular EO qubits connected at their vertices, forming a larger triangle. Figure 5.1(e) presents three spins arranged linearly in one DFS qubit, while two DFS qubits are arranged in a triangular configuration.

Figure 5.1(f) illustrates three triangular DFS qubits arranged more efficiently; two DFS qubits (cyan and yellow) have two possible exchange pulses with the third DFS qubit (red) and one possible exchange pulse with each other. Figure 5.1(g) shows each triangular DFS qubit having two possible exchange pulses with the adjacent two DFS qubits.

# References

[1] Abaach, S., Faqir, M., and El Baz, M. (2022). Long-range entanglement in quantum dots with fermi-hubbard physics. *Phys. Rev. A*, 106:022421.

[2] Abaach, S., Mzaouali, Z., and El Baz, M. (2023). Long distance entanglement and high-dimensional quantum teleportation in the fermi–hubbard model. *Sci. Rep.*, 13(1):964.

[3] Acuna, E., Broz, J. D., Shyamsundar, K., Mei, A. B., Feeney, C. P., Smetanka, V., Davis, T., Lee, K., Choi, M. D., Boyd, B., Suh, J., Ha, W. D., Jennings, C., Pan, A. S., Sanchez, D. S., Reed, M. D., and Petta, J. R. (2024). Coherent control of a triangular exchange-only spin qubit.

[4] Amico, L., Fazio, R., Osterloh, A., and Vedral, V. (2008). Entanglement in many-body systems. *Rev. Mod. Phys.*, 80:517–576.

[5] Andrews, R. W., Jones, C., Reed, M. D., Jones, A. M., Ha, S. D., Jura, M. P., Kerckhoff, J., Levendorf, M., Meenehan, S., Merkel, S. T., Smith, A., Sun, B., Weinstein, A. J., Rakher, M. T., Ladd, T. D., and Borselli, M. G. (2019). Quantifying error and leakage in an encoded si/sige triple-dot qubit. *Nat. Nanotechnol.*, 14(8):747–750.

[6] Anfossi, A., Giorda, P., and Montorsi, A. (2007). Entanglement in extended hubbard models and quantum phase transitions. *Phys. Rev. B*, 75:165106.

[7] Bacon, D., Kempe, J., Lidar, D. A., and Whaley, K. B. (2000). Universal fault-tolerant quantum computation on decoherence-free subspaces. *Phys. Rev. Lett.*, 85:1758–1761.

[8] Bakker, M. A., Mehl, S., Hiltunen, T., Harju, A., and DiVincenzo, D. P. (2015). Validity of the single-particle description and charge noise resilience for multielectron quantum dots. *Phys. Rev. B*, 91:155425.

[9] Barenco, A., Bennett, C. H., Cleve, R., DiVincenzo, D. P., Margolus, N., Shor, P., Sleator, T., Smolin, J. A., and Weinfurter, H. (1995). Elementary gates for quantum computation. *Phys. Rev. A*, 52:3457–3467.

[10] Barnes, E., Kestner, J. P., Nguyen, N. T. T., and Das Sarma, S. (2011). Screening of charged impurities with multielectron singlet-triplet spin qubits in quantum dots. *Phys. Rev. B*, 84:235309.

[11] Barnes, E., Nichol, J. M., and Economou, S. E. (2019). Stabilization and manipulation of multispin states in quantum-dot time crystals with heisenberg interactions. *Phys. Rev. B*, 99:035311.

[12] Baruffa, F., Stano, P., and Fabian, J. (2010). Spin-orbit coupling and anisotropic exchange in two-electron double quantum dots. *Phys. Rev. B*, 82:045311.

[13] Bauer, B., Bravyi, S., Motta, M., and Chan, G. (2020). Quantum algorithms for quantum chemistry and quantum materials science. *Chemical Reviews*.

[14] Benito, M., Croot, X., Adelsberger, C., Putz, S., Mi, X., Petta, J. R., and Burkard, G. (2019). Electric-field control and noise protection of the flopping-mode spin qubit. *Phys. Rev. B*, 100:125430.

[15] Benjamin, S. C. and Bose, S. (2004). Quantum computing in arrays coupled by "always-on" interactions. *Phys. Rev. A*, 70:032314.

[16] Bennett, C. H. and DiVincenzo, D. P. (2000). Quantum information and computation. *Nature*, 404(6775):247–255.

[17] Bluhm, H., Foletti, S., Neder, I., Rudner, M., Mahalu, D., Umansky, V., and Yacoby, A. (2011). Dephasing time of gaas electron-spin qubits coupled to a nuclear bath exceeding 200 $\mu$s. *Nat. Phys.*, 7(2):109–113.

[18] Blume-Kohout, R. and Viola, L. (2002). A robust, high-fidelity quantum memory. *Phys. Lett. A*, 298(3):198–202.

[19] Borjans, F., Croot, X. G., Mi, X., Gullans, M. J., and Petta, J. R. (2019). Phonon-mediated decay of singlet-triplet qubits in double quantum dots. *Nature*, 577:195–198.

[20] Borsoi, F., Hendrickx, N. W., John, V., Meyer, M., Motz, S., van Riggelen, F., Sammak, A., de Snoo, S. L., Scappucci, G., and Veldhorst, M. (2024). Shared control of a 16 semiconductor quantum dot crossbar array. *Nat. Nanotechnol.*, 19(1):21–27.

[21] Botzem, T., McNeil, R. P. G., Mol, J.-M., Schuh, D., Bougeard, D., and Bluhm, H. (2016). Quadrupolar and anisotropy effects on dephasing in two-electron spin qubits in gaas. *Nat. Commun.*, 7(1).

[22] Braunstein, S. L. and van Loock, P. (2005). Quantum information with continuous variables. *Rev. Mod. Phys.*, 77:513–577.

[23] Brown, K., Munro, W., and Kendon, V. (2010). Using quantum computers for quantum simulation. *Entropy*.

[24] Burkard, G., Ladd, T. D., Pan, A., Nichol, J. M., and Petta, J. R. (2023). Semiconductor spin qubits. *Rev. Mod. Phys.*, 95:025003.

[25] Burkard, G., Loss, D., and DiVincenzo, D. P. (1999). Coupled quantum dots as quantum gates. *Phys. Rev. B*, 59:2070–2078.

[26] Buterakos, D. and Das Sarma, S. (2021). Spin-valley qubit dynamics in exchange-coupled silicon quantum dots. *PRX Quantum*, 2:040358.

[27] Buterakos, D. and Das Sarma, S. (2023). Certain exact many-body results for hubbard model ground states testable in small quantum dot arrays. *Phys. Rev. B*, 107:014403.

[28] Byrd, M. S. and Lidar, D. A. (2002). Empirical memory effects in decoherence-free subspaces. *Phys. Rev. A*, 65(1):012322.

[29] Byrnes, T., Kim, N. Y., and Yamamoto, Y. (2008). Quantum simulation with interacting photons. *Nat. Phys.*, 4(11):803–810.

[30] Calabrese, P. and Cardy, J. (2004). Entanglement entropy and quantum field theory. *J. Stat. Mech.: Theory Exp.*, 2004(06):P06002.

[31] Calderbank, A. R. and Shor, P. W. (1996). Good quantum error-correcting codes exist. *Phys. Rev. A*, 54(2):1098–1105.

[32] Calderon-Vargas, F. A. and Kestner, J. P. (2015). Directly accessible entangling gates for capacitively coupled singlet-triplet qubits. *Phys. Rev. B*, 91:035301.

[33] Carollo, A., Spagnolo, B., Dubkov, A. A., and Valenti, D. (2019). J. stat. mech.: Theory exp. 2019(9):094010.

[34] Carollo, A., Valenti, D., and Spagnolo, B. (2020). Phys. rep. 838:1–72.

[35] Cerfontaine, P., Botzem, T., Ritzmann, J., Humpohl, S. S., Ludwig, A., Schuh, D., Bougeard, D., Wieck, A. D., and Bluhm, H. (2020). Closed-loop control of a gaas-based singlet-triplet spin qubit with 99.5*Nat. Commun.*, 11(1).

[36] Chan, G. X., Kestner, J. P., and Wang, X. (2021). Charge noise suppression in capacitively coupled singlet-triplet spin qubits under magnetic field. *Phys. Rev. B*, 103:L161409.

[37] Chan, G. X. and Wang, X. (2019). Suppression of leakage for a charge qubit in triangular triple quantum dots. *Adv. Quantum Technol.*, 2(12):1900072.

[38] Chan, G. X. and Wang, X. (2022a). Microscopic theory of a magnetic-field-tuned sweet spot of exchange interactions in multielectron quantum-dot systems. *Phys. Rev. B*, 105:245409.

[39] Chan, G. X. and Wang, X. (2022b). Robust entangling gate for capacitively coupled few-electron singlet-triplet qubits. *Phys. Rev. B*, 106:075417.

[40] Chan, G. X. and Wang, X. (2022c). Sign switching of superexchange mediated by a few electrons in a nonuniform magnetic field. *Phys. Rev. A*, 106:022420.

[41] Child, T., Sheekey, O., Lüscher, S., Fallahi, S., Gardner, G. C., Manfra, M., Mitchell, A., Sela, E., Kleeorin, Y., Meir, Y., and Folk, J. (2022). Entropy measurement of a strongly coupled quantum dot. *Phys. Rev. Lett.*, 129:227702.

[42] Choi, J., Zhou, H., Knowles, H., Landig, R., and Choi, S. (2020). Robust dynamic hamiltonian engineering of many-body spin systems. *Physical Review X*.

[43] Choquette, A., Di Paolo, A., Barkoutsos, P. K., Sénéchal, D., Tavernelli, I., and Blais, A. (2021). Quantum-optimal-control-inspired ansatz for variational quantum algorithms. *Phys. Rev. Res.*, 3:023092.

[44] Ciftja, O. (2013). Understanding electronic systems in semiconductor quantum dots. *Physica Scripta*.

[45] Climente, J. I., Bertoni, A., Goldoni, G., Rontani, M., and Molinari, E. (2007). Magnetic field dependence of triplet-singlet relaxation in quantum dots with spin-orbit coupling. *Phys. Rev. B*, 75:081303.

[46] Cruz, P. M. Q. and Murta, B. (2024). Shallow unitary decompositions of quantum fredkin and toffoli gates for connectivity-aware equivalent circuit averaging. *APL Quantum.*, 1(1):016105. Accessed: 2024-06-21.

[47] Daley, A. (2014). Quantum trajectories and open many-body quantum systems. *Adv. Phys.*

[48] Danon, J. and Veldhorst, M. (2018). Exchange-only singlet-only spin qubit. *Phys. Rev. B*, 98(11):115307.

[49] D'Arrigo, A., Falci, G., and Paladino, E. (2016). High-fidelity two-qubit gates via dynamical decoupling of local $1/f$ noise at the optimal point. *Phys. Rev. A*, 94:022303.

[50] Das, K., Lehmann, J., and Loss, D. (2011). Quantum dots for quantum information processing: Controlling and exploiting the quantum dot environment. *Annu. Rev. Condens.*, 2(1):153–185.

[51] Das Sarma, S., Wang, X., and Yang, S. (2011). Hubbard model description of silicon spin qubits: Charge stability diagram and tunnel coupling in si double quantum dots. *Phys. Rev. B*, 83:235314.

[52] de Arquer, F. G., Talapin, D., Klimov, V., and Arakawa, Y. (2021). Semiconductor quantum dots: Technological progress and future challenges. *Science*.

[53] de Keijzer, R., Tse, O., and Kokkelmans, S. (2023). Pulse based variational quantum optimal control for hybrid quantum computing. *Quantum*, 7:908.

[54] Dehollain, J. P., Mukhopadhyay, U., Michal, V. P., Wang, Y., Wunsch, B., Reichl, C., Wegscheider, W., Rudner, M. S., Demler, E., and Vandersypen, L. M. K. (2020). Nagaoka ferromagnetism observed in a quantum dot plaquette. *Nature*, 579(7800):528–533.

[55] DEL TITOLO, T. P. I. C. and DI DOTTORE, D. R. (1999). Electronic states in semiconductor quantum dots.

[56] Deng, K. and Barnes, E. (2020). Interplay of exchange and superexchange in triple quantum dots. *Phys. Rev. B*, 102:035427.

[57] Deng, K., Calderon-Vargas, F. A., Mayhall, N. J., and Barnes, E. (2018). Negative exchange interactions in coupled few-electron quantum dots. *Phys. Rev. B*, 97:245301.

[58] Department of Computer Science and Technology, University of Cambridge (2020). Quantum computing. Accessed: June 22, 2024.

[59] Deutsch, D. and Jozsa, R. (1992). Rapid solution of problems by quantum computation. *Proceedings of the Royal Society of London. Series A: Mathematical and Physical Sciences*, 439(1907):553–558.

[60] Dial, O. E., Shulman, M. D., Harvey, S. P., Bluhm, H., Umansky, V., and Yacoby, A. (2013). Charge noise spectroscopy using coherent exchange oscillations in a singlet-triplet qubit. *Phys. Rev. Lett.*, 110:146804.

[61] Diepen, C. V., Eendebak, P., and Buijtendorp, B. (2018). Automated tuning of inter-dot tunnel coupling in double quantum dots. *Applied Physics Letters*.

[62] Dijkema, J., Xue, X., Harvey-Collard, P., Rimbach-Russ, M., de Snoo, S. L., Zheng, G., Sammak, A., Scappucci, G., and Vandersypen, L. M. K. (2023). Two-qubit logic between distant spins in silicon.

[63] DiVincenzo, D. P., Bacon, D., Kempe, J., Burkard, G., and Whaley, K. B. (2000). Universal quantum computation with the exchange interaction. *Nature*, 408(6810):339–342.

[64] Doherty, A. C. and Wardrop, M. P. (2013). Two-qubit gates for resonant exchange qubits. *Phys. Rev. Lett.*, 111:050503.

[65] Donnelly, M., Keizer, J., Pye, D., and Kiczynski, M. (2024a). A quantum materials simulator based on coulomb-confined quantum dots. *Bulletin of the American Physical Society*.

[66] Donnelly, M. B., Rowlands, J., Kranz, L., Hsueh, Y. L., Chung, Y., Timofeev, A. V., Geng, H., Singh-Gregory, P., Gorman, S. K., Keizer, J. G., Rahman, R., and Simmons, M. Y. (2024b). Noise correlations in a 1d silicon spin qubit array.

[67] Duan, L.-M. and Guo, G.-C. (1998). Phys. rev. a. 57:737–741.

[68] Dutta, P. and Horn, P. M. (1981). Low-frequency fluctuations in solids: $\frac{1}{f}$ noise. *Rev. Mod. Phys.*, 53:497–516.

[69] Eisert, J., Cramer, M., and Plenio, M. B. (2010). Area laws for the entanglement entropy. *Rev. Mod. Phys.*, 82:277–306.

[70] Ekert, A. and Jozsa, R. (1996). Quantum computation and shor's factoring algorithm. *Rev. Mod. Phys.*, 68(3):733–753.

[71] Ercan, H., Coppersmith, S., and Friesen, M. (2021). Strong electron-electron interactions in si/sige quantum dots. *Phys. Rev. B*.

[72] Ercan, H. E., Anderson, C. R., Coppersmith, S. N., Friesen, M., and Gyure, M. F. (2023). Multielectron dots provide faster rabi oscillations when the core electrons are strongly confined.

[73] Fedele, F., Chatterjee, A., Fallahi, S., Gardner, G. C., Manfra, M. J., and Kuemmeth, F. (2021). Simultaneous operations in a two-dimensional array of singlet-triplet qubits. *PRX Quantum*, 2:040306.

[74] Fei, J., Hung, J.-T., Koh, T. S., Shim, Y.-P., Coppersmith, S. N., Hu, X., and Friesen, M. (2015). Characterizing gate operations near the sweet spot of an exchange-only qubit. *Phys. Rev. B*, 91:205434.

[75] Feng, M., Yoneda, J., Huang, W., Su, Y., Tanttu, T., Yang, C. H., Cifuentes, J. D., Chan, K. W., Gilbert, W., Leon, R. C. C., Hudson, F. E., Itoh, K. M., Laucht, A., Dzurak, A. S., and Saraiva, A. (2023). Control of dephasing in spin qubits during coherent transport in silicon. *Phys. Rev. B*, 107:085427.

[76] Feng, M., Zaw, L. H., and Koh, T. S. (2021). Two-qubit sweet spots for capacitively coupled exchange-only spin qubits. *npj Quantum Inf.*, 7(1):112.

[77] Fernández-Fernández, D., Ban, Y., and Platero, G. (2024). Hole flying qubits in quantum dot arrays.

[78] Ferraro, E., Fanciulli, M., and Michielis, M. D. (2017). Controlled-not gate sequences for mixed spin qubit architectures in a noisy environment. *Quantum Inf. Process.*, 16(11):277.

[79] Ferreira, D. L. B., Maciel, T. O., Vianna, R. O., and Iemini, F. (2022). Quantum correlations, entanglement spectrum, and coherence of the two-particle reduced density matrix in the extended hubbard model. *Phys. Rev. B*, 105:115145.

[80] Fong, B. H. and Wandzura, S. M. (2011). Universal quantum computation and leakage reduction in the 3-qubit decoherence free subsystem.

[81] Fonseca, M. E., Fanchini, F. F., de Lima, E. F., and Castelano, L. K. (2023). Effectiveness of the krotov method in controlling open quantum systems.

[82] Foulk, N. L. and Das Sarma, S. (2023). Realizable time crystal of four silicon quantum dot qubits. *Phys. Rev. B*, 107:125420.

[83] Friesen, M., Joynt, R., and Eriksson, M. A. (2017). Decoherence-free subspaces for quantum computing in semiconductor quantum dots. *Phys. Rev. Lett.*, 98(23):230503.

[84] Fölsch, S., Martínez-Blanco, J., and Yang, J. (2014). Quantum dots with single-atom precision. *Nat. Nanotechnol.*

[85] Gamble, J. K., Friesen, M., Coppersmith, S. N., and Hu, X. (2012). Two-electron dephasing in single si and gaas quantum dots. *Phys. Rev. B*, 86:035302.

[86] Gaudreau, L., Granger, G., Kam, A., Aers, G., Studenikin, S., Zawadzki, P., Pioro-Ladriere, M., Wasilewski, Z., and Sachrajda, A. (2012). Coherent control of three-spin states in a triple quantum dot. *Nat. Phys.*, 8(1):54–58.

[87] Geier, S., Thaicharoen, N., Hainaut, C., and Franz, T. (2021). Floquet hamiltonian engineering of an isolated many-body spin system. *Science*.

[88] Georgescu, I., Ashhab, S., and Nori, F. (2014). Quantum simulation. *Rev. Mod. Phys.*

[89] Giovannetti, V., Lloyd, S., and Maccone, L. (2006). Quantum metrology. *Phys. Rev. Lett.*, 96(1):010401.

[90] Goerz, M., Basilewitsch, D., Gago-Encinas, F., Krauss, M. G., Horn, K. P., Reich, D. M., and Koch, C. (2019). Krotov: A python implementation of krotov's method for quantum optimal control. *SciPost Physics*, 7(6):080. IF: 5.5 Q1.

[91] Gonzalez-Zalba, M. F., de Franceschi, S., Charbon, E., Meunier, T., Vinet, M., and Dzurak, A. S. (2021). Scaling silicon-based quantum computing using cmos technology. *Nat. Electron.*, 4(12):872–884.

[92] Greenberger, D. M., Horne, M. A., and Zeilinger, A. (1989). Going beyond bell's theorem. In *Bell's Theorem, Quantum Theory and Conceptions of the Universe*, pages 69–72. Springer.

[93] Gross, C. and Bloch, I. (2017). Quantum simulations with ultracold atoms in optical lattices. *Science*, 357(6355):995–1001.

[94] Grover, L. K. (1996). A fast quantum mechanical algorithm for database search. In *Proceedings of the 28th Annual ACM Symposium on Theory of Computing*, pages 212–219. ACM.

[95] Grover, T. (2014). Certain general constraints on the many-body localization transition. *Phys. Rev. Lett.*, 112(20):2014.

[96] Gu, S.-J., Deng, S.-S., Li, Y.-Q., and Lin, H.-Q. (2004). Entanglement and quantum phase transition in the extended hubbard model. *Phys. Rev. Lett.*, 93:086402.

[97] Guarcello, C., Valenti, D., Carollo, A., and Spagnolo, B. (2015). Entropy. 17(5):2862–2875.

[98] Haferkamp, J., Faist, P., Kothakonda, N. B. T., Eisert, J., and Halpern, N. Y. (2022). Linear growth of quantum circuit complexity. *Nat. Phys.*, 18(5):528–532.

[99] Hanson, R., Kouwenhoven, L. P., Petta, J. R., Tarucha, S., and Vandersypen, L. M. (2007). Spins in few-electron quantum dots. *Rev. Mod. Phys.*, 79(4):1217.

[100] Hartmann, M. and Brandao, F. (2008). Quantum many-body phenomena in coupled cavity arrays. *Laser & Photonics Reviews*.

[101] Harvey, S. et al. (2022). Quantum dots / spin qubits. *arXiv preprint arXiv:2204.04261*.

[102] Harvey-Collard, P., Jacobson, N. T., Bureau-Oxton, C., Jock, R. M., Srinivasa, V., Mounce, A. M., Ward, D. R., Anderson, J. M., Manginell, R. P., Wendt, J. R., Pluym, T., Lilly, M. P., Luhman, D. R., Pioro-Ladrière, M., and Carroll, M. S. (2019). Spin-orbit interactions for singlet-triplet qubits in silicon. *Phys. Rev. Lett.*, 122:217702.

[103] Hastings, M. B. (2010). Random quantum systems. *Phys. Rev. Lett.*, 104(15):157201.

[104] Hausler, W. and Egger, R. (2009). Artificial atoms in interacting graphene quantum dots. *Phys. Rev. B*.

[105] He, G., Chan, G. X., and Wang, X. (2023). Theory on electron–phonon spin dephasing in gaas multi-electron double quantum dots. *Adv. Quantum Technol.*, 6(3):2200074.

[106] Head-Marsden, K., Flick, J., and Ciccarino, C. (2020). Quantum information and algorithms for correlated quantum matter. *Chemical Reviews*.

[107] Heinz, I., Borjans, F., Curry, M., Kotlyar, R., Luthi, F., Mądzik, M. T., Mohiyaddin, F. A., Bishop, N., and Burkard, G. (2024a). Fast quantum gates for exchange-only qubits using simultaneous exchange pulses.

[108] Heinz, I., Mills, A. R., Petta, J. R., and Burkard, G. (2024b). Analysis and mitigation of residual exchange coupling in linear spin-qubit arrays. *Phys. Rev. Res.*, 6:013153.

[109] Hensgens, T., Fujita, T., Janssen, L., Li, X., Van Diepen, C. J., Reichl, C., Wegscheider, W., Das Sarma, S., and Vandersypen, L. M. K. (2017). Quantum simulation of a fermi–hubbard model using a semiconductor quantum dot array. *Nature*, 548(7665):70–73.

[110] Hepp, S., Jetter, M., and Portalupi, S. (2019). Semiconductor quantum dots for integrated quantum photonics. *Adv. Quantum Technol.*

[111] Hickman, G. T., Wang, X., Kestner, J. P., and Das Sarma, S. (2013). Dynamically corrected gates for an exchange-only qubit. *Phys. Rev. B*, 88:161303.

[112] Higginbotham, A. P., Kuemmeth, F., Hanson, M. P., Gossard, A. C., and Marcus, C. M. (2014). Coherent operations and screening in multielectron spin qubits. *Phys. Rev. Lett.*, 112:026801.

[113] Hoffman, S., Schrade, C., Klinovaja, J., and Loss, D. (2016). Universal quantum computation with hybrid spin-majorana qubits. *Phys. Rev. B*, 94:045316.

[114] Horodecki, R., Horodecki, P., Horodecki, M., and Horodecki, K. (2009). Quantum entanglement. *Rev. Mod. Phys.*, 81:865–942.

[115] Hu, X. (2011). Two-spin dephasing by electron-phonon interaction in semiconductor double quantum dots. *Phys. Rev. B*, 83:165322.

[116] Hu, X. and Das Sarma, S. (2001). Spin-based quantum computation in multi-electron quantum dots. *Phys. Rev. A*, 64:042312.

[117] Hu, X. and Das Sarma, S. (2006). Charge-fluctuation-induced dephasing of exchange-coupled spin qubits. *Phys. Rev. Lett.*, 96:100501.

[118] Hu, X., Zhang, F., Li, Y., and Long, G. (2021). Optimizing quantum gates within decoherence-free subspaces. *Phys. Rev. A*, 104:062612.

[119] Huang, P. (2021). Dephasing of exchange-coupled spins in quantum dots for quantum computing. *Adv. Quantum Technol.*, 4(11):2100018.

[120] Hubbard, J. (1963). Electron correlations in narrow energy bands. *Proc. R. Soc. A.*, 276(1365):238–257.

[121] Hung, J.-T., Fei, J., Friesen, M., and Hu, X. (2014). Decoherence of an exchange qubit by hyperfine interaction. *Phys. Rev. B*, 90:045308.

[122] Hur, K. L., Henriet, L., Petrescu, A., and Plekhanov, K. (2016). Many-body quantum electrodynamics networks: Non-equilibrium condensed matter physics with light. *Comptes Rendus Physique*.

[123] Hyart, T. and Lado, J. L. (2022). Non-hermitian many-body topological excitations in interacting quantum dots. *Phys. Rev. Res.*, 4:L012006.

[124] Iemini, F., Maciel, T. O., and Vianna, R. O. (2015). Entanglement of indistinguishable particles as a probe for quantum phase transitions in the extended hubbard model. *Phys. Rev. B*, 92:075423.

[125] Imada, M., Fujimori, A., and Tokura, Y. (1998). Rev. mod. phys. 70:1039–1263.

[126] Ivanova-Rohling, V. N., Rohling, N., and Burkard, G. (2024). Discovery of an exchange-only gate sequence for cnot with record-low gate time using reinforcement learning.

[127] Iyer, S., Oganesyan, V., Refael, G., and Huse, D. A. (2013). Many-body localization in a quasiperiodic system. *Phys. Rev. B*, 87:134202.

[128] J., A., Adedoyin, A., Ambrosiano, J., Anisimov, P., Casper, W., Chennupati, G., Coffrin, C., Djidjev, H., Gunter, D., Karra, S., Lemons, N., Lin, S., Malyzhenkov, A., Mascarenas, D., Mniszewski, S., Nadiga, B., O'malley, D., Oyen, D., Pakin, S., Prasad, L., Roberts, R., Romero, P., Santhi, N., Sinitsyn, N., Swart, P. J., Wendelberger, J. G., Yoon, B., Zamora, R., Zhu, W., Eidenbenz, S., Bärtschi, A., Coles, P. J., Vuffray, M., and Lokhov, A. Y. (2022). Quantum algorithm implementations for beginners. *ACM Trans. Quantum Comput.*, 3(4).

[129] Johnson, N. (1995). Quantum dots: few-body, low-dimensional systems. *J. Phys.: Condens. Matter.*

[130] Jorgensen, M. (2011). Many-body approaches to quantum dots. *duo.uio.no*.

[131] Kambhampati, P. (2011). Unraveling the structure and dynamics of excitons in semiconductor quantum dots. *Accounts of Chemical Research*.

[132] Kanaar, D. W., Wolin, S., Güngördü, U., and Kestner, J. P. (2021). Single-tone pulse sequences and robust two-tone shaped pulses for three silicon spin qubits with always-on exchange. *Phys. Rev. B*, 103:235314.

[133] Kandel, Y. P., Qiao, H., Fallahi, S., Gardner, G. C., Manfra, M. J., and Nichol, J. M. (2021). Adiabatic quantum state transfer in a semiconductor quantum-dot spin chain. *Nat. Commun.*, 12(1):2156.

[134] Karnieli, A. and Fan, S. (2023). Jaynes-cummings interaction between low-energy free electrons and cavity photons. *Sci. Adv.*

[135] Karrasch, C., Bardarson, J. H., and Moore, J. E. (2012). Finite-temperature dynamical density matrix renormalization group and the drude weight of spin-1/2 chains. *Phys. Rev. Lett.*, 108(22):227206.

[136] Kempe, J., Bacon, D., DiVincenzo, D. P., and Whaley, K. B. (2001a). Encoded universality from a single physical interaction.

[137] Kempe, J., Bacon, D., Lidar, D. A., and Whaley, K. B. (2001b). Theory of decoherence-free fault-tolerant universal quantum computation. *Phys. Rev. A*, 63:042307.

[138] Khaetskii, A. V., Loss, D., and Glazman, L. (2002). Electron spin decoherence in quantum dots due to interaction with nuclei. *Phys. Rev. Lett.*, 88(18):186802.

[139] Khordad, R. (2024). The singlet–triplet transition of two interacting electrons in a frost–musulin quantum dot. *Optical and Quantum Electronics*.

[140] Kiczynski, M., Gorman, S., Geng, H., and Donnelly, M. (2022). Engineering topological states in atom-based semiconductor quantum dots. *Nature*.

[141] Kim, C., Nichol, J., Jordan, A., and Franco, I. (2022). Analog quantum simulation of the dynamics of open quantum systems with quantum dots and microelectronic circuits. *PRX Quantum*.

[142] Kiyama, H., Yoshimi, K., Kato, T., Nakajima, T., Oiwa, A., and Tarucha, S. (2021). Preparation and readout of multielectron high-spin states in a gate-defined gaas/algaas quantum dot. *Phys. Rev. Lett.*, 127:086802.

[143] Kleemans, N., Van Bree, J., Govorov, A., and Keizer, J. (2010). Many-body exciton states in self-assembled quantum dots coupled to a fermi sea. *Nat. Phys.*

[144] Kloeffel, C. and Loss, D. (2013). Prospects for spin-based quantum computing in quantum dots. *Annu. Rev. Condens.*

[145] Koch, C. P., Boscain, U., Calarco, T., Dirr, G., Filipp, S., Glaser, S. J., Kosloff, R., Montangero, S., Schulte-Herbrüggen, T., Sugny, D., and Wilhelm, F. K. (2022). Quantum optimal control in quantum technologies. strategic report on current status, visions and goals for research in europe. *EPJ Quantum Technology*, 9(1):19.

[146] Kollath, C., Läuchli, A. M., and Altman, E. (2007). Quench dynamics and nonequilibrium phase diagram of the bose-hubbard model. *Phys. Rev. Lett.*, 98(18):180601.

[147] Kornich, V., Kloeffel, C., and Loss, D. (2014). Phonon-mediated decay of singlet-triplet qubits in double quantum dots. *Phys. Rev. B*, 89:085410.

[148] Kornich, V., Kloeffel, C., and Loss, D. (2018). Phonon-assisted relaxation and decoherence of singlet-triplet qubits in si/sige quantum dots. *Quantum*, 2:70.

[149] Kouwenhoven, L. P., Austing, D. G., and Tarucha, S. (2001). Few-electron quantum dots. *Reports on Progress in Physics*, 64(6):701.

[150] Kouwenhoven, L. P., Marcus, C. M., McEuen, P. L., Tarucha, S., Westervelt, R. M., and Wingreen, N. S. (1997a). Electron transport in quantum dots. *Mesoscopic electron transport*, pages 105–214.

[151] Kouwenhoven, L. P., Oosterkamp, T., Danoesastro, M., Eto, M., Austing, D., Honda, T., and Tarucha, S. (1997b). Excitation spectra of circular, few-electron quantum dots. *Science*, 278(5344):1788–1792.

[152] Krummheuer, B., Axt, V. M., and Kuhn, T. (2002). Theory of pure dephasing and the resulting absorption line shape in semiconductor quantum dots. *Phys. Rev. B*, 65(19):195313.

[153] Krzywda, J. A. and Cywiński, L. (2021). Interplay of charge noise and coupling to phonons in adiabatic electron transfer between quantum dots. *Phys. Rev. B*, 104:075439.

[154] Kwiat, P. G., Mattle, K., Weinfurter, H., Zeilinger, A., Sergienko, A. V., and Shih, Y. (2000). Experimental entanglement distillation and 'hidden'non-locality. *Phys. Rev. Lett.*, 75(24):4337–4341.

[155] Ladd, T. D., Jelezko, F., Laflamme, R., Nakamura, Y., Monroe, C., and O'Brien, J. L. (2010). Quantum computers. *Nature*, 464(7285):45–53.

[156] Lagoin, C., Bhattacharya, U., Grass, T., Chhajlany, R. W., Salamon, T., Baldwin, K., Pfeiffer, L., Lewenstein, M., Holzmann, M., and Dubin, F. (2022). Extended bose–hubbard model with dipolar excitons. *Nature*, 609(7927):485–489.

[157] Laird, E. A., Taylor, J. M., DiVincenzo, D. P., Marcus, C. M., Hanson, M. P., and Gossard, A. C. (2010). Coherent spin manipulation in an exchange-only qubit. *Phys. Rev. B*, 82:075403.

[158] Lan, K., Du, Q., Kang, L., Tang, X., Jiang, L., Zhang, Y., and Cai, X. (2020). Dynamics of an open double quantum dot system via quantum measurement. *Phys. Rev. B*, 101:174302.

[159] Langrock, V. and DiVincenzo, D. P. (2020). A reset-if-leaked procedure for encoded spin qubits.

[160] Laucht, A., Villas-Bôas, J., Stobbe, S., and Hauke, N. (2010). Mutual coupling of two semiconductor quantum dots via an optical nanocavity. *Phys. Rev. B*.

[161] Le, N. H., Fisher, A. J., Curson, N. J., and Ginossar, E. (2020). Topological phases of a dimerized fermi–hubbard model for semiconductor nano-lattices. *npj Quantum Inf.*, 6(1):24.

[162] Lehtonen, O., Sundholm, D., and Vänskä, T. (2008). Computational studies of semiconductor quantum dots. *Physical Chemistry Chemical Physics*.

[163] Leon, R. C. C., Yang, C. H., Hwang, J. C. C., Camirand Lemyre, J., Tanttu, T., Huang, W., Huang, J. Y., Hudson, F. E., Itoh, K. M., Laucht, A., Pioro-Ladrière, M., Saraiva, A., and Dzurak, A. S. (2021). Bell-state tomography in a silicon many-electron artificial molecule. *Nat. Commun.*, 12(1).

[164] Levy, J. (2002). Universal quantum computation with spin-1/2 pairs and heisenberg exchange. *Phys. Rev. Lett.*, 89(14):147902.

[165] Li, X. and Wang, C. (2023). Efficient quantum algorithms for quantum optimal control.

[166] Lidar, D. A., Chuang, I. L., and Whaley, K. B. (1998). Decoherence-free subspaces for quantum computation. *Phys. Rev. Lett.*, 81(12):2594.

[167] Loss, D. and DiVincenzo, D. P. (1998). Quantum computation with quantum dots. *Phys. Rev. A*, 57:120–126.

[168] Luitz, D. J., Plat, X., and Alet, F. (2015). Many-body localization edge in the random-field heisenberg chain. *Phys. Rev. B*, 91(8):081103.

[169] Lundberg, T., Li, J., Hutin, L., Bertrand, B., Ibberson, D. J., Lee, C.-M., Niegemann, D. J., Urdampilleta, M., Stelmashenko, N., Meunier, T., Robinson, J. W. A., Ibberson, L., Vinet, M., Niquet, Y.-M., and Gonzalez-Zalba, M. F. (2020). Spin quintet in a silicon double quantum dot: Spin blockade and relaxation. *Phys. Rev. X*, 10:041010.

[170] Madzik, M. T., Ladd, T. D., Hudson, F. E., Itoh, K. M., Jakob, A. M., Johnson, B. C., McCallum, J. C., Jamieson, D. N., Dzurak, A. S., Laucht, A., and Morello, A. (2020). Controllable freezing of the nuclear spin bath in a single-atom spin qubit. *Sci. Adv.*, 6(27).

[171] Magann, A. B., Arenz, C., Grace, M. D., Ho, T.-S., Kosut, R. L., McClean, J. R., Rabitz, H. A., and Sarovar, M. (2021). From pulses to circuits and back again: A quantum optimal control perspective on variational quantum algorithms. *PRX Quantum*, 2:010101.

[172] Mahan, G. D. (2000). *Many-Particle Physics*. Kluwer Academic/Plenum Publishers, 3rd edition.

[173] Malinowski, F. K., Martins, F., Smith, T. B., Bartlett, S. D., Doherty, A. C., Nissen, P. D., Fallahi, S., Gardner, G. C., Manfra, M. J., Marcus, C. M., and Kuemmeth, F. (2018). Spin of a multielectron quantum dot and its interaction with a neighboring electron. *Phys. Rev. X*, 8:011045.

[174] Malinowski, F. K., Martins, F., Smith, T. B., Bartlett, S. D., Doherty, A. C., Nissen, P. D., Fallahi, S., Gardner, G. C., Manfra, M. J., Marcus, C. M., and Kuemmeth, F. (2019). Fast spin exchange across a multielectron mediator. *Nat. Commun.*, 10(1).

[175] Martins, F., Malinowski, F. K., Nissen, P. D., Barnes, E., Fallahi, S., Gardner, G. C., Manfra, M. J., Marcus, C. M., and Kuemmeth, F. (2016). Noise suppression using symmetric exchange gates in spin qubits. *Phys. Rev. Lett.*, 116:116801.

[176] Martins, F., Malinowski, F. K., Nissen, P. D., Fallahi, S., Gardner, G. C., Manfra, M. J., Marcus, C. M., and Kuemmeth, F. (2017). Negative spin exchange in a multielectron quantum dot. *Phys. Rev. Lett.*, 119:227701.

[177] Maune, B. M., Borselli, M. G., Huang, B., Ladd, T. D., Deelman, P. W., Holabird, K. S., Kiselev, A. A., Alvarado-Rodriguez, I., Ross, R. S., Schmitz, A. E., et al. (2012). Coherent singlet-triplet oscillations in a silicon-based double quantum dot. *Nature*, 481(7381):344–347.

[178] Mavadia, S., Frey, V., Sastrawan, J., Dona, S., and Biercuk, M. J. (2017). Prediction and real-time compensation of qubit decoherence via machine learning. *Nat. Commun.*, 8(1):1–6.

[179] Medford, J., Beil, J., Taylor, J. M., Rashba, E. I., Lu, H., Gossard, A. C., and Marcus, C. M. (2013). Quantum-dot-based resonant exchange qubit. *Phys. Rev. Lett.*, 111:050501.

[180] Mehl, S. and DiVincenzo, D. P. (2013). Noise-protected gate for six-electron double-dot qubit. *Phys. Rev. B*, 88:161408.

[181] Mehl, S. and DiVincenzo, D. P. (2014). Inverted singlet-triplet qubit coded on a two-electron double quantum dot. *Phys. Rev. B*, 90:195424.

[182] Mehl, S. and DiVincenzo, D. P. (2015). Simple operation sequences to couple and interchange quantum information between spin qubits of different kinds. *Phys. Rev. B*, 92:115448.

[183] Mills, A., Guinn, C., Gullans, M., Sigillito, A., Feldman, M., Nielsen, E., and Petta, J. (2021). Two-qubit silicon quantum processor with operation fidelity exceeding 99%. *arXiv preprint arXiv:2111.11937*.

[184] Monz, T. et al. (2011). 14-qubit entanglement: Creation and coherence. *Phys. Rev. Lett.*, 106(13):130506.

[185] Mortemousque, P., Chanrion, E., Jadot, B., and Thiney, V. (2021). Coherent control of individual electron spins in a two-dimensional quantum dot array. *Nat. Nanotechnol.*

[186] Morzhin, O. V. and Pechen, A. N. (2019). Krotov method for optimal control of closed quantum systems. *Russian Mathematical Surveys*, 74(5):851.

[187] Mozyrsky, D., Kogan, S., Gorshkov, V. N., and Berman, G. P. (2002). Time scales of phonon-induced decoherence of semiconductor spin qubits. *Phys. Rev. B*, 65:245213.

[188] Mądzik, M. T., Asaad, S., Youssry, A., Joecker, B., Rudinger, K. M., Nielsen, E., Young, K. C., Proctor, T. J., Baczewski, A. D., Laucht, A., Schmitt, V., Hudson, F. E., Itoh, K. M., Jakob, A. M., Johnson, B. C., Jamieson, D. N., Dzurak, A. S., Ferrie, C., Blume-Kohout, R., and Morello, A. (2022). Precision tomography of a three-qubit donor quantum processor in silicon. *Nature*, 601(7893):348–353.

[189] Nakajima, T., Delbecq, M. R., Otsuka, T., Amaha, S., Yoneda, J., Noiri, A., Takeda, K., Allison, G., Ludwig, A., Wieck, A. D., et al. (2018). Coherent transfer of electron spin correlations assisted by dephasing noise. *Nat. Commun.*, 9(1):1–8.

[190] Neder, I., Marquardt, F., Heiblum, M., Mahalu, D., and Umansky, V. (2007). Controlled dephasing of electrons by non-gaussian shot noise. *Nat. Phys.*, 3(8):534–537.

[191] Neyens, S. F., MacQuarrie, E., Dodson, J., Corrigan, J., Holman, N., Thorgrimsson, B., Palma, M., McJunkin, T., Edge, L., Friesen, M., Coppersmith, S., and Eriksson, M. (2019). Measurements of capacitive coupling within a quadruple-quantum-dot array. *Phys. Rev. Appl.*, 12:064049.

[192] Nielsen, M. A. and Chuang, I. L. (2010). *Quantum Computation and Quantum Information*. Cambridge University Press.

[193] Noh, C. and Angelakis, D. (2016). Quantum simulations and many-body physics with light. *Reports on Progress in Physics*.

[194] Obata, T., Pioro-Ladrière, M., Tokura, Y., Shin, Y., and Kubo, T. (2010). Coherent manipulation of individual electron spin in a double quantum dot integrated with a micromagnet. *Phys. Rev. B*.

[195] Ota, Y., Iwamoto, S., Kumagai, N., and Arakawa, Y. (2009). Impact of electron-phonon interactions on the optical properties of quantum dots. *Phys. Rev. Lett.*, 103(6):066404.

[196] Pal, A. and Huse, D. A. (2010). Many-body localization phase transition. *Phys. Rev. B*, 82:174411.

[197] Pal, A., Rashba, E. I., and Halperin, B. I. (2014). Driven nonlinear dynamics of two coupled exchange-only qubits. *Phys. Rev. X*, 4:011012.

[198] Palma, G. M., antti Suominen, K., and Ekert, A. (1996). Proc. r. soc. london a. 452(1946):567–584.

[199] Patel, A. A., Strack, P., and Sachdev, S. (2018). Critical strange metal from fluctuating gauge fields in a solvable random model. *Phys. Rev. Lett.*, 119(27):2018.

[200] Penc, P., Moca, C. P., Örs Legeza, Prosen, T., Zaránd, G., and Werner, M. A. (2024). Loss-induced quantum information jet in an infinite temperature hubbard chain.

[201] Petersson, K. D., Petta, J. R., Lu, H., and Gossard, A. C. (2010a). Charge and spin state readout of a double quantum dot coupled to a resonator. *Nature*, 490(7420):380–383.

[202] Petersson, K. D., Petta, J. R., Lu, H., and Gossard, A. C. (2010b). Quantum coherence in a one-electron semiconductor charge qubit. *Phys. Rev. Lett.*, 105:246804.

[203] Petroff, P., Lorke, A., and Imamoglu, A. (2001). Epitaxially self-assembled quantum dots. *Physics Today*.

[204] Petta, J. R. et al. (2021). Semiconductor spin qubits. *arXiv preprint arXiv:2112.08863*.

[205] Petta, J. R., Johnson, A. C., Taylor, J. M., Laird, E. A., Yacoby, A., Lukin, M. D., Marcus, C. M., Hanson, M. P., and Gossard, A. C. (2005). Coherent manipulation of coupled electron spins in semiconductor quantum dots. *Science*, 309(5744):2180–2184.

[206] Pham, D. N., Bharadwaj, S., and Ram-Mohan, L. R. (2020). Tuning spatial entanglement in interacting two-electron quantum dots. *Phys. Rev. B*, 101:045306.

[207] Philips, S. G. J., Madzik, M. T., Amitonov, S. V., de Snoo, S. L., Russ, M., Kalhor, N., Volk, C., Lawrie, W. I. L., Brousse, D., Tryputen, L., Wuetz, B. P., Sammak, A., Veldhorst, M., Scappucci, G., and Vandersypen, L. M. K. (2022). Universal control of a six-qubit quantum processor in silicon. *Nature*, 609(7929):919–924.

[208] Porod, W., Lent, C., Bernstein, G., and Orlov, A. (1999). Quantum-dot cellular automata: computing with coupled quantum dots. *International Journal of Electronics*.

[209] Potts, H., Josefi, J., Chen, I.-J., Lehmann, S., Dick, K. A., Leijnse, M., Reimann, S. M., Bengtsson, J., and Thelander, C. (2021). Symmetry-controlled singlet-triplet transition in a double-barrier quantum ring. *Phys. Rev. B*, 104:L081409.

[210] Qiao, H., Kandel, Y. P., Deng, K., Fallahi, S., Gardner, G. C., Manfra, M. J., Barnes, E., and Nichol, J. M. (2020). Coherent multispin exchange coupling in a quantum-dot spin chain. *Phys. Rev. X*, 10:031006.

[211] Qiao, H., Kandel, Y. P., Fallahi, S., Gardner, G. C., Manfra, M. J., Hu, X., and Nichol, J. M. (2021). Long-distance superexchange between semiconductor quantum-dot electron spins. *Phys. Rev. Lett.*, 126:017701.

[212] Rachel, S. and Hur, K. L. (2010). Topological insulators and mott physics from the hubbard model. *Phys. Rev. B*, 82(7):075106.

[213] Raussendorf, R. and Briegel, H. J. (2001). A one-way quantum computer. *Phys. Rev. Lett.*, 86(22):5188–5191.

[214] Reed, M. D., Maune, B. M., Andrews, R. W., Borselli, M. G., Eng, K., Jura, M. P., Kiselev, A. A., Ladd, T. D., Merkel, S. T., Milosavljevic, I., Pritchett, E. J., Rakher, M. T., Ross, R. S., Schmitz, A. E., Smith, A., Wright, J. A., Gyure, M. F., and Hunter, A. T. (2016). Reduced sensitivity to charge noise in semiconductor spin qubits via symmetric operation. *Phys. Rev. Lett.*, 116:110402.

[215] Reilly, D. J., Taylor, J. M., Petta, J. R., Marcus, C. M., Hanson, M. P., and Gossard, A. C. (2008). Suppressing spin qubit dephasing by nuclear state preparation. *Science*, 321(5890):817–821.

[216] Reiter, D. E., Kuhn, T., and Axt, V. M. (2019). The role of phonons for exciton and biexciton generation in an optically driven quantum dot. *J. Phys.: Condens. Matter.*, 31(10):103002.

[217] Ridley, B. K. (1999). *Quantum Processes in Semiconductors*. Oxford University Press, 4th edition.

[218] Rinaldi, R. (1998). Artificial atoms in magnetic field: Electronic and optical properties. *International Journal of Modern Physics B*.

[219] Rohling, N. and Burkard, G. (2016). Optimizing electrically controlled echo sequences for the exchange-only qubit. *Phys. Rev. B*, 93:205434.

[220] Rontani, M., Cavazzoni, C., Bellucci, D., and Goldoni, G. (2006). Full configuration interaction approach to the few-electron problem in artificial atoms. *J. Chem. Phys.*, 124(12):124102.

[221] Roszak, K., Filip, R., and Novotnỳ, T. (2015). Decoherence control by quantum decoherence itself. *Sci. Rep.*, 5(1):1–10.

[222] Russ, M. and Burkard, G. (2015a). Asymmetric resonant exchange qubit under the influence of electrical noise. *Phys. Rev. B*, 91:235411.

[223] Russ, M. and Burkard, G. (2015b). Long distance coupling of resonant exchange qubits. *Phys. Rev. B*, 92:205412.

[224] Russ, M. and Burkard, G. (2017). Three-electron spin qubits. *J. Phys.: Condens. Matter.*, 29(39):393001.

[225] Russ, M., Petta, J. R., and Burkard, G. (2018). Quadrupolar exchange-only spin qubit. *Phys. Rev. Lett.*, 121:177701.

[226] Sala, A. and Danon, J. (2017). Exchange-only singlet-only spin qubit. *Phys. Rev. B*, 95:241303.

[227] Sala, A., Qvist, J. H., and Danon, J. (2020). Highly tunable exchange-only singlet-only qubit in a gaas triple quantum dot. *Phys. Rev. Research*, 2:012062.

[228] Scarlino, P., Kawakami, E., Ward, D. R., Savage, D. E., Lagally, M. G., Coppersmith, S. C., Eriksson, M. A., and Vandersypen, L. M. K. (2017). Toward high-fidelity coherent electron spin transport in a gaas double quantum dot. *Phys. Rev. B*, 95(16):165429.

[229] Schlosshauer, M. (2007). *Decoherence and the Quantum-to-Classical Transition*. Springer, Berlin, Heidelberg.

[230] Schollwöck, U. (2011). The density-matrix renormalization group in the age of matrix product states. *Ann. Phys.*, 326(1):96–192.

[231] Schulz, M., Hooley, C. A., Moessner, R., and Pollmann, F. (2019). Stark many-body localization. *Phys. Rev. Lett.*, 122:040606.

[232] Serina, M., Kloeffel, C., and Loss, D. (2017). Long-range interaction between charge and spin qubits in quantum dots. *Phys. Rev. B*, 95:245422.

[233] Setiawan, F., Hui, H.-Y., Kestner, J. P., Wang, X., and Sarma, S. D. (2014). Robust two-qubit gates for exchange-coupled qubits. *Phys. Rev. B*, 89:085314.

[234] Shi, Z., Simmons, C. B., Prance, J. R., Gamble, J. K., Koh, T. S., Shim, Y.-P., Hu, X., Savage, D. E., Lagally, M. G., Eriksson, M. A., Friesen, M., and Coppersmith, S. N. (2012). Fast hybrid silicon double-quantum-dot qubit. *Phys. Rev. Lett.*, 108:140503.

[235] Shim, Y.-P. and Tahan, C. (2016a). Charge-noise-insensitive gate operations for always-on, exchange-only qubits. *Phys. Rev. B*, 93:121410.

[236] Shim, Y.-P. and Tahan, C. (2016b). Single-qubit gates in two steps with rotation axes in a single plane. *Phys. Rev. B*, 93(12):121410.

[237] Shor, P. W. (1997). Polynomial-time algorithms for prime factorization and discrete logarithms on a quantum computer. *SIAM Journal on Computing*, 26(5):1484–1509.

[238] Shulman, M. D., Dial, O. E., Harvey, S. P., Bluhm, H., Umansky, V., and Yacoby, A. (2012). Demonstration of entanglement of electrostatically coupled singlet-triplet qubits. *Science*, 336(6078):202–205.

[239] Smet, M. D., Matsumoto, Y., Zwerver, A.-M. J., Tryputen, L., de Snoo, S. L., Amitonov, S. V., Sammak, A., Samkharadze, N., Önder Gül, Wasserman, R. N. M., Rimbach-Russ, M., Scappucci, G., and Vandersypen, L. M. K. (2024). High-fidelity single-spin shuttling in silicon.

[240] Song, C. et al. (2019). Generation of multi-qubit entangled states on a programmable superconducting processor. *Nature*, 574(7779):622–625.

[241] Spethmann, M., Bosco, S., Hofmann, A., Klinovaja, J., and Loss, D. (2024). High-fidelity two-qubit gates of hybrid superconducting-semiconducting singlet-triplet qubits. *Phys. Rev. B*, 109:085303.

[242] Srinivasa, V., Xu, H., and Taylor, J. M. (2015). Tunable spin-qubit coupling mediated by a multielectron quantum dot. *Phys. Rev. Lett.*, 114:226803.

[243] Stafford, C. A. and Das Sarma, S. (1994). Collective coulomb blockade in an array of quantum dots: A mott-hubbard approach. *Phys. Rev. Lett.*, 72(21):3590.

[244] Steane, A. (1996). Error correcting codes in quantum theory. *Phys. Rev. Lett.*, 77(5):793–797.

[245] Stepanenko, D., Bonesteel, N. E., DiVincenzo, D. P., Burkard, G., and Loss, D. (2003). Spin-orbit coupling and time-reversal symmetry in quantum gates. *Phys. Rev. B*, 68:115306.

[246] Stepanenko, D., Rudner, M., Halperin, B. I., and Loss, D. (2012). Singlet-triplet splitting in double quantum dots due to spin-orbit and hyperfine interactions. *Phys. Rev. B*, 85:075416.

[247] Struck, T., Hollmann, A., Schauer, F., Fedorets, O., Schmidbauer, A., Sawano, K., Riemann, H., Abrosimov, N. V., Łukasz Cywiński, Bougeard, D., and Schreiber, L. R. (2020). Low-frequency spin qubit energy splitting noise in highly purified $^{28}$si/sige. *npj Quantum Inf.*, 6(1):40.

[248] Sun, B., Brecht, T., Fong, B. H., Akmal, M., Blumoff, J. Z., Cain, T. A., Carter, F. W., Finestone, D. H., Fireman, M. N., Ha, W., Hatke, A. T., Hickey, R. M., Jackson, C. A. C., Jenkins, I., Jones, A. M., Pan, A., Ward, D. R., Weinstein, A. J., Whiteley, S. J., Williams, P., Borselli, M. G., Rakher, M. T., and Ladd, T. D. (2024). Full-permutation dynamical decoupling in triple-quantum-dot spin qubits. *PRX Quantum*, 5:020356.

[249] Suter, D. and Alvarez, G. A. (2016). Colloquium: Protecting quantum information against environmental noise. *Rev. Mod. Phys.*, 88(4):041001.

[250] Szabo, A., Szabó, A., and Ostlund, N. (1982). *Modern Quantum Chemistry: Introduction to Advanced Electronic Structure Theory*. Macmillan.

[251] Szulakowska, L. (2020). *Electron-electron interactions and optical properties of two-dimensional nanocrystals*. PhD thesis, University of Ottawa.

[252] Takakura, T., Noiri, A., Obata, T., Otsuka, T., Yoneda, J., Yoshida, K., and Tarucha, S. (2014). Single to quadruple quantum dots with tunable tunnel couplings. *Applied Physics Letters*, 104(11):113109.

[253] Takeda, K., Noiri, A., Nakajima, T., Kobayashi, T., and Tarucha, S. (2022). Quantum error correction with silicon spin qubits. *Nature*, 608(7924):682–686.

[254] Tariq, B. and Hu, X. (2022). Impact of the valley orbit coupling on exchange gate for spin qubits in silicon. *npj Quantum Inf.*, 8(1):53.

[255] Tarucha, S., Honda, T., Austing, D., and Tokura, Y. (1998). Electronic states in quantum dot atoms and molecules. *Physica E: low-dimensional systems and nanostructures*.

[256] Taylor, J. M., Engel, H.-A., Dür, W., Yacoby, A., Marcus, C. M., Zoller, P., and Lukin, M. D. (2005). Fault-tolerant architecture for quantum computation using electrically controlled semiconductor spins. *Nat. Phys.*, 1(3):177–183.

[257] Taylor, J. M., Petta, J. R., Johnson, A. C., Yacoby, A., Marcus, C. M., and Lukin, M. D. (2007). Relaxation, dephasing, and quantum control of electron spins in double quantum dots. *Phys. Rev. B*, 76:035315.

[258] Taylor, J. M., Srinivasa, V., and Medford, J. (2013). Electrically protected resonant exchange qubits in triple quantum dots. *Phys. Rev. Lett.*, 111:050502.

[259] Taylor, J. R., Foulk, N. L., and Das Sarma, S. (2024). Assessing quantum dot swap gate fidelity using tensor network methods. *Phys. Rev. B*, 109:165403.

[260] Terhal, B. M. (2015). Quantum error correction for quantum memories. *Rev. Mod. Phys.*, 87(2):307–346.

[261] Thorgrimsson, B., Fung, H. F., Hwang, S. J., Jarausch, D. D., Eyck, G. A. T., Croke, J. R. W., Foote, R. H., Ward, D. R., Savage, D. E., Lagally, M. G., Coppersmith, S. N., Eriksson, M. A., and Friesen, M. (2017). Phonon-assisted relaxation and decoherence of singlet-triplet qubits in si/sige quantum dots. *Phys. Rev. Lett.*, 119(4):046802.

[262] Throckmorton, R. E. and Das Sarma, S. (2022). Effects of leakage on the realization of a discrete time crystal in a chain of singlet-triplet qubits. *Phys. Rev. B*, 106:245419.

[263] Tisdale, W. and Zhu, X. (2011). Artificial atoms on semiconductor surfaces. *PNAS*.

[264] Vahapoglu, E., Slack-Smith, J. P., Leon, R. C. C., Lim, W. H., Hudson, F. E., Day, T., Cifuentes, J. D., Tanttu, T., Yang, C. H., Saraiva, A., Abrosimov, N. V., Pohl, H.-J., Thewalt, M. L. W., Laucht, A., Dzurak, A. S., and Pla, J. J. (2022). Coherent control of electron spin qubits in silicon using a global field. *npj Quantum Inf.*, 8(1):126.

[265] Valenti, D., Carollo, A., and Spagnolo, B. (2018). Phys. rev. a. 97:042109.

[266] Valenti, D., Magazzù, L., Caldara, P., and Spagnolo, B. (2015). Phys. rev. b. 91:235412.

[267] van Beest, M. (2015). Fock-darwin states for an elliptical spin-orbit coupled quantum well. *Bachelor Thesis*.

[268] van Diepen, C. J., Hsiao, T.-K., Mukhopadhyay, U., Reichl, C., Wegscheider, W., and Vandersypen, L. M. K. (2021). Quantum simulation of antiferromagnetic heisenberg chain with gate-defined quantum dots. *Phys. Rev. X*, 11:041025.

[269] Van Dyke, J. S., Kandel, Y. P., Qiao, H., Nichol, J. M., Economou, S. E., and Barnes, E. (2021). Protecting quantum information in quantum dot spin chains by driving exchange interactions periodically. *Phys. Rev. B*, 103:245303.

[270] van Meter, J. R. and Knill, E. (2019). Approximate exchange-only entangling gates for the three-spin-$1/2$ decoherence-free subsystem. *Phys. Rev. A*, 99:042331.

[271] van Riggelen, F., Lawrie, W. I. L., Russ, M., Hendrickx, N. W., Sammak, A., Rispler, M., Terhal, B. M., Scappucci, G., and Veldhorst, M. (2022). Phase flip code with semiconductor spin qubits. *npj Quantum Inf.*, 8(1):124.

[272] Veldhorst, M., Eenink, H. G. J., Yang, C. H., and Dzurak, A. S. (2017). Silicon cmos architecture for a spin-based quantum computer. *Nat. Commun.*, 8(1):1766.

[273] Verstraete, F., Murg, V., and Cirac, J. I. (2008). Matrix product states, projected entangled pair states, and variational renormalization group methods for quantum spin systems. *Adv. Phys.*, 57(2):143–224.

[274] Viola, L., Knill, E., and Lloyd, S. (1999). Dynamical decoupling of open quantum systems. *Phys. Rev. Lett.*, 82(12):2417.

[275] Volk, C., Chatterjee, A., Ansaloni, F., Marcus, C. M., and Kuemmeth, F. (2019). Fast charge sensing of si/sige quantum dots via a high-frequency accumulation gate. *Nano Letters*, 19(8):5628–5633.

[276] Vukmirović, N. and Bruder, C. (2011). Quantum dot arrays: Simulation of electronic structure and electron transport. *Phys. Rev. B*, 83(16):165411.

[277] Wang, C.-A., John, V., Tidjani, H., Yu, C. X., Ivlev, A., Déprez, C., van Riggelen-Doelman, F., Woods, B. D., Hendrickx, N. W., Lawrie, W. I. L., Stehouwer, L. E. A., Oosterhout, S., Sammak, A., Friesen, M., Scappucci, G., de Snoo, S. L., Rimbach-Russ, M., Borsoi, F., and Veldhorst, M. (2024). Operating semiconductor quantum processors with hopping spins.

[278] Wang, H., Zhan, A., Xu, Y., and Chen, H. (2017). Quantum many-body simulation using monolayer exciton-polaritons in coupled-cavities. *J. Phys.: Condens. Matter.*

[279] Wang, K., Dahan, R., Shentcis, M., and Kauffmann, Y. (2020). Coherent interaction between free electrons and a photonic cavity. *Nature*.

[280] Wang, K., Xu, G., Gao, F., Liu, H., Ma, R., and Zhang, X. (2022a). Ultrafast coherent control of a hole spin qubit in a germanium quantum dot. *Nat. Commun.*

[281] Wang, P.-C., Wang, Y.-H., Chen, C., Zhang, J., Li, W., Nan, N., Wang, J.-N., Yang, J.-T., Laref, A., and Xiong, Y.-C. (2022b). Frustration-controlled quantum phase transition between multiple singular two-stage kondo behaviors in a tetrahedral quadruple quantum dot structure. *Phys. Rev. B*, 105:075430.

[282] Wang, X., Rana, S., Thompson, J. K., and Vuckovic, J. (2015). Cavity qed based on collective magnetic dipolar interactions: Spin dynamics. *Phys. Rev. B*, 91:165117.

[283] Wang, X., Yang, S., and Das Sarma, S. (2011). Quantum theory of the charge-stability diagram of semiconductor double-quantum-dot systems. *Phys. Rev. B*, 84:115301.

[284] Watson, T. F., Philips, S. G. J., Kawakami, E., Ward, D. R., Scarlino, P., Veldhorst, M., Savage, D. E., Lagally, M. G., Friesen, M., Coppersmith, S. N., Eriksson, M. A., and Vandersypen, L. M. K. (2018). A programmable two-qubit quantum processor in silicon. *Nature*, 555(7698):633–637.

[285] Weidenmuller, H. (2001). Quantum dots and the many-body problem. *International Journal of Modern Physics B*.

[286] Weinstein, A. J., Reed, M. D., Jones, A. M., Andrews, R. W., Barnes, D., Blumoff, J. Z., Euliss, L. E., Eng, K., Fong, B. H., Ha, S. D., Hulbert, D. R., Jackson, C. A. C., Jura, M., Keating, T. E., Kerckhoff, J., Kiselev, A. A., Matten, J., Sabbir, G., Smith, A., Wright, J., Rakher, M. T., Ladd, T. D., and Borselli, M. G. (2023). Universal logic with encoded spin qubits in silicon. *Nature*, 615(7954):817–822.

[287] Wilson-Rae, I. and Imamoglu, A. (2002). Quantum dot cavity-qed in the presence of strong electron-phonon interactions. *Phys. Rev. B*, 65(23):235311.

[288] Wolfe, M. A., Calderon-Vargas, F. A., and Kestner, J. P. (2017). Robust operating point for capacitively coupled singlet-triplet qubits. *Phys. Rev. B*, 96:201307.

[289] Yamachika, R. (2009). *Probing atomic-scale properties of organic and organometallic molecules by scanning tunneling spectroscopy*. PhD thesis, University of California.

[290] Yan, W., Liu, Q., Wang, C., Yang, X., and Yao, T. (2014). Realizing ferromagnetic coupling in diluted magnetic semiconductor quantum dots. *Journal of the American Chemical Society*.

[291] Yang, S., Wang, X., and Das Sarma, S. (2011a). Generic hubbard model description of semiconductor quantum-dot spin qubits. *Phys. Rev. B*, 83:161301.

[292] Yang, S.-S., Feng, W., and Kouwenhoven, L. P. (2011b). Quantum dots and electron transport. *Chinese Physics B*, 20(5):058101.

[293] Yang, X.-C., Chan, G. X., and Wang, X. (2018). Tunable charge qubit based on barrier-controlled triple quantum dots. *Phys. Rev. A*, 98:032334.

[294] Yoneda, J., Huang, W., Feng, M., Yang, C. H., Chan, K. W., Tanttu, T., Gilbert, W., Leon, R., Hudson, F., Itoh, K., et al. (2021). Coherent spin qubit transport in silicon. *Nat. Commun.*, 12(1):1–9.

[295] Yoneda, J., Takeda, K., Otsuka, T., Nakajima, T., Delbecq, M. R., Allison, G., Honda, T., Kodera, T., Oda, S., Hoshi, Y., et al. (2018). A quantum-dot spin qubit with coherence limited by charge noise and fidelity higher than 99.9%. *Nat. Nanotechnol.*, 13(2):102–106.

[296] Yue, Z., Levine, H., Srinivasa, V., Choi, S., Kestner, J. P., Tyryshkin, A. M., Lyon, S. A., Economou, S. E., and Petta, J. R. (2022). Quantum simulation of extended hubbard models with dopant-based quantum dots. *Nat. Phys.*, 18(11):1324–1329.

[297] Zajac, D. M., Sigillito, A. J., Russ, M., Borjans, F., Taylor, J. M., Burkard, G., and Petta, J. R. (2018). Resonantly driven cnot gate for electron spins. *Science*, 359(6374):439–442.

[298] Zanardi, P. and Rasetti, M. (1997). Noiseless quantum codes. *Phys. Rev. Lett.*, 79(17):3306.

[299] Zenkevich, E., Stupak, A., and Göhler, C. (2015). Tuning electronic states of a cdse/zns quantum dot by only one functional dye molecule. *ACS Nano*.

[300] Zeuch, D. and Bonesteel, N. E. (2016). Simple derivation of the fong-wandzura pulse sequence. *Phys. Rev. A*, 93:010303.

[301] Zeuch, D. and Bonesteel, N. E. (2020). Efficient two-qubit pulse sequences beyond cnot. *Phys. Rev. B*, 102:075311.

[302] Zeuch, D., Cipri, R., and Bonesteel, N. E. (2014). Analytic pulse-sequence construction for exchange-only quantum computation. *Phys. Rev. B*, 90:045306.

[303] Zhang, C., Yang, X.-C., and Wang, X. (2016). Benchmarking of dynamically corrected gates for the exchange-only spin qubit in a $1/f$ noise environment. *Phys. Rev. A*, 94:042323.

[304] Zhang, H.-K., Liu, S., and Zhang, S.-X. (2024). Absence of barren plateaus in finite local-depth circuits with long-range entanglement. *Phys. Rev. Lett.*, 132:150603.

[305] Zhang, L.-L. and Gong, W.-J. (2017). Transport properties in a non-hermitian triple-quantum-dot structure. *Phys. Rev. A*, 95:062123.

[306] Zhang, L.-L., Li, Z.-Z., Zhan, G.-H., Yi, G.-Y., and Gong, W.-J. (2019a). Eigenenergies and quantum transport properties in a non-hermitian quantum-dot chain with side-coupled dots. *Phys. Rev. A*, 99:032119.

[307] Zhang, X., Li, H.-O., Cao, G., Xiao, M., Guo, G.-C., and Guo, G.-P. (2018). Semiconductor quantum computation. *Natl. Sci. Rev.*, 6(1):32–54.

[308] Zhang, X., Morozova, E., Rimbach-Russ, M., Jirovec, D., Hsiao, T.-K., Fariña, P. C., Wang, C.-A., Oosterhout, S. D., Sammak, A., Scappucci, G., Veldhorst, M., and Vandersypen, L. M. K. (2023). Universal control of four singlet-triplet qubits.

[309] Zhang, X.-M., Wei, Z., Asad, R., Yang, X.-C., and Wang, X. (2019b). When does reinforcement learning stand out in quantum control? a comparative study on state preparation. *npj Quantum Inf.*, 5(1):85.

[310] Zhang, Z., Chang, K., and Peeters, F. (2008). Tuning of energy levels and optical properties of graphene quantum dots. *Phys. Rev. B*.

[311] Zhao, X. and Hu, X. (2018). Toward high-fidelity coherent electron spin transport in a gaas double quantum dot. *Sci. Rep.*, 8(1):1–15.

[312] Zhu, H., Yang, Y., Wu, K., and Lian, T. (2016). Charge transfer dynamics from photoexcited semiconductor quantum dots. *Annual Review of Physical Chemistry*.

[313] Zhu, X. (2004). Electronic structure and electron dynamics at molecule–metal interfaces: implications for molecule-based electronics. *Surface Science Reports*.

[314] Zwolak, J. P. and Taylor, J. M. (2023). Colloquium: Advances in automation of quantum dot devices control. *Rev. Mod. Phys.*, 95:011006.

# Publications

- G. He, G. X. Chan, X. Wang. **(2023)** Theory on Electron-Phonon Spin Dephasing in GaAs Multi-Electron Double Quantum Dots. *Advanced Quantum Technologies* 6(3), 2200074. doi: 10.1002/qute.202200074
- G. He, X. Wang. **(2024)** Exploring Entanglement Spectrum and Phase Diagram in Multi-Electron Quantum Dot Chains. *arXiv preprint*. doi: 10.48550/ARXIV.2405.06083 Under Review
- Constructing Three-Qubit Gate Pulse Sequences in Exchange-Only Spin System. Preparing